%% file: DQM_ijmpe.tex
\documentclass[12pt]{ws-ijmpe}
\usepackage[super,compress]{cite}
\expandafter\let\csname equation*\endcsname\relax

\expandafter\let\csname endequation*\endcsname\relax

\usepackage{amsmath,slashed}

\usepackage{cite}
\usepackage{epsfig}
\usepackage{epsf}
\usepackage{graphicx}
\usepackage{psfrag}
\usepackage[T1]{fontenc}
\usepackage[latin1]{inputenc}
\usepackage[T1]{fontenc}
\usepackage{color}

\usepackage{amssymb}

\usepackage{bm}
\usepackage{latexsym}
\usepackage{subfigure}
\usepackage{needspace}
\usepackage{pslatex}
\usepackage{txfonts}
\usepackage{caption2,hyperref}

\usepackage[OMLmathrm,OMLmathbf]{isomath}

\input{mycommands}

\begin{document}

\title{Dynamics of QCD Matter -- current status}

%\author{Amaresh Jaiswal$^1$, Najmul Haque$^1$ (editors)}

\author{ 
Amaresh Jaiswal$^1$\footnote{a.jaiswal@niser.ac.in}~, 
Najmul Haque$^1$\footnote{nhaque@niser.ac.in}~,
Aman Abhishek$^2$,
Raktim Abir$^3$,
Aritra Bandyopadhyay$^4$,
Khatiza Banu$^3$,
Samapan Bhadury$^1$,
Sumana Bhattacharyya$^5$,
Trambak Bhattacharyya$^6$,
Deeptak Biswas$^5$,
H. C. Chandola$^7$,
Vinod Chandra$^8$,
Bhaswar Chatterjee$^9$,
Chandrodoy Chattopadhyay$^{10}$,
Nilanjan Chaudhuri$^{11,~12}$,
Aritra Das$^{13,~12}$, 
Arpan Das$^2$,
Santosh K. Das$^{14}$,
Ashutosh Dash$^1$,
Kishan Deka$^1$,
Jayanta Dey$^{15}$,
Ricardo L. S. Farias$^4$,
Utsab Gangopadhyaya$^{11}$,
Ritesh Ghosh$^{13}$,
Sabyasachi Ghosh$^{15}$,
Snigdha Ghosh$^{13}$,
Ulrich Heinz$^{10}$,
Sunil Jaiswal$^{16}$,
Guruprasad Kadam$^{17}$,
Pallavi Kalikotay$^{18}$,
Bithika Karmakar$^{13}$,
Gast\~ao Krein$^{19}$,
Avdhesh Kumar$^{20}$,
Deepak Kumar$^{2,~8}$,
Lokesh Kumar$^{21}$,
Manu Kurian$^8$,
Soumitra Maity$^5$,
Hiranmaya Mishra$^2$,
Payal Mohanty$^1$,
Ranjita K. Mohapatra$^{22}$,
Arghya Mukherjee$^{13}$,
Munshi G. Mustafa$^{13}$,
Subrata Pal$^{16}$,
H. C. Pandey$^{23}$,
Mahfuzur Rahaman$^{11}$,
Ralf Rapp$^{24}$,
Deependra Singh Rawat$^7$,
Sutanu Roy$^1$,
Victor Roy$^1$,
Kinkar Saha$^{25}$,
Nihar R. Sahoo$^{26}$,
Subhasis Samanta$^1$,
Sourav Sarkar$^{11,~12}$,
Sarthak Satapathy$^{15}$,
Fernando E. Serna$^{19}$,
Mariyah Siddiqah$^3$,
Pracheta Singha$^5$,
V. Sreekanth$^{27}$,
Sudipa Upadhaya$^{11}$,
Nahid Vasim$^3$,
Dinesh Yadav$^7$
(authors)\footnote{
The contributors on this author list have contributed only to those sections of the report, which they cosign with their name. Only those have collaborated together, whose names appear together in the header of a given section.}}

\address{
$^1$ National Institute of Science Education and Research, HBNI, Jatni 752050, Odisha, India\\
$^2$ Theory Division, Physical Research Laboratory, Navrangpura, Ahmedabad 380 009, India\\
$^3$ Department of Physics, Aligarh Muslim University, Aligarh (U.P.)-202002, India\\
$^4$ Departamento de F\'isica, Universidade Federal de Santa Maria, Santa Maria, RS, 97105-900, Brazil\\
$^5$ Center for Astroparticle Physics \& Space Science, Bose Institute, EN-80, Sector-5, Bidhan Nagar, Kolkata-700091, India\\
$^6$ University of Cape Town, Rondebosch 7701, Cape Town, South Africa. Presently at Bogoliubov Laboratory of Theoretical Physics, Joint Institute for Nuclear Research, Dubna, 141980, Moscow region, Russian Federation\\
$^7$ Department of Physics (UGC-Centre of Advanced Study), Kumaun University, Nainital, India\\
$^8$ Indian Institute of Technology Gandhinagar, Gandhinagar 382 355, Gujarat, India\\
$^9$ Department of Physics, Indian Institute of Technology Roorkee, Roorkee 247 667, India\\
$^{10}$ Department of Physics, The Ohio State University, Columbus, Ohio 43210-1117, USA\\
$^{11}$ Variable Energy Cyclotron Centre, 1/AF Bidhan Nagar, Kolkata 700 064, India\\
$^{12}$ Homi Bhabha National Institute, Training School Complex, Anushaktinagar, Mumbai - 400085, India\\
$^{13}$ Saha Institute of Nuclear Physics, 1/AF Bidhan Nagar, Kolkata 700064, India\\
$^{14}$ School of Physical Sciences, Indian Institute of Technology Goa, Ponda-403401, Goa, India\\
$^{15}$ Indian Institute of Technology Bhilai, GEC Campus, Sejbahar, Raipur 492015, Chhattisgarh, India\\
$^{16}$ Department of Nuclear and Atomic Physics, Tata Institute of Fundamental Research, Mumbai 400005, India\\
$^{17}$ Department of Physics, Shivaji University, Kolhapur, Maharashtra-416004, India\\
$^{18}$ Department of Physics, Kazi Nazrul University, Asansol - 713340, West Bengal, India\\
$^{19}$ Instituto de F\'isica Te\'orica, Universidade Estadual Paulista, Rua Dr. Bento Teobaldo Ferraz, 271 - Bloco II, 01140-070 S\~ao Paulo, SP, Brazil\\
$^{20}$ Institute of Nuclear Physics Polish Academy of Sciences, PL-31-342 Krak\'ow, Poland\\
$^{21}$ Department of Physics, Panjab University, Chandigarh, India-160014\\
$^{22}$ Department of Physics, Indian Institute of Technology Bombay, Mumbai, 400076, India\\
$^{23}$ Birla Institute of Applied Sciences, Bhimtal, India\\
$^{24}$ Texas A \& M University, Department of Physics and Astronomy and Cyclotron Institute,  College Station, TX 77843-3366, USA\\
$^{25}$ Department of Physics, University of Calcutta, 92, A. P. C. Road, Kolkata - 700009, India\\
$^{26}$ Shandong University, Qingdao, China\\
$^{27}$ Department of Sciences, Amrita School of Engineering, Coimbatore, Amrita Vishwa Vidyapeethom, India
}

\maketitle

\begin{abstract}

In this article, there are 18 sections discussing various current topics in the field of relativistic heavy-ion collisions and related phenomena, which will serve as a snapshot of the current state of the art. 

Section 1 reviews experimental results of some recent light-flavored particle production data from ALICE collaboration. Other sections are mostly theoretical in nature.

Very strong but transient magnetic field created in relativistic heavy-ion collisions could have important observational consequences. This has generated a lot of theoretical activity in the last decade. Sections 2, 7, 9, 10 and 11 deal with the effects of the magnetic field on the properties of the QCD matter. More specifically, Sections 2, discusses mass of $\pi^0$ in the linear sigma model coupled to quarks at zero temperature. In Section 7, one-loop calculation of the anisotropic pressure is discussed in presence of strong magnetic field. In Section 9, chiral transition and chiral susceptibility in the NJL model is discussed for a chirally imbalanced plasma in the presence of magnetic field using a Wigner function approach. Sections 10 discusses electrical conductivity and Hall conductivity of hot and dense hadron gas within Boltzmann approach and Section 11 deals with electrical resistivity of quark matter in presence of magnetic field. There are several unanswered questions about the QCD phase diagram. Sections 3, 11 and 18 discuss various aspects of the QCD phase diagram and phase transitions.

Recent years have witnessed interesting developments in foundational aspects of hydrodynamics and their application to heavy-ion collisions. Sections 12, 15, 16 and 17 of this article probe some aspects of this exciting field. In Section 12, analytical solutions of viscous Landau hydrodynamics in 1+1D is discussed. Section 15 deals with derivation of hydrodynamics from effective covariant kinetic theory. Sections 16 and 17 discusses hydrodynamics with spin and analytical hydrodynamic attractors, respectively.

Transport coefficients together with their temperature- and density-dependence, are essential inputs in hydrodynamical calculations. Sections 5, 8 and 14 deal with calculation/estimation of various transport coefficients (shear and bulk viscosity, thermal conductivity, relaxation times, etc.) of quark matter and hadronic matter.

Sections 4, 6 and 13 deals with interesting new developments in the field. Section 4 discusses color dipole gluon distribution function at small transverse momentum in the form of a series of Bells polynomials. Section 6 discusses the properties of Higgs boson in the quark gluon plasma using Higgs-quark interaction and calculate the Higgs decays into quark and anti-quark, which shows a dominant on-shell contribution in the bottom-quark channel. Section 13 discusses modification of coalescence model to incorporate viscous corrections and application of this moedel to study hadron production from a dissipative quark-gluon plasma.

\end{abstract}

\newpage
\tableofcontents

\markboth{Dynamics of QCD Matter -- current status}{Dynamics of QCD Matter -- current status}

%Preface
%\input{preface.tex}

\newpage

%
\input{Lokesh/lokesh.tex}
%
\newpage
\input{Aritra/aritra.tex}

\newpage
\input{Ashutosh/Ashutosh.tex}

\newpage
\input{Mariyah/mariyah.tex}

\newpage
\input{Pallavi/pallavi.tex}

\newpage
\input{Sarthak/sarthak.tex}

\newpage

\input{Ritesh/ritesh.tex}

\newpage

\input{Sabya/sabya.tex}

%
\newpage
%
\input{Arpan/arpan.tex}
\newpage
\input{Ranjita/ranjita.tex}
%
\newpage

\input{Jayanta/jayanta.tex}

%
\newpage

\input{Deeptak/deeptak.tex}
\newpage
%
\input{Sumana/sumana.tex}

%
\newpage
%
\input{Sreekanth/sreekanth.tex}
\newpage
\input{Samapan/samapan.tex}
\newpage

\input{Avdhesh/avdhesh.tex}
\newpage

\input{Sunil/sunil.tex}
%
\newpage

\input{Deependra/deependra.tex}
%
\newpage

%Acknowledgments
\input{acknow.tex}

\newpage
%

%\section*{References}

\end{document}

%% file: mycommands.tex
\newcommand{\lsim}{~{\buildrel < \over {_\sim}}~}
\newcommand{\gsim}{~{\buildrel > \over {_\sim}}~}

\renewcommand{\d}{{\rm d}}

\newcommand{\vp}{\vec p}
\newcommand{\FB}[1]{\left(#1\right)}
\newcommand{\ep}{\epsilon}

\newcommand{\sF}{\scriptscriptstyle{f}}
\newcommand{\sB}{\scriptscriptstyle{B}}
\newcommand{\sperp}{\scriptscriptstyle{\perp}}
\newcommand{\shp}{\shortparallel}

\def\({\left(}
\def\){\right)}
\newcommand{\nn}{\nonumber}

\def\lsim{\raise0.3ex\hbox{$<$\kern-0.75em\raise-1.1ex\hbox{$\sim$}}}
\def\gsim{\raise0.3ex\hbox{$>$\kern-0.75em\raise-1.1ex\hbox{$\sim$}}}

\def\a{\alpha}

\def\del{\partial}

\def\eps{\varepsilon}

\def\lsim{\mathrel{\rlap{\lower4pt\hbox{\hskip1pt$\sim$}}
		\raise1pt\hbox{$<$}}}         %less than or approx. symbol
\def\gsim{\mathrel{\rlap{\lower4pt\hbox{\hskip1pt$\sim$}}
		\raise1pt\hbox{$>$}}}         %greater than or approx. sy
\def\lessim{\lower.5ex\hbox{$\; \buildrel < \over \sim \;$}}

\def\be{\begin{equation}}
	\def\ee{\end{equation}}
\newcommand{\bel}[1]{\begin{eqnarray}\label{#1}}
	\newcommand{\eel}{\end{eqnarray}}
\def\barr{\begin{array}}
	\def\earr{\end{array}}
\def\beq{\begin{eqnarray}}
	\def\eeq{\end{eqnarray}}
\def\bfig{\begin{figure}}
	\def\efig{\end{figure}}

\newcommand{\f}[2]{\frac{#1}{#2}}

\newcommand{\p}{\partial}
\newcommand{\rf}[1]{Eq.~(\ref{#1})}

\newcommand{\rfn}[1]{(\ref{#1})}

\def\a{\alpha}
\def\b{\beta}
\def\g{\gamma}
\def\d{\delta}

\def\LR{\left(} % round
\def\RR{\right)}
\def\HP{\hphantom{\alpha}} % horizontal

\newcommand{\sh}[1]{\sinh#1}
\newcommand{\ch}[1]{\cosh#1}

\def\GLW{{\rm GLW}}

\def\nU{n_{(0)}}
\def\eU{\varepsilon_{(0)}}
\def\PU{P_{(0)}}

\def\Cv{{\boldsymbol C}}
\usepackage{amssymb,csquotes}
\usepackage{graphicx}
\usepackage{amsmath}
\usepackage{subfigure}
\usepackage{bbold}
\usepackage{yfonts}
\usepackage{placeins}
\usepackage{bm} 
\usepackage{nicefrac}
\usepackage{slashed}
\def\be{\begin{equation}}
	\def\ee{\end{equation}}
\def\barr{\begin{array}}
	\def\earr{\end{array}}
\def\beq{\begin{eqnarray}}
	\def\eeq{\end{eqnarray}}
\def\bfig{\begin{figure}}
	\def\efig{\end{figure}}

\def\a{\alpha}
\def\b{\beta}
\def\g{\gamma}
\def\d{\delta}

\def\LR{\left(} % round
\def\RR{\right)}
\def\HP{\hphantom{\alpha}} % horizontal

\def\GLW{{\rm GLW}}

\def\nU{n_{(0)}}
\def\eU{\varepsilon_{(0)}}
\def\PU{P_{(0)}}

\newcommand{\lab}[1]{\label{#1}}

\def\pmu{p^\mu}

\def\pv{{\boldsymbol p}}

\def\Cv{{\boldsymbol C}}

\def\piv{{\boldsymbol \pi}}
\def\omnL{\omega_{\mu\nu}}
\def\omnU{\omega^{\mu\nu}}

\def\omnUD{\tilde {\omega}^{\mu\nu}}

\def\fplusrsxp{f^+_{rs}(x,p)}

\def\fminusrsxp{f^-_{rs}(x,p)}

\def\SmunuU{{\Sigma}^{\mu\nu}}

\def\S0iU{{\Sigma}^{0i}} 
 
\def\SmnU{{\Sigma}^{\mu\nu}}

\def\ubarrp{{\bar u}_r(p)}

\def\usp{u_s(p)}

\def\urp{u_r(p)}

\def\vbarsp{{\bar v}_s(p)}

\def\vrp{v_r(p)}
\def\bmu{\beta_\mu}

\def\n0{n_{(0)}}
\def\e0{\varepsilon_{(0)}}
\def\P0{P_{(0)}}

\def\omnL{\omega_{\mu\nu}}
\def\omnU{\omega^{\mu\nu}}

\def\omnUD{\tilde {\omega}^{\mu\nu}}

\def\pmu{p^\mu}

\def\Wxk{{\cal W}(x,k)}
\def\Weqxk{{\cal W}_{\rm eq}(x,k)}
\def\Weqpxk{{\cal W}^{+}_{\rm eq}(x,k)}
\def\Weqmxk{{\cal W}^{-}_{\rm eq}(x,k)}
\def\Weqpmxk{{\cal W}^{\pm}_{\rm eq}(x,k)}
\def\Feqpmxk{{\cal F}^{\pm}_{\rm eq}(x,k)}
\def\Peqpmxk{{\cal P}^{\pm}_{\rm eq}(x,k)}

\def\be {\begin{equation}}
\def\ee {\end{equation}}
\def\bea {\begin{eqnarray}}
\def\eea {\end{eqnarray}}
\def\bc {\begin{center}}
	\def\ec {\end{center}}
\def\nn {\nonumber}
\def\eps {\epsilon}
\def\gm {\gamma}

\def\mn {\mu\nu}
\def\({\left(}
\def\){\right)}
\def\[{\left[}
\def\]{\right]}
\def\sp {\shortparallel}
\newcommand \Tr{\operatorname{\text{Tr}}}

\def\slashed{\slash\!\!\!\!}
\def\slashedl{\slash\!\!\!}

\newcommand{\ks}{k \!\!\!\! /}
\newcommand{\qs}{q \!\!\!\! /}

%% file: Lokesh/lokesh.tex
%\title{Overview of Recent Results from ALICE on Particle Production}
\section{Recent Selected Results from ALICE on Particle Production}

\textit{Lokesh Kumar} 

\bigskip

%\date{\today}

%\begin{abstract}
{\small
We present a selection of recent results on light-flavored particle
production from ALICE experiment. The results are presented on the
charge particle multiplicity, 
 average
transverse momentum, kinetic freeze-out parameters, enhancement of
strangeness production, suppression of resonance yields in central
nucleus-nucleus collisions, and first experimental observation of spin-orbit coupling in
high-energy heavy-ion collisions, latest estimation of hypertriton
lifetime from ALICE. Comparison of these results among
several collision systems
such as pp, p-Pb, Pb-Pb, and Xe-Xe  at various
center-of-mass energies is presented. 
}
%\end{abstract}

\bigskip

%\pacs{25.75.-q,25.75.Gz,25.75.Bh}  
%\keywords{Nuclei, anit-nuclei, coalescence, di-hadron correlations, and jets.}
%\maketitle

\subsection{Introduction}
The ALICE experiment at the Large Hadron Collider has collected a large
amount of data for various systems and energies.
In view of the large data sample collected in small systems, it is
possible to study the  
multiplicity dependence of various observables. 
Many interesting observations have come out of ALICE
by comparing results from of small systems pp and p-Pb with large systems Pb-Pb
and Xe-Xe.  We
present a selection of recent results, mostly on light-flavored
particle production. 

\subsection{Collectivity}\label{Sec:collect}
%%%%%%%%%%%%%%%%%%%%%%%%%%%%%%%%%%%%%%%%%%%%%
\begin{figure}[!b]
\begin{center}
%\rotatebox{0}{\resizebox{\columnwidth}{!}{
%        \includegraphics{Figures/2019-06-06-LFmeanPt_logx_0.pdf}}}
%\includegraphics[width =
%12cm]{Figures/2019-06-06-LFmeanPt_logx_0.pdf}
\includegraphics[width = 9cm]{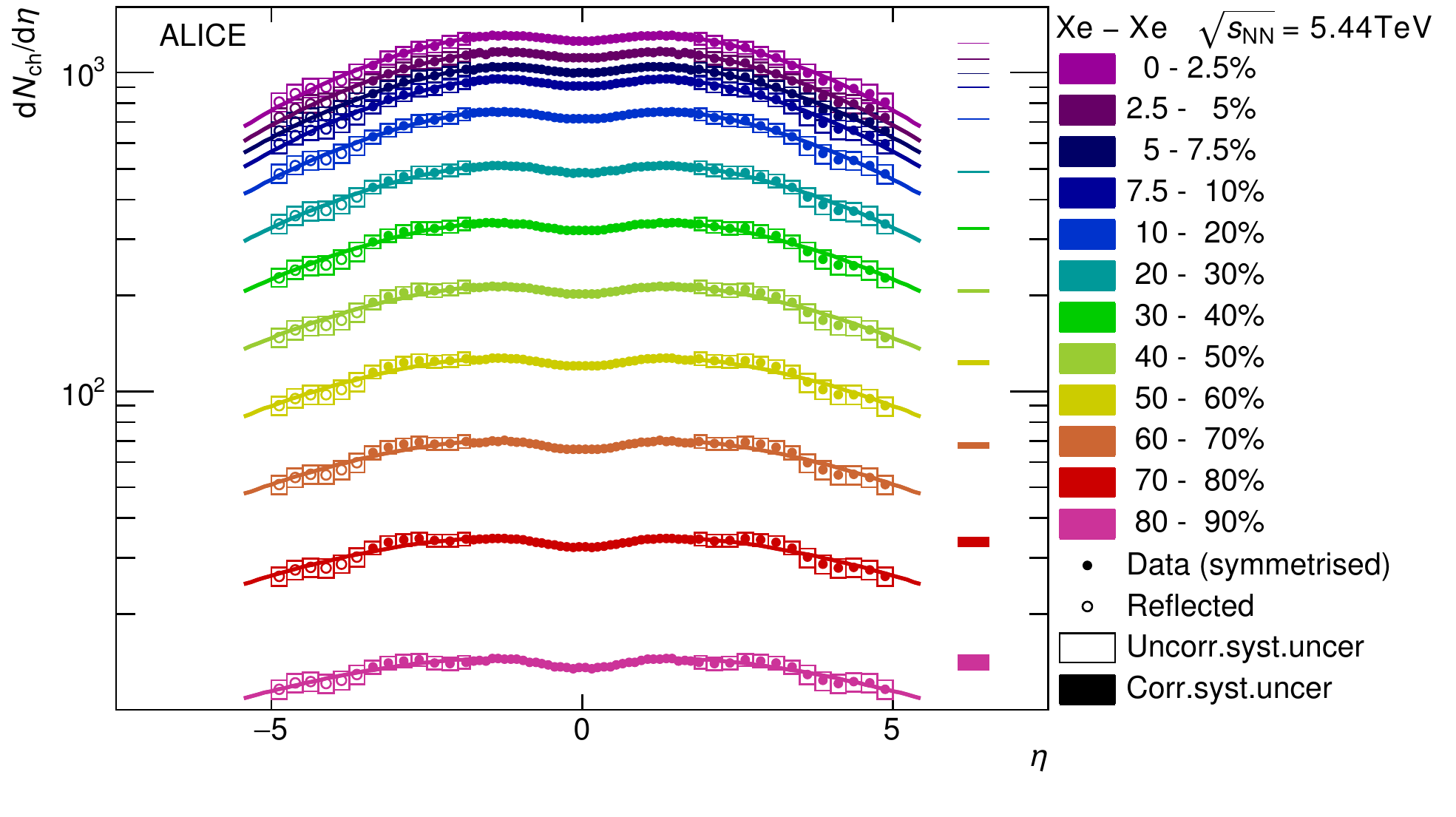}
\caption{Charged particle pseudorapidity density for various centrality
  classes over a broad range of $\eta$ in Xe-Xe collisions at $\sqrt{s_{NN}} =$ 5.44 TeV~\cite{Acharya:2018hhy}. 
}\label{Fig:eta}
\end{center}
\end{figure}
%%%%%%%%%%%%%%%%%%%%%%%%%%%%%%%%%%%%%%%%%%%%%
Figure~\ref{Fig:eta} shows the new results on charged particle pseudorapidity density for various centrality
  classes over a broad range of $\eta$ in Xe-Xe collisions at
  $\sqrt{s_{NN}} =$ 5.44 TeV~\cite{Acharya:2018hhy}. 
The data are presented for 12 centrality classes. At midrapidity the $\langle dN_{\rm{ch}}/d\eta
  \rangle$ is about 1302 $\pm$ 17 for Xe-Xe at $\sqrt{s_{NN}} =$ 5.44 TeV. 
The charged particle multiplicity has been measured for small systems as well as
  Pb-Pb collisions and it is observed that energy dependence behavior of
  midrapidity $\langle dN_{\rm{ch}}/d\eta \rangle/(0.5 \langle
  N_{\rm{part}} \rangle$) is different for small systems and large
  systems. %At similar $N_{part}$, the $\langle dN_{\rm{ch}}/d\eta \rangle/(0.5 \langle  N_{part} \rangle$) for Xe-Xe is larger compared to Pb-Pb.
The $\langle dN_{\rm{ch}}/d\eta \rangle/(0.5 \langle
  N_{\rm{part}} \rangle$) does not scale with number of participant
  nucleons $\langle  N_{\rm{part}} \rangle$, however, it scales
  approximately with number of   wounded constituent quarks $\langle
  N_{\rm{q-part}} \rangle$ calculated using quark-Glauber parameterization~\cite{Loizides:2016djv}.

%%%%%%%%%%%%%%%%%%%%%%%%%%%%%%%%%%%%%%%%%%%%%
\begin{figure}[tbh]
\begin{center}
%\rotatebox{0}{\resizebox{\columnwidth}{!}{
%        \includegraphics{Figures/2019-06-06-LFmeanPt_logx_0.pdf}}}
%\includegraphics[width =
%12cm]{Figures/2019-06-06-LFmeanPt_logx_0.pdf}
\includegraphics[width = 13cm]{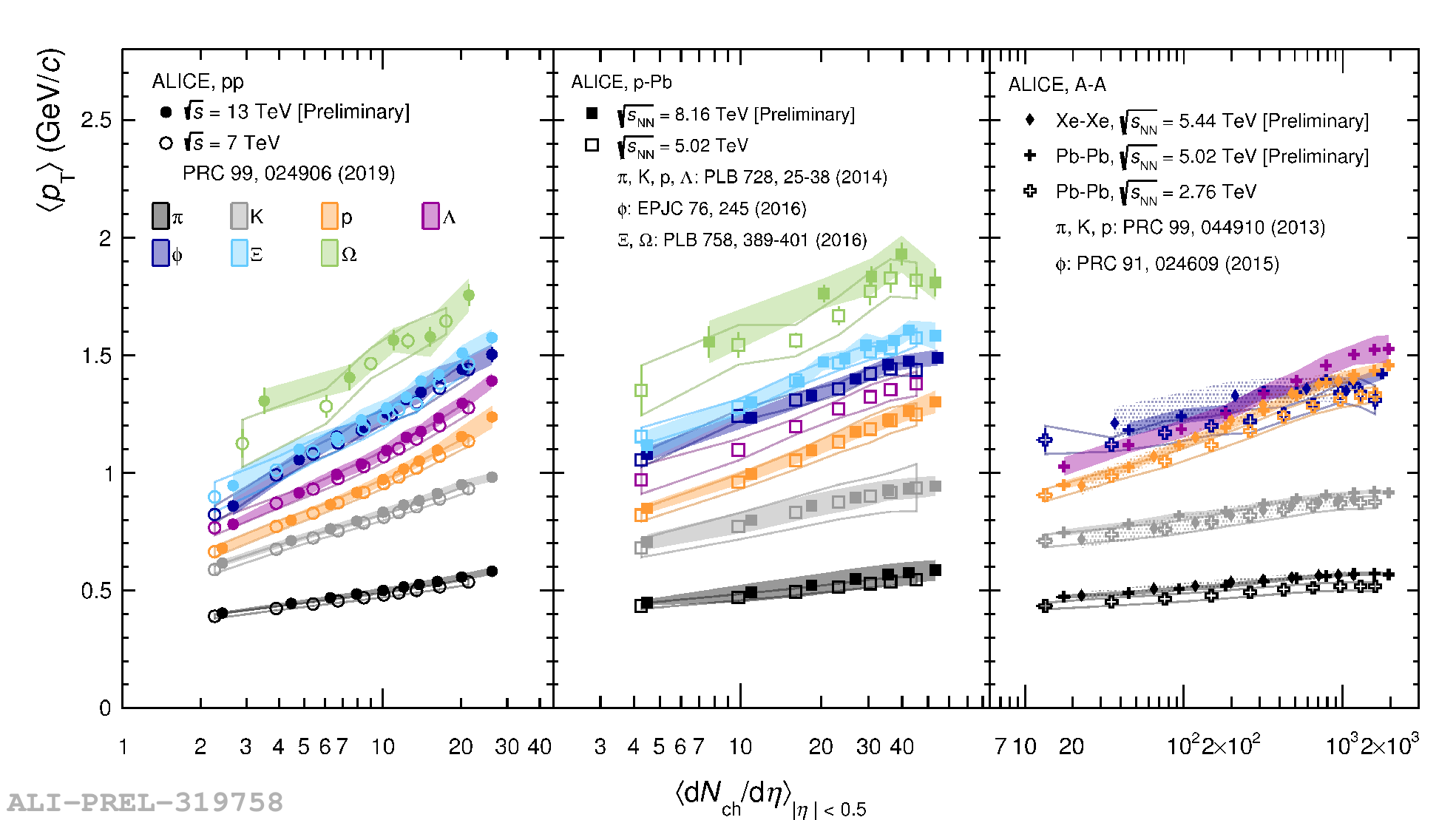}
\caption{Average transverse momentum $\langle p_T \rangle$ of
  identified hadrons as a function of $\langle dN_{\rm{ch}}/d\eta
  \rangle_{|\eta|<0.5}$ in different collision systems pp, p-Pb,
  Xe-Xe, and Pb-Pb at various center of mass energies~\cite{Knospe:2018mek}. 
}\label{Fig:meanpt}
\end{center}
\end{figure}
%%%%%%%%%%%%%%%%%%%%%%%%%%%%%%%%%%%%%%%%%%%%%
\begin{figure}[tbh]
\begin{center}
%\rotatebox{0}{\resizebox{\columnwidth}{!}{
%\includegraphics{Figures/2019-07-07-With13TeV.pdf}}}
\includegraphics[width = 9cm]{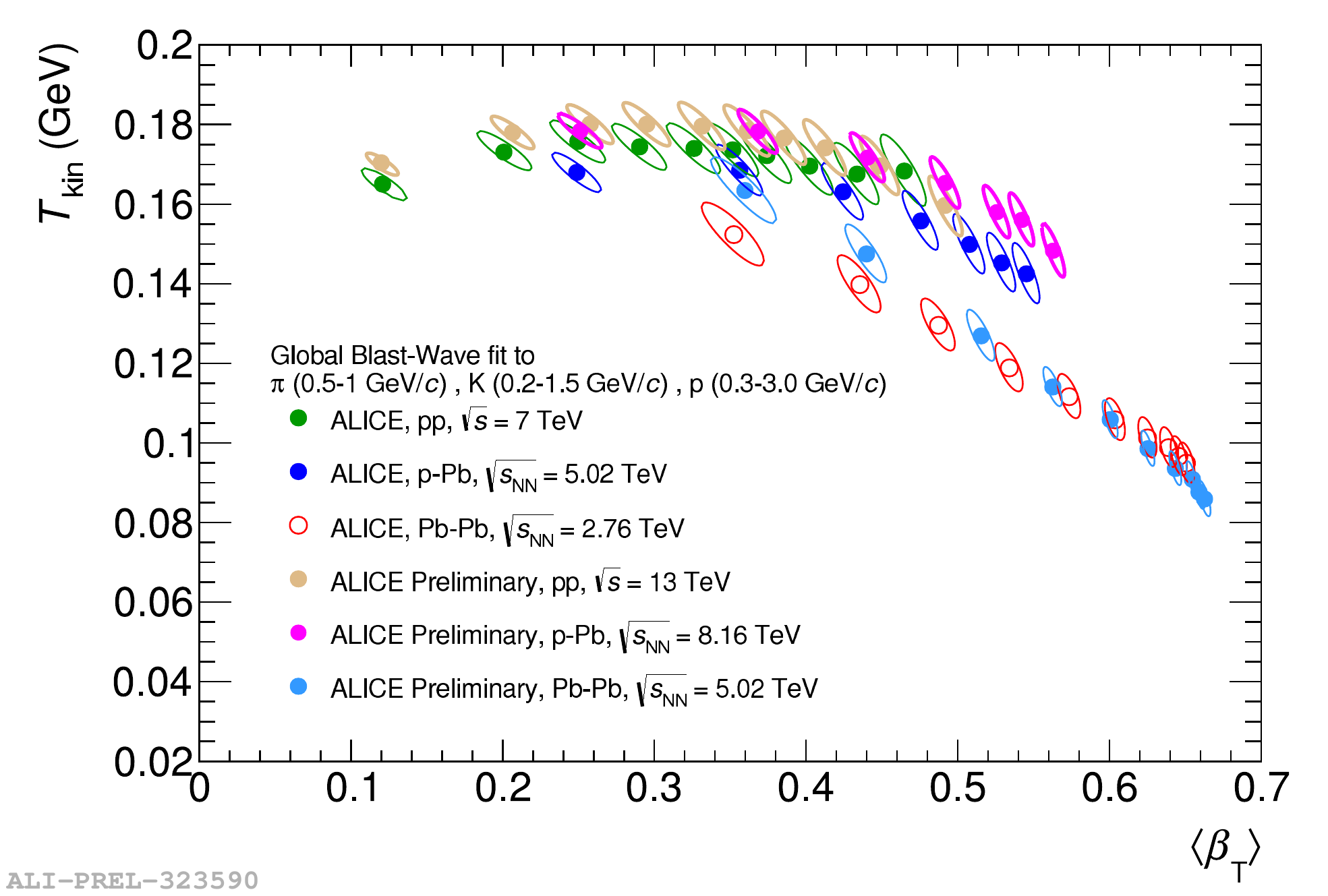}
\caption{The extracted kinetic freeze-out parameters using blast-wave
  model for various collision systems and energies~\cite{Abelev:2013vea,Adam:2016dau}. 
}\label{Fig:kfo}
\end{center}
\end{figure}
%%%%%%%%%%%%%%%%%%%%%%%%%%%%%%%%%%%%%%%%%%%%%
Figure~\ref{Fig:meanpt} shows the average transverse momentum $\langle p_T \rangle$ of
  identified hadrons plotted as a function of $\langle dN_{\rm{ch}}/d\eta
  \rangle_{|\eta|<0.5}$ in different collision systems pp, p-Pb,
  Xe-Xe, and Pb-Pb at various center of mass energies~\cite{Knospe:2018mek}.
In general, the $\langle p_T \rangle$ increases with increasing
multiplicity for all systems and energies.
 It is observed that for
  central A-A collisions, the $\langle p_T \rangle$ increases with
  mass of hadrons. This is referred to as the mass ordering and is consistent
  with hydrodynamical behavior. As can be seen, the proton and $\phi(1020)$ meson
  having similar masses have same $\langle p_T \rangle$ values in
  central A-A collisions.
 However, in peripheral A-A, p-Pb, and pp collisions, the mass ordering
 seems to be violated for $\phi(1020)$ mesons where it is observed that its
 $\langle p_T \rangle$ even exceeds those of protons and
 $\Lambda$. It is also observed that the increase in $\langle p_T
 \rangle$ with $\langle dN_{\rm{ch}}/d\eta  \rangle_{|\eta|<0.5}$ is
 faster in light systems than in heavy-ions.

Figure~\ref{Fig:kfo} shows the extracted kinetic freeze-out parameters using blast-wave
  model for various collision systems and energies~\cite{Abelev:2013vea,Adam:2016dau}. Blast-wave model
  is a hydrodynamical based model which assumes that the system is expanding
  radially with common radial flow velocity and undergoing common
  freeze-out. 
Simultaneous blast wave fits are performed on the transverse momentum
spectra of pions, kaons, and protons. 
The fit parameters, kinetic freeze-out
  temperature $T_{\rm{kin}}$ and average transverse flow velocity $\langle
  \beta_T \rangle$ are plotted in Fig.~\ref{Fig:kfo} for various
  multiplicity classes. The multiplicity increases from left to right
  in the shown figure. It is observed that for heavy-ions A-A 
  collisions the $T_{\rm{kin}}$ decreases with multiplicity while $\langle
  \beta_T \rangle$  increases. There is 
  is no clear energy dependence of the freeze-out parameters. For
  small systems pp and p-Pb, the $T_{\rm{kin}}$ remains constant while
  $\langle \beta_T \rangle$ increases rapidly with multiplicity. At
  similar multiplicity values, the $\langle \beta_T \rangle$ is larger
  for small systems.

\subsection{Strangeness Production}
%%%%%%%%%%%%%%%%%%%%%%%%%%%%%%%%%%%%%%%%%%%%%
\begin{figure}[tbh]
\begin{center}
%\rotatebox{0}{\resizebox{\columnwidth}{!}{
%        \includegraphics{Figures/2019-08-08-img_ToPionRatios_w_pPb8.pdf}}}
\includegraphics[width = 7cm]{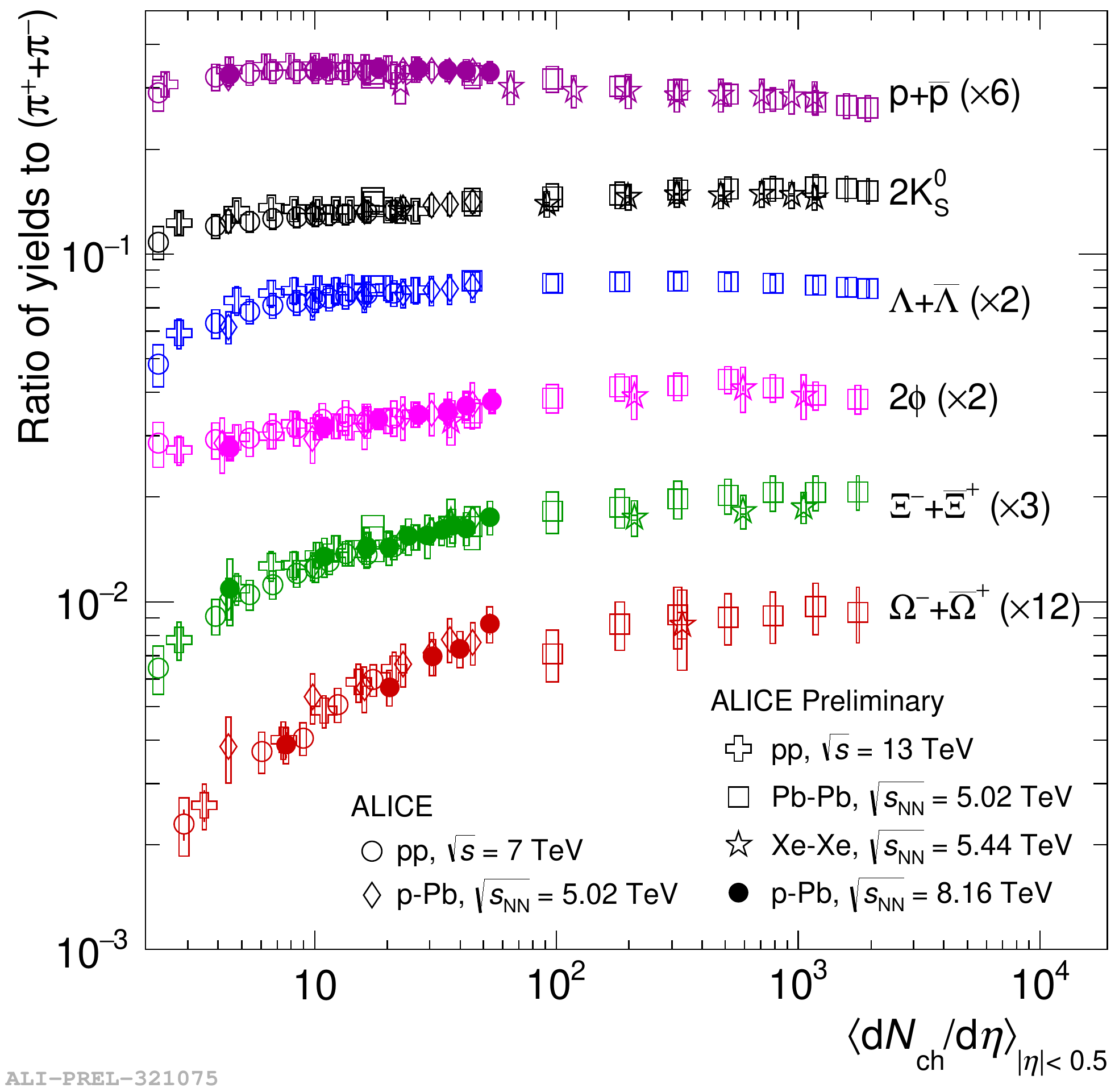}
\caption{Ratios of various particle yields to pion yield as a function of $\langle dN_{\rm{ch}}/d\eta
  \rangle_{|\eta|<0.5}$ for small systems pp and p-Pb, and for
  large systems A-A at various center-of-mass energies~\cite{ALICE:2017jyt,Adam:2015vsf,Abelev:2013haa}. 
}\label{Fig:strange}
\end{center}
\end{figure}
%%%%%%%%%%%%%%%%%%%%%%%%%%%%%%%%%%%%%%%%%%%%%
Figure~\ref{Fig:strange} shows the ratios of various particle yields to pion yield as a function of $\langle dN_{\rm{ch}}/d\eta  \rangle_{|\eta|<0.5}$ for small systems pp and p-Pb, and for
  large systems A-A at various center-of-mass energies~\cite{ALICE:2017jyt,Adam:2015vsf,Abelev:2013haa}. The ratios
  evolve smoothly as a function of multiplicity. There is no energy
  dependence neither the system-size dependence observed. This
  suggests that the particle production is driven by the charged
  particle multiplicity. It is observed that the ratios that involve 
  strange particles increase with increasing multiplicity and
  saturate for heavy-ions. Thus, there is a enhancement
  in strange particle yields as a function of multiplicity. Strangeness
  enhancement has been predicted as a signature of QGP in heavy-ion
  collisions~\cite{Koch:1986ud}. However, for small systems, we also observe the
  enhancement in strange particle yields. It is observed that the
  particle with more strangeness content exhibits larger
  enhancement. It is further noted that $\phi(1020)$ also exhibits
  strangeness enhancement though its a ``hidden strangeness'' state
  and has total strangeness zero. It is observed that all particles
  with open strangeness undergo canonical suppression in small systems
  but $\phi(1020)$ does not~\cite{Acharya:2018orn}. %More investigations have been
                                %done using Monte-Carlo models and
                                %Core-Corno to understand the
                                %enhancement of $\phi(1020)$.
 The investigations based on model
  calculations are ongoing to understand this observation. Recent
  studies with $\phi(1020)$ meson using statistical thermal model
  in small systems and various ratios involving $\phi(1020)$ suggest that the
  ``effective strangeness'' of $\phi(1020)$ meson is 1-2 units~\cite{Sharma:2018owb,Knospe:2018mek}.

\subsection{Resonance Production}
\begin{figure}[tbh]
\begin{center}
%\rotatebox{0}{\resizebox{\textwidth}{!}{
%        \includegraphics{Figures/2019-05-31-2019-05-29-ResonancePlot_v1.pdf}}}
\includegraphics[width = 9cm]{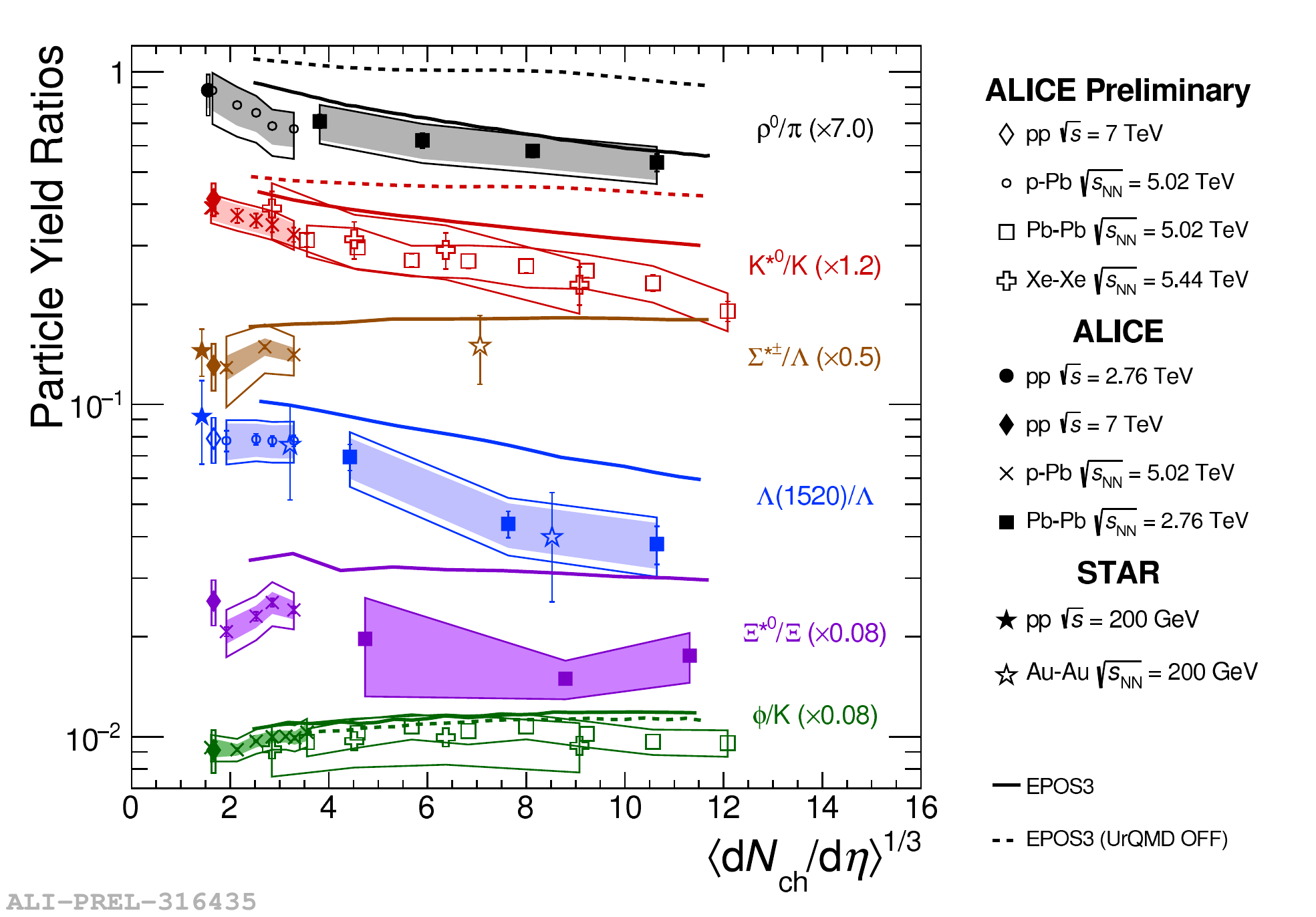}
  \caption{Ratios of particle yields involving short and long-lived
    resonance particles as a function of $\langle dN_{\rm{ch}}/d\eta
    \rangle^{1/3}$ for various systems and energies~\cite{Knospe:2018mek}. Results are compared with EPOS model~\cite{Knospe:2015nva}.}
\label{Fig:reson}
\end{center}
\end{figure}
Figure~\ref{Fig:reson} shows the ratios of particle yields involving short and long-lived
    resonance particles as a function of $\langle dN_{\rm{ch}}/d\eta
    \rangle^{1/3}$ for various systems and energies~\cite{Knospe:2018mek}. Results
    are compared with EPOS model~\cite{Knospe:2015nva}. The resonance particles are
    reconstructed through their hadronic decay channel. The yields of resonances are
    affected by the medium through re-scattering and re-generation
    processes. In re-scattering, the resonance particles that decay in
    the hadronic phase are not reconstructed due the re-scattering of
    their decay daughters in the hadronic phase. There may be also
    re-generation of resonance particles due to the pseudo-elastic
    scattering. At the kinetic freeze-out stage, the resonance yields
    depend on various factors that include chemical freeze-out
    temperature, lifetime of hadronic phase, resonance particle
    lifetime, and scattering cross-section of decay products. In
    Fig.~\ref{Fig:reson} it is observed that yields of short lived resonances
    such as $\rho(770)^0$, $K^*(892)^0$, and $\Lambda(1520)$ decrease
    as a function of multiplicity. The lifetimes of these particles are
    1.3 fm/$c$, 4.2 fm/$c$, and 12.6 fm/$c$, respectively, hence the
    decrease in their yields as a function of multiplicity is
    consistent with the fact that the yields might have been reduced due
    to re-scattering in the hadronic phase. The $\phi(1020)$ yield
    remains constant as a function of multiplicity. Since its lifetime
    is about 46.2 fm/$c$, this suggests that the $\phi(1020)$ meson
    decays after the hadronic phase and is not affected by
    re-scattering or re-generation processes. It is also observed that
    all these ratios, too, do not depend on the system-size and energy
    but only depend on the multiplicity. The EPOS model with UrQMD to
    describe the hadronic scattering effects describe the centrality
    dependence of ratios in heavy-ions, while turning-off the
    UrQMD results in poorer description.

\subsection{Spin Alignment}\label{Sec:spin}
\begin{figure}[tbh]
\begin{center}
%\rotatebox{0}{\resizebox{\columnwidth}{!}{
%        \includegraphics{Figures/RhoVsPtKStar_ppandpbpbandkshort_withProdPlane.pdf}}}
\includegraphics[width = 8cm]{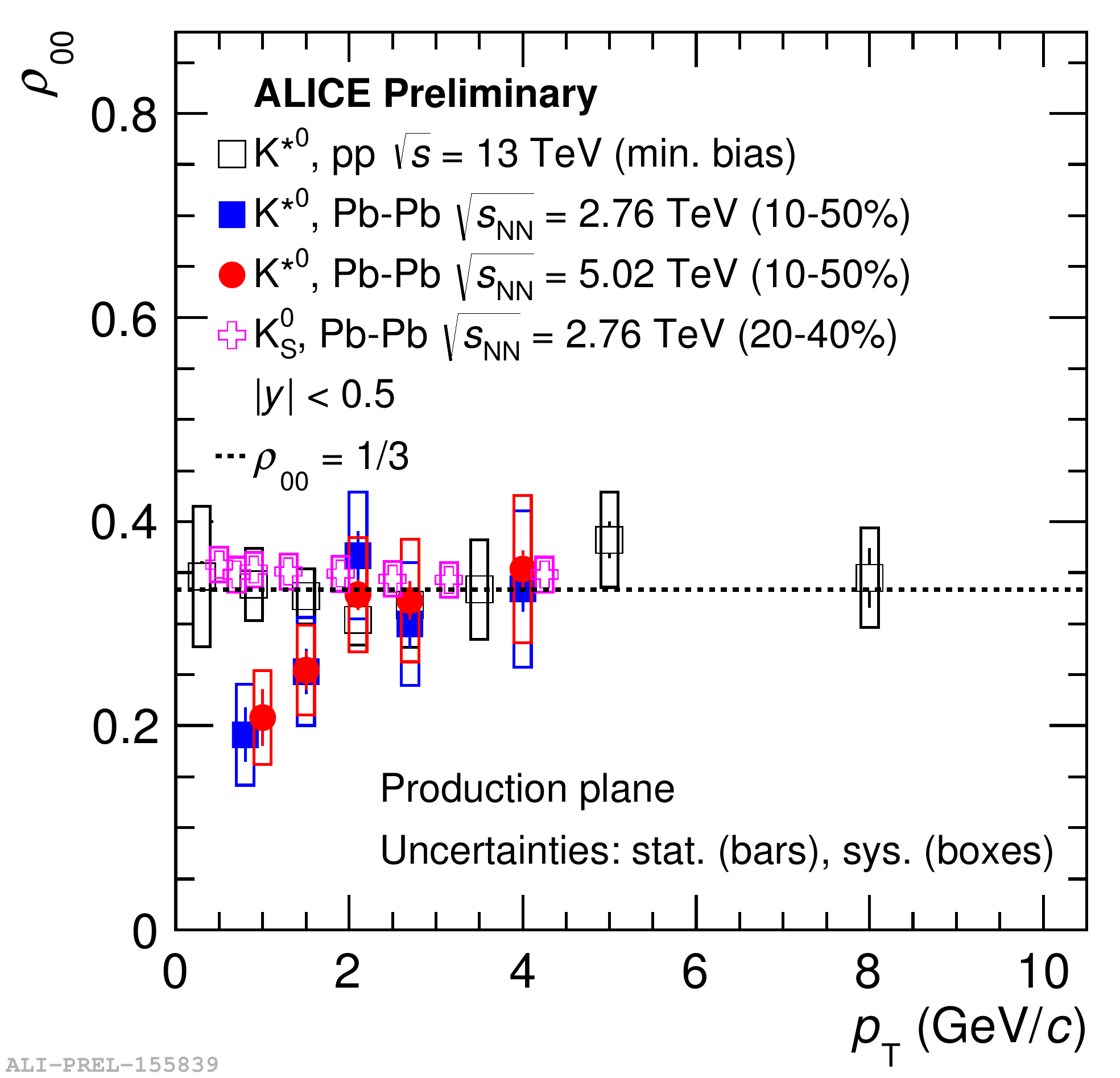}
\caption{The $\rho_{00}$ values as a function of $p_T$ for $K^{*0}$ in
  pp collisions at $\sqrt{s}=$ 13 TeV and Pb-Pb collisions
  $\sqrt{s_{NN}}=$ 2.76 and 5.02 TeV, and $K^0_S$ in Pb-Pb collisions
  at  $\sqrt{s_{NN}}=$ 2.76 TeV at 
  mid-rapidity corresponding to the production plane.  
}
\label{Fig:spin}
\end{center}
\end{figure}

In high-energy heavy-ion collisions with non-zero impact parameter, a
large angular momentum ($\sim10^5 \hbar$) and magnetic field ($10^{14}$ T) is expected to be created~\cite{Becattini:2007sr,Kharzeev:2007jp}. The deconfined state of quarks and
gluons, called the Quark Gluon Plasma (QGP) is also created in high-energy
heavy-ion collisions. In the large angular momentum the 
spin-orbit coupling of quantum chromodynamics (QCD) could lead to
polarization of quarks and hence net-polarization of spin 1 vector
mesons along the direction of the angular momentum~\cite{Liang:2004xn,Yang:2017sdk}.
The spin alignment is studied through the angular distribution of decay
daughters of the vector mesons with respect to the quantization
axis. The quantization axis is perpendicular to the production plane
of the vector meson,
defined by the momentum of the momentum of the vector meson and the
beam direction, or normal to the reaction plane of the system,
defined by impact parameter and the beam direction. The angular
distribution is given by~\cite{Fano:1957zz} 
\begin{equation}
\frac{dN}{d \cos \theta^*} \propto \left[1 - \rho_{00} + \cos^2 \theta^*
(3\rho_{00} - 1)\right], 
\end{equation}
where the $\rho_{00}$ is the zeroth element of the 3$\times$3
spin-density matrix~\cite{Yang:2017sdk}. It is the probability of finding a vector
meson in the spin state of zero out of the possible spin states of $-1,
0$, and $1$. If there is no polarization, all spin states are expected to
be equally probable leading to $\rho_{00}=1/3$. Thus, any deviation of
$\rho_{00}$ value from the 1/3 would lead to non-uniform angular
distribution preferring a spin state. Figure~\ref{Fig:spin}  shows the $\rho_{00}$ values as a function of $p_T$ for $K^{*0}$ in
  pp collisions at $\sqrt{s}=$ 13 TeV and Pb-Pb collisions
  $\sqrt{s_{NN}}=$ 2.76 and 5.02 TeV, and $K^0_S$ in Pb-Pb collisions
  at  $\sqrt{s_{NN}}=$ 2.76 TeV at 
  mid-rapidity corresponding to the production plane.  
It is observed that $\rho_{00}<1/3$ for $K^{*0}$ in Pb-Pb collisions
at both energies. As expected, the $\rho_{00}=1/3$ for  $K^{*0}$ (and
$\phi$ meson)  in pp collisions and for spin 0 state $K^0_S$. The
results are consistent between event and production planes. The
results suggest the first
experimental observation of the spin-orbital interaction in heavy-ion collisions.

\begin{figure}[tbh]
\begin{center}
%\rotatebox{0}{\resizebox{\columnwidth}{!}{
%        \includegraphics{Figures/2018-Jun-01-LifetimeCollection_2018_2body.pdf}}}
%        \includegraphics[width =
%        12cm]{Figures/2018-Jun-01-LifetimeCollection_2018_2body.pdf}
        \includegraphics[scale=.55]{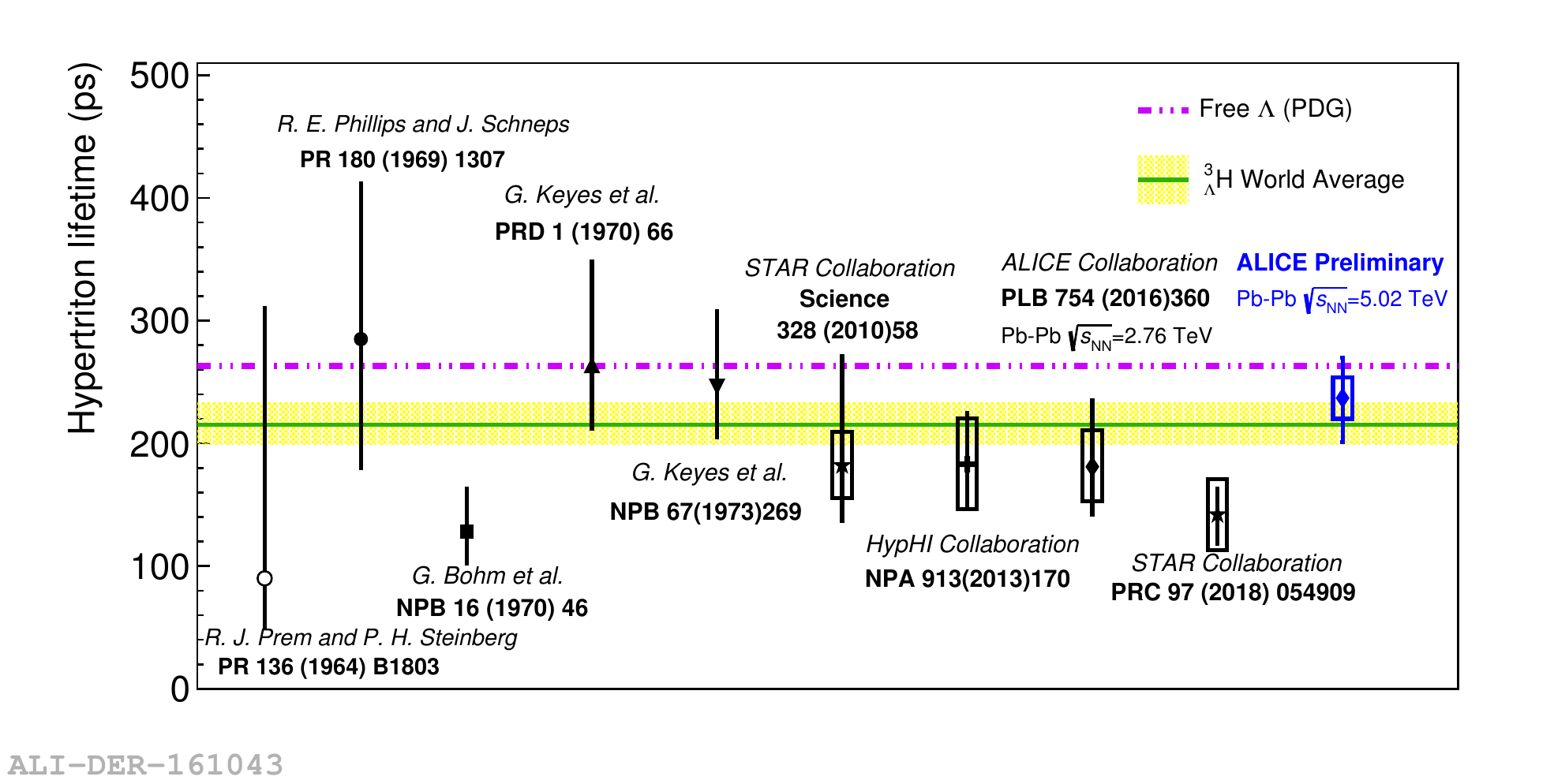}
\caption{Comparison of new hypertriton lifetime measurement results  from ALICE at $\sqrt{s_{NN}}=$ 5.02 TeV with previously published results.}
\label{Fig:hyper}
\end{center}
\end{figure}
There have been lot of efforts in estimating the lifetime of
hypertriton $^3_\Lambda \rm{H}$ which is a bound state of proton, neutron
and $\Lambda$ and is a lightest hypernucleus. Very small $\Lambda$
binding energy has led to the hypothesis that  $^3_\Lambda \rm{H}$ is lower
than free $\Lambda$. Figure~\ref{Fig:hyper} shows the comparison of
new hypertriton lifetime measurement results  from ALICE at
$\sqrt{s_{NN}}=$ 5.02 TeV with previously published results. The new
results from ALICE are obtained from the full statistics data of Pb-Pb
5.02 TeV and the $^3_\Lambda \rm{H}$ are reconstructed through the two-body
decay $^3_\Lambda \rm{H} \rightarrow\ ^3\rm{He} + \pi$. The new ALICE results are
consistent with both free $\Lambda$ and world average.

\subsection{Summary}
In summary, we have presented a selected recent results on particle
production from ALICE. The latest results on charged particle
multiplicity confirm the violation of scaling of number of participant
nucleons.  The $\langle p_T \rangle$ of identified hadrons increases
with multiplicity for both small and large systems. Mass ordering is
observed for heavy-ions but seems to be broken for small systems. The
kinetic freeze-out parameters are extracted for small and large
systems. The extracted $\langle \beta_T \rangle$ increases with
increasing multiplicity for all systems. At
  similar multiplicity values, the $\langle \beta_T \rangle$ is larger
  for small systems. The strangeness enhancement as a function of
  multiplicity is also observed for the first time in small
  systems. The short-lived resonance particles yield decreases with
  increasing multiplicity suggesting re-scattering effect in
  hadronic phase. The results on spin alignment studies suggest the first experimental
  observation of spin-orbit coupling in heavy-ion
  collisions. The latest hypertriton lifetime measurement from ALICE is
  consistent with free $\Lambda$ and the world average.
  
%I would like to thank the organizers of the DQM2019 for inviting me to
%discuss these results. The discussions 
%with Bedangadas Mohanty and Natasha Sharma are acknowledged. The
%support from the SERB Grant No. ECR/2016/000109 is acknowledged and
%would like to thank the ALICE collaboration.

%\section{Summary}\label{Sec:Conclusions}

% \section{Acknowledgements}

%\bibliography{Bibliography,b2_paper}   
%\bibliography{ref}   

%\end{document}

%% file: Aritra/aritra.tex
\section{Study of neutral pion mas in presence of a magnetic field  in the linear sigma model coupled to quarks}
\label{Aritra}

{\it Aritra Das and Najmul Haque}

\bigskip

{\small
In the framework of linear sigma model coupled to quark, we calculate the neutral pion mass in the presence of an external arbitrary magnetic field at zero temperature. A non-monotonic behavior of pion mass as a function of magnetic field is found.
Existing weak-field result has also been reproduced.}

\bigskip

\subsection{Introduction}
	In heavy-ion collisions experiments, a very strong anisotropic magnetic field ($\sim$ $10^{19}$ Gauss) is generated in peripheral collisions perpendicular to the reaction plane due to the relative motion of the colliding ions~\cite{Skokov:2009qp}. In the interior of dense astrophysical objects like compact stars, magnetars~\cite{Duncan:1992hi} and also in the early universe, magnetic field is also involved. The effects of such magnetic fields on fundamental particles cannot be neglected and the detailed study of the effects on the elementary particles is essential at fundamental levels.
	
	The linear sigma model (LSM) is one of the simplest model in pre-QCD era. It was originally proposed by Gell-Mann and L{\'e}vy to study phenomena such as pion-nucleon interaction. The addition of light quarks to the LSM Lagrangian density has given more flexibility to the existing model and it is called linear sigma model coupled to quark (LSMq). 
	
	In this proceedings contribution, we discuss the $\pi^0$-mass in the presence of an arbitrary magnetic field using LSMq.
	
\subsection{Linear sigma model coupled to quarks}
	The Lagrangian of the model is written as
	\begin{align}
	\mathcal{L}&=\underbrace{\frac{1}{2}(\partial_{\mu}\sigma)^2+\frac{1}{2}(\partial_{\mu}\mathbold{\mathbold{\pi}})^2+\frac{a^2}{2}(\sigma^2+\mathbold{\pi}^2)-\frac{\lambda}{4}(\sigma^2+\mathbold{\pi}^2)^2}_{\text{LSM part}} \nn\\ &+\underbrace{i\bar{\psi}\gamma^{\mu}\partial_{\mu}\psi-g\bar{\psi}(\sigma+i\gamma_5\mathbold{\tau}\cdot\mathbold{\pi})\psi}_{\text{quark part}} 
	%\label{unb_lag}
	\end{align}
	 The charged and neutral pion fields are usually defined as 
	\bea
	\pi^{\pm} = \frac{1}{\sqrt{2}}\left(\pi^{1}\pm i \pi^2\right), \qquad \pi^0 = \pi^{3}. 
	\eea
	\\
	 $\sigma$ is the sigma meson of LSM, and  $\psi$ is the $u,d$ quark doublet as
	\bea
	\psi = \begin{pmatrix}
		u \\ d
	\end{pmatrix},
	\eea
	$\mathbold\tau = (\tau^1,\tau^2,\tau^3)$ represents the Pauli spin matrices; $a^2$ is the mass parameter of the theory and we take $a^2<0$ in symmetry unbroken state. Finally, $\lambda$ is the coupling within $\sigma$-$\sigma$, $\pi$-$\pi$, $\sigma$-$\pi$; $g$ represents the coupling between degrees of freedom (DOFs) of LSM with that of quarks.
	
	When $a^2>0$, the symmetry is broken. After symmetry breaking, the Lagrangian takes the form
	\bea
	\mathcal{L} &=& \bar{\psi}(i\gamma^{\mu}\partial_{\mu}-M_f)\psi+\frac{1}{2}(\partial_{\mu}\sigma)^2+\frac{1}{2}(\partial_{\mu}\mathbold{\mathbold{\pi}})^2 - \frac{1}{2}M^2_{\sigma}\sigma^2-\frac{1}{2}M^2_{\pi}\mathbold{\pi}^2 \nn\\
	&-&g\bar{\psi}(\sigma+i\gamma_5\mathbold{\tau}\cdot\mathbold{\pi})\psi -V(\sigma,\pi)-V_{\rm tree}(v), \label{eq:lag_ssb}
	\eea
	with 
	\bea
	V(\sigma,\pi) &=& \lambda v \sigma(\sigma^2+\mathbold{\pi}^2)+\frac{\lambda}{4} (\sigma^2+\mathbold{\pi}^2)^2, \label{eq:pot_V} \\
	V_{\text{tree}}(v)&=&-\frac{1}{2}a^2v^2+\frac{1}{4}\lambda v^4 .\label{eq:pot_tree} 
	\eea
	In unbroken state, the masses of quarks, sigma and three pions are given by 
	\bea
	M_{\sF} &=& gv, \nn\\
		M^2_{\sigma} &=& 3 \lambda v^2 - a^2, \nn\\
	M^2_{\pi} &=& \lambda v^2 - a^2.
\label{eq:tree_masses}
	\eea
	To incorporate non vanishing pion mass we add a term $\mathcal{L}_{ESB}=\frac{1}{2}m_{\pi}^2v (\sigma + v)$ to the Lagrangian density that is obtained after symmetry breaking and as a result the masses are modified to 
	\bea
	M_{\sF}(v^{\prime}_0) &=& g \left(\frac{a^2+m^2_{\pi}}{\lambda}\right)^{1/2}, \nn\\
		M^2_{\sigma}(v^{\prime}_0) &=& 2a^2+3m^2_{\pi}, \nn\\
	M^2_{\pi}(v^{\prime}_0) &=& m^{2}_{\pi}.
 \label{eq:mass_modified}
	\eea 
	
\subsection{Background magnetic field}
We consider a homogeneous,  time-independent background magnetic field in $z$-direction as $\vec{\mathcal{B}} = B\, \hat{\mathbold{z}}$ and the corresponding four-potential is $\mathcal{A}^{\mu}=\displaystyle \frac{B}{2}(0,-y,x,0)$. The four-derivative $\partial_{\mu}$ is replaced by covariant four-derivative $D_{\mu}=\partial_{\mu}+i\mathcal{Q}\mathcal{A}^{\mu}$  for the charged DOFs (quarks and charged pions). Here $\mathcal{Q}=q_{\sF}$ for quark of flavor $f$ and $\mathcal{Q}=e$ for $\pi^{\pm}$, respectively. 
	 
\subsection{One-loop pion self-energy} \label{sec:pi0}
\begin{figure}[tbh]
		\centering
		\includegraphics[scale=0.5]{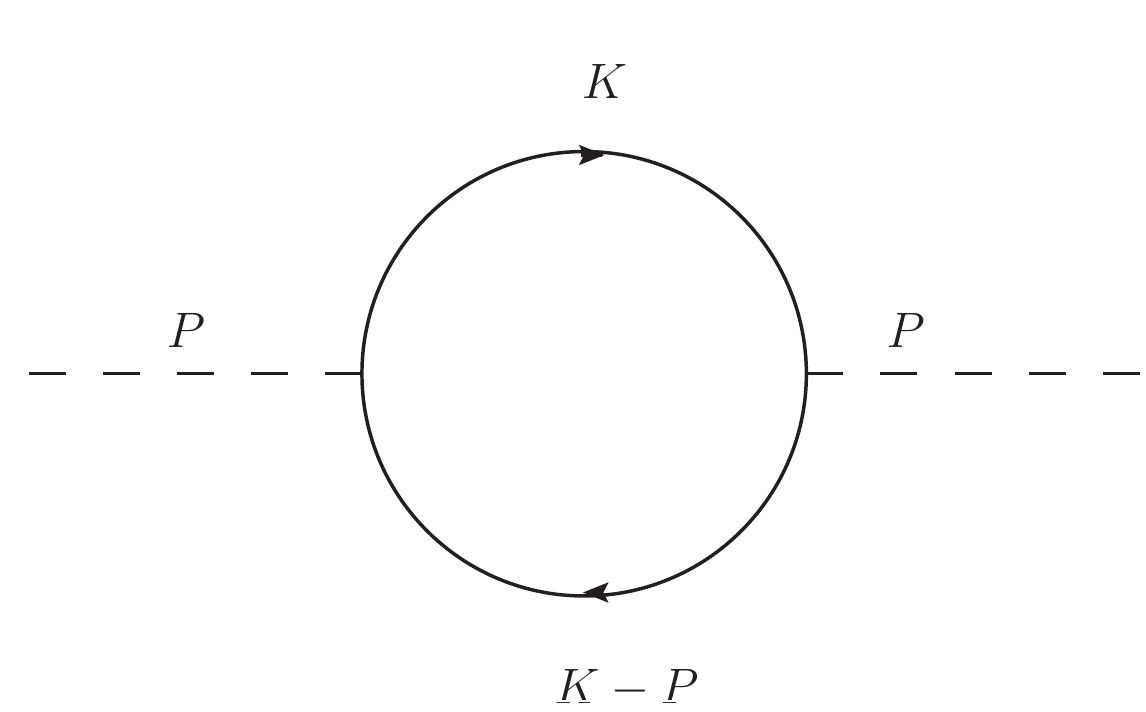}
		\caption{Feynman diagram for the $\pi^{0}$ self-energy containing quark-antiquark loop }
		\label{aritra_fig:pi_quark}
	\end{figure}

	The neutral pion self-energy has the following four contributions:
	\bea
	\Pi(B, P)=\Pi_{f \overline{f}}(B, P)+\Pi_{\pi^{ \pm}}(B)+\Pi_{\pi^{0}}+\Pi_{\sigma}. \label{eq:sigma_con}
	\eea
	The one-loop diagram for quark-antiquark contribution $\Pi_{f \overline{f}}(B, P)$ is depicted in Fig.~\ref{aritra_fig:pi_quark}, whereas that for charged pion contribution $\Pi_{\pi^{ \pm}}(B)$ is depicted in Fig.~\ref{aritra_fig:pi_pi}. Note that for the last two terms $[\Pi_{\pi^{0}}\ \mbox{and}\ \Pi_{\sigma}]$, there are no magnetic corrections as the particles in the loop are chargeless.
	
	\subsubsection{Pion to quark-antiquark loop}
	\label{sec:Piff}
	The expression for the pion self-energy with a quark loop in presence of magnetic field reads
	\bea
	\Pi_{\sF{\bar\sF}}(B,P)
	&=&i\sum_{\sF}g^2\int\frac{d^4K}{(2\pi)^4}\text{Tr}[\gamma_5 iS_{\sF}^{\sB}(K) \gamma_5 iS_{\sF}^{\sB}(K-P)], \label{eq-ff}
	\eea
	where $S_{\sF}^{\sB}(K)$ is the quark propagator given as
	\bea
iS_{\sF}^{\sB}(K) &=& \int_{0}^{\infty} ds\,\exp\left[is\left\{K_{\shortparallel}^{2}+K^2_{\sperp}\frac{\tan(|q_{\sF}B|s)}{|q_{\sF}B|s}-M_{\sF}^2+i\epsilon\right\}\right]\nn\\ &\times&\Big[\left(\slashed{\!K}_{\shortparallel}+M_{\sF}\right)\big\{1+\text{sgn}(q_{\sF}B)\tan(|q_{\sF}B|s)\gamma^1\gamma^2\big\} +\ \slashed{\!K}_{\sperp}\sec^2(|q_{\sF}B|s)\Big], \label{eq:full_sf}
	\eea
	where $\text{sgn}$ is the sign-function. Now, to carry out the loop-momentum integration over $K$, we switch from Minkowski to Euclidean space-time by replacement $k^0\rightarrow ik^0_E$ and also with the additional substitution $( s\rightarrow -is,\ \ t\rightarrow -it)$ as in Ref.~\cite{Alexandre:2000jc}. The subscript $E$ represents the momentum components in Euclidean spacetime.
	Thus, after integration over the four-momentum, the expression for $\Pi_{\sF\bar{\sF}}$ can be written in terms of two proper-time integrations as
    \begin{align}
    \Pi_{\sF\bar{\sF}}(B,P) &= \sum_{\sF}\frac{g^2}{4\pi^2}\int\limits_{0}^{\infty}ds\,dt\,\frac{|q_{\sF}B|}{(s+t)}
    e^{-\left\{M^2_{\sF}(s+t)+(p_E^{\shp})^2\frac{st}{s+t}+\frac{(p_E^{\sperp})^2}{|q_{\sF}B|}\frac{\sinh(|q_{\sF}B|s)\sinh(|q_{\sF}B|t)}{\sinh[|q_{\sF}B|(s+t)]}\right\}} \nn\\
    %%%%%
    &\hspace{.5cm}\times\Bigg[\frac{1+M^2_{\sF}(s+t)-(p_E^{\shp})^2\frac{st}{s+t}}{\(s+t\)\tanh\(|q_fB|(s+t)\)}-\frac{|q_fB|}{\sinh^2\(|q_fB|(s+t)\)} \nn\\
    &\hspace{2cm}\times\left(1-\frac{(p^{\sperp}_E)^2}{|q_fB|}\frac{\sinh(|q_{\sF}B|s)\sinh(|q_{\sF}B|t)}{\sinh[|q_{\sF}B|(s+t)]}\right)\Bigg]. \label{eq:full_mom_Piff}
    \end{align}
    
    \subsubsection{Charged pion loop}
    \label{sec:Pipm}
    \begin{figure}[tbh]
    	\centering
    	\includegraphics[scale=0.5]{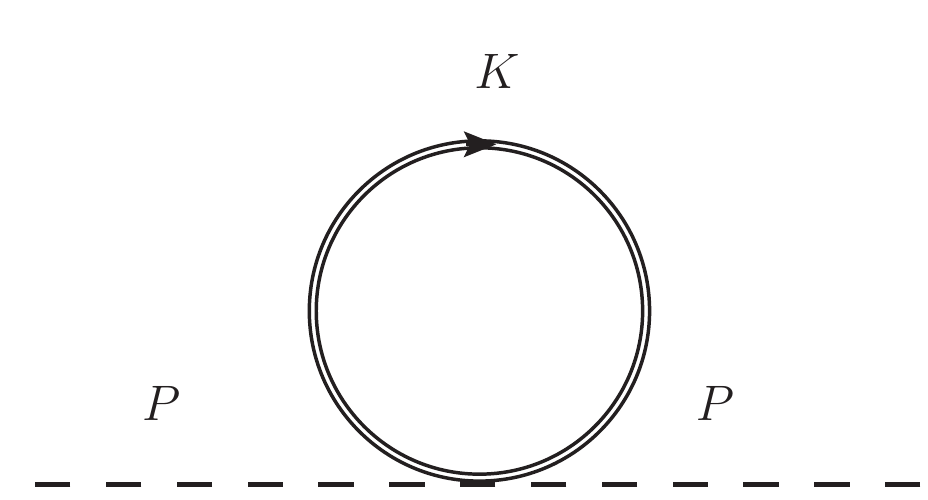}
    	\caption{Feynman diagram for one loop charged pion contribution to the $\pi^{0}$ self-energy}
    	\label{aritra_fig:pi_pi}
    \end{figure}
    The tadpole diagram, shown in Fig.~\ref{aritra_fig:pi_pi}, reads
    \begin{align}
    \Pi_{\pi_{\pm}}(B)=\frac{\lambda}{4}\int\frac{d^4K}{(2\pi)^4}iD_{B}(K). \label{eq:pi_pi}
    \end{align}
    Charged pion propagator $D_B(K)$ in presence of magnetic field is given by
    \begin{align}
    iD_B(K) =\int\limits_{0}^{\infty}\frac{ds}{\cos(|eB| s)}e^{i\left(K^2_{\shp}+K^2_{\sperp}\frac{\tan(|eB|s)}{|eB|s}-m_{\pi}^2\right)s}. \label{eq:scalar_prop}
    \end{align}
     After momentum integration, we are left with the expression of $\Pi_{\pi_{\pm}}(B)$ given as
    \begin{align}
    \Pi_{\pi_{\pm}}(B) = \frac{\lambda}{4}\frac{|eB|}{16\pi^2}\int\limits_{0}^{\infty} \frac{ds}{s} \frac{e^{-sm^2_{\pi}}}{\sinh(|eB|s)}. \label{eq:pi_pi_complete}
    \end{align}
 %%%%%%%%   
 \subsection{Pion Mass} \label{sec:Pi_mass}
    %\subsection{Exact solution}
We need to solve the equation
\begin{equation}
p^2_0-|\mathbold{p}|^2-m^2_{\pi}-\text{Re}[\Pi(B,P)]=0 \label{eq:ex_disp}
\end{equation}
in the limit $\mathbold{p}\rightarrow 0$ and $p_0=M_{\pi}(B)$ to obtain modified pion mass $M_{\pi}(B)$. The self-energy of $\pi^0$ has four contributions out of which $\Pi_{f \overline{f}}(B, P)$ and $\Pi_{\pi_{\pm}}(B)$ will contribute to magnetic field correction as mentioned in Eq.~\eqref{eq:sigma_con}.
    The total self-energy $\Pi(B,p_0,\mathbold{p}=\mathbold{0})$ can be written as 
    %%%
    \begin{align}
    &\Pi(B,p_0)=\sum_f \frac{g^2}{4\pi^2}\int\limits_{0}^{\infty}ds\,dt\,\frac{|q_{\sF}B|}{(s+t)}e^{-(s+t)M^2_{\sF}-\frac{s t}{s+t}(p^0_E)^2}\nn\\
    &\hspace{0.2cm}\times\Bigg[\frac{1+M^2_{\sF}(s+t)-\frac{s t}{s+t}(p^0_E)^2}{(s+t)\tanh\(|q_{\sF}B|(s+t)\)} -\frac{|q_{\sF}B|}{\sinh^2\(|q_{\sF}B|(s+t)\)} -\frac{M_{\sF}^2(s+t)-\frac{s t}{s+t}(p^0_E)^2}{|q_{\sF}B|\(s+t\)^2}\Bigg] \nn\\
    &\hspace{2.2cm} + \frac{\lambda}{4}\frac{1}{16\pi^2}\int\limits_{0}^{\infty}ds\frac{e^{-sm^2_{\pi}}}{s}\left[\frac{|eB|}{\sinh(|eB|s)}-\frac{1}{s}\right]. \label{eq:Pi_ex_st}
    \end{align}
    %%%%%
    We can make a variable change from $(s,t)$ to $(u,v$~\cite{Alexandre:2000jc,Tsai:1974ap} in Eq.~\eqref{eq:Pi_ex_st} as
    \begin{align}
    s=\frac{1}{2}u(1-v),\qquad t=\frac{1}{2}u(1+v).
    \end{align} 
  This leads Eq.~\eqref{eq:ex_disp} to 
\bea
M_{\pi}^2(B) &=&  (v_0^{\prime})^2\lambda-a^2 -  \sum_{f} \frac{g^2}{4\pi^2}\int\limits_{0}^{\infty}du\int\limits_{-1}^{1}dv\,\frac{|q_{\sF}B|\, }{2}e^{-u\left[M_{\sF}^2-\frac{1}{4}(1-v^2)M_{\pi}^2(B)\right]}\nn\\
&\times&\left[\left\lbrace\frac{1+uM^2_{\sF}+\frac{1}{4}u(1-v^2)M^2_{\pi}(B)}{u\tanh(|q_{\sF}B|u)} - \frac{|q_{\sF}B|}{\sinh^{2}(|q_{\sF}B|u)}\right\rbrace \right. \nn\\
&&\left.-\frac{M_{\sF}^2+\frac{1}{4}(1-v^2)M^2_{\pi}(B)}{u}\right]-\frac{\lambda}{4}\frac{1}{16\pi^2}\int\limits_{0}^{\infty}du\frac{e^{-u\,m^2_{\pi}}}{u}\left[\frac{|eB|}{\sinh(|eB|u)}-\frac{1}{u}\right].
\label{MpiB}
\eea
 Equation~\eqref{MpiB} for the magnetic-field dependent neutral pion mass is incomplete; one also needs to incorporate the one-loop magnetic-field correction to the boson self-coupling $\lambda$, the fermion coupling $g$ and the minimum of the potential $v_0'$. 
 
 Now, the effective fermion mass becomes
 \bea
 M_{\sF,\rm{eff}} &= g_{\rm eff}\, v_0^{B},
 \label{eff_quantities} 
 \eea
 where $g_{\rm eff}$ represents magnetic field dependent one-loop effective fermion vertex whereas $v_0^{B}$ is the the magnetic-field-dependent minimum of the potential after symmetry breaking. 
 
 Using Eq.~\eqref{eff_quantities} and replacing $g_{\rm eff}$ with the other effective quantities, Eq.~\eqref{MpiB} becomes
 \bea
 M_{\pi}^2(B) &=&  (v_0^{B})^2\lambda_{\text{eff}}-a^2 -  \sum_{f} \frac{1}{4\pi^2}\frac{1}{(v_0^{B})^2}\int\limits_{0}^{\infty}du\int\limits_{-1}^{1}dv\,\frac{|q_{\sF}B|\, M_{\sF,\rm{eff}}^2}{2}e^{-u\left[M_{\sF}^2-\frac{1}{4}(1-v^2)M_{\pi}^2(B)\right]}\nn\\
 &\times&\left[\left\lbrace\frac{1+uM^2_{\sF}+\frac{1}{4}u(1-v^2)M^2_{\pi}(B)}{u\tanh(|q_{\sF}B|u)}  -\frac{|q_{\sF}B|}{\sinh^{2}(|q_{\sF}B|u)}\right\rbrace-\frac{M_{\sF}^2+\frac{1}{4}(1-v^2)M^2_{\pi}(B)}{u}\right]\nn\\
 &-&\frac{\lambda_{\rm eff}}{4}\frac{1}{16\pi^2}\int\limits_{0}^{\infty}du\frac{e^{-u\,m^2_{\pi}}}{u}\left[\frac{|eB|}{\sinh(|eB|u)}-\frac{1}{u}\right].
 \label{MpiB_eff}
 \eea
 %%%%%
  The expression of effective self-coupling $(\lambda_{\rm eff})$ can be obtained from the following vertex diagrams.
   \begin{figure}[tbh]
  	\centering
  	\includegraphics[scale=0.5]{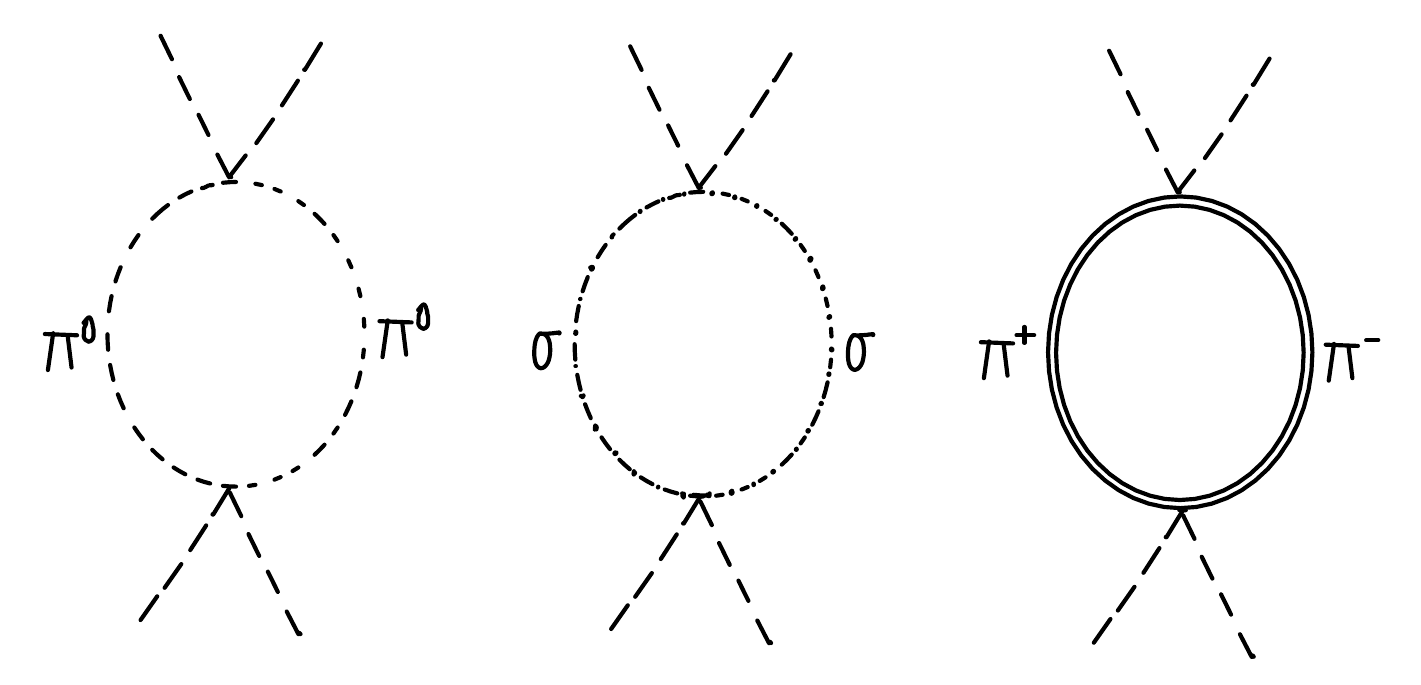}
  	\caption{One-loop corrections to the self-coupling $\lambda$. The dashed line denotes $\pi^{0}$, double line denotes $\pi^{\pm}$, dashed-dotted line denotes $\sigma$-meson.}
  	\label{fig:pi_vertex}
  \end{figure}

  The effective self-coupling $\lambda_{\rm eff}$ to one-loop order is obtained from Fig.~\ref{fig:pi_vertex} as
    	\bea
    	\lambda_{\rm eff}&=&\lambda + \frac{3\lambda^2}{8\pi^2}\int\limits_{-1}^{1}dv\int\limits_{0}^{\infty}du\,e^{-u\,\left\{m^2_{\pi}+\frac{1}{4}(1-v^2)(p_E^0)^2\right\}} \times\ \left[\frac{|eB|}{2\sinh\big(|eB|u\big)}-\frac{1}{2u}\right].
    	\eea
 %   \end{widetext}
  The other effective magnetic-field-dependent quantities, namely,  $v_0^{B}$ and $M_{\sF,\rm{eff}}$ can be found in appendices of Ref.~\cite{Das:2019ehv}.

 \begin{figure}[tbh]
 	\centering
 	\subfigure{
 		\includegraphics[scale=0.3]{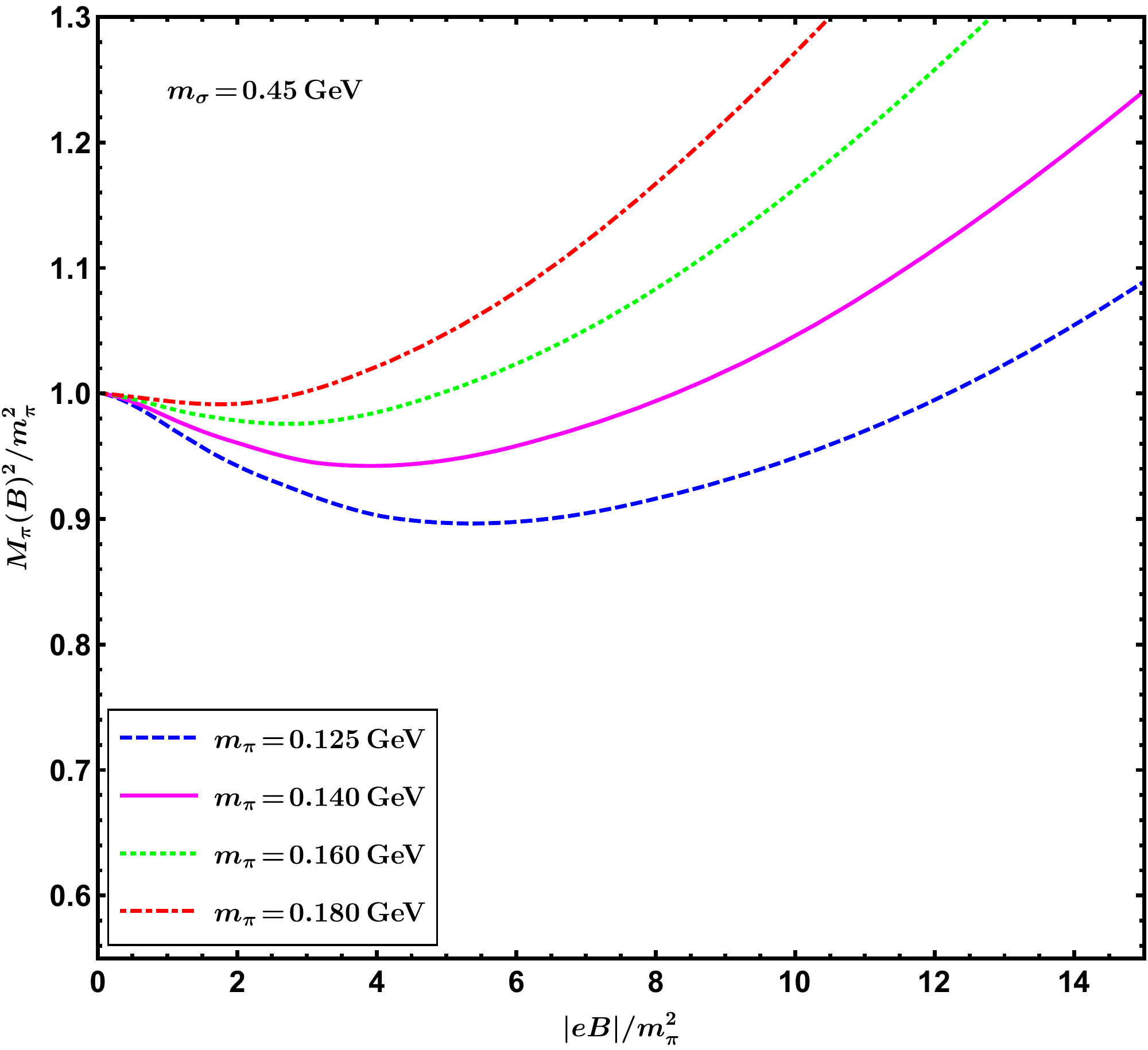}}
 	\subfigure{
 		\includegraphics[scale=0.3]{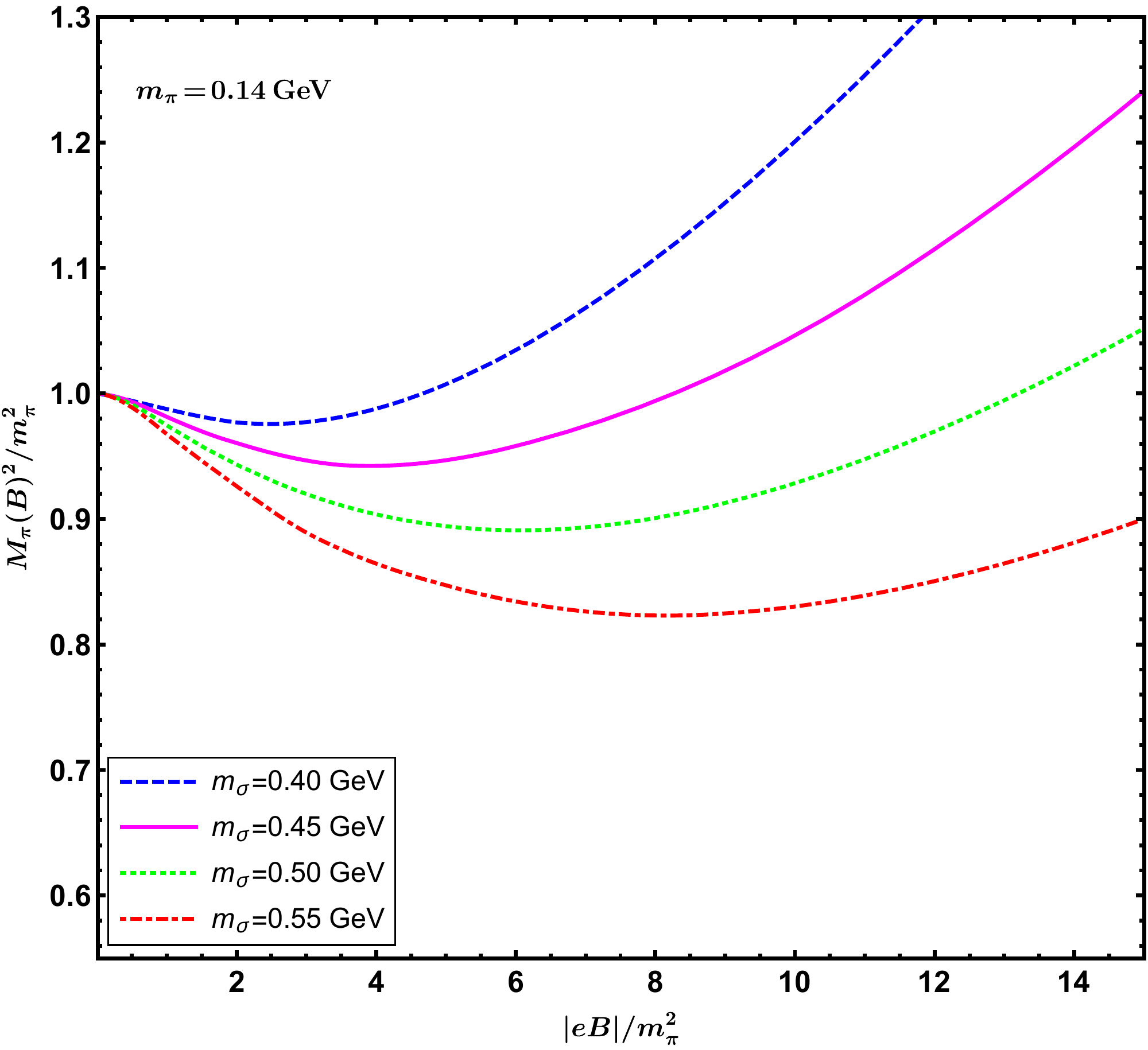}}
 	\caption{Figure shows magnetic field dependence of neutral pion mass for a fixed $m_{\sigma}=0.45\,\,\text{GeV}$ with $m_{\pi}=0.125,\, 0.140,\, 0.160,\, 0.180 \,\text{GeV} $ (left panel) and for a fixed $m_{\pi}=0.14\,\, \text{GeV}$ with $m_{\pi}=0.40,\, 0.45,\, 0.50,\, 0.55\,\, \text{GeV} $ (right panel).}
 	\label{fig:pion_mass_exact_full}
 \end{figure}  
 
  Solving Eq.~\eqref{MpiB_eff} numerically, we get $\pi^0$ mass as shown in Fig.~\ref{fig:pion_mass_exact_full} in which a non-monotonic behavior with magnetic field is observed~\cite{Das:2019ehv}. It \emph{decreases} with increasing magnetic field at weak magnetic field~\cite{Ayala:2018zat} but in large values of magnetic field it starts to increase.
    
\subsection{Conclusion and outlook} \label{sec:con}
In conclusion, we have studied effect of external magnetic field to the mass of the neutral pion mass under the framework of LSMq. The calculation is performed taking into account one-loop self-coupling of pions $ \lambda_{\text{eff}}$, one-loop effective fermion\label{key} mass $M_{\sF,\rm{eff}}$ and one-loop effective minimum of the potential $v_0^{B}$. When the strength of magnetic field is increased, we get a non-monotonic behavior. Our result also qualitatively agrees with LQCD studies as in Ref.~\cite{Bali:2017ian} up to a moderate strength of the magnetic field. Looking to the future the present calculation can be extended to the case of astrophysical objects where the baryon density and also the magnetic field are very high. Nevertheless, using LSMq model, we can qualitatively capture essential features that is obtained by much more involving and rigorous studies.

%% file: Ashutosh/Ashutosh.tex
\section{Flow correlations as a measure of phase transition}
\label{Ashutosh}

{\it Ashutosh Dash and Victor Roy}

\bigskip

{\small
In the present exploratory study,  using a hydrodynamic model, we study the imprint of two different equation of state: one with crossover transition and other with first order phase transition, on the flow correlations developed in the medium. We find that the normalized symmetric cummulants between different flow harmonics are sensitive to the nature of phase transition. 
}

\subsection{Introduction}
It is well known that at low temperature ($T$) and  baryon chemical potential ($\mu_B$), nuclear matter is in a state of confined color neutral hadrons while at high temperature or high baryon chemical potential, nuclear matter is in a state of deconfined matter of quarks and gluons called the quark gluon plasma (QGP). Nuclear matter at  at small baryon chemical potential  and finite temperature  is believed to undergo a crossover transition from the hadronic phase to the QGP phase
and a first order phase transition at relatively larger $\mu_B$ and the first order phase transition line terminates at a critical point \cite{Gavai:2004sd}.

The  present study aims to find a unique observable which connects QCD
Equation of State (EoS) and the experimental data of heavy-ion collisions using hydrodynamical model. We find the linear/Pearson correlation (defined later) of initial geometric asymmetry of colliding nuclei to the corresponding flow coefficient 
(particularly the second-order flow coefficient $v_{2}$, Eq.~\ref{Eq:vn}) is a unique observable which can differentiate 
between EoS with a first-order phase transition to that with a crossover transition irrespective of the 
initial condition used.  It has been known that the event averaged $v_{2},$ and the eccentricity
of the averaged initial state, $\epsilon_{2}$, Eq.~\ref{Eq:eccen} are approximately linearly correlated \cite{Niemi:2012aj},
and the  Pearson correlation is quite insensitive to the shear viscosity of the fluid and the initial condition used \cite{Niemi:2012aj}, 
which makes it a robust observable to disentangle between the two different EoSs.
\subsection*{Results and Discussion}

 \begin{figure}
 \centering
\includegraphics[width =0.45\textwidth]{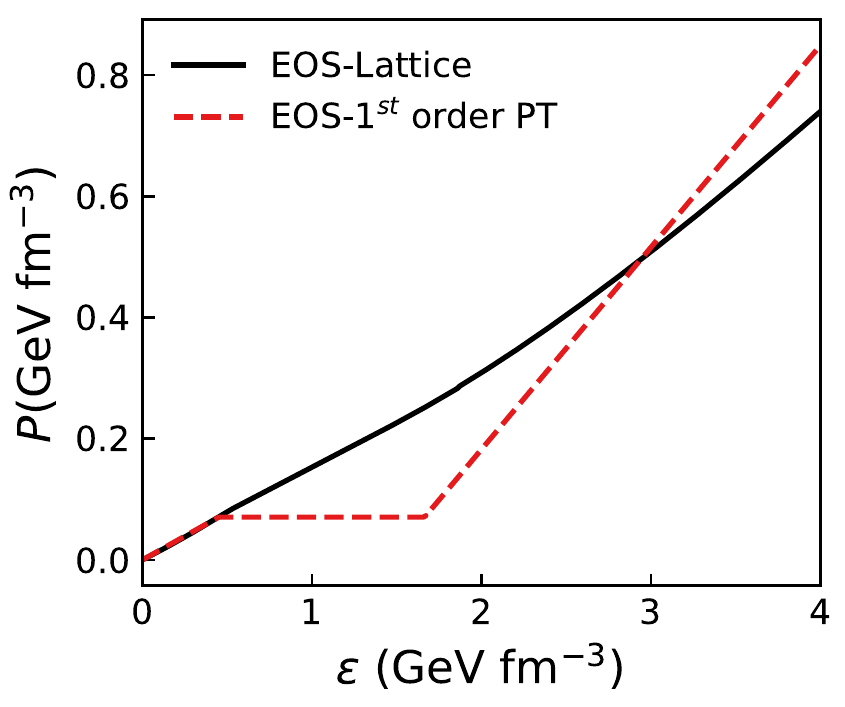}
\caption{ Equation of state with a cross-over transition (solid black line) and with first order phase transition (dashed red line), at $\mu_B=0$ MeV.}
\label{figEOS}
\end{figure}
In the present work, we will be using two kinds of  EoSs (shown in Fig.~\ref{figEOS}) \cite{Dash:2019fsm}~:\\
(i) A parameterized EoS (EoS Lattice) which has a cross-over transition between
high temperature QGP phase obtained from lattice QCD and a hadron resonance gas below the crossover temperature.\\
(ii) An EoS (EoS $1^{st}$ order PT) connecting a non-interacting massless QGP gas at high temperature to a hadron resonance gas at low temperatures through a first order phase transition.  
The bag constant $B$ is a parameter adjusted  to yield a critical temperature $T_c=164~$MeV.

Similarly, we consider here two initial conditions, where the initial energy density $\varepsilon(x,y)$ is obtained at initial time $\tau_{0}=0.6$ fm from the MC-Glauber and the Trento model using Gaussian smearing,
\begin{equation}\label{Eq:energy}
\varepsilon(x, y)=\kappa\sum_{i=1}^{N_{\text {WN}}}\exp \left(\frac{-\left(\vec{r}-\vec{r}_{i}\right)^{2}}{\left(2 \sigma^{2}\right)}\right),
\end{equation}
 where $\vec{r}_i=(x_{i}, y_{i})$ are the spatial coordinates of either wounded  nucleons (initial condition $\varepsilon_{WN}$) or 
 binary collisions (initial condition $\varepsilon_{BC}$). 
 $\kappa$ is a normalization constant fixed to provide the observed multiplicity of pions at $\sqrt{s_{NN}}=62.4$ GeV and  $\sigma=0.7$ fm  is the spatial scale of a wounded nucleon or a binary collision. 
 The initial geometry/anisotropy of the overlap zone of two colliding nucleus is quantified in terms of coefficients $\epsilon_{n} $ 
 \begin{equation}\label{Eq:eccen}
\epsilon_{n} e^{i n \Phi_{n}}=-\frac{\int d x d y r^{n} e^{i n \phi} \varepsilon(x, y)}{\int d x d y r^{n} \varepsilon(x, y)}.
\end{equation}
where $\phi$ is the azimuthal angle in position space and $\varepsilon(x, y)$ is as defined in Eq.~\ref{Eq:energy}. The final azimuthal momentum
 anisotropy is characterized in terms of the coefficients $v_{n}$ and is defined as Fourier expansion of the single particle azimuthal distribution
 \begin{eqnarray}\label{Eq:vn}
\frac{\mathrm{d} N}{\mathrm{d} \phi_p} \propto 1+2 \sum_{n=1}^{\infty} v_{n}^{\mathrm{obs}} \cos n\left(\phi_p-\Psi_{n}^{\mathrm{obs}}\right)
\end{eqnarray}
where $\phi_p$ is the azimuthal angle in momentum space and $\Psi_{n}^{\mathrm{obs}}$ is the event plane angle. In order to quantify the  linear correlation we use Pearson's correlation coefficient which is defined as
  \begin{figure}
  \centering
\includegraphics[width =0.45\textwidth]{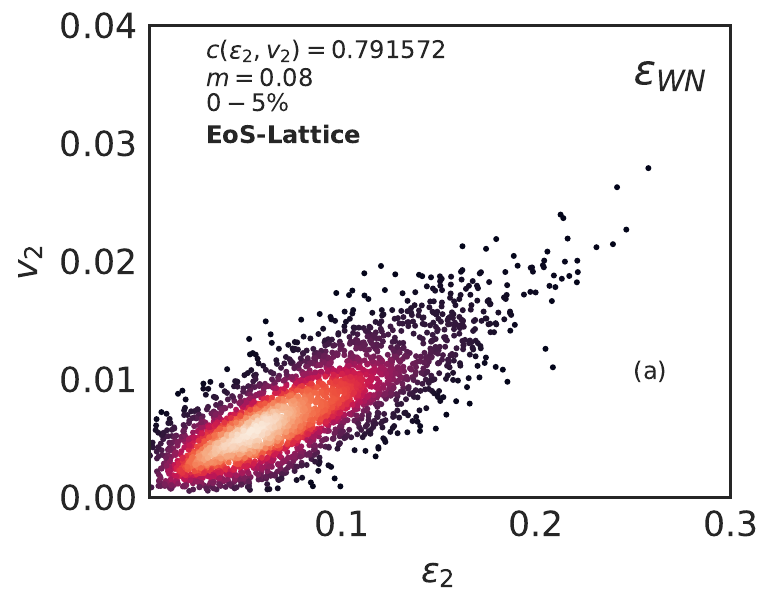}
\includegraphics[width =0.45\textwidth]{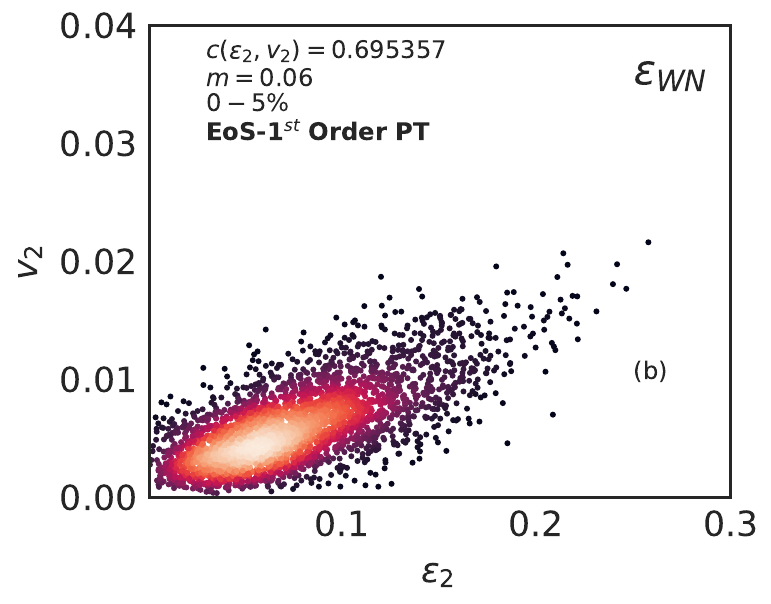}
\caption{\label{fig1} (a) Event-by-event distribution of $v_{2}$ vs $\epsilon_{2}$  for $0-5\%$ Au+Au collisions at $\sqrt{s_{NN}}=200$ GeV. 
    (b) Same as left panel but for EoS with first order phase transition.}
\end{figure}
\begin{equation}
c(x, y)=\left\langle\frac{\left(x-\langle x\rangle_{\mathrm{ev}}\right)\left(y-\langle y\rangle_{\mathrm{ev}}\right)}{\sigma_{x} \sigma_{y}}\right\rangle_{\mathrm{ev}},
\end{equation}
where $\sigma_x$ and $\sigma_y$ are the standard deviations of the quantities $x$ and $y$. A value of $1(-1)$ implies that a linear (anti-linear) 
correlation between $x$ and $y$.  A value of $0$ implies that there is no linear correlation between the variables. 

For centrality $0-5\%$ as shown in Fig.~\ref{fig1},
using two different EoS we found $\sim 15\%$ decrease in $c(\epsilon_{2},v_{2})$ for first order phase transition compared to a crossover transition,  
which clearly indicates that  $c(\epsilon_{2},v_{2})$ can be treated as a good signal of phase 
transition in the nuclear matter.
However, the initial eccentricities $\epsilon_{n}$ are not accessible in real experiments (and are model dependent) which makes $c(v_{n},v_{m})$ more interesting. Nevertheless, instead of $c(v_{n},v_{m})$, a better experimental observable would rather be \textit{ normalized symmetric cummulants} (NSC) defined as
 \begin{equation}
 \mathrm{NSC}(m,n)=\frac{\left\langle v_{m}^{2} v_{n}^{2}\right\rangle-\left\langle v_{m}^{2}\right\rangle\left\langle v_{n}^{2}\right\rangle}{\left\langle v_{m}^{2}\right\rangle\left\langle v_{n}^{2}\right\rangle}.
 \end{equation}
 \begin{figure}
\includegraphics[width =0.48\textwidth]{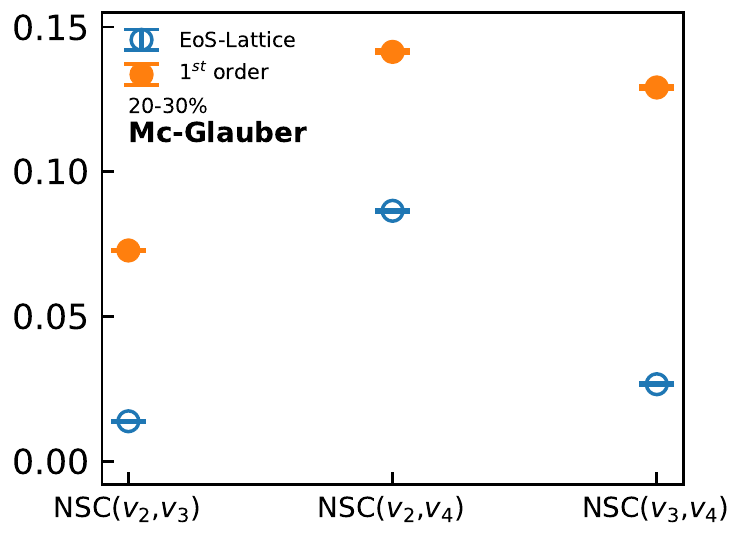}
\includegraphics[width =0.48\textwidth]{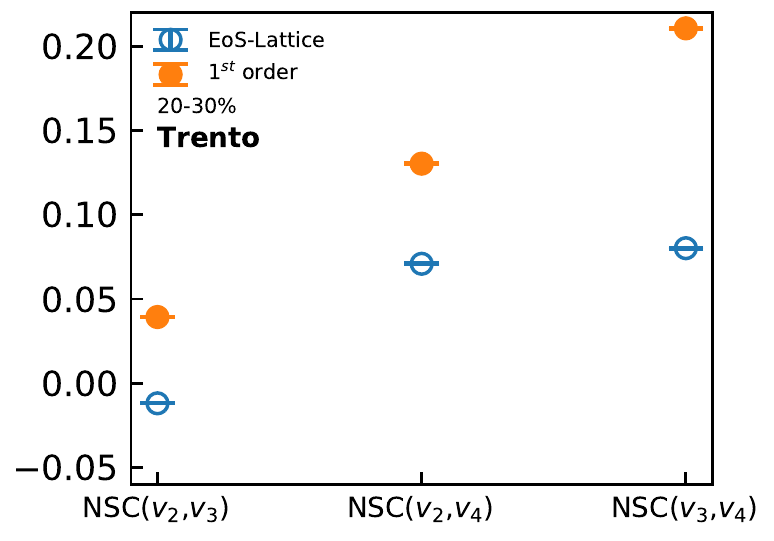}
\caption{ (a) Normalised symmetric cummulants $NSC(m,n)$ for EoS-Lattice (solid red circles), and first order phase transition (open blue circle) for $20-30\%$ collision centrality.  (b) same as left panel but for Trento model. Error bars are statistical.}
\label{fig2}
\end{figure}
The results of $\mathrm{NSC}(m,n)$ two different initial condition is shown in Fig.~\ref{fig2}.  We found in the mid central collisions always show higher values of $\mathrm{NSC}(m,n)$ for the EoS with first order phase transition than 
crossover transition irrespective of the initial conditions. For example we can calculate $\mathrm{NSC}(m,n)$ from available
experimental data for various $\sqrt{s_{NN}}$ and pinpoint the energies where $\mathrm{NSC}(m,n)$ shows an enhancement. These observations may be attributed to very different evolutionary dynamics of the system for the  
two different EoS, as the speed  of sound becomes zero in first-order phase transition hence the  linear/non-linear coupling of 
$\epsilon_{n}$ - $v_{n}$ and $v_{n}$ - $v_{m}$ is different in the two scenario.

%% file: Mariyah/mariyah.tex
\section{Color dipole distribution at small transverse momentum} %Force line breaks with

%\title{Probing fusion incompleteness in $^{20}$Ne + $^{165}$Ho reaction at E$_{lab}$ $\approx$ 4-7 MeV/A} %Force line breaks with \\
%\title{Systematic of fusion incompleteness in $\alpha$ cluster projectile: $^{20}$Ne + $^{165}$Ho reaction a case study} %Force line breaks with \\

\textit{Mariyah Siddiqah, Nahid Vasim, Khatiza Banu, Raktim Abir, Trambak Bhattacharyya}

\bigskip

{\small
We derive analytical results for unintegrated color dipole gluon distribution function at small
transverse momentum
in the form of a series of Bells polynomials. Interestingly, when resumming the series in leading log accuracy, the results show striking similarity with the Sudakov form factor when one identifies the coupling term with a constant that stems from the saddle point condition along the saturation line.
}

\bigskip

\subsection{Introduction}\label{mariyah_sec:level1}

Parton distribution function (PDF) encodes the nonperturbative structure of hadrons by providing information about the probability distribution of partons with longitudinal momentum fraction $x$ at some resolution scale $Q^2$  inside the hadron or nucleus. PDFs are the source of attraction for numerous dedicated experimental and theoretical efforts. 
PDFs are universal as they can be  extracted from one experiment and used in some other scattering process at some  other resolution scale. They  play a central role in QCD predictions.

In order to know transverse momentum distribution of  quarks and gluons inside the hadron/nucleus it is necessary to consider some other distribution function. In this context Transverse  momentum dependent (TMDs) parton  distributions  or unintegrated parton distribution (UPDFs) functions are objects of interest. They not only provide the information about the longitudinal momentum distributions but also gives the information about the transverse momentum distributions of partons within the hadron/nucleus.
% transverse momentum distribution of  quarks and gluons inside the hadron/nucleus,  we need to consider other parton distribution function known as  transverse momentum dependent (TMDs) parton  distributions  or unintegrated parton distribution (UPDFs) functions \cite{Bacchetta:2016ccz}.
%  Transverse  momentum dependent (TMDs) parton  distributions  or unintegrated parton distribution (UPDFs) functions  provide the information about the longitudinal as well as the transverse   momentum distribution 
 Thus  providing a more detailed information on the internal structure of protons 
\cite{Bacchetta:2016ccz}.

TMDs have recently attracted a huge amount of interest and are fully investigated at the current and future facilities including JLAB 12 GeV upgrade, RHIC and planed electron-ion collider(EIC). Recently unpolarised quark TMD from global data analysis has been extracted from the TMD factorised formullas derived from the semi-inclusive deep inelastic scattering and Drell-Yan and Z-boson production in proton-proton(pp) collisions.

The deep inelastic scattering experiments at HERA also provide intense indications that there exists a novel, yet unexplored, saturation regime  in high energy limit of QCD which  corresponds  the small values of Bjorken-$x$. In this regime the gluon cascade occupy all the phase space  available to them to such an extent that the fusion of newly emited gluons starts, leading to the gluon saturation. A dynamical scale gets generated due to this QCD self regulation mechanism known as saturation scale $Q_s$ \cite{Gribov:1984tu}. At  this scale the gluon splitting balances gluon recombination.

In last few years lot of  efforts have been done in connecting TMDs and small-$x$ saturation physics. Like PDFs, TMDs are also non-perturbative quantities and can be extracted from experiment using the same factorization approach but they are not universal as their operator definitions are process dependent. This process dependence of UPDFs is related to different choices of gauge links. The future and past gauge links correspond to final and initial state interactions, respectively.  Depending on these gauge links there are various UPDFs, but only two of them are universal and all other more complicated UPDFs can be seen as the convolution of these two gluon distribution functions \cite{Dominguez:2011wm}. The two different UPDFs are:  Weizs\"{a}cker-Williams  (WW)  gluon distribution and dipole gluon (DP)  distribution function \cite{Dominguez:2011wm,Buffing:2013eka,Hatta:2016dxp,Xiao:2017yya}.

\subsection{Gluon distributions}
Weizs\"{a}cker-Williams  gluon distribution can be directly probed in the quark-anti-quark jet correlation in deep inelastic scattering while the dipole gluon distribution  can be probed in the direct photon-jet correlation in pA collisions. In the light-cone gauge with the proper boundary conditions the gauge links in the definition of WW gluon distribution disappears completely, indicating that WW gluon distribution can be interpreted as the genuine gluon density.  While as on the other hand, the dipole gluon distribution does not have any such interpretation as the gauge link dependence always remains in its definition, thus it is defined as the the Fourier transform of the color dipoles.   
 
The operator definition of Weizs\"{a}cker-Williams gluon distribution is, 
 \begin{eqnarray}
 &&x G^{WW}\left(x,k_{\bot}\right)=2\int\frac{d\xi^{-}d^2\xi_{\bot}}{(2\pi)^3P^{+}}e^{ixP^{+}\xi^{-}-ik_{\bot}.\xi_{\bot}}\nonumber\\
 &&
 ~\langle P|{\rm Tr}\left[F^{+i}(\xi^{-},\xi_{\bot})~\mathcal{U}^{[+]\dagger}~F^{+i}(0,0)~\mathcal{U}^{[+]}\right]|P\rangle~, 
 \label{WW-Operator}
 \end{eqnarray}
 whereas the operator definition of color dipole gluon distribution in the fundamental representation is, 
 \begin{eqnarray}
 &&x G^{DP}\left(x,k_{\bot}\right)=2\int\frac{d\xi^{-}d^2\xi_{\bot}}{(2\pi)^3P^{+}}e^{ixP^{+}\xi^{-}-ik_{\bot}.\xi_{\bot}}\nonumber\\
 &&
 ~\langle P|{\rm Tr}\left[F^{+i}(\xi^{-},\xi_{\bot})~\mathcal{U}^{[-]\dagger}~F^{+i}(0,0)~\mathcal{U}^{[+]}\right]|P\rangle ~.
 \label{DP-Operator}
\end{eqnarray}
In both the definitions $F^{\mu\nu}$ is gluon field strength tensor $F^{\mu\nu}_a$ 
and the gauge links involved are,   
%
%
%  \begin{eqnarray}
%  \mathcal{U}^{[+]}&=&U^n\left[0,\infty;0_{\bot}\right]U^t\left[\infty;0_{\bot},   \infty_\perp\right]
%                    U^t\left[\infty;\infty_{\bot},\xi_\perp\right]U^n\left[\infty;\xi^-,\xi_\perp\right] \\
%  \mathcal{U}^{[-]}&=&U^n\left[0,-\infty;0_{\bot}\right]U^t\left[-\infty;0_{\bot},\infty_\perp\right]
%                    U^t\left[\infty;\infty_{\bot},\xi_\perp\right]U^n\left[-\infty;\xi^-,\xi_\perp\right]
%  \end{eqnarray}
%
\begin{eqnarray}
 \cal{U}^{[+]} &=&
 U^n\left[0^{-},0_{\bot};\infty^{-},0_{\bot}\right]U^t\left[\infty^{-},0_{\bot};\infty^{-},\infty_\perp\right]\nonumber\\
 &&
                   U^t\left[\infty^{-},\infty_{\bot};\infty^{-},\xi_\perp\right]U^n\left[\infty^{-},\xi_{\bot};\xi^-,\xi_\perp\right]~, \\
\cal{U}^{[-]}&=&U^n\left[0^{-},0_{\bot};-\infty^{-},0_{\bot}\right]U^t\left[-\infty^{-},0_{\bot};-\infty^{-},\infty_\perp\right]\nonumber\\
 &&
                   U^t\left[-\infty^{-},\infty_{\bot};-\infty^{-},\xi_\perp\right]U^n\left[-\infty^{-},\xi_{\bot};\xi^-,\xi_\perp\right]~, 
 \end{eqnarray}
where the longitudinal ($U^{n}$) and transverse ($U^{t}$) gauge links are defined as, 
 \begin{eqnarray}
  U^n\left[a^{-},x_{\bot};b^{-},x_{\bot}\right]&=&\mathcal{P} \exp\left[ig~\int_{a^{-}}^{b^{-}}dx^{-} 
   A^{+}\left(0,x^{-},x_{\bot}\right)\right]      ~, \nonumber\\   
  U^t\left[x^{-},a_{\bot};x^{-},b_{\bot}\right]&=&\mathcal{P} \exp\left[ig~\int_{a_{\bot}}^{b_{\bot}}dx_{\bot} .~
   A_{\bot}\left(0,x^{-},x_{\bot}\right)\right] ~.\nonumber 
 \end{eqnarray}

Both these gluon distributions in the McLerran-Venugopalan model for a larger nucleus shows a dramatic behaviour as a function of $k_{\perp}$. For the larger values of $k_{\perp}$, both the  WW gluon distribution and DP gluon distribution  is proportional to $Q^2/k_{\perp}^2$ while at smaller values of $k_{\perp}$ WW gluon distribution  
is proportional to $\ln Q^2/k_{\perp}^2$ and DP gluon distribution is proportional to $k_{\perp}^2$. 
\begin{figure}[!t]
\centering
% \cente
  \includegraphics[width=9cm, height=9.5cm,angle=270]{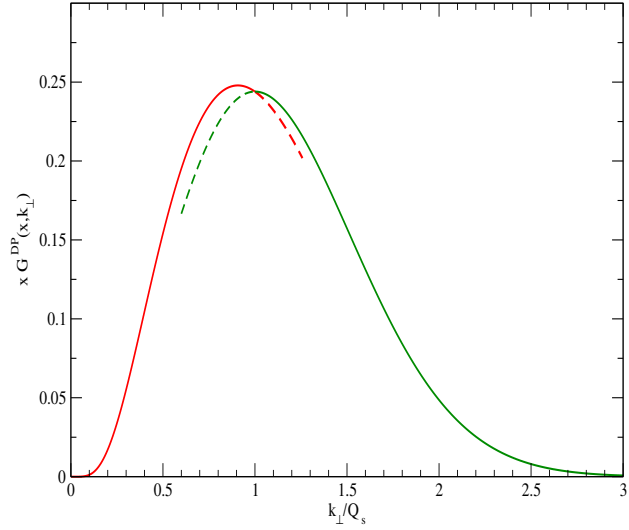}
\caption{The unintegrated dipole gluon distribution  $xG^{DP}(r_\perp,Y)$ plotted as function of $\xi=k_{\perp}/Q_s(Y)$ for nucleus of typical radius $\sim 7$ fm at $\alpha_s \sim 0.1$.
The tail of the green curve, has been tweaked by scaling down to match at $\xi =1$, about to follow power law fall.}
 \label{SexyFigure1}
\end{figure}
 
\subsection{Results and discussions}
Recently we have derived the analytical results of color dipole distribution function at small transverse momentum in series of Bells polynomial \cite{Abir:2018hvk}. 
% \item The dipole gluon distribution can be probed directly in the photon jet correlation measurement in $pA$ collisions while as Weizsacker-williams can be directly measured in the DIS dijet production and 
 We consider the Levin-Tuchin (LT) solution \cite{Levin:1999mw,Levin:2000mv} of the leading order Balitsky Kovchegov equation in the black disc limit. Interestingly,
when resuming the series in leading log accuracy, the results showing up striking similarity with the Sudakov form factor with role play of coupling is being done by a constant
$(\tau = 0.2)$ that stems from the saddle point
condition along the saturation line. The key result of our study, unintegrated dipole gluon 
distribution at small transverse momentum, is as follows,
\begin{equation}
xG^{DP}(x,k_{\perp})
\approx -\frac{S_{\perp}N_c \tau}{\pi^3\alpha_s}\ln\left(\frac{k_{\perp}^{2}}
{4Q_{s}^{2}}\right)\exp\left[-\tau\ln^2\left(\frac{k_{\perp}^{2}}{4Q_{s}^{2}}\right)\right]\nonumber
\end{equation}
The result (as shown in Fig.~\ref{SexyFigure1}) indicates that at small transverse momentum, $xG^{DP} (x, k_\perp) $is not actually 
proportional to $k_{\perp}^{2}$ as
previously anticipated, rather it is proportional to $\ln(k_{\perp}^{2}/4 Q_s^2)$
times the double log soft factor.

%% file: Pallavi/pallavi.tex
\section{Viscous coefficients and thermal conductivity of a $\pi K N$ gas mixture in the medium}

\textit{Pallavi Kalikotay, Nilanjan Chaudhuri, Snigdha Ghosh, Utsab Gangopadhyaya, Sourav Sarkar}

\bigskip

{\small
The temperature and density dependence of the relaxation times, thermal conductivity, shear viscosity and bulk viscosity for a hot and dense gas consisting of pions, kaons and nucleons have been evaluated in the kinetic theory approach. The in-medium cross-sections for $\pi\pi$, $\pi K$ and $\pi N$ scatterings were obtained by using complete propagators for the exchanged $\rho$, $\sigma$, $K^*$ and $\Delta$ excitations derived using thermal field theoretic techniques. Significant deviations have been observed when compared with corresponding calculations using vacuum cross-sections usually employed in the literature. The value of the specific shear viscosity $\eta/s$ is found to agree well with available estimates.	
}

\bigskip

%\end{abstract}
%\maketitle
%\vspace{-1cm}

\subsection{Introduction}
The effects of dissipation on the dynamical  evolution of matter produced in relativistic heavy ion collisions have been a much discussed topic in recent times. Dissipative phenomena are generally studied by considering small deviations from equilibrium  at the microscopic level. Transport coefficients  such as shear and bulk viscosity and thermal conductivity are estimated considering the transport of momenta and heat among the constituents. Collisions among constituents are responsible for the transport of momenta, heat etc. within the system and so the scattering cross-section is the principal dynamical input in transport equations where it appears in the collision integral. It is thus necessary that the relaxation time which quantifies the time scale of approach to equilibrium should be evaluated using in-medium scattering cross-sections in order to obtain a more realistic estimate of the transport coefficients. We have considered a hadron gas mixture consisting of Pions- the most abundant hadron gas produced in HIC, Kaons- the next abundant species and Nucleons - for introducing finite baryon density.

% % % % % % % % % % % % % % % % % % % % % % % % % %
Using the kinetic theory approach the expressions for thermal conductivity $\lambda$, shear viscosity $\eta$ and bulk viscosity $\zeta$ is found to be
\begin{eqnarray}
&&\lambda =\frac{1}{3T^2} \sum_{k=1}^{N} \int\frac{d^3 p_k}{(2\pi)^3}\frac{g_k \tau_k}{E_k^2} p_k^2  (p_k^\nu u_\nu -h_k)^2  f_k^{(0)}(1\pm f_k^{(0)}),\label{lambda1} \nonumber\\
&&\eta=\frac{1}{15 T}\sum_{k=1}^{N} \int\frac{d^3 p_k}{(2\pi)^3}\frac{g_k \tau_k}{E_{p_k}^2}|\vec{p}_k|^4
f_k^{(0)}(1\pm f_k^{(0)}), \label{eta1} \nonumber\\
&&\zeta =\frac{1}{T}\sum_{k=1}^{N}\int \frac{d^3 p_k}{(2\pi)^3}\frac{g_k \tau_k}{E_{p_k}^2}Q_k^2f_k^{(0)} (1\pm f_k^{(0)})~.\nonumber \label{zeta1}
\end{eqnarray}

%\iffalse
	
	\begin{figure}[!ht]
		\begin{center}
			\includegraphics[angle=-90, scale=0.162]{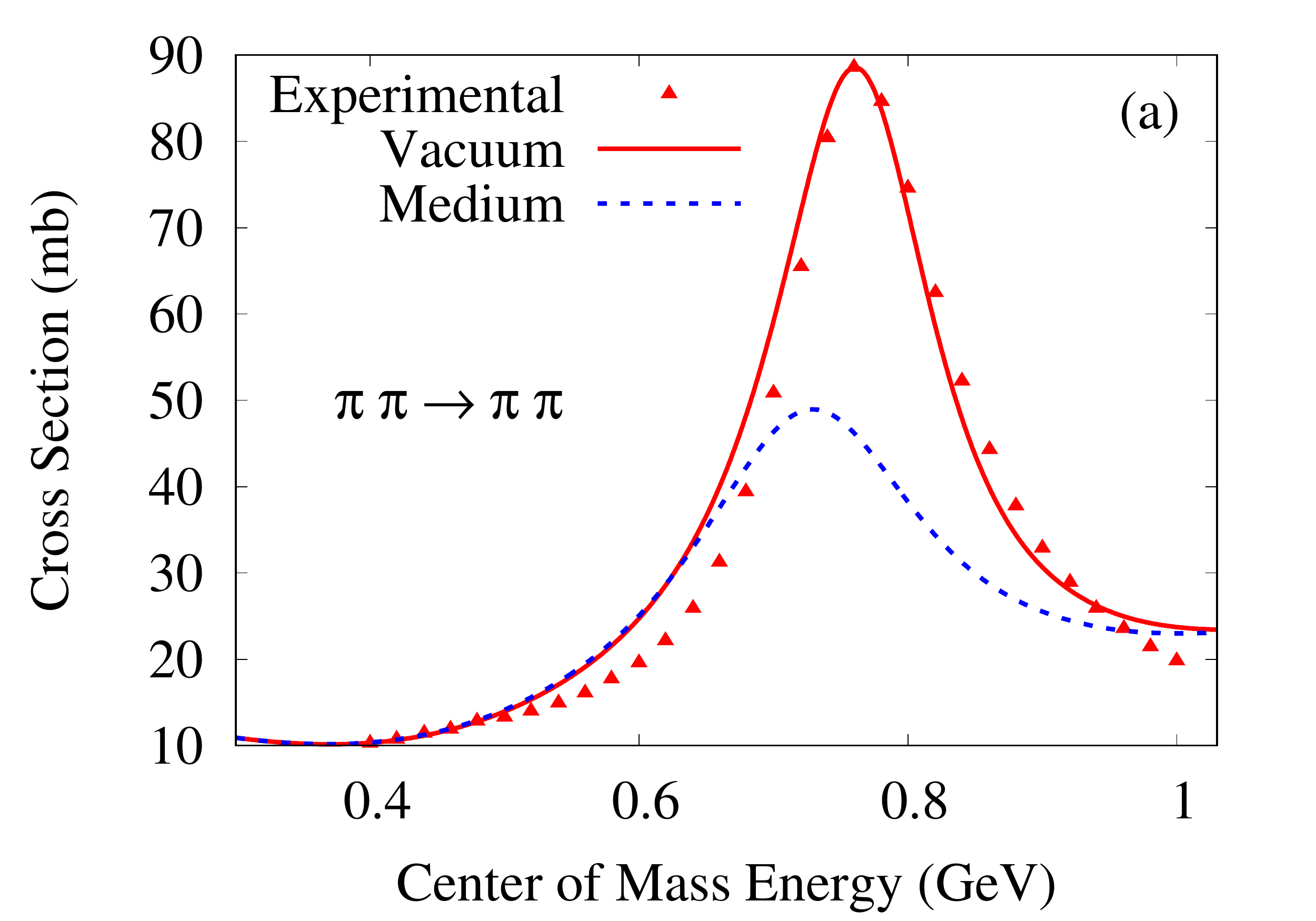}	\includegraphics[angle=-90, scale=0.162]{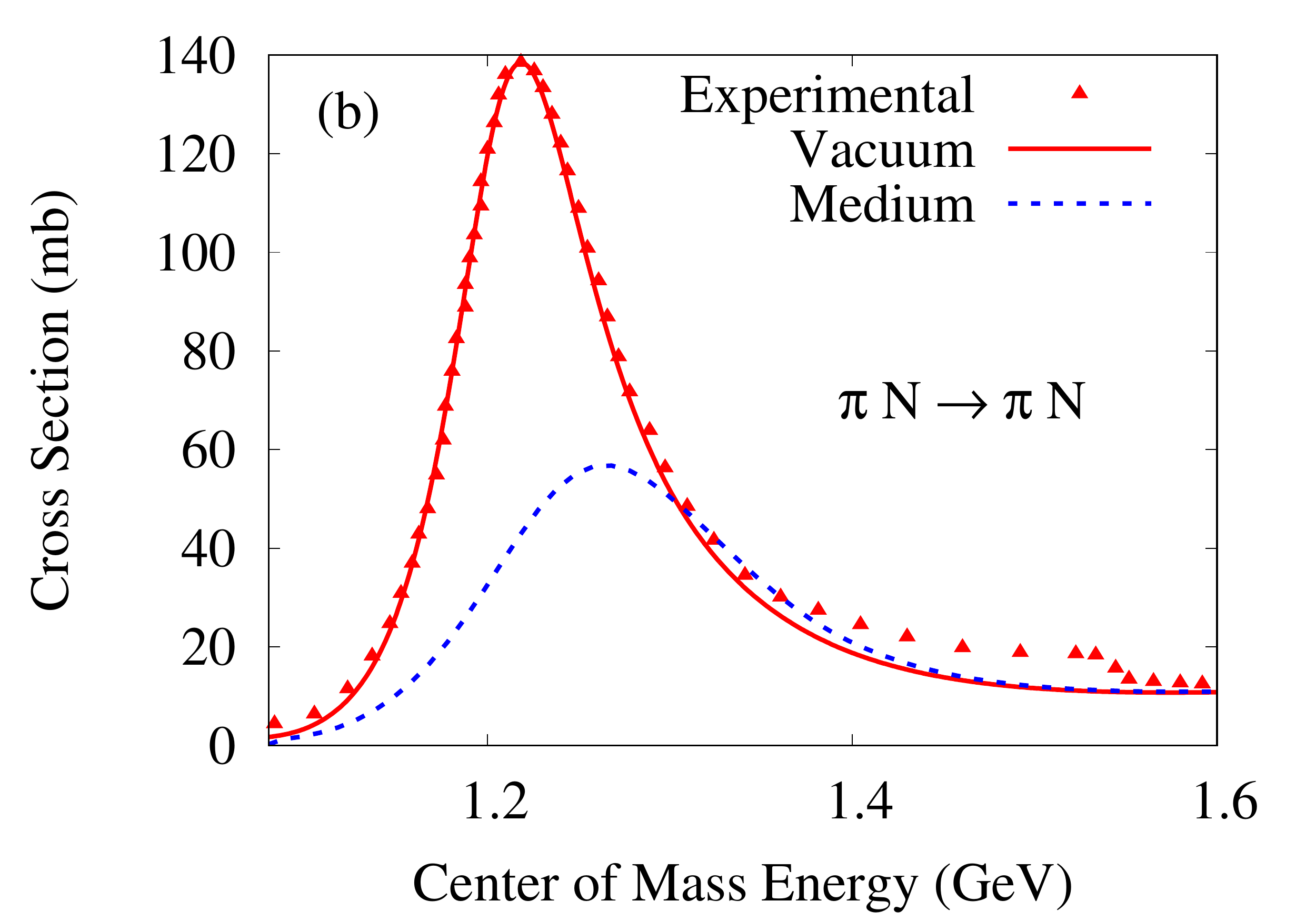}	\includegraphics[angle=-90, scale=0.162]{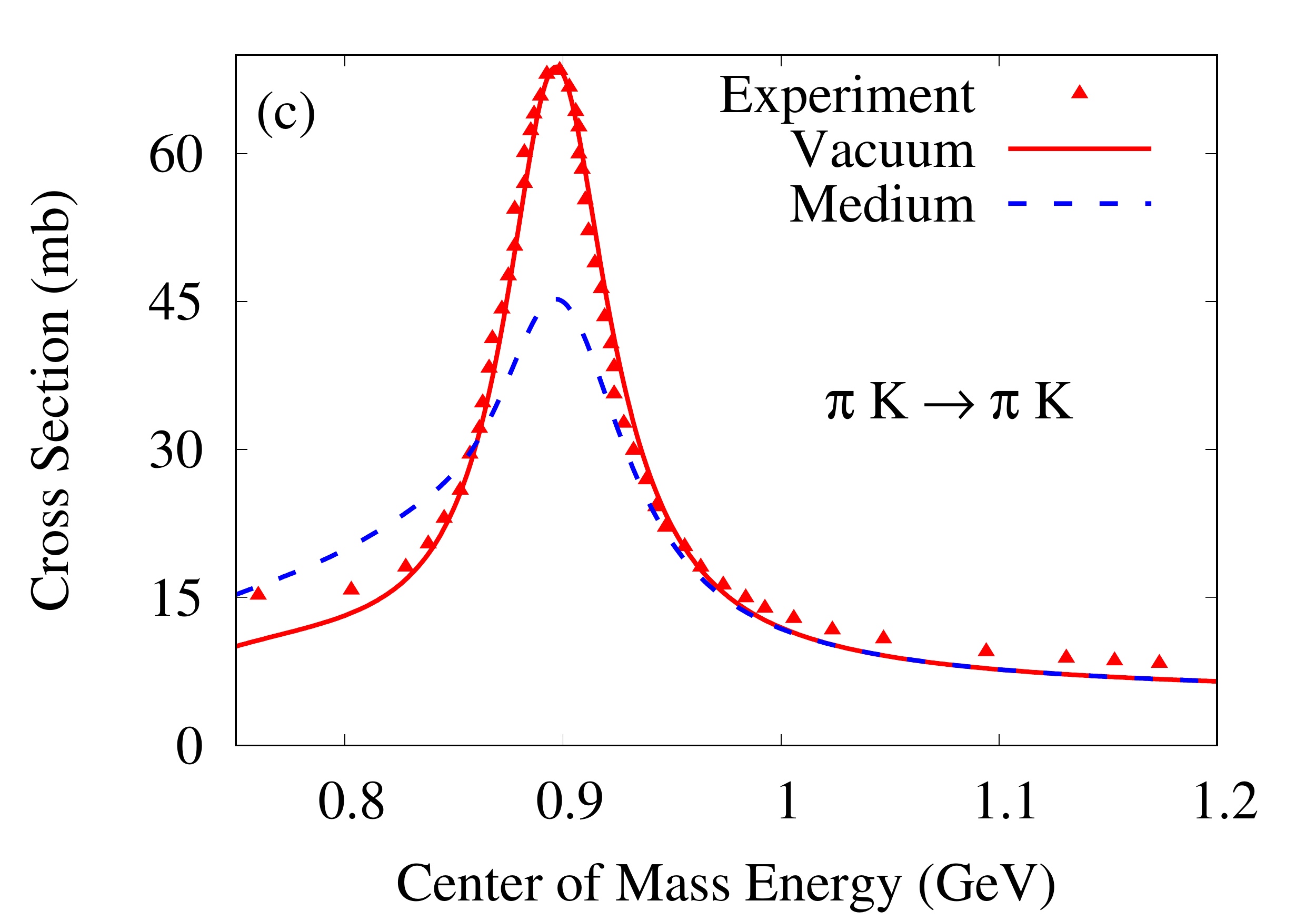}
		\end{center}
		\caption{The (a)$ \pi \pi \rightarrow\pi \pi $, (b)$ \pi N\rightarrow\pi N $ and (c) $ \pi K \rightarrow \pi K $ elastic scattering cross section as a function of centre of	mass energy compared among experiment, vacuum and medium corresponds to $ T = 160 $ MeV and $ \mu_N=200 $ MeV. Experimental data have been taken from Ref.~\cite{Prakash:1993bt} }
		\label{cross_sec}
	\end{figure}

%\fi

We will now start to discuss the results of our work. Fig.(\ref{cross_sec}) shows the elastic scattering cross sections for $\pi\pi\rightarrow\pi\pi$, 
$\pi N\rightarrow\pi N$ and $\pi K\rightarrow\pi K$. The thermal medium has the effect of suppressing the cross section at the resonance energy, which shows to be about $ 50-70 \% $  at $ T=160$ MeV.

We have calculated all the results for three different set of values of pion, nucleon and kaon chemical potential, the choice of these sets have been tabulated below.
\begin{center}
	\begin{tabular}{|c|c|c|c|}
		\hline 
		Chemical potential  & $\mu_\pi$ & $\mu_k$ & $\mu_N$ \\
		\hline
		Set 1 & 0 & 0 & 0\\ 
		Set 2 & 50 & 100 & 200\\ 
		Set 3 & 100 & 200 & 500\\ 
		\hline
		
	\end{tabular}
\end{center}

% % % % % % % % % % % % % % % % % % % % % % % % % % %

%\iffalse
\begin{figure}[!ht]
	\begin{center}
		\includegraphics[angle=-90, scale=0.162]{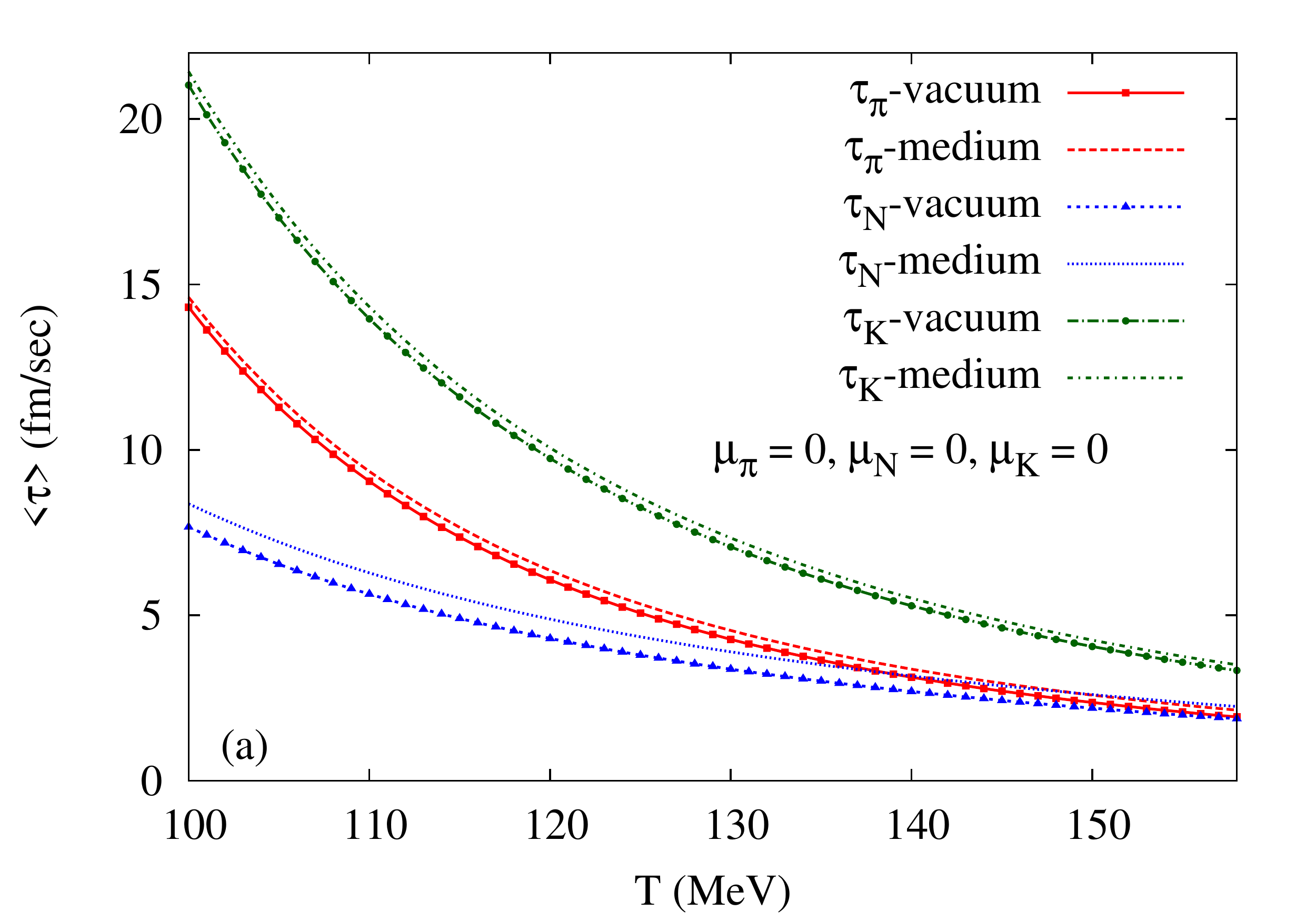}  
		\includegraphics[angle=-90, scale=0.162]{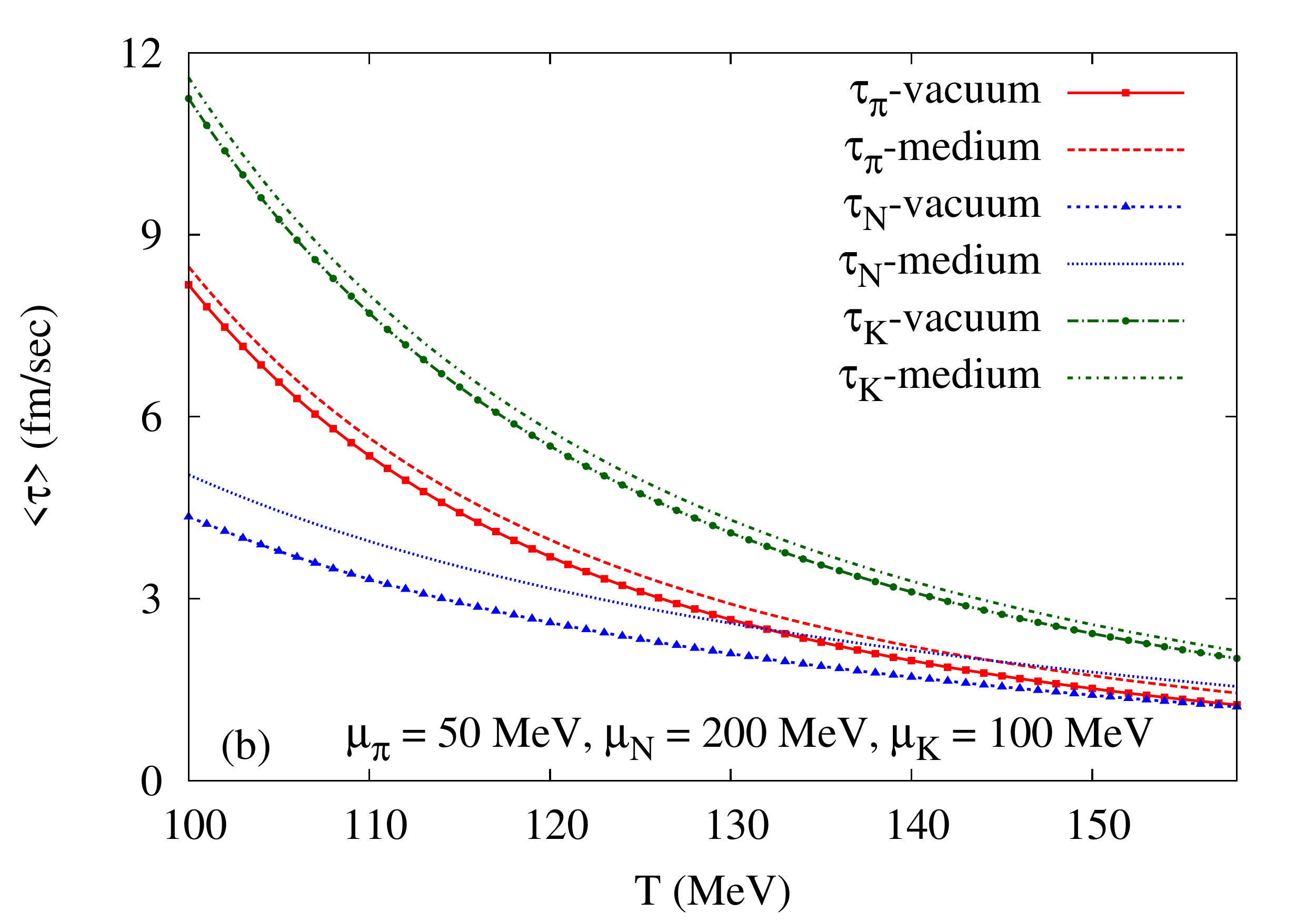} 	
		\includegraphics[angle=-90, scale=0.162]{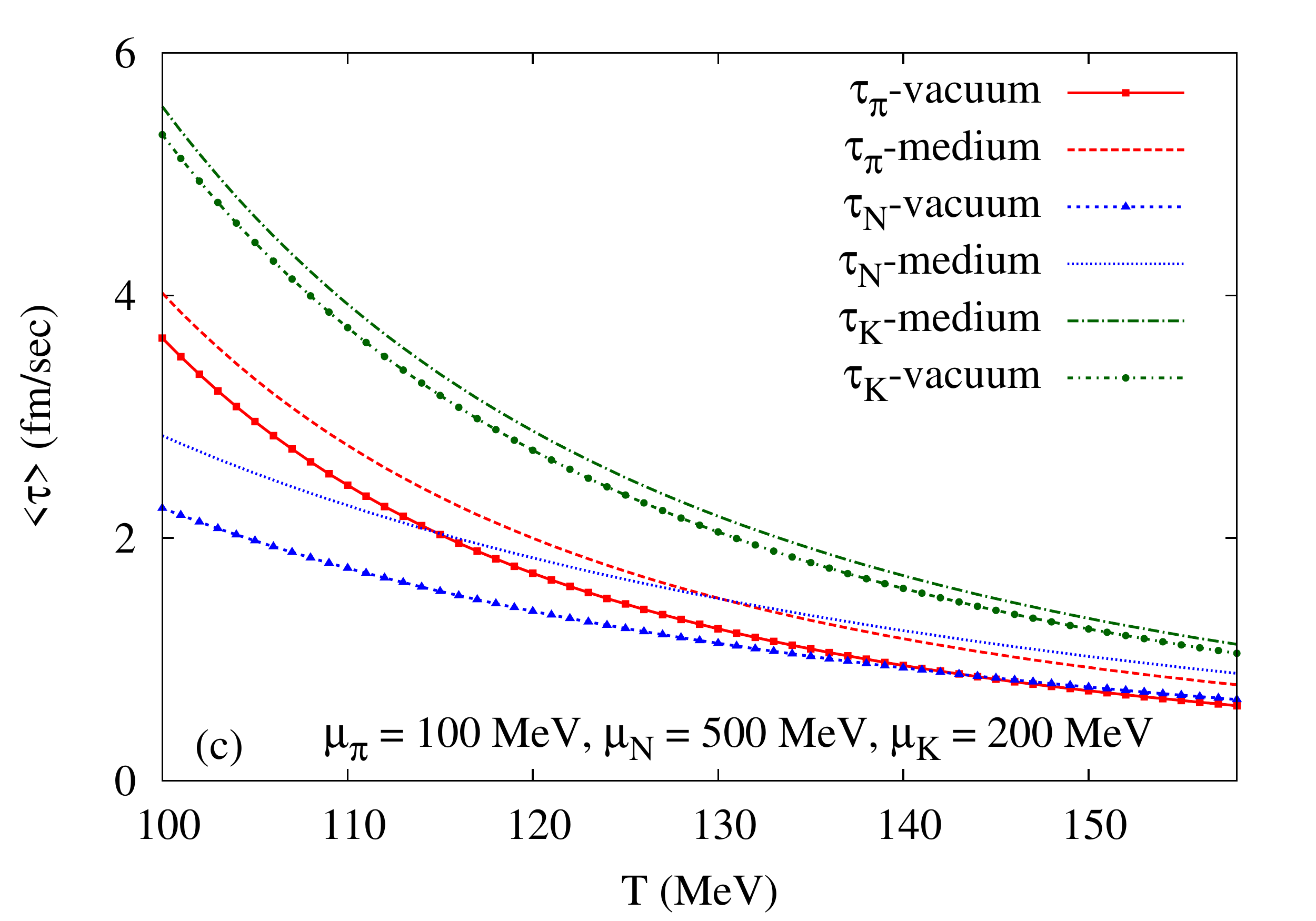}
		\includegraphics[angle=-90, scale=0.162]{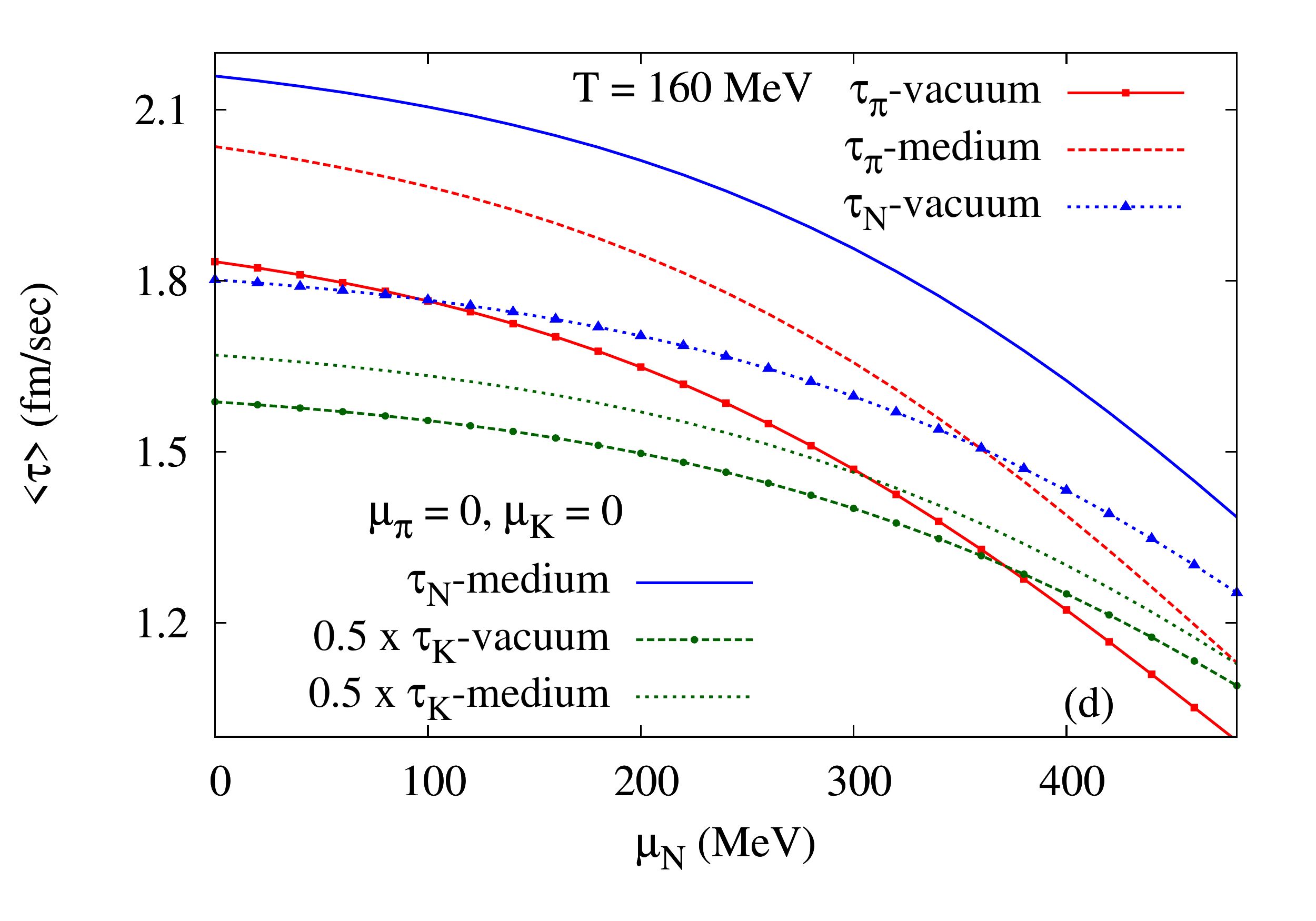}  
		\includegraphics[angle=-90, scale=0.162]{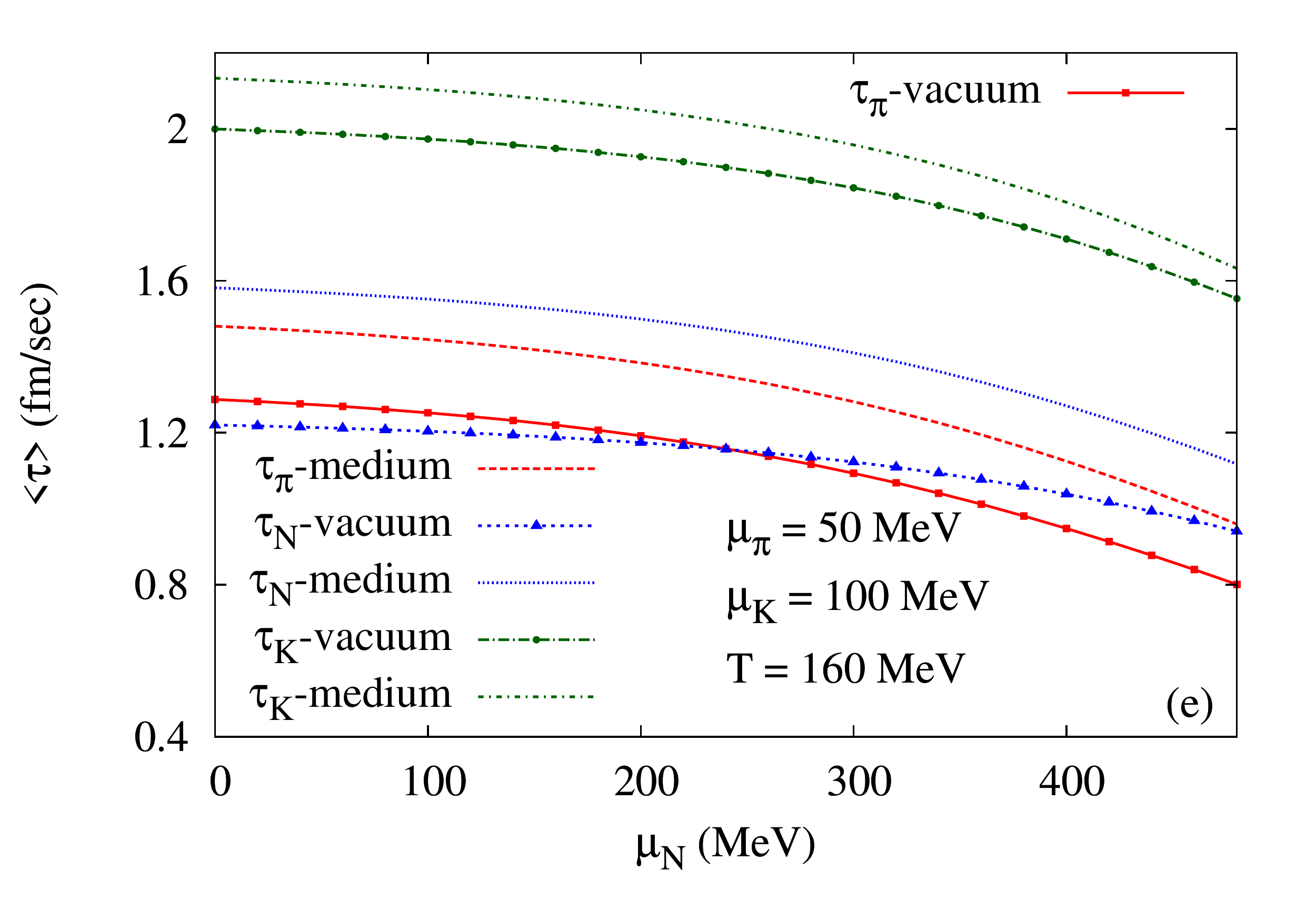} 	
		\includegraphics[angle=-90, scale=0.162]{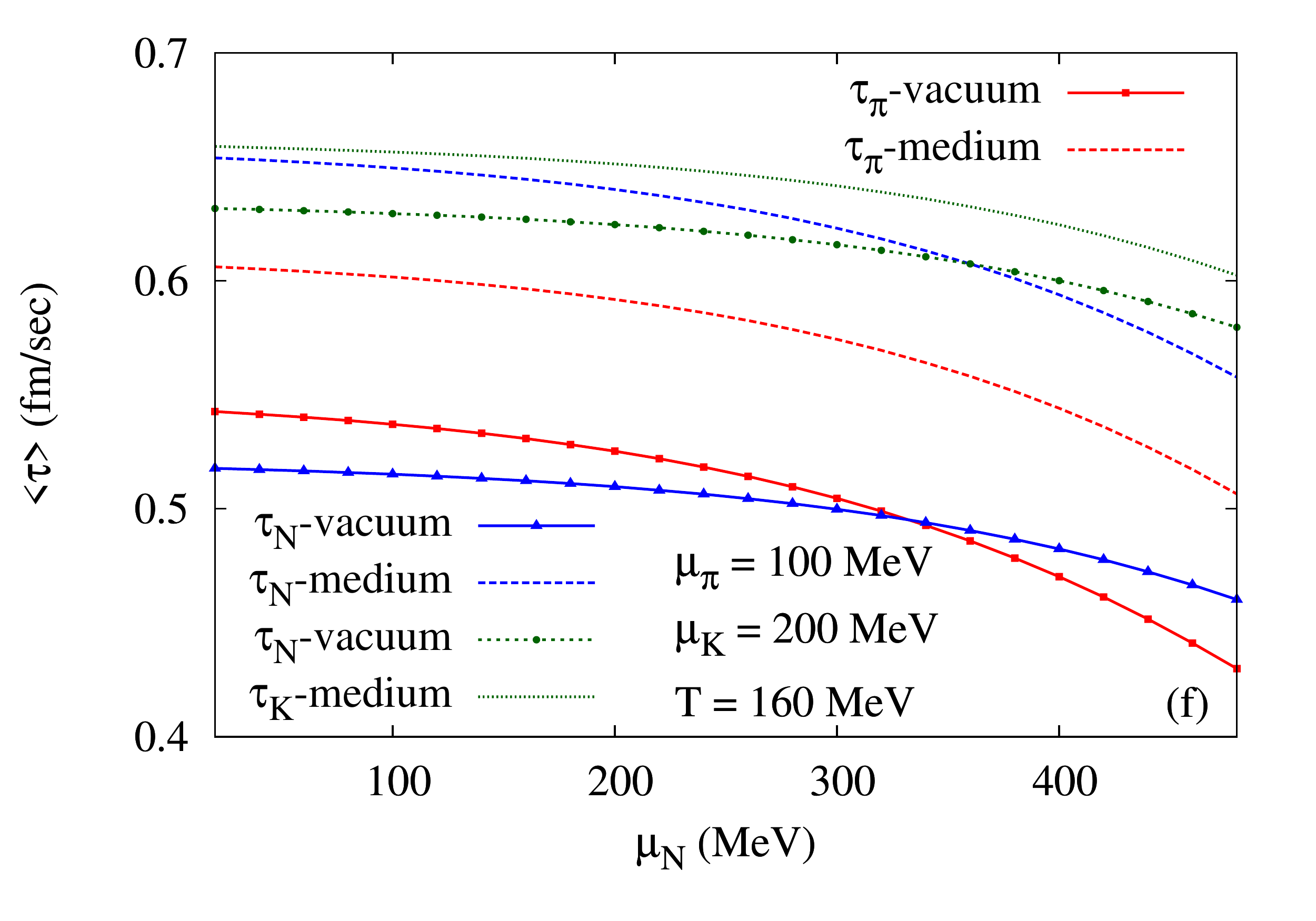}
	\end{center}
	\caption{Momentum averaged relaxation time of pions, nucleons and kaons in a pion-nucleon-kaon hadronic gas as function of temperature for (a) Set 1, (b) Set 2 and (c) Set 3 of chemical potentials of individual components and as function of baryonic density at $ T=160 $ MeV for (d) $ \mu_\pi = 0,~\mu_K=0 $ (e) $ \mu_\pi = 50~{\rm MeV} ,~\mu_K=100~{\rm MeV} $ (f) $ \mu_\pi = 100~{\rm MeV},~\mu_K= 200~{\rm MeV} $  }
	\label{Fig_tau_vs_T}
\end{figure}
%\fi
Fig.(\ref{Fig_tau_vs_T}) shows the average relaxation times of $\pi$, K and N in $\pi$KN system as a function of temperature and baryon chemical potential. With the increase in temperature number density of the system increases. As relaxation time is inversely related to number density of the system hence relaxation time decreases with increase in temperature. Also with the increase of temperature cross section (in-medium cross section) gets suppressed as shown in Fig.~(\ref{cross_sec}) hence the magnitude of relaxation time increases. When baryonic density increases the number of particle available for collision increases hence the system relaxes faster. This explains the decreasing relaxation time with increase in baryonic chemical potential.
\vspace{-0.0cm}
% % % % % % % % % % % % % % % % % % % % % % % % % % % %
\subsection{Thermal conductivity}
\vspace{-.0cm}

%\iffalse
	\begin{figure}[!ht]
		\begin{center}
			\includegraphics[angle=-90, scale=0.162]{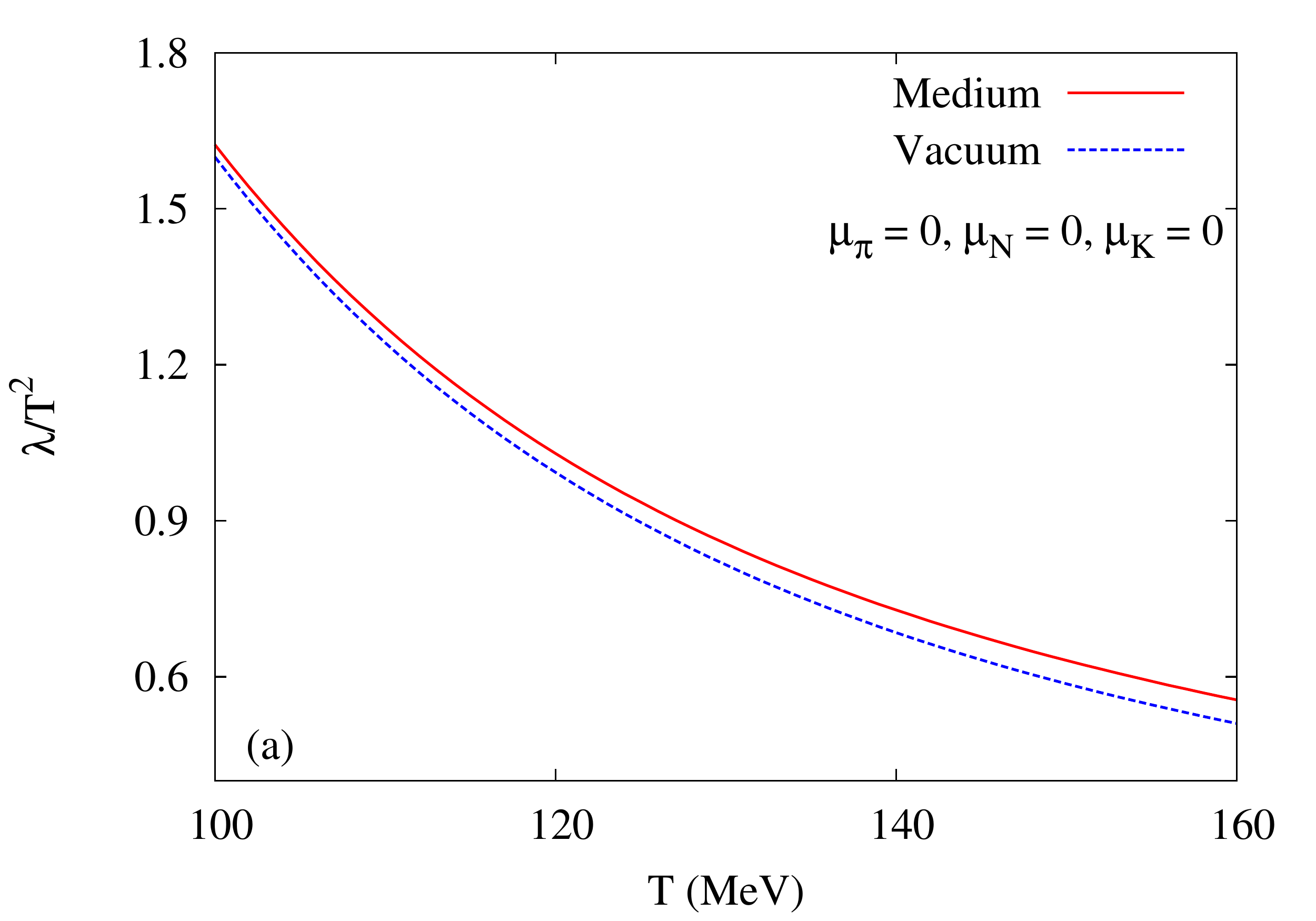}  
			\includegraphics[angle=-90, scale=0.162]{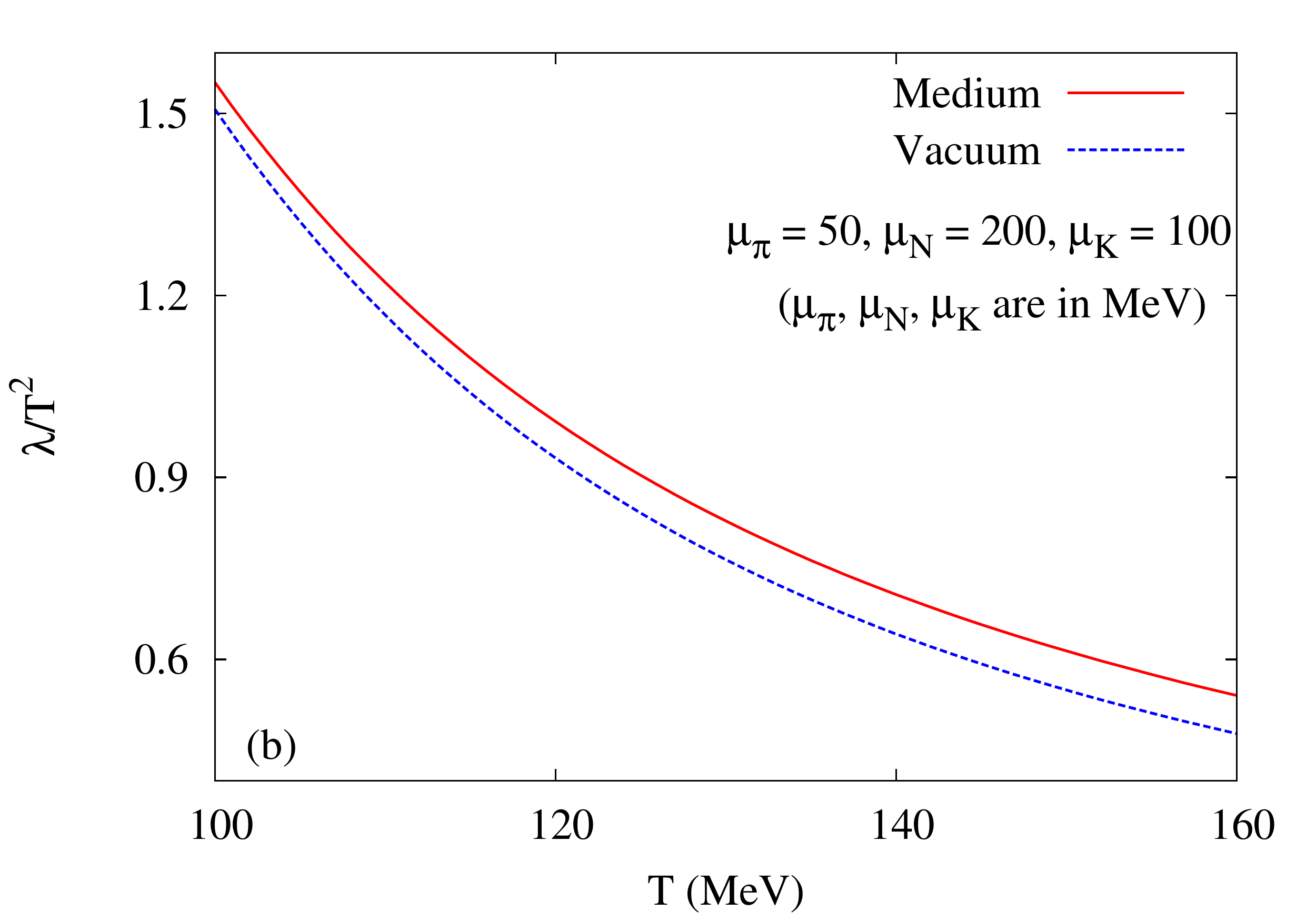} 	
			\includegraphics[angle=-90, scale=0.162]{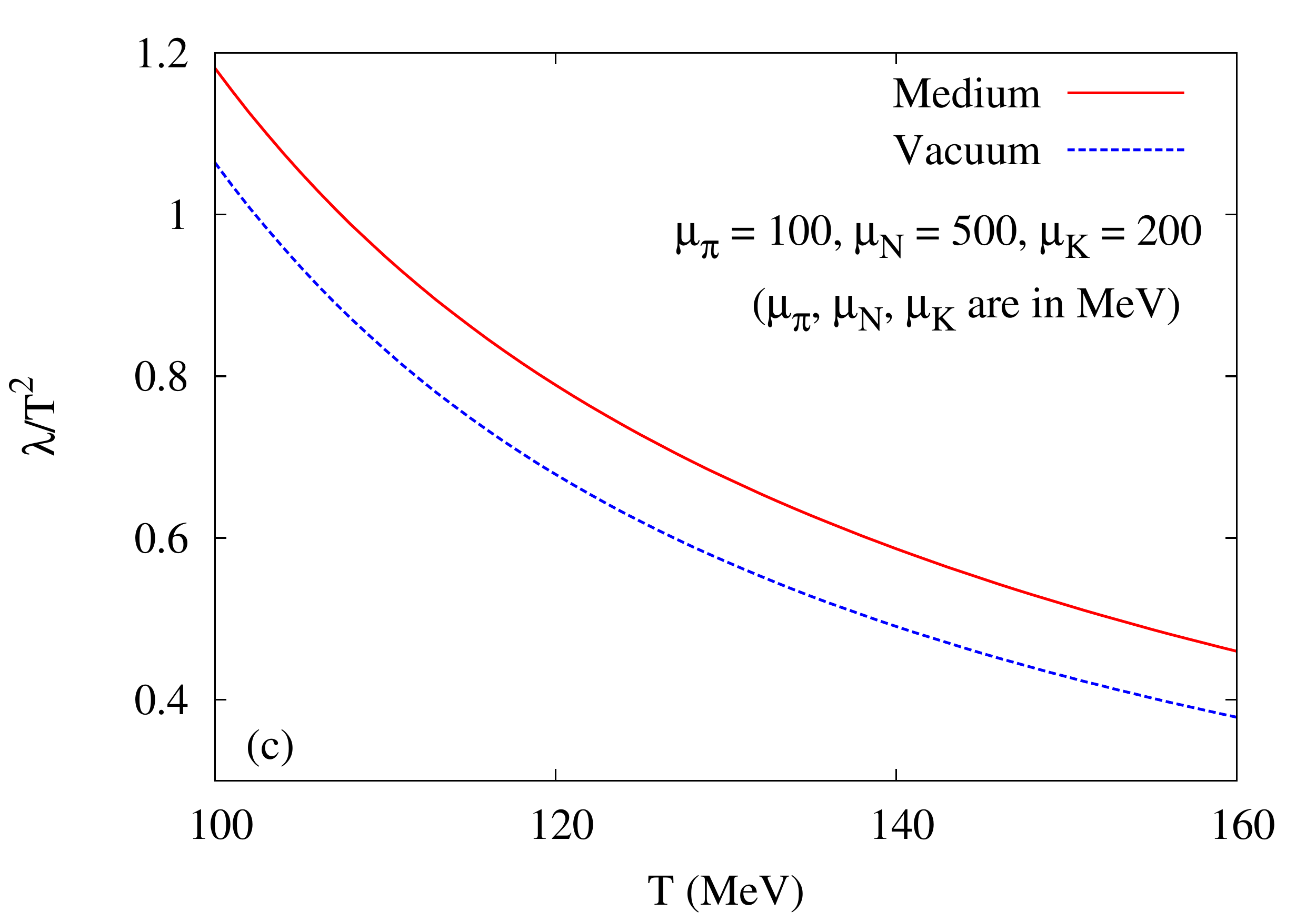}
		\end{center}
		\caption{$\lambda/T^{2}$ as a function of temperature for different set of chemical potential of individual components. (a) Set 1, (b) Set 2 and (c) Set 3.   }
		\label{Fig_lambda_vs_T}
	\end{figure}
%\fi

Plot of $\lambda/T^2 $ as a function of temperature is shown for different sets of chemical potential in Fig.(\ref{Fig_lambda_vs_T}). The figure shows a decrease in its magnitude with the increase of temperature. The decrease in relaxation time with increase of temperature causes $\lambda/T^2$ to decrease with temperature. The medium effects increases the magnitude of $\lambda/T^2$. The increasing chemical potential causes $\lambda/T^2$ to decrease which is because of the increase in relaxation time brought down by the increase in density of nucleons and kaons.
% % % % % % % % % % % % % % % % % % % % % % % % % % % % % % % % %
%\iffalse 
 \begin{figure}[!ht]
	\begin{center}
		\includegraphics[angle=-90, scale=0.162]{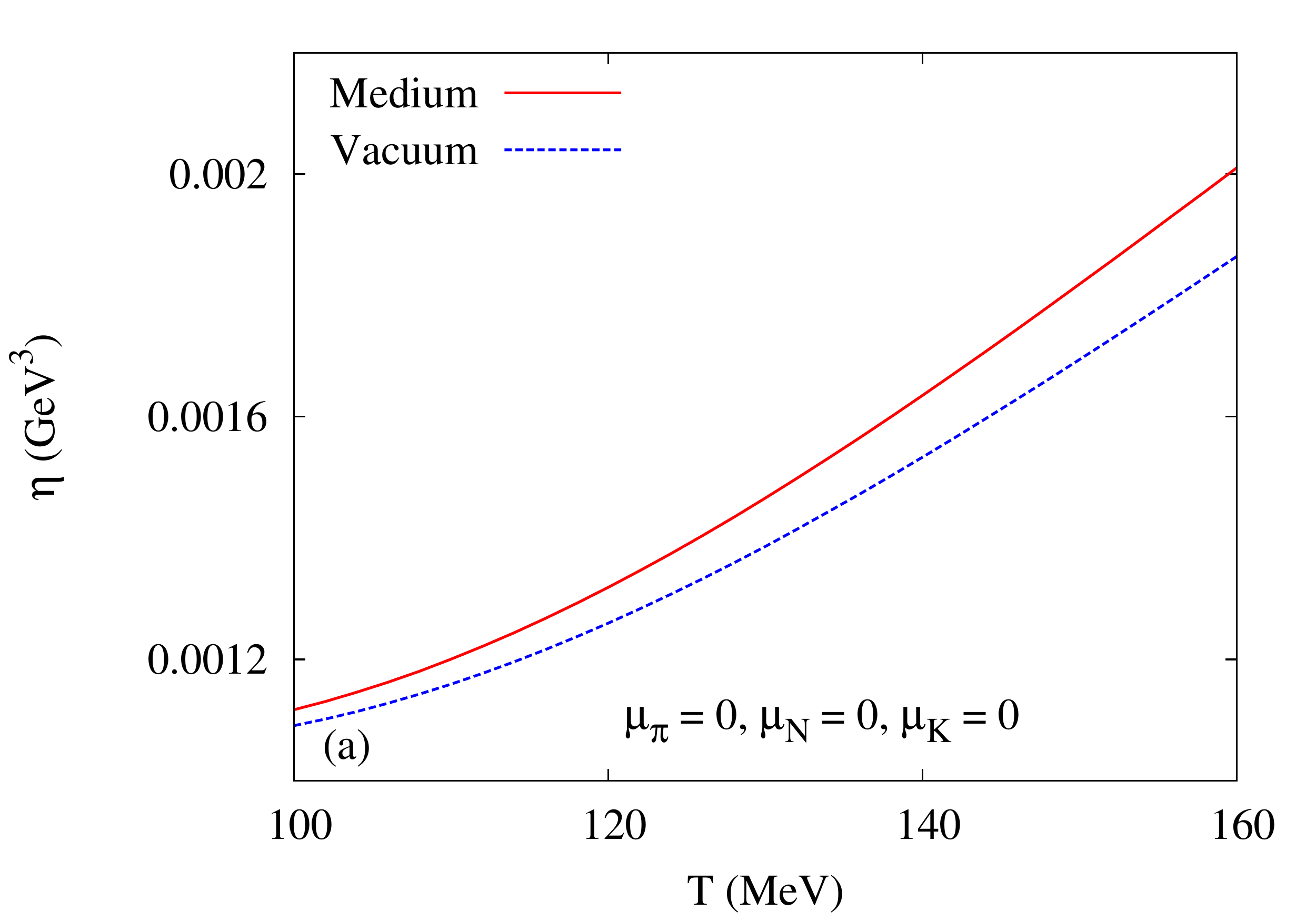}  
		\includegraphics[angle=-90, scale=0.162]{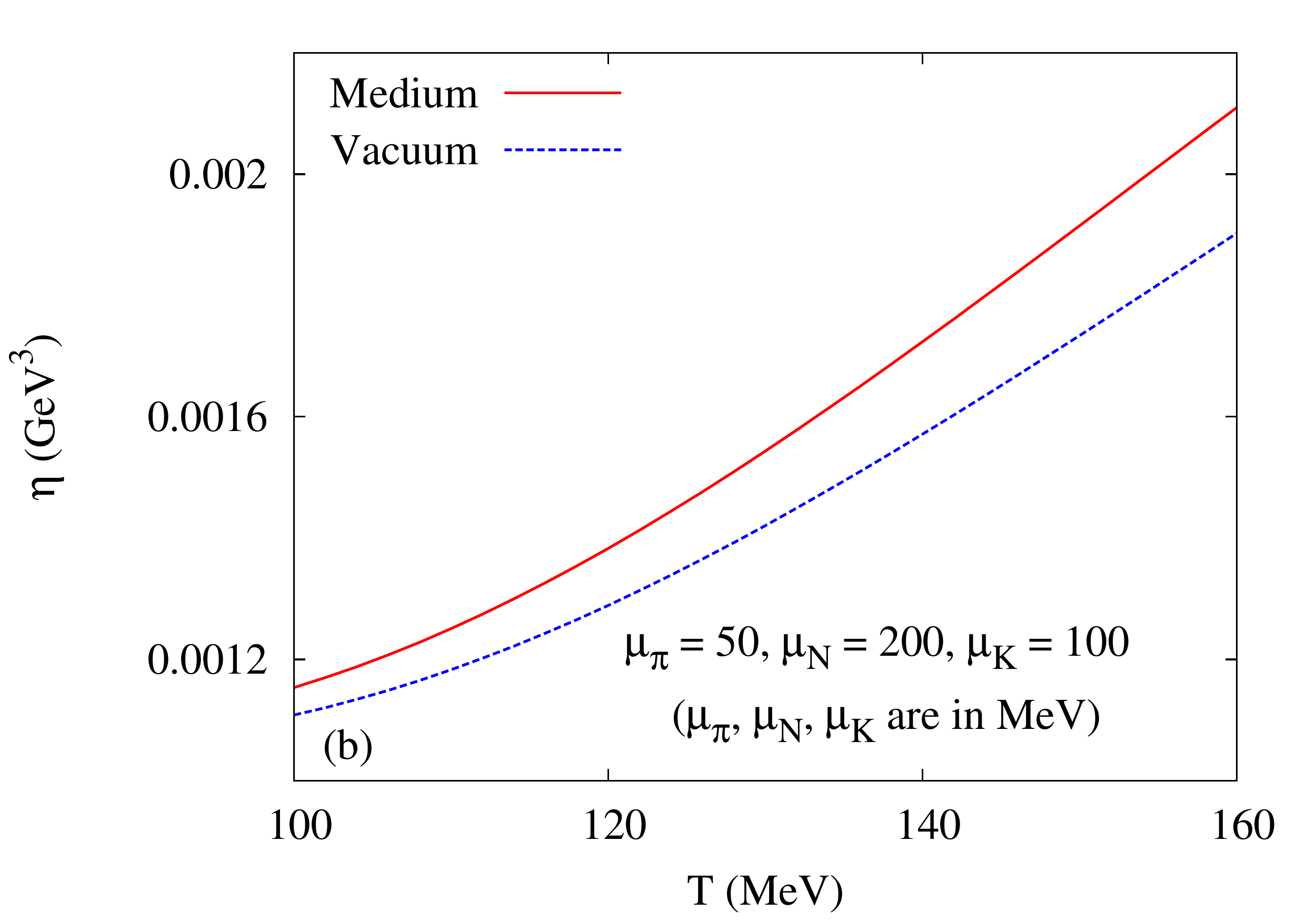} 	
		\includegraphics[angle=-90, scale=0.162]{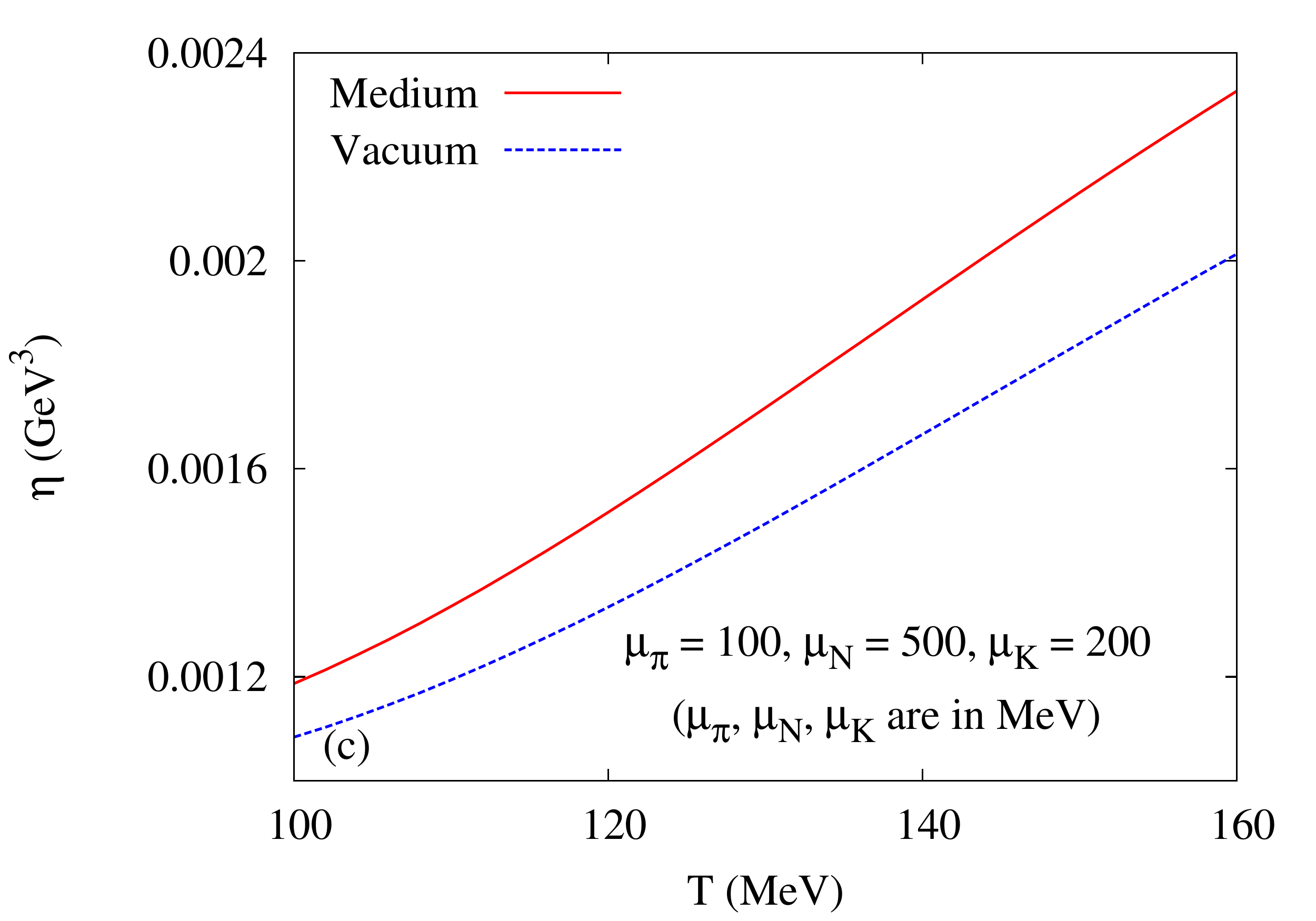}\\
		\includegraphics[angle=-90, scale=0.162]{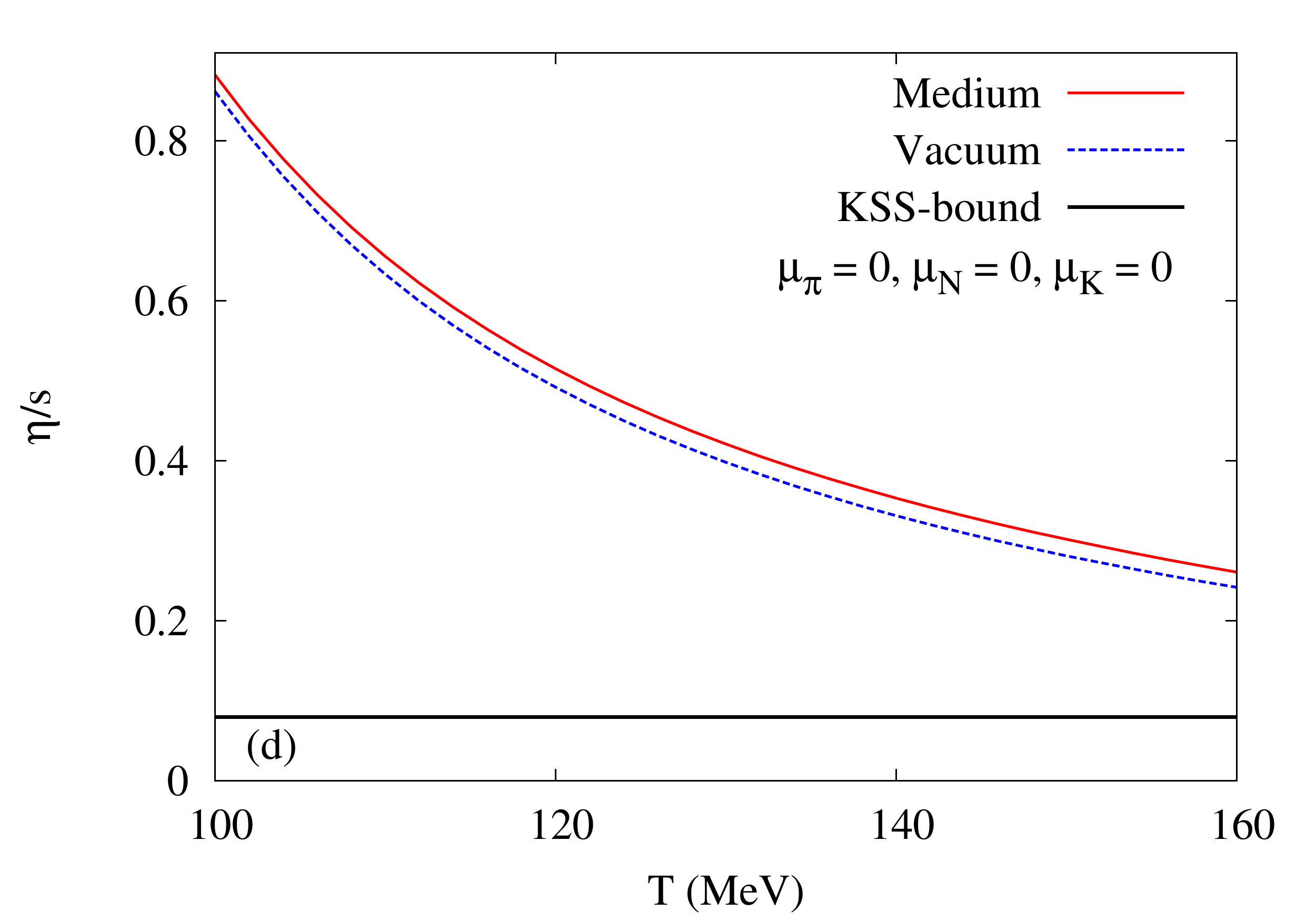}  
		\includegraphics[angle=-90, scale=0.162]{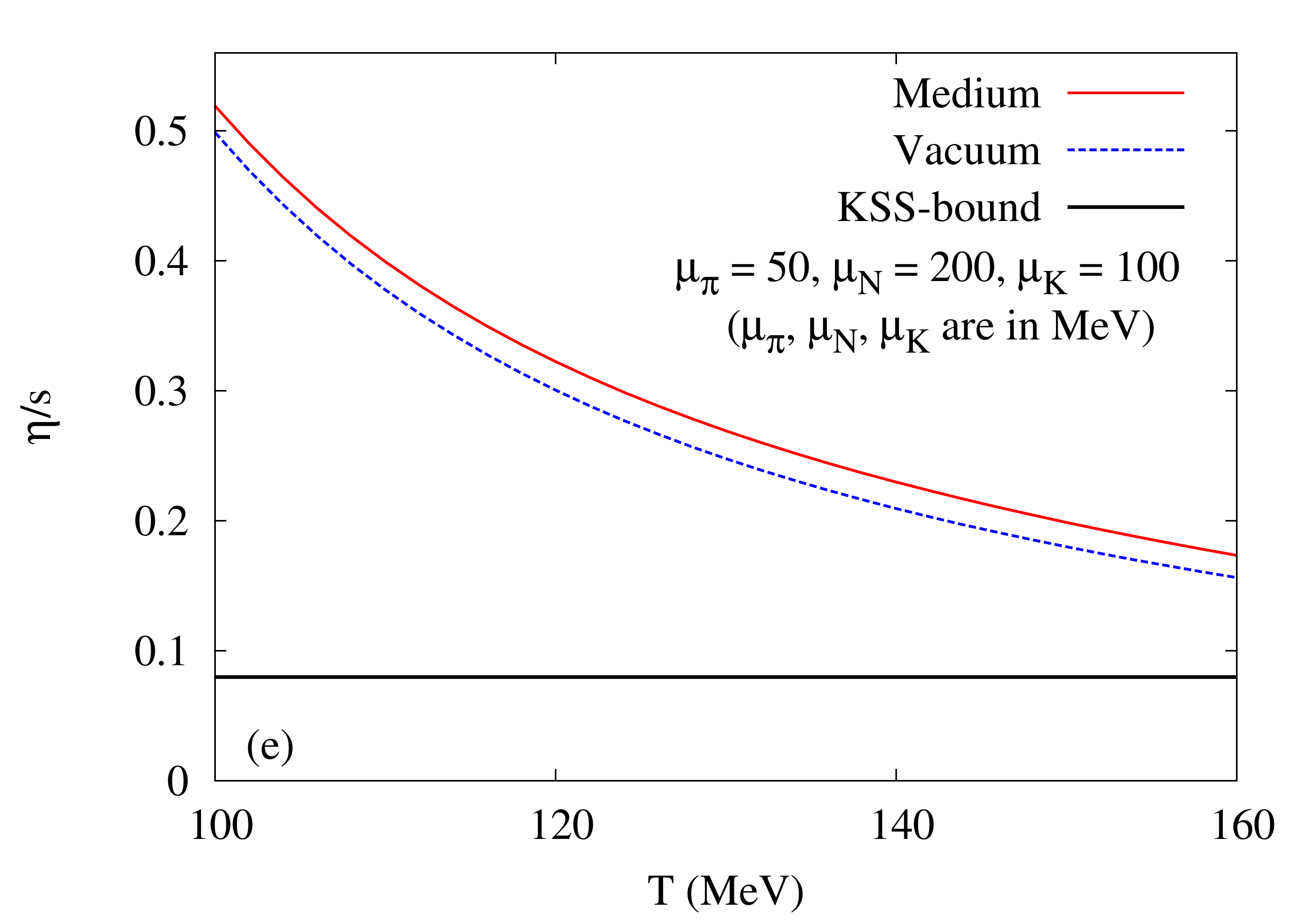} 	
		\includegraphics[angle=-90, scale=0.162]{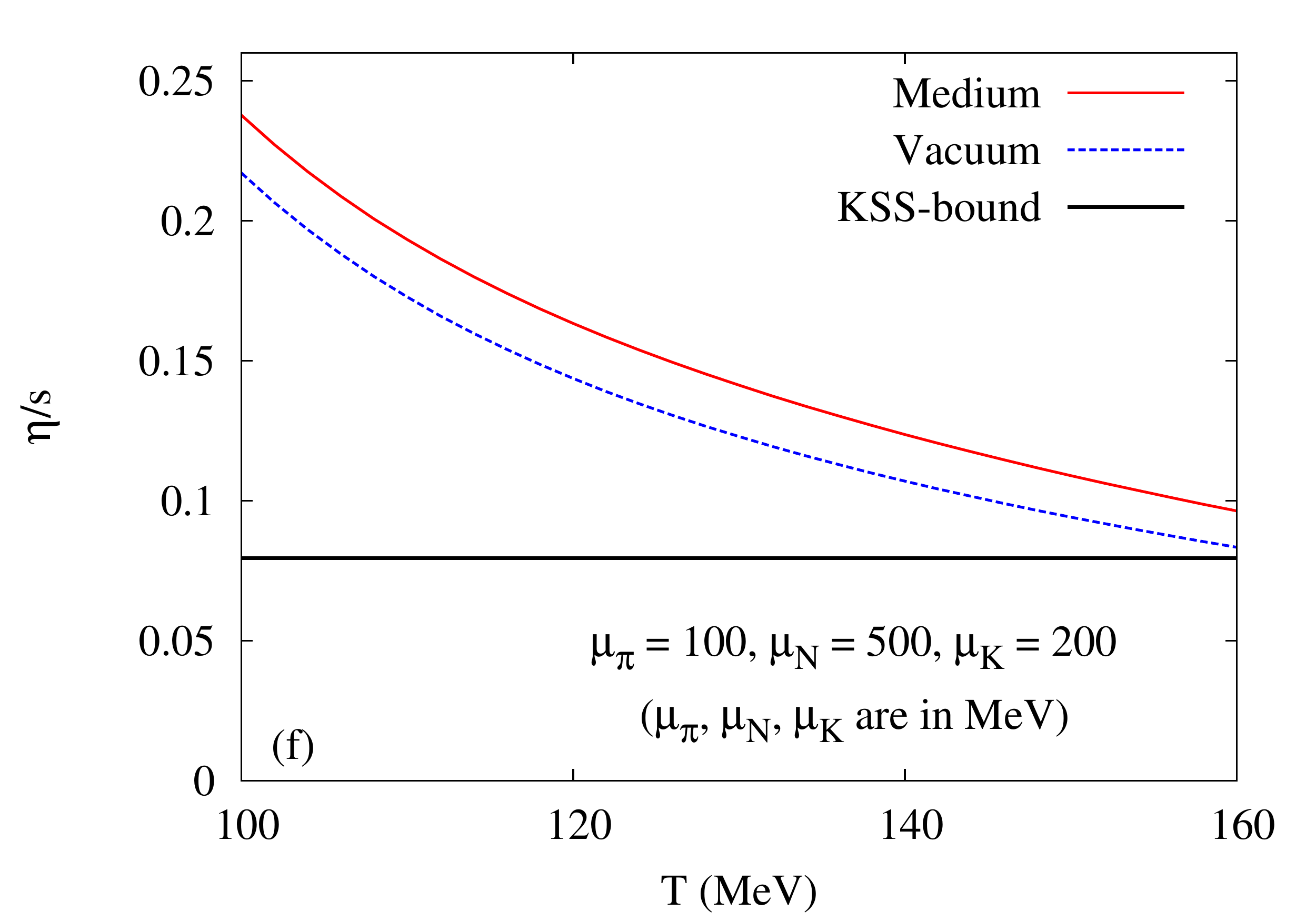}
	\end{center}
	\caption{Shear viscosity($ \eta $) and  Specific shear viscosity($ \eta/s $)  vs Temperature (T)for a pion-kaon-nucleon hadronic gas for different set of chemical potential of individual components with and without including medium effects. }
	\label{Fig_eta_vs_T}
\end{figure}
%\fi

Variation of $\eta$ and $\eta/s$ with temperature is shown in Fig.(\ref{Fig_eta_vs_T}). Plots (a),(b) and (c)
 shows increase in $\eta$ with increase in temperature which is due to the increase in density. $\eta/s$ decreases with increase in temperature due to the increase in entropy density with increase in temperature. Entropy density increases with increase in chemical potential thus decreasing $\eta/s$. Here $\eta/s$ respects the KSS bound. Due to the medium effects both $\eta$ and $\eta/s$ increases in magnitude.

% % % % % % % % % % % % % % % % % % % % % % % %

%\iffalse

%\begin{widetext}
\begin{figure}[!ht]
	\begin{center}
		\includegraphics[angle=-90, scale=0.162]{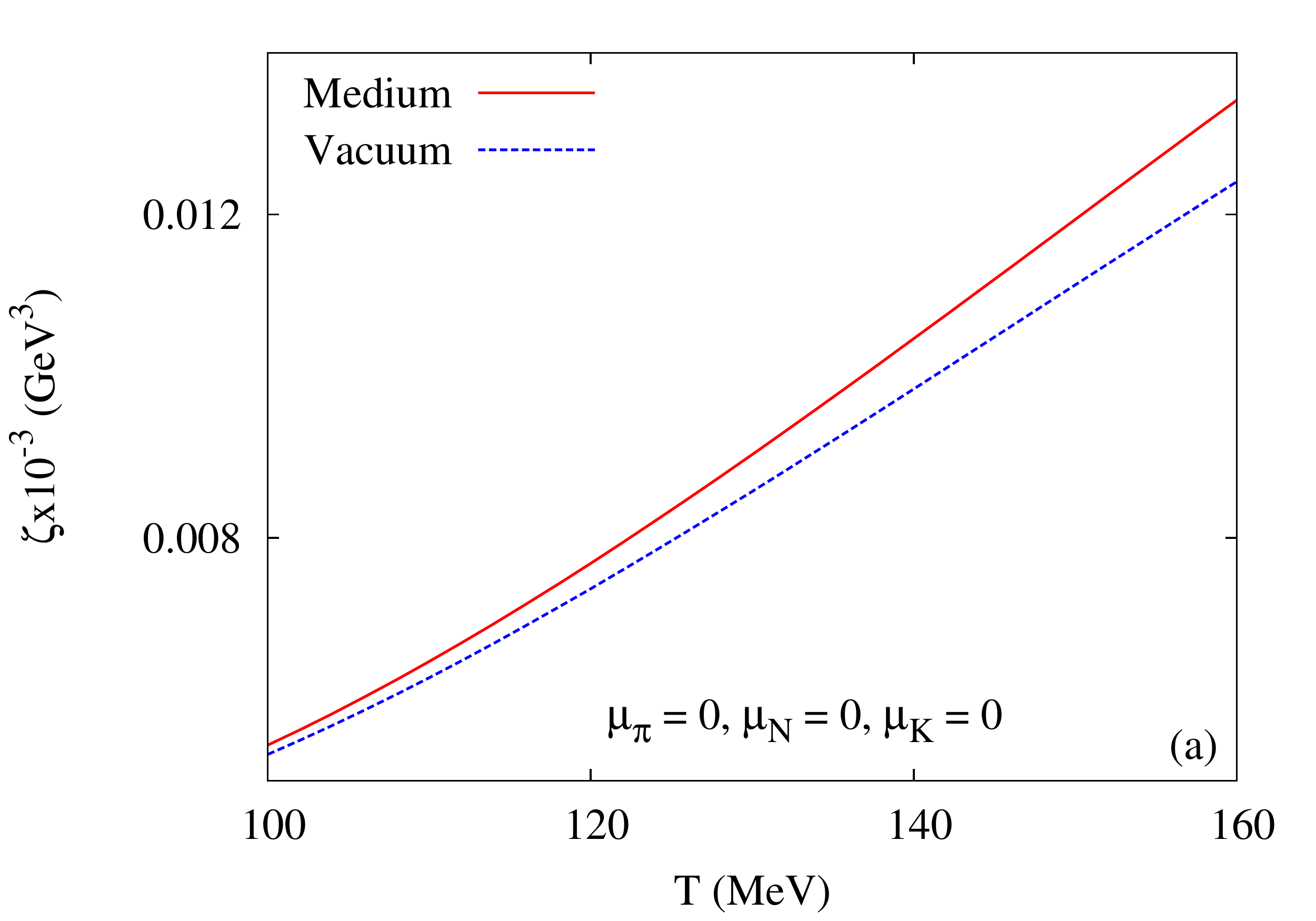}  
		\includegraphics[angle=-90, scale=0.162]{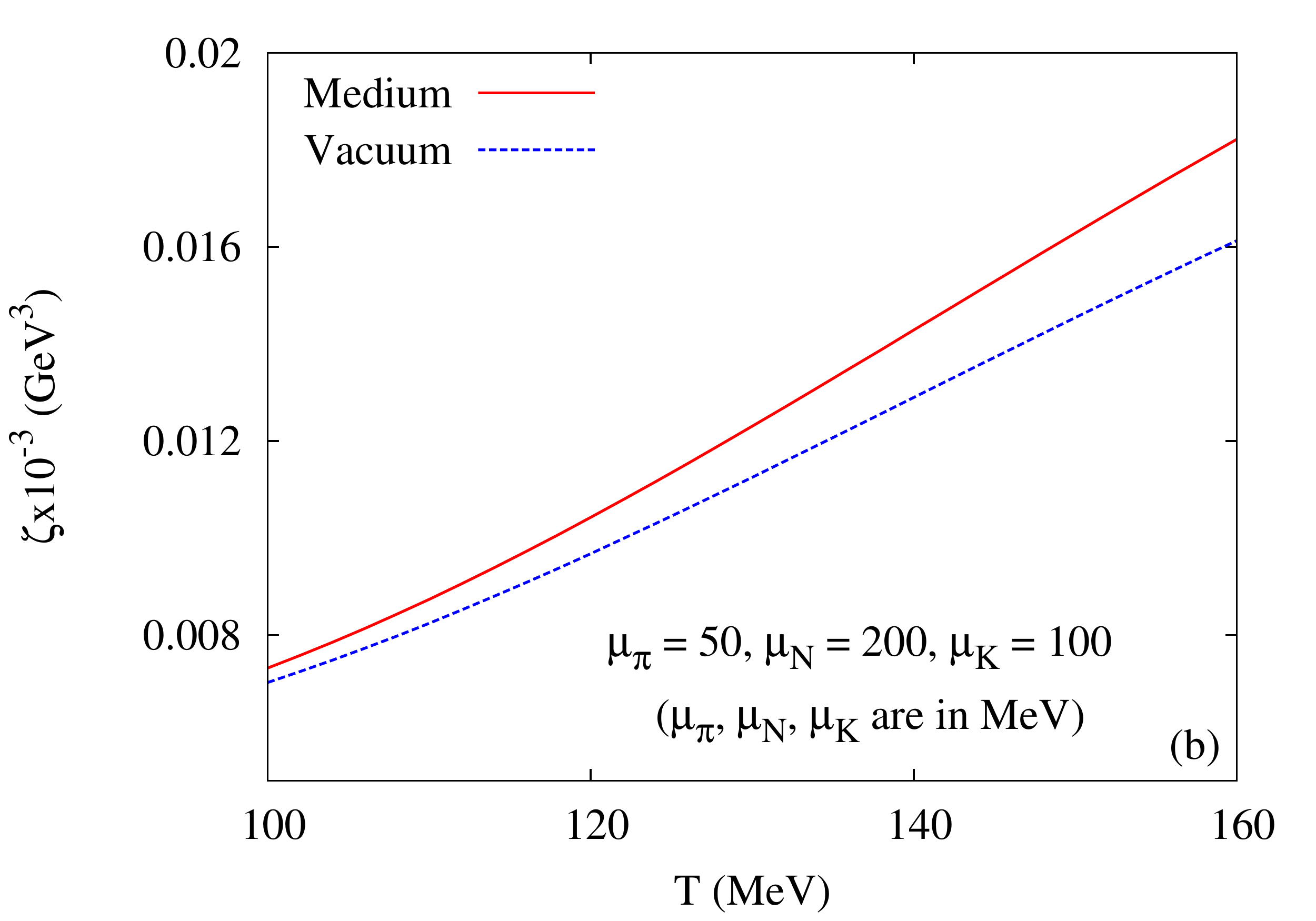} 	
		\includegraphics[angle=-90, scale=0.162]{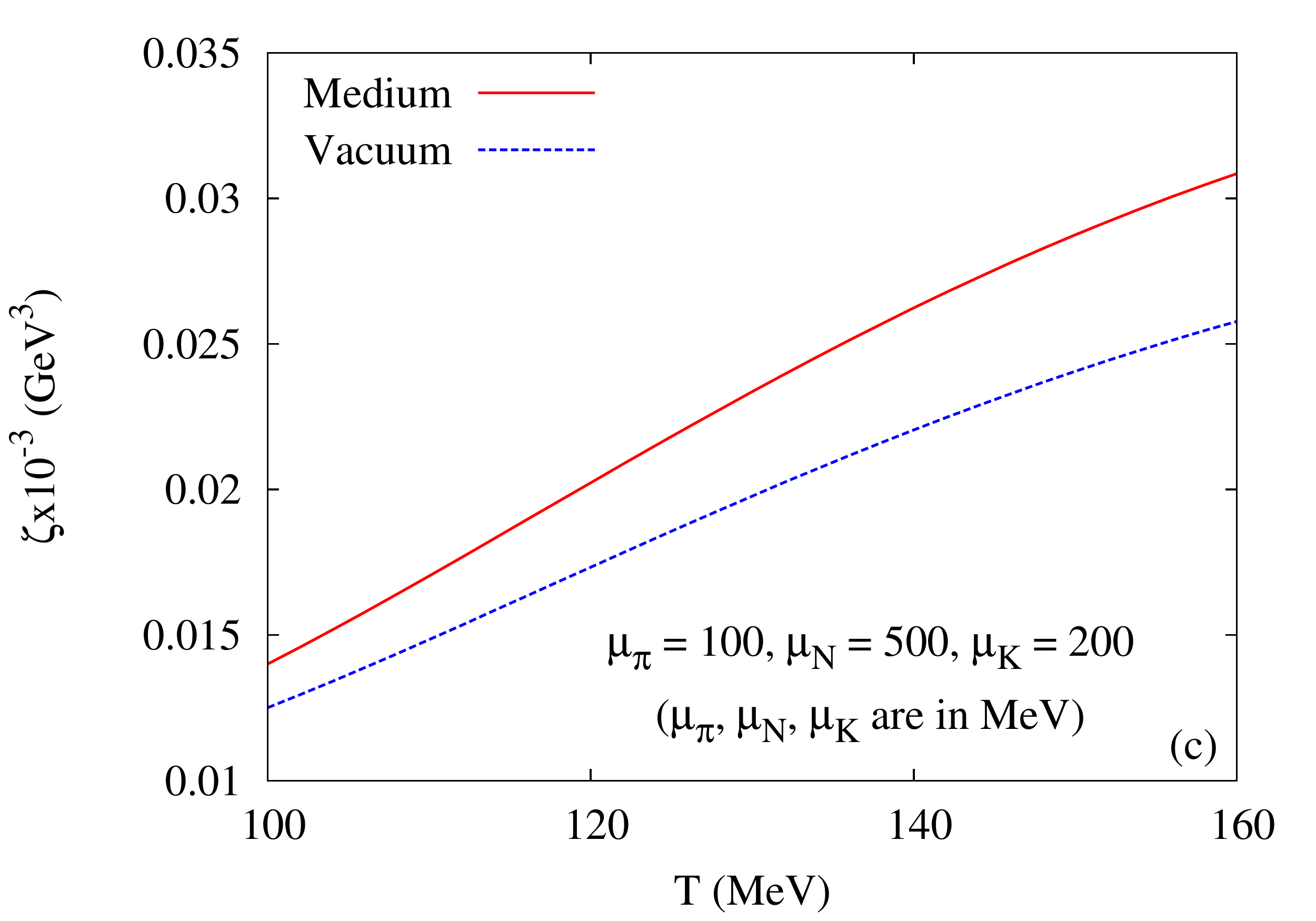}
		\\
		\includegraphics[angle=-90, scale=0.162]{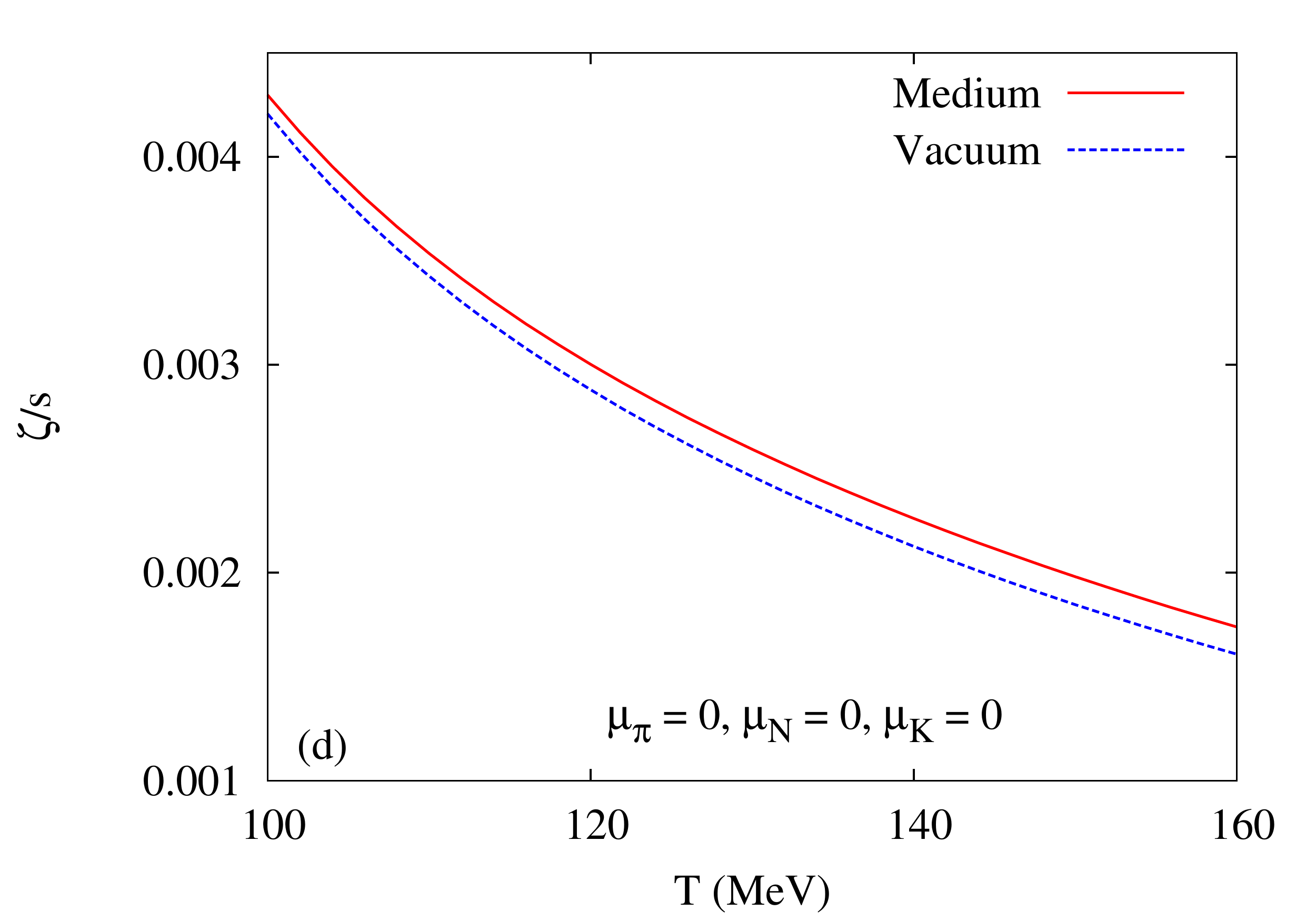}  
		\includegraphics[angle=-90, scale=0.162]{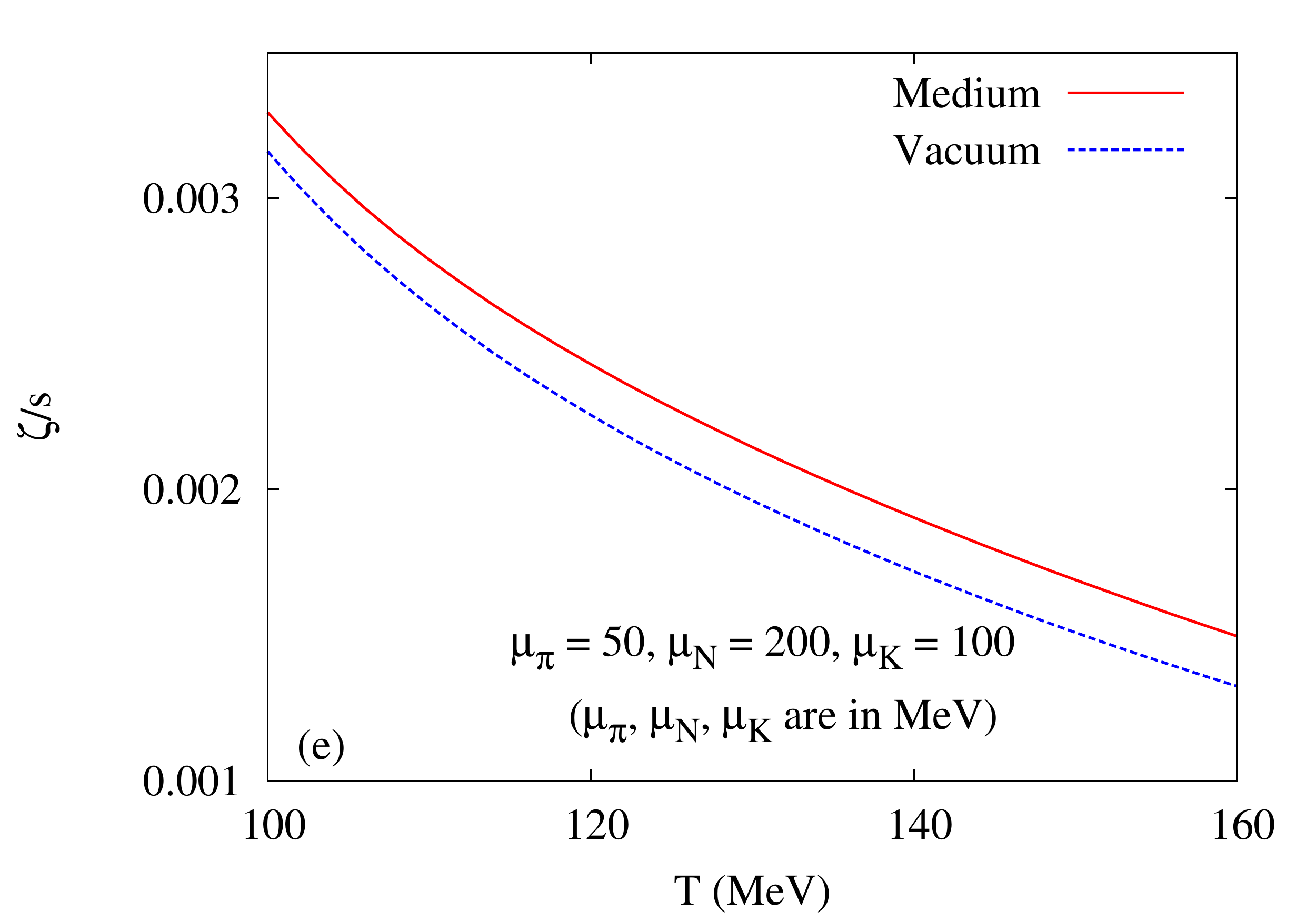} 	
		\includegraphics[angle=-90, scale=0.162]{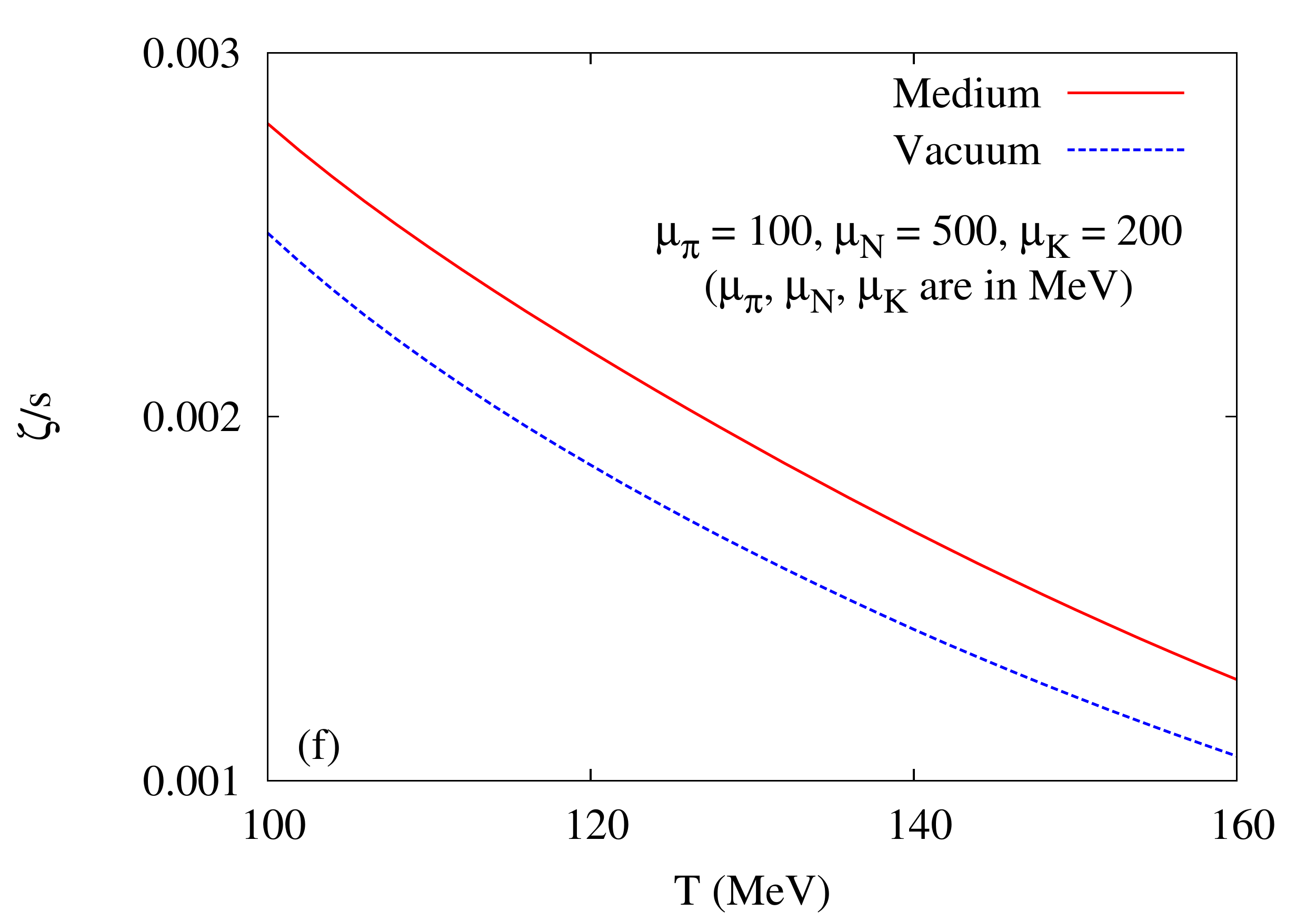}
	\end{center}
	\caption{Bulk viscosity($ \zeta $) and its entropy ratio ($\zeta/s$) vs temperature ($T$) for different set of chemical potentials with and without including medium effects  }
	\label{Fig_zetabys_vs_T}
\end{figure}
%\end{widetext}
%\fi

Fig.~(\ref{Fig_zetabys_vs_T})) shows plot of $\zeta$ and $\zeta/s$ with temperature. The trend in the plots can be explained in similar lines done for $\eta$ and $\eta/s$. The medium effects are 
visible here as well.

% % % % % % % % % % % % % % % % % % % % % % % % % % % %

%	\iffalse
%\begin{widetext}
 \begin{figure}[!ht]
	\begin{center}
		\includegraphics[angle=0, scale=0.18]{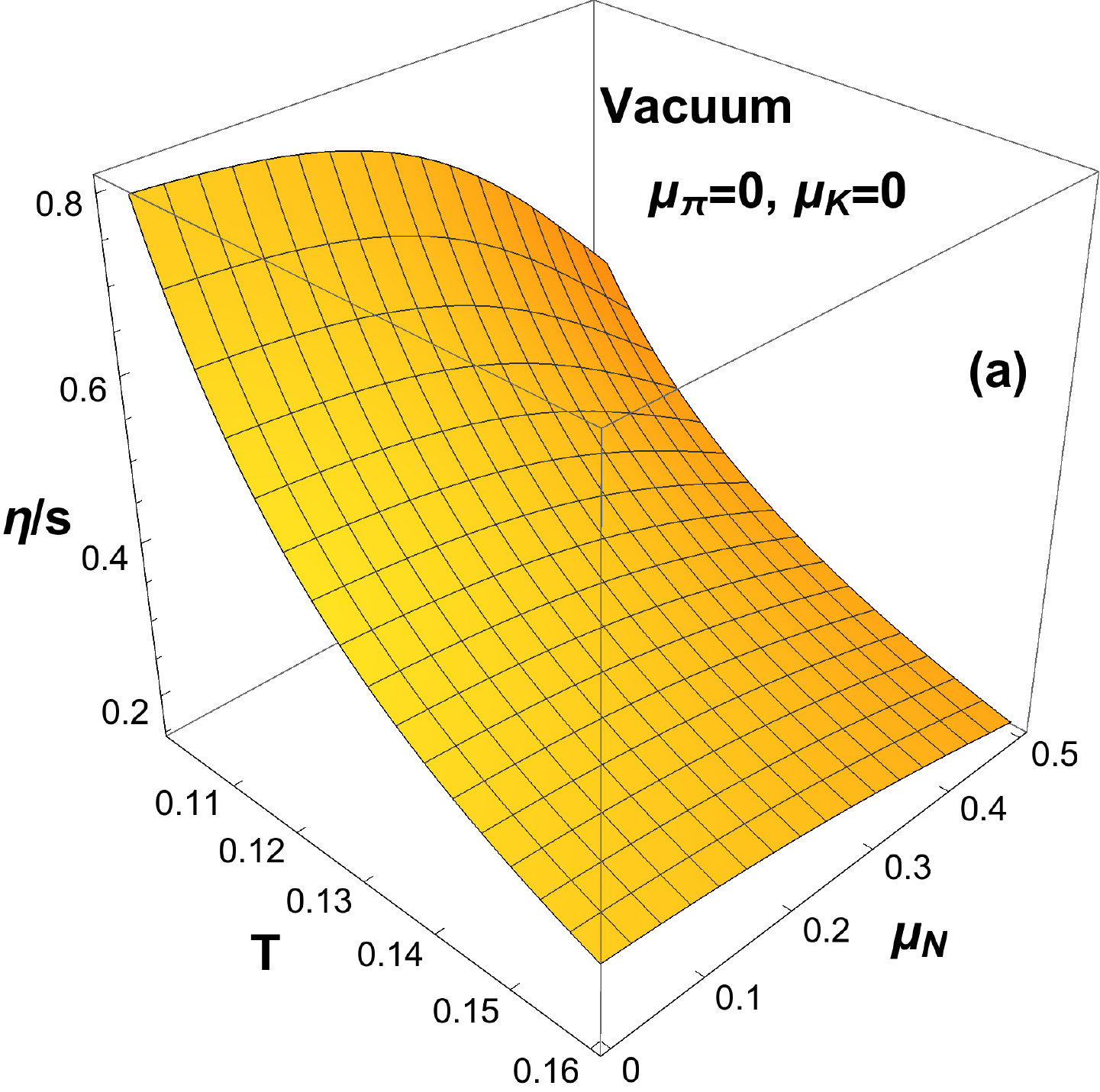}  ~~~~~ \includegraphics[angle=0, scale=0.18]{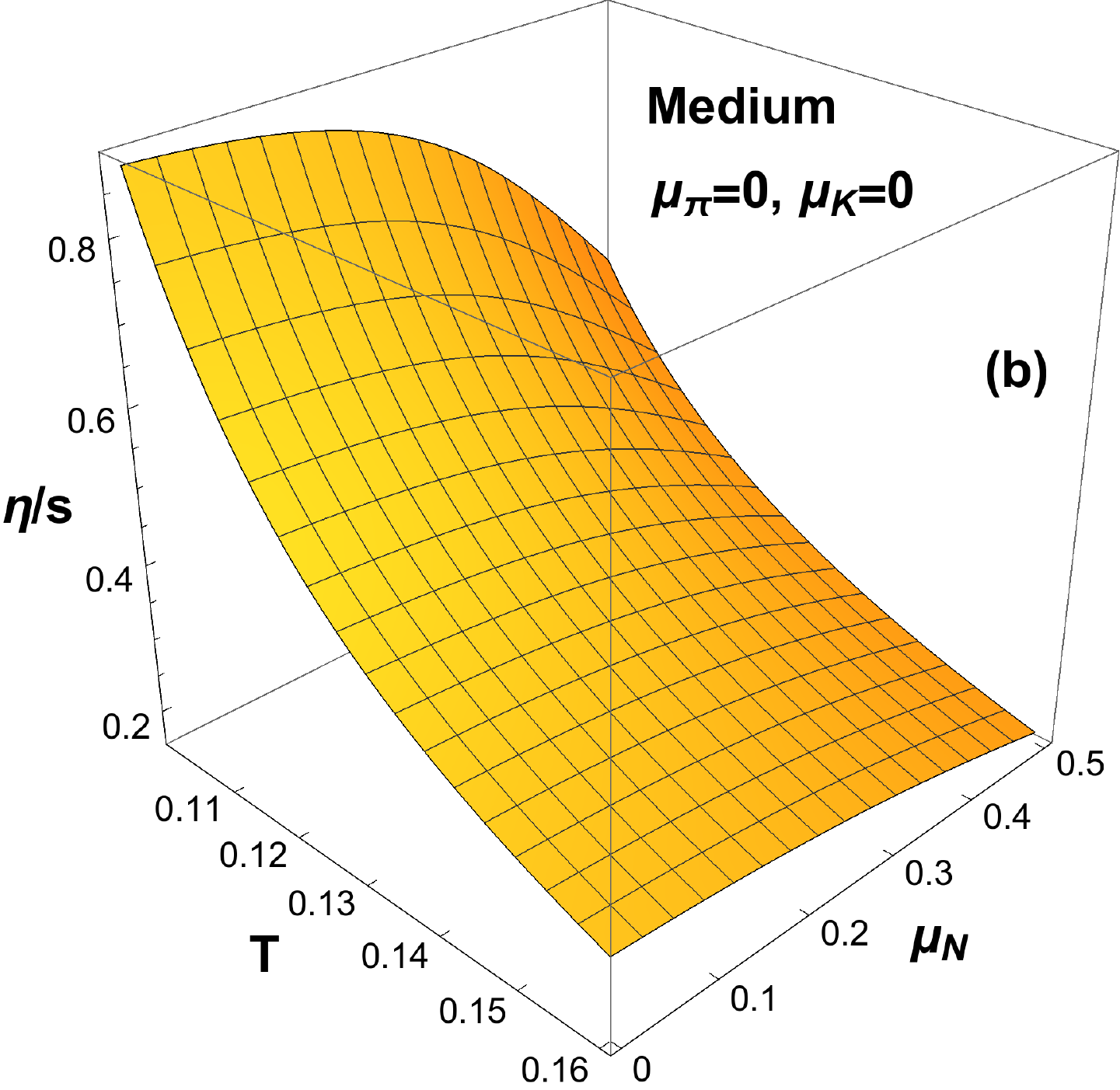}
		\includegraphics[angle=0, scale=0.18]{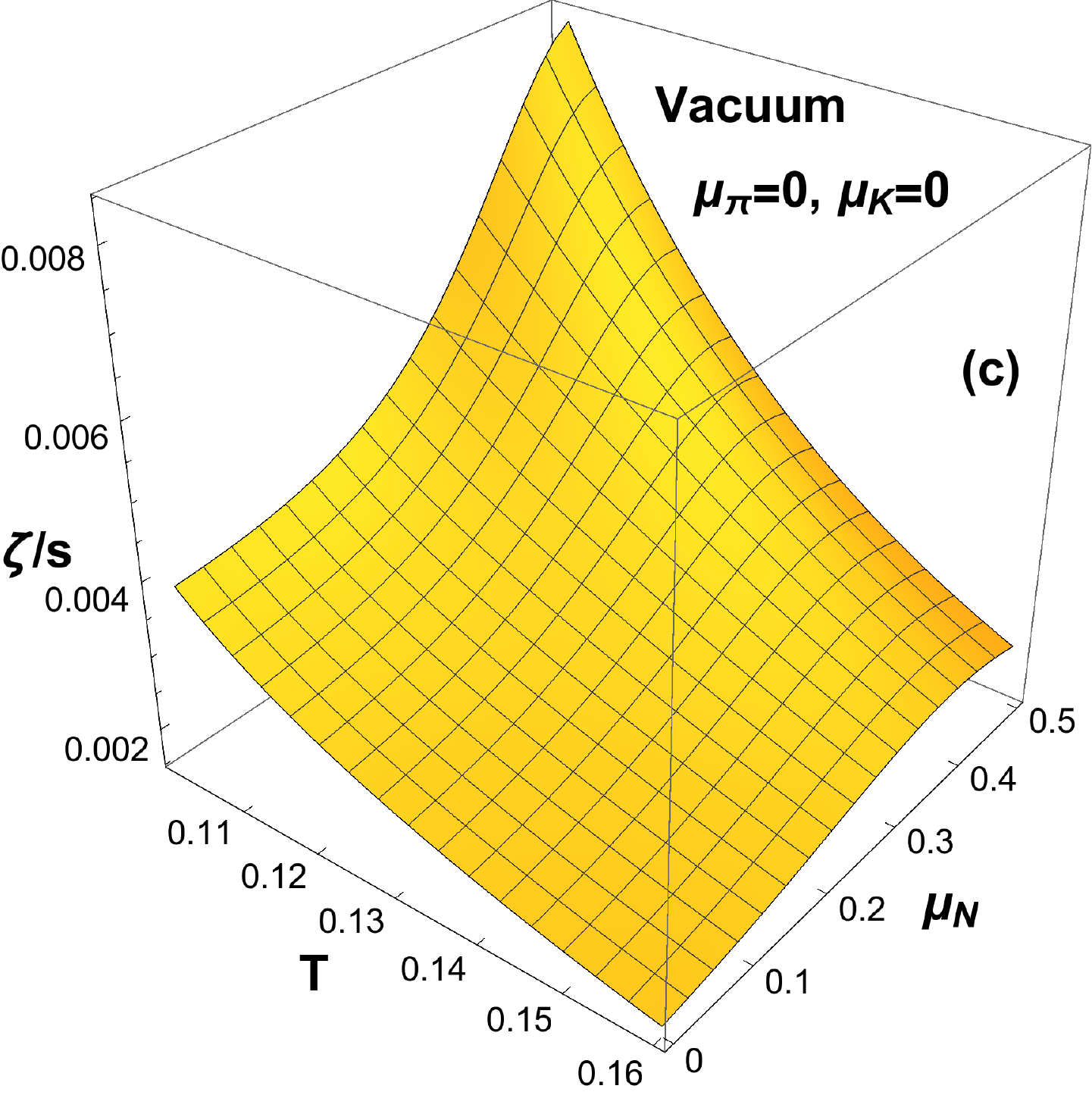} ~~~~~ \includegraphics[angle=0, scale=0.18]{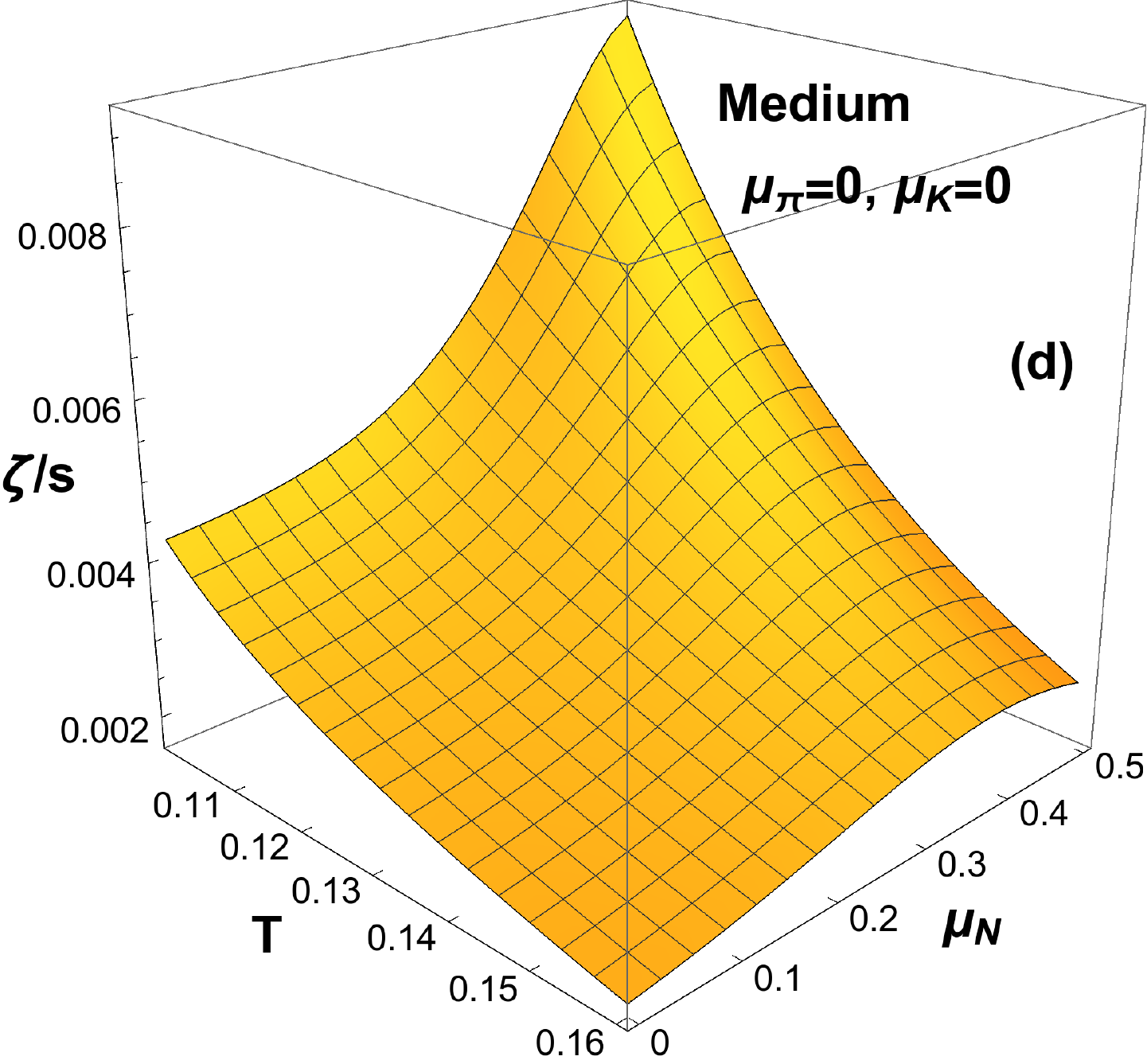}      
	\end{center}
	\caption{ Shear viscosity to entropy density ration ($\eta/s$) as a function of temperature and nucleon chemical potential at 
		$\mu_\pi = \mu_K = 0$ with (a) vacuum and (b) in-medium cross sections. 
		Bulk viscosity to entropy density ration ($\zeta/s$) as a function of temperature and nucleon chemical potential at 
		$\mu_\pi = \mu_K = 0$ with (c) vacuum and (d) in-medium cross sections.  }
	\label{Fig:3D}
\end{figure}
%\fi
%\end{widetext}

Variation of $\eta/s$ and $\zeta/s$ with temperature and baryon chemical potential has been studied using both vacuum and in-medium cross sections for $\mu_\pi=$0 and $\mu_K=$0 in Fig.(\ref{Fig:3D}). It is seen from the figure that $\eta/s$ decreases with increasing $\mu_N$ whereas $\zeta/s$ increases with increase in $\mu_N$.

% % % % % % % % % % % % % % % % % % % % % % % % % % % % % % % % % %

%\iffalse
\begin{figure}[tbh]
	\begin{center}
		\includegraphics[angle=-90, scale=0.195]{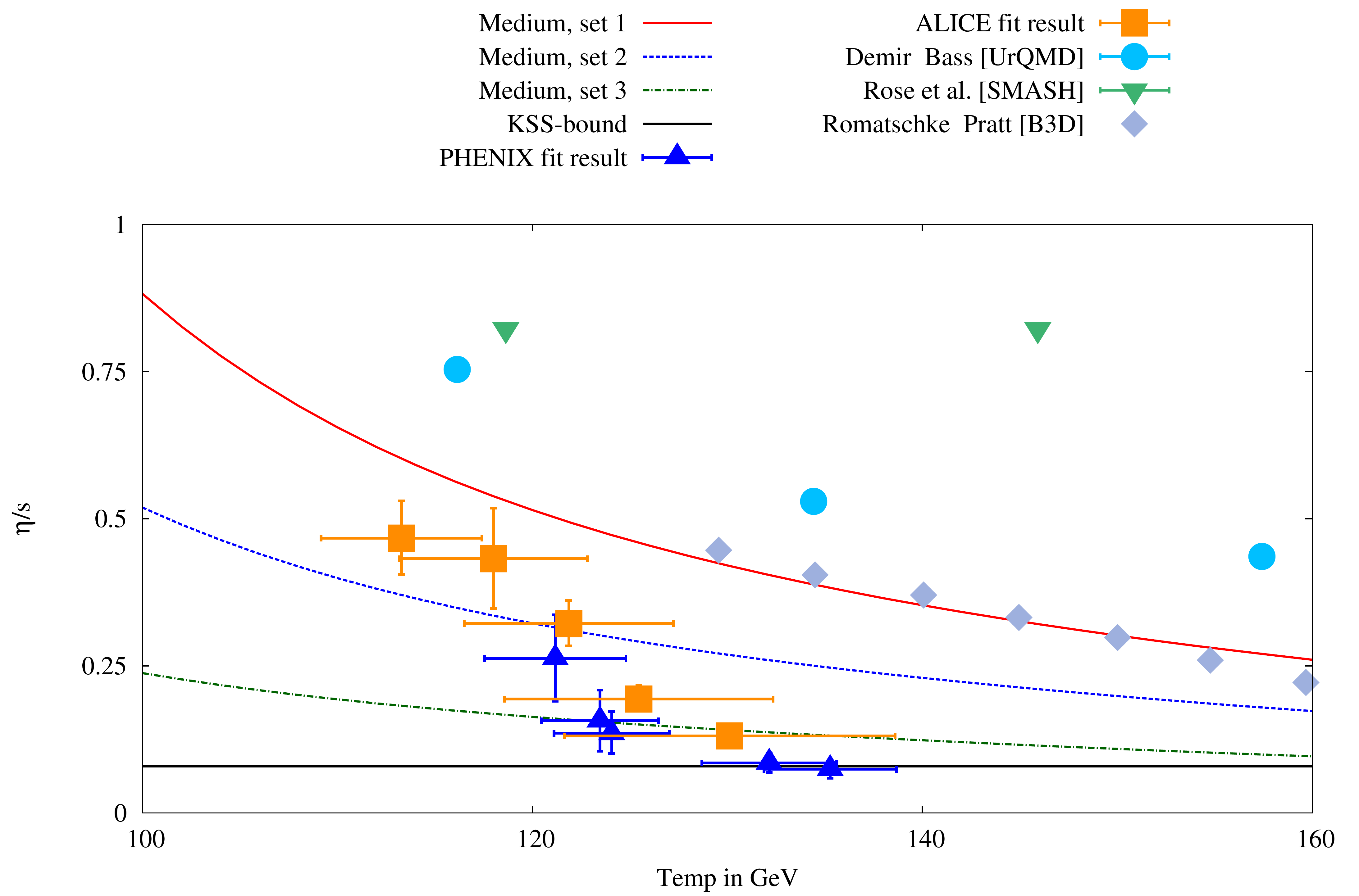}  
	\end{center}
	\caption{The result obtained in this paper compared to various data of the specific shear viscosity $\eta/s$ as a function of temperature
		available in the literature. A line of KSS bound has been drawn as a reference. }
	\label{Fig_etabysexp_vs_T}
\end{figure}
%\fi
Fig.(\ref{Fig_etabysexp_vs_T}) shows the comparison of $\eta/s$ from our work with other datas in literature. From the figure we see that our results are well within the range. We also found that $\eta/s$ calculated for $\mu_\pi=0$, $\mu_k=0$ and $\mu_N=0$ shows a good agreement with the data obtained by~\cite{Romatschke:2014gna}. 
% % % % % % % % % % % % % % % % % % % % % % % % % % %
% % % % % % % % % % % % % % % % % % % % % % % % % % %
\subsection{Summary and Discussions}
\vspace{-0.0cm}
In this work we have considered a hot and dense hadronic gas mixture consisting of pions, kaons and nucleons 
which are the most important components of the system produced during the later stages of heavy ion collisions. We have endeavored  to present a systematic study of the relaxation times, viscous coefficients and thermal conductivity for a  system consisting 
only of pions, a system of pions and kaons and finally for a pion-kaon-nucleon system using the Boltzmann transport equation which has been linearised using the Enskog expansion. The key ingredient is the use of in-medium cross-sections which were obtained using one-loop corrected thermal propagators 
in the matrix elements for $\pi\pi$, $\pi K$ and $\pi N$ scattering. The suppression of the in-medium cross-sections at 
finite temperature and density were reflected in the enhancement of relaxation times. This in turn results in a 
significant modification of the temperature dependence of the viscous coefficients. In particular, 
the value of $\eta/s$ in the medium was found to be in good agreement with those found in the literature. These results may have significant effects on the evolution of the hot/dense hadronic matter produced in the 
later stages of heavy ion collisions.

%% file: Sarthak/sarthak.tex
\section{Higgs propagation in Quark Gluon Plasma}
% Force line breaks with \\

\textit{Sarthak Satapathy, Sabyasachi Ghosh, Santosh K. Das, Ralf Rapp, Nihar R. Sahoo}

\bigskip

{\small
We are studying the properties of Higgs boson in the quark gluon plasma (QGP). 
From the Higgs-quark interaction Lagrangian density, we calculate the Higgs decays into quark and anti-quark, which shows a dominant on-shell contribution in the bottom-quark channel.
A large thermal suppression of the in-medium correction  to the Higgs width is found in a straightforward thermal-field theory
 calculation. Alternatively, an operator product expansion has been adopted in a recent calculation 
%investigation for getting a non-negligeable thermal width  of Higgs. 
In the present project we aim at building a unified picture, including both decay and scattering diagrams, thereby also being able to include the impact of non-equilibrium effects.
% from light quark to heavy quark to Higgs.
In the future, these interactions will be implemented into transport simulations to estimate the nuclear suppression factor of Higgs in QGP as formed in high-energy collisions of heavy nuclei.
}

\bigskip

%\end{abstract}
%
%\maketitle

\subsection{Introduction}
In the standard model, the Higgs boson, whose mass is measured at around 125 GeV~\cite{Djouadi:2005gi}, 
has a very small decay width of about 4 MeV. With respect to typical typical tie scales in Quantum Chromodynamics (QCD) 
$\sim$ 1 fm/c, the mean life-time of the Higgs (50 fm/c) is quite large, whereas the lifetime of the quark gluon plasma (QGP) created in ultrarelativistic heavy-ion 
collisions is about 10 fm/c. This leads to a rather intriguing hierarchy of time scales for the interaction of the Higgs in the medium formed in heavy-ion collision. Indeed, it has been conjectured as a possibly relevant topic in the discussion
of the  Future Circular Collider (FCC), 
where these interactions might have non-negligible consequences. Inspired by recent work on Higgs boson suppression in the QGP~\cite{dEnterria:2018bqi} and on its thermal 
width~\cite{Ghiglieri:2019lzz}, we are interested in 
finite-temperature calculation of the Higgs boson spectral function and its phenomenological connections. 

In the following section (\ref{sec:FD}), we will first carry out a quantum field-theoretical calculation of Higgs boson decay width through quark anti-quark channels and confirm that a straightforward application gives almost no width enhancement relative to the vacuum. We will then discuss an alternative calculation of the thermal width correction calculation through the operator product expansion (OPE) methodology as well as a study of
thermal scattering and suppression of Higgs in QGP, following along the lines of Ref.~\cite{Ghiglieri:2019lzz}.
In Sec.~(\ref{sec:sum}) we give a brief summary an indicate a future strategy of our ongoing work. 
% Finding the touching overlapping zone between the GeV scale
% electro-weak phase transition and MeV scale quark-hadron phase transition is also our matter of
% interest.

\subsection{Framework and Discussion}
\label{sec:FD}
 \begin{figure}
	\centering
	\includegraphics[scale=0.9]{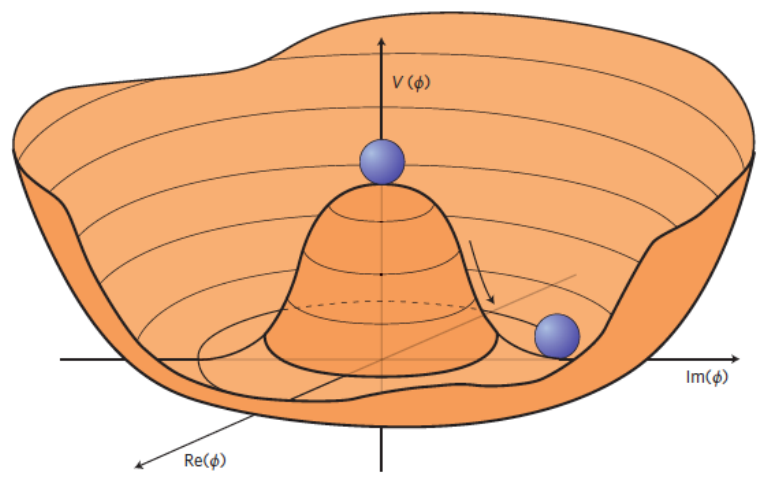}
	\caption{A sketch of mexican-hat type potential, $V(\phi)$, having degenerate vacua, one of which
	is chosen during the spontaneous symmetry breaking in the Higgs mechanism.}
	\label{fig:Mhat}
 \end{figure}
The particles in the standard model acquire mass through the Higgs mechanism. 
The carriers of the weak interactions are the 
$W^{\pm} $and $Z^0$ bosons, which is a $SU(2)$ gauge theory, and 
the electromagnetic interactions is carried by the photon( $\gamma$),
which is a $U(1)$ gauge theory. Electromagnetic and weak interactions are unified
within a $SU(2)_L\times U(1)_Y$ symmetry to form the Electroweak theory. In addition to 
this, the standard model contains the strong interactions mediated by gluons which are
based on a $SU(3)$ gauge theory, QCD. Thus the complete gauge group of the 
standard model is $SU(3)\times SU(2)\times U(1)$.

Spontaneous symmetry breaking is the essence governing Higgs mechanism 
which is triggered by a mexican-hat potential for a complex scalar field theory as shown in Fig~(\ref{fig:Mhat}). The Electroweak Lagrangian is constructed to allow for the Higgs mechanism giving rise to interaction terms fermions and bosons with a Higgs field. 
% These terms are responsible for describing
% decay and scattering of particles in electroweak theory. (Mexican Hat Potential Figure)

Here we are interested in Higgs boson's coupling to particles in the QCD sector. 
The interaction Lagrangian term for Higgs decaying to quark and anti-quark is given by
\bea
\mathcal{L}_{Hq\bar{q}} = -\frac{m_q H\psi_q\bar{\psi}_q}{v}~,
\eea
where $v =246$\,GeV is the vacuum expectation value of the Higgs field and $m_q$ is the bare mass of the various quarks. To calculate Higgs decay width in vacuum, we evaluate the quark-anti-quark loop diagram of Higgs boson, given by 
\bea
\Pi(q) = \int \frac{d^4q}{(2\pi)^4} L(k,q)D_kD_{k-q}~,
\eea
where 
\bea
L(k,q) &=& \left(\frac{-im_q}{v}\right)^2Tr[(\ks +m)(\ks -\qs+m)] 
\nn\\
&=& -\frac{4m_q^2}{v^2}\Big[k^2-\frac{q^2}{2}+m^2\Big]
\eea
and $D_k$ , $D_{k-q}$ are the scalar terms of the fermionic propagator given by 
\be
D_{k, k-q} = \frac{1}{(k, k-q)^2 - m_q^2} \ .
\ee
The decay width $\Gamma_H$ can be obtained from imaginary part of the vacuum self-energy,
\be
\Gamma(q) = \frac{\rm{Im}\Pi(q)}{q}=N_c\frac{m_q^2q}{8\pi v^2}\left[1-\frac{4m_q^2}{q^2}\right]^{3/2}~,
\ee
where $q$ is the 4-momentum of Higgs boson and $N_c$ is color degeneracy factor.  

% &&\ln\Big |\frac{e^{\beta(\omega_k^+-\mu)}-1}{e^{\beta(\omega_k^--\mu)}-1}\Big | +
% \frac{1}{\beta} \nn \ln\Big |\frac{e^{\beta(\omega_k^--\mu)}+1}{e^{\beta(\omega_k^-+\mu)}+1}\Big | \Big]\\
%
\begin{figure}
	\centering
	\includegraphics[scale=0.3]{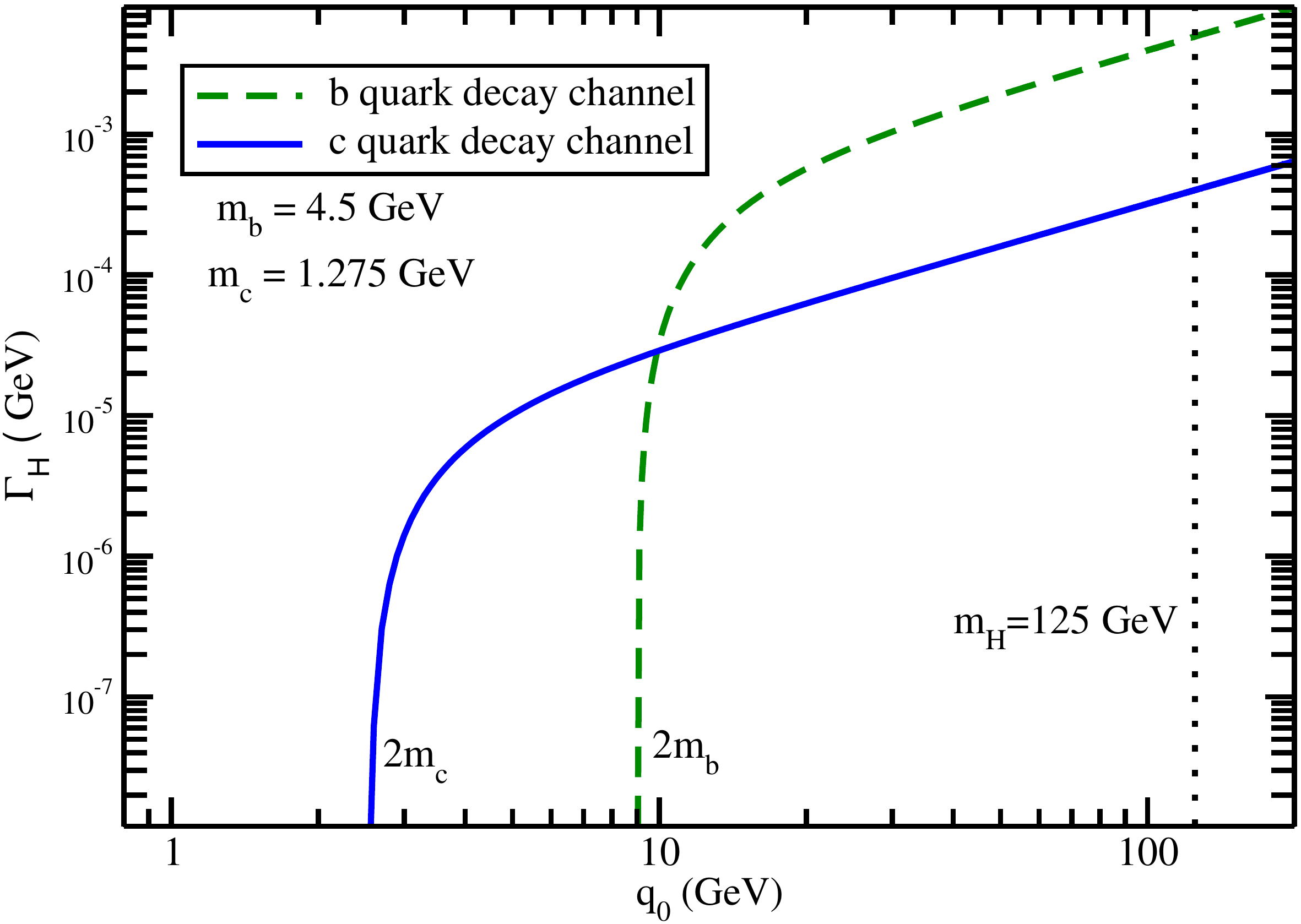}
	\caption{Off-mass shell distribution of vacuum widths for the $H\rightarrow b{\bar b}$ 
	and $H\rightarrow c{\bar c}$ decay channels.}
	\label{fig:Higgs_Decay_Width}
 \end{figure}
The off-mass shell distributions of the Higgs boson decay widths $\Gamma_{b\overline{b}}$ and $\Gamma_{c\overline{c}}$ for 
charm- and bottom-quark decay channels are shown in Fig.~(\ref{fig:Higgs_Decay_Width}),
where their unitarity cut thresholds of $2m_c$ and $2m_b$ can be seen very distinctly.
The dotted vertical line indicates the pole mass
of the Higgs, which marks the on-shell values of $\Gamma_{b\overline{b}}$ and $\Gamma_{c\overline{c}}$,
where former is the dominant contribution with a value of approximately 4\,MeV. 
The light-quark channels are rather suppressed, by factors of $\frac{m_{u, d, s}^2}{m_{b}^2}$; the quark masses are the bare ones, with $m_{u, d}\simeq 5$\,MeV, much smaller than heavy quark masses $m_{c, b}$ (recall that, while the phase space is ultrelativistic in all cases, the respective coupling constants are proportional to the masses).
 
The thermal corrections to the decay of Higgs boson are given by the expression of the finite-temperatures self-energies, 
\bea
\mbox{Im}\left(\Pi^T\right) = \frac{m_q^2q^2}{8\pi v^2 |\overrightarrow{q}|}\left[1-\frac{4m_q^2}{q^2}\right]\left(\frac{1}{\beta}  
\ln\Bigg |\frac{e^{\beta(\omega_k^+-\mu)}-1}{e^{\beta(\omega_k^--\mu)}-1}\Bigg | +
\frac{1}{\beta}  \ln\Bigg |\frac{e^{\beta(\omega_k^--\mu)}+1}{e^{\beta(\omega_k^-+\mu)}+1}\Bigg | \right)
\label{Pi_T}
\eea
where $\omega_k^{\pm} = \frac{1}{2}[q_0 \pm |\overrightarrow{q}| ]$. 
These corrections are very small compared to the vacuum decay width 
because of the $v^2$ term figures in the denominator and the large $M_H$ value figures in the logarithmic term in Eq~.(\ref{Pi_T}) and thus is contributes rather little.

A non-trivial evaluation of the decay of Higgs to quarks and gluons in a hot QCD medium has recently been recently in 
Ref.~\cite{Ghiglieri:2019lzz} using an alternative technique -- the operator product expansion(OPE).  
Pioneering work in Ref.~\cite{CaronHuot:2009ns} has utilized the use OPE technique~\cite{Wilson:1969zs, Wilson:1972ee} to study the asymptotic behavior of different spectral functions  and stress-energy tensors at finite temperature in the 
high-energy time-like region, $\omega >> T$,  and their thermal corrections in Euclidean Yang-Mills theory.
The Euclidean current-current correlator relates the spectral function with the Euclidean Green's function.
Thermal corrections to the decay rate depend on thermal corrections to the spectral 
function~\cite{CaronHuot:2009ns,Moore:2008ws,Laine:2010tc}. 
Reference \cite{Ghiglieri:2019lzz} makes use of the technique of Ref.~\cite{CaronHuot:2009ns} for obtaining a pertinent result for the Higgs boson. 
%Then  the stress-energy tensors for gluons and fermions as well %as the 
%trace of the stress-energy tensor are calculated in Ref-~%\cite{CaronHuot:2009ns}. 
The Euclidean OPE coefficients are calculated by taking the 
Euclidean current-current correlator given by
\bea
G_E(q) = \int d^4x e^{-iqx} \Big<J(x)J(0)\Big>~,
\eea
evaluated at $q=(0,0,0,q_E)$. 

The dispersion relation which relates Euclidean green's function to the spectral function 
is~\cite{Ghiglieri:2019lzz,CaronHuot:2009ns}  
\bea
G_E(q_E) &=& P(q_E) 
+ \int_{-\infty}^{+\infty} \frac{d\omega}{(2\pi)(\omega-iq_E)}\rho_J(\omega)~,
\eea
whose bulk channel is useful for calculating decay widths, as  has been done for the 
problem at hand in Ref.~\cite{Ghiglieri:2019lzz}. The
asymptotic expansion of $\rho_J(K)$ for large time-like $K$ has 
then been obtained from matching term-by-term to the OPE of $G_E(q_E)$ for large space-like $q_E$.

Leading thermal corrections to spectral functions in QCD are proportional to $T^4$, which is a standard result in perturbative QCD. 
%The estimated temperature that can be reached in heavy-ion collisions is of the order of 1 GeV. 
To apply OPE one has to then distinguish the kinematic regions based on the validity of the OPE technique. A detailed analysis of OPE applied to the Higgs in the QGP has been carried out in Ref.~\cite{Ghiglieri:2019lzz}, 
and the thermal correction to the decay width of Higgs to quark and anti-quark pairs (in particular bottom) has been obtained as
\bea
\delta \Gamma_{H\rightarrow b\overline{b}}  
= -\Gamma^{vac}_{H\rightarrow b\overline{b}}\, \alpha_s\frac{T^4}{m_H^4}\frac{128\pi^3}{135}~,
\eea
where $\Gamma^{vac}_{H\rightarrow b\overline{b}}\approx N_c\frac{m_q^2 m_H}{8\pi v^2}$ is 
the on-shell vacuum decay width of Higgs the boson to bottom quarks by ignoring the 
multiplicative factor of $\Big(1-\frac{m_q^2}{m_H^2} \Big)^{3/2}$.
For all partial decay widths into $q\overline{q}$ pairs the thermal correction is of ${\cal O}( \alpha_s \frac{T^4}{m_H^4} )$ to the vacuum decay width. 

%is a $\mathcal{O}\Big(\alpha_s\frac{T^4}{m_H^4}\Big)$ correction 

In another recent work~\cite{dEnterria:2018bqi} on the interactions of Higgs in quark-gluon matter, the scattering amplitude is employed for estimating the Higgs adsorption in QGP. 
Here, the cross-sections for Higgs-parton scattering 
has been analyzed and reproduced by a power-law fit of the form
\bea
\sigma_{Hgq} (\sqrt{s}) &=& K.A[\mu b]\Big((\sqrt(s)-m_H)/[GeV] \Big)^{-n}~,
\nn\\
\eea
with an amplitude $A = 2\mu b$ and $n=3$. They have taken a $K=3$ factor to
map higher order corrections in
Higgs-parton scattering. This has been obtained through $N^3LO/LO$ ratio of the
$gg \rightarrow H + X$ production cross-section, featuring the same diagrams. 
They have made use of thermal mass prescriptions for partons 
in medium giving finite Higgs-parton scattering ratio of the order of $\mu b$.
By using this in-medium cross-section, they obtain a non-negligible suppression of Higgs in QGP. However, when also including virtual corrections, they find a large cancellation which results in an essentially negligible final result compatible with that of Ref.~\cite{Ghiglieri:2019lzz}.

\subsection{Summary and Future Plan}
\label{sec:sum}
In the present article we have first given a brief survey of the Higgs boson connection to the quark-gluon plasma. Starting with the spontaneous symmetry breaking mechanism of the standard-model Mexican-hat potential within the QCD sector, our interest has been focused on the Higgs coupling to quarks as described by the interaction Lagrangian density. After illustrating the off-shell mass distribution function of Higgs boson going to quark and anti-quark decay channels in vacuum, we have addressed its thermal-field theoretical correction, which turns out to be very small. This is in line with previous works by Ghiglieri and Wiedemann adopting an operator product expansion methodology, and by d'Enterria and Loizides using a fitted cross section including virtual corrections. Based on these existing investigations, we have attempted a unified description, which includes both decay and scattering both diagrams, as a work in progress. Realizing the suppressed thermal correction, we also plan to develop a non-equilibrium spectral function of Higgs from QGP via an intermediate non-equilibrium mechanism of diffusing heavy quarks, whose number in heavy-ion collisions is usually much larger then the equilibrium value.
The rough sketch would be to obtain non-equilibrium Higgs properties from non-equilibrium heavy quarks in an equilibrium light-quark and gluon bath. After developing the spectral form, our next step is to implement it into a transport approach
for revisiting the nuclear suppression factor of the Higgs in heavy-ion collisions.

%% file: Ritesh/ritesh.tex
\section{Anisotropic pressure of deconfined Hot QCD matter in presence of strong magnetic field within one loop approximation}
	
\textit{Ritesh Ghosh, Bithika Karmakar, Aritra Bandyopadhyay, Najmul Haque, Munshi G. Mustafa}	
	
\bigskip

{\small	
We constructed general structure of fermion self-energy in strong magnetic field and obtained dispersion relation by calculating one loop fermion self-energy. We obtained analytic expression for anisotropic pressure and magnetization of a strongly magnetized hot QCD matter created in heavy-ion collisions considering the general structure of the two point functions of both quarks and gluons(within one-loop approximation) using hard thermal loop approximation for the heat bath.  The obtained anisotropic pressure may be useful for a magnetohydrodynamics description of a hot and dense deconfined QCD matter produced in heavy-ion collisions.
}

\bigskip

\subsection{\label{sec:level1}Introduction}
	A new hot and dense state of quarks and gluons is created in relativistic heavy ion collisions(HIC) in RHIC at BNL and LHC at CERN in recent times. This new state known as QGP can be explained by non-abelian gauge theory of QCD which is the theory of strong interaction of quarks and gluons. This theory explains a phase transition from confined state in low energy to deconfined state of quarks and gluons(QGP) in high energy. 
	 It is believed that such QGP state was created in early universe after few microseconds of big bang and exists in core of neutron star where matter density is much higher than normal matter density. Upcoming experiments are to be performed in FAIR at GSI and NICA at Dubna to explore more. 
	  In recent years study of non-central collisions says that high magnetic field can be generated in direction perpendicular to reaction plane due to the spectator particles~\cite{Fukushima:2012vr}. Strength of magnetic field decreases very fast from (30-10) $m_{\pi}^2$ to (1-2) $m_{\pi}^2$ in about (4-5) fm/c~\cite{McLerran:2013hla}. So one can work in two different regions: one is strong magnetic field limit ($q_fB>T^2$) and other is weak magnetic field limit ($q_fB<T^2$). 
	  
As EoS has phenomenological importance for studying hot and dense QCD matter we computed the EoS within the strong limit. We work in lowest Landau levels (LLL) with scale hierarchy ($q_fB>T^2>m_f^2$) as in strong field limit magnetic field pushes the higher Landau levels (HLL) to infinity compared to LLL~\cite{Bandyopadhyay:2016fyd}.

	\subsection{\label{sec:level2}Quarks in strong magnetic field}
	\subsubsection{General structure}
	 Presence of heat bath breaks the Lorentz (boost) invariance, whereas the presence of magnetic field breaks the rotational invariance of the system. So one needs to construct a manifestly covariant structure of the self-energy. We have external fermion momentum $P^{\mu}$. We have worked in rest frame of heat bath $u^{\mu}=(1,0,0,0)$. As we are considering non-central HIC, we are taking background magnetic field in z-direction $n_{\mu}=\frac{1}{2 B}\epsilon_{\mu\nu\rho\lambda}u^{\nu}F^{\rho\lambda}=(0,0,0,1)$, where $F^{\mn}$ is the electromagnetic field tensor.
	
	The fermion self-energy is a $4\times4$ matrix as well as Lorentz scalar. General structure should be made of basis matrices $\{\bf{I},\gamma_\mu,\gamma_{5},\gamma_{\mu}\gamma_5,\sigma_{\mu\nu}\}$. As we are working in strong magnetic field limit($q_fB\gg T^2$), we confine ourselves in LLL where transverse component of fermion momentum $P_\perp=0$. General structure of fermion self-energy in LLL can be written as~\cite{Karmakar:2019tdp}
	\bea
	\Sigma(p_0,p_3)=a\slashed{u}+b\slashed{n}+c\gm_5\slashed{u}+d\gamma_5\slashed{n},
	\eea
	where 	
	\bea
	a&=&\frac{1}{4}\Tr[\Sigma\ \slashed\! u]\,, \hspace{1cm}	b = -\frac{1}{4}\Tr[\Sigma \ \slashed\!n] ,\nn \\
	c&= &\frac{1}{4}\Tr[ \gamma_5 \Sigma\slashedl\! u]\,\, \rm{and} \,\,\,	d=-\frac{1}{4}\Tr[ \gamma_5 \nn \Sigma\ \slashed\!n].
	\eea
%%%%%%%%%%%%%%%%%%%%%%	
	\subsubsection{One loop quark self energy in strong field}
\begin{center}
		\begin{figure}[!ht]
			\begin{center}
				\includegraphics[scale=0.4]{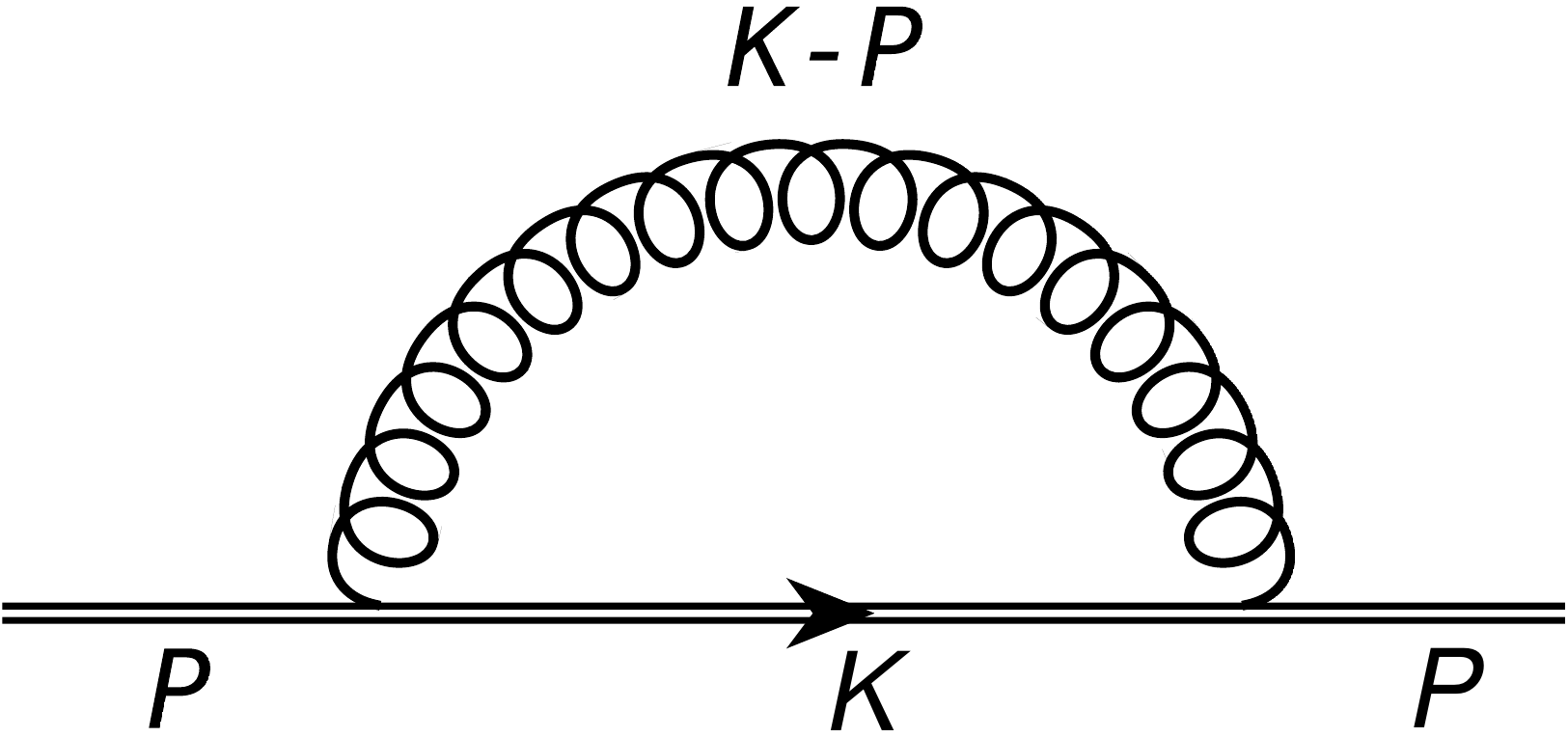} 
				\caption{Self-energy diagram for a quark in a strong magnetic field approximation. 
					The double line indicates the modified quark propagator in presence of strong magnetic 
					field.}
				\label{quark_se}
			\end{center}
		\end{figure}
	\end{center}

One-loop quark self-energy in Feynman gauge can be written from Fig.~\ref{quark_se} as
\bea
\Sigma(P) &=& -ig^2C_F\int\frac{d^4K}{(2\pi)^4}\gamma_\mu S(K) \gamma^\mu 
\Delta(K-P),
\label{sfa_self}	
\eea
	where the unmodified  gluonic propagator is given as
	\bea
	\Delta(K-P) = \frac{1}{(K-P)_\sp^2-(K-P)_\perp^2}
	\eea  
	 and modified fermion propagator in LLL is given by
	\bea
	iS(K)=ie^{-{k_\perp^2}/{q_fB}}~~\frac{\slashed{\!K}_\sp+m_f}{
		K_\sp^2-m_f^2}(1-i\gamma_1\gamma_2).
	\label{prop_sfa}
	\eea
	Here we used $(K-P)_\sp^2=(k_0-p_0)^2-(k_3-p_3)^2$ and $(K-P)_\perp^2=(k_1-p_1)^2+(k_2-p_2)^2$.

\subsubsection{Effective propagator and dispersion relation}  
\begin{center}
	\begin{figure}[!ht]
		\begin{center}
			\includegraphics[scale=0.6]{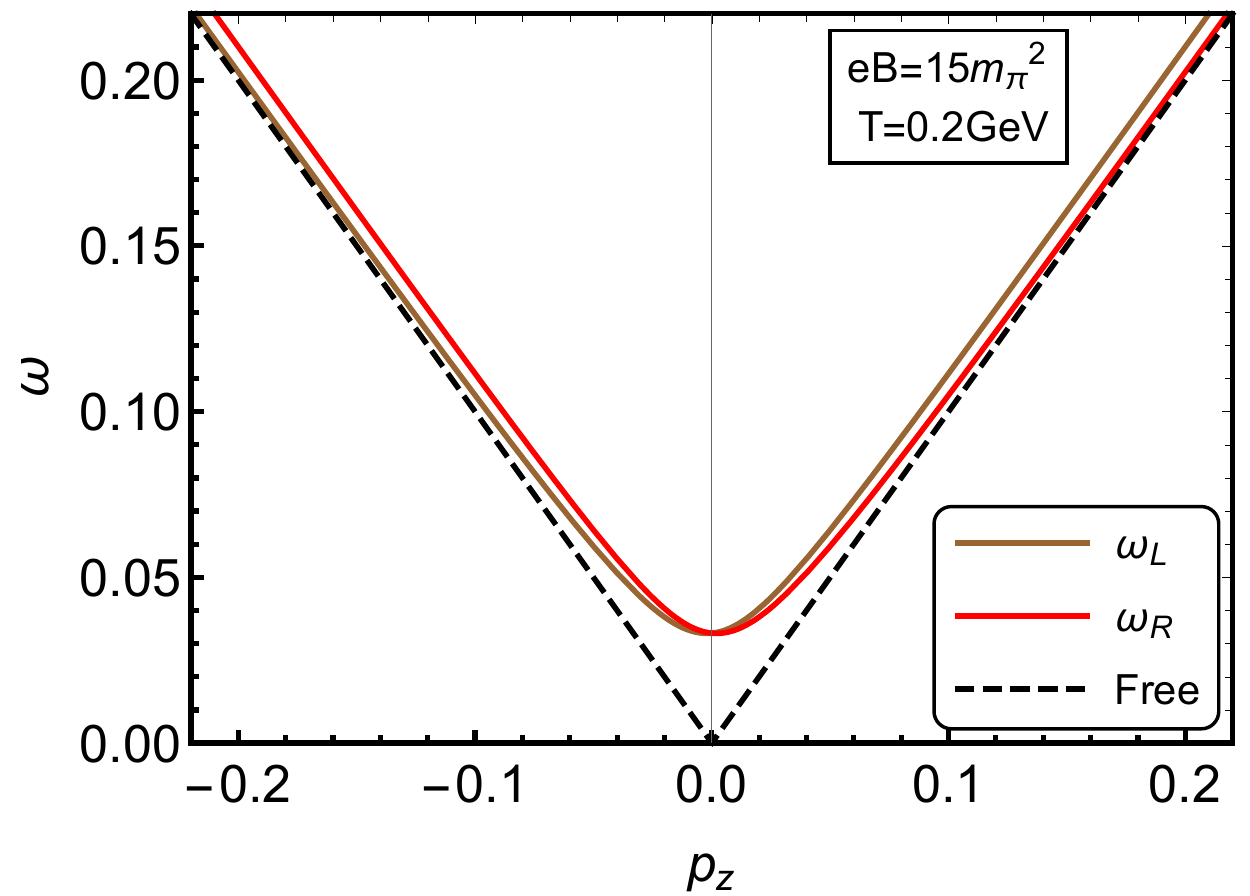} 
			\caption{Dispersion relation of fermion in presence of strong magnetic field}
			\label{quark_disp}
		\end{center}
	\end{figure}
\end{center}

In LLL, the effective fermion propagator can be written as 
\bea
S_{\text{eff}}(P_{\sp})=\frac{1}{\slashed{\!P_{\sp}}+\Sigma}.
\eea
Using chiral projectors effective propagator can also be written as,
\bea
{S}_{\text{eff}}(P_{\sp})=\mathcal{P}_R\frac{\slashed{\!R}}{R^2}\mathcal{P}_L+\mathcal{P}_L\frac{\slashed{\!L}}{L^2}\mathcal{P}_R, \label{fermion_eff_S}
\eea
where $\mathcal{P}_R=\frac{1}{2}(1+\gamma_5)$ and $\mathcal{P}_L=\frac{1}{2}(1-\gamma_5)$.
We have,
\begin{align}
&L^2=(p_0+(a+c))^2-\big(p_3-(b+d)\big)^2,\\
&R^2=(p_0+(a-c))^2-\big(p_3-(b-d)\big)^2.
\end{align}
We find the expression of form factors from one loop fermion self-energy and then obtain
dispersion curves by solving $R^2=0$ and $L^2=0$ from Eq.~(\ref{fermion_eff_S}). 
There are four modes, two comes from $L^2=0$ and two from $R^2=0$. In LLL only two modes are allowed~\cite{Das:2017vfh}: one $L$-mode 
with energy $\omega_L$ of a positively charged fermion having spin up and another one from $R$-mode 
with energy $\omega_R$ of a negatively charged fermion having spin down. These two modes 
are plotted in Fig.~\ref{quark_disp}. At high $p_z$ both the mode of dispersion resembles free dispersion mode. We also note that the reflection symmetry is broken 
in presence of magnetic field~\cite{Das:2017vfh}.

\subsection{Gluon and quark free-energy in a strongly magnetized hot medium}
General structure of gauge boson two-point function in strong magnetic field and one loop dispersion relation is obtained in paper~\cite{Karmakar:2018aig}. We have calculated hard and soft contribution (considering soft gluon momentum $P\sim gT$) of gluon free energy upto $\mathcal{O}(g^4)$ in hard thermal loop approximation. We also calculate quark free-energy upto $\mathcal{O}(g^4)$. Analytic expressions of quark and gluon free-energy can be found in \cite{Karmakar:2019tdp}.
	
\subsection{Anisotropic pressure in strong magnetic field }
So total one loop free energy of deconfined QCD matter can be written as 
\bea
F=F_q+F_g^{hard}+F_g^{soft}+F_0+\Delta \mathcal{E}^0_T+\Delta \mathcal{E}^B_T,
\eea
where $F_q$, $F_g^{hard}$, $F_g^{soft}$ are respectively quark free-energy and hard and soft contribution of gluon self energy. From one-loop calculation, different kind of divergences arises of $\mathcal{O}[\frac{1}{\eps}]$ and these are renormalized by adding the last three counterterms in the $\overline{MS}$ renormalization scheme. 
In presence of strong magnetic field space becomes anisotropic and we get different pressures~\cite{PerezMartinez:2007kw} for direction parallel and perpendicular to magnetic field. Longitudinal and transverse pressures are defined as 
\bea
P_{z}=-F ,\,\,\, P_{\perp}=-F-eB\cdot M=P_z-eB\cdot M,\label{trans_pressure}
\eea
where  the magnetization per unit volume $M=-\frac{\partial(F)}{\partial (eB)}$.

\subsection{Results and conclusions}	
\begin{center}
\begin{figure}[!ht]
\begin{center}
\includegraphics[scale=0.49]{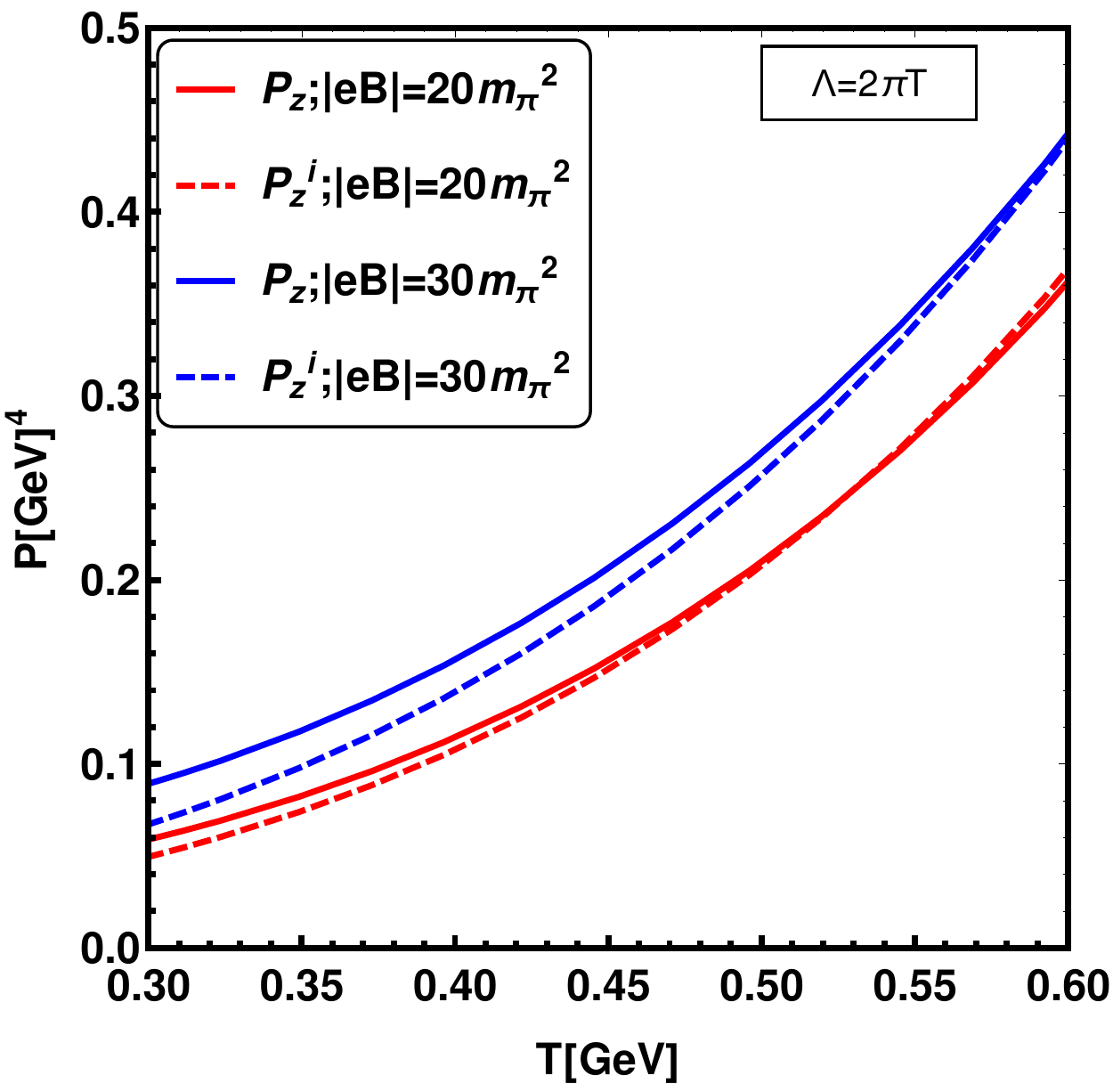}
\includegraphics[scale=0.49]{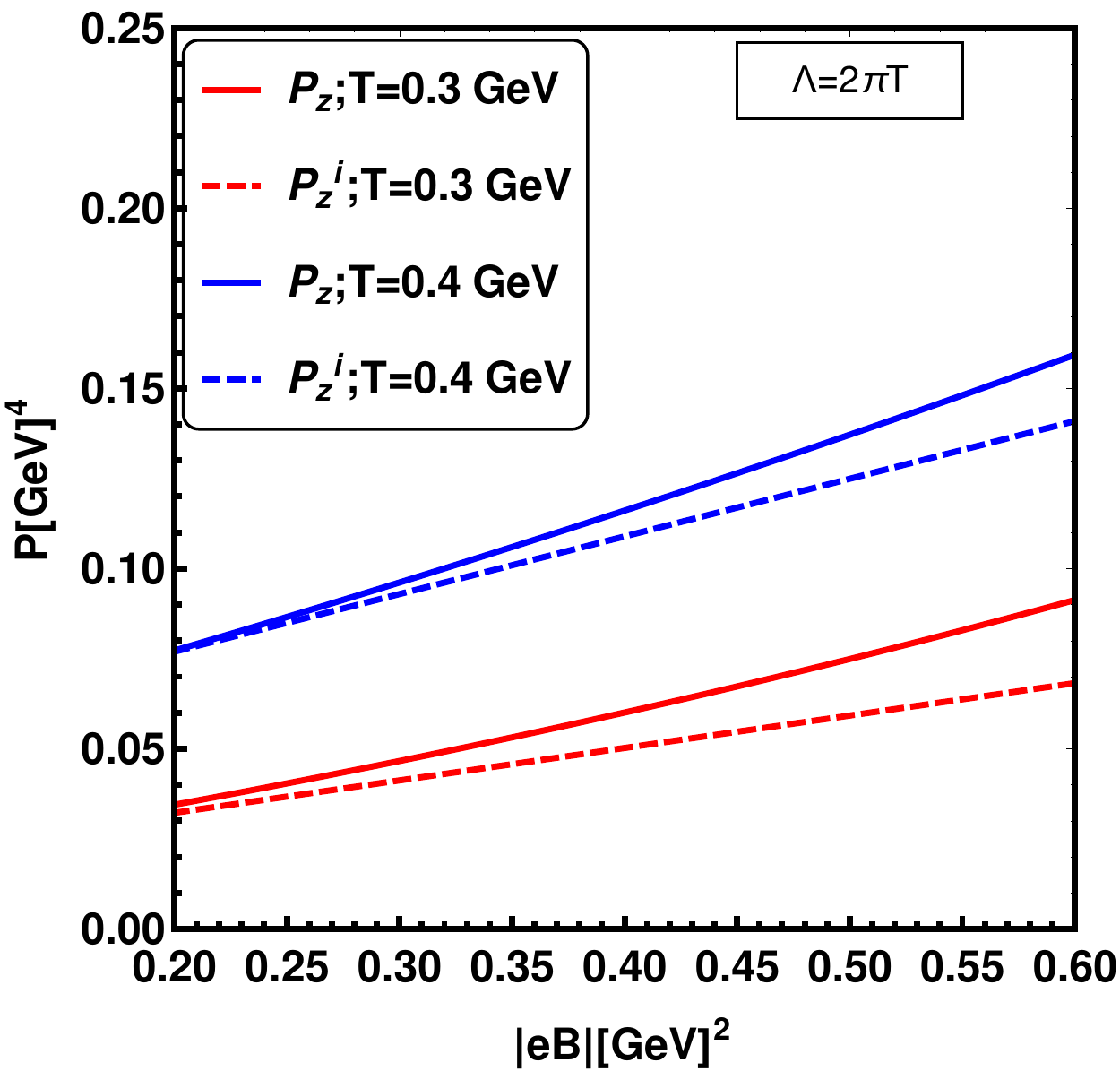}
\caption{Variation of one-loop longitudinal pressure  as a function of temperature for different value of magnetic field (left panel)  and as a function of magnetic field  at different temperature (right panel) for number of quark flavor $N_f=3$ and the central value of the  renormalization scale, $\Lambda=2\pi T$.  Dashed curves represent ideal longitudinal pressure.}
\label{1loop_long_pressure_T_eB}
\end{center}
\end{figure}
\end{center}

\begin{center}
\begin{figure}[!ht]
\begin{center}
\includegraphics[scale=0.49]{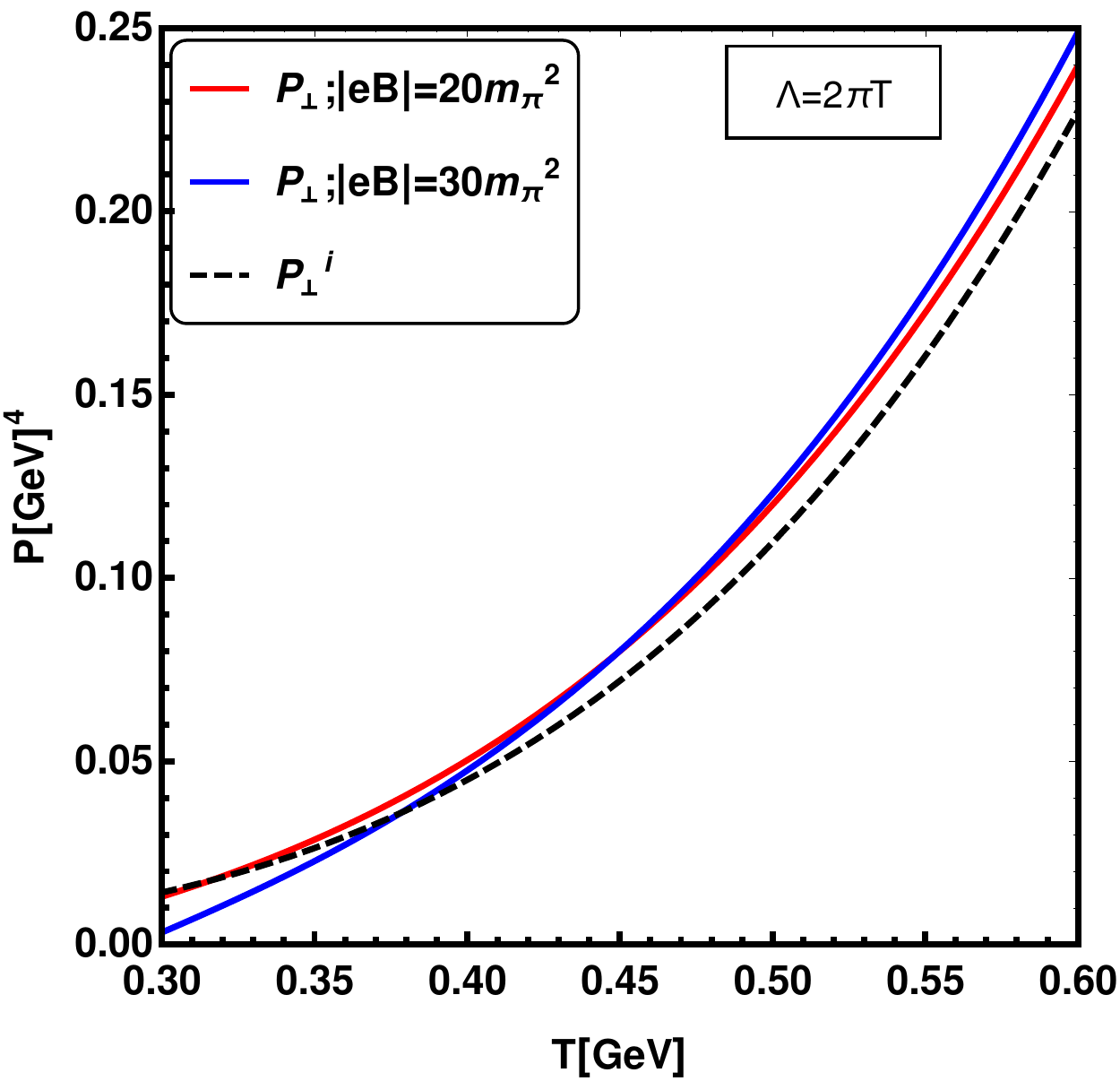}
\includegraphics[scale=0.49]{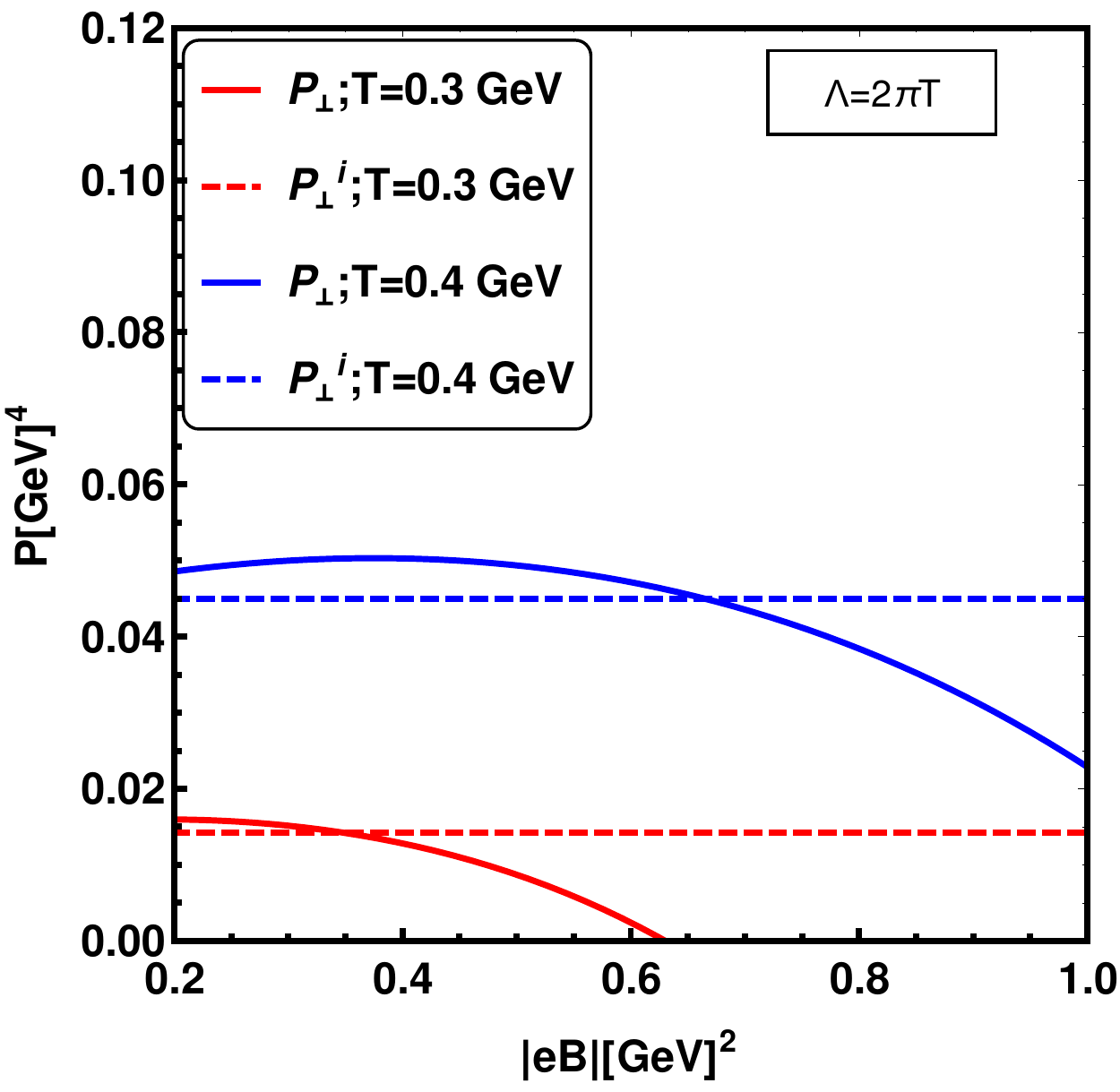}
\caption{Variation of the  one-loop transverse pressure as a function of temperature for various magnetic fields (left panel) and as a function of magnetic field for different temperatures (right panel). Dashed curves represent ideal transverse pressure.}
\label{trans_ideal_unscaled_T_eB}
\end{center}
\end{figure}
\end{center}

From Fig.~\ref{1loop_long_pressure_T_eB} we can see that one-loop pressure increases with the increase in temperature and field strength, respectively. However, the one-loop interacting pressure is higher than that of ideal~\cite{Karmakar:2019tdp} one in both panels. From Fig.~\ref{trans_ideal_unscaled_T_eB} we can see that the one loop transverse pressure increases with temperature showing similar nature as longitudinal pressure 
(left panel of Fig.~\ref{1loop_long_pressure_T_eB}) but lower in magnitude. Dashed lines represent transverse ideal 
pressure which is independent of magnetic field. For a given high value of the magnetic field, the pressure starts with a lower value than that of ideal gas particularly at low $T$ and then a crossing takes place. 

%\begin{widetext}
%\end{widetext}
This can also be understood from the right panel where the transverse pressure is displayed as a function of magnetic field for two different temperatures. Here also the dashed lines represent the ideal transverse pressure which is independent of magnetic field. The transverse pressure for interacting case is given in Eq.~\eqref{trans_pressure} as $P_{\perp}=P_z-eB \cdot M$. Now for a given temperature its variation is very slow	(or almost remain unaltered) with lower value of the magnetic field because there is a competition between $P_z$ and $eBM$. Due to increase of the magnetization $M$  with magnetic field the transverse pressure, $P_\perp$ tends to decrease, falls below ideal gas value  and  may even go to negative values for low $T$ at large value of magnetic field. This is an indication that the system may shrink in the transverse direction.
	
Finally we can conclude that due to the presence of strong background magnetic field one gets different pressures in direction parallel and perpendicular to magnetic field. Both the pressures are calculated analytically by calculating the magnetization of the system. This anisotropic pressure can be useful for magnetohydrodynamics description of hot deconfined QCD. We have calculated one loop HTL perturbation theory upto $\mathcal{O}(g^4)$; $ \mathcal{O}(g^2)$ and $\mathcal{O}(g^4)$ are incomplete and this result can be improved by higher order loop calculation.

%% file: Sabya/sabya.tex
\section{On the microscopic estimated values of transport coefficients for quark and hadronic matter}

\textit{Sabyasachi Ghosh, Subhasis Samanta, Kinkar Saha, Snigdha Ghosh, Fernando E. Serna, Mahfuzur Rahaman, Aman Abhishek, Guruprasad Kadam, Pracheta Singha, Sudipa Upadhaya, Soumitra Maity, Sumana Bhattacharyya, Arghya Mukherjee, Payal Mohanty, Bhaswar Chatterjee}

\bigskip

%\begin{abstract}
{\small
From a long list of microscopic calculations of transport coefficients of quark and hadronic matter,
few selective references are chosen and their estimated values are tabulated.
Through this catalogue-type draft on microscopic calculations of transport coefficients, we have pointed out
a particular investigation series, which has identified
three possible sources - (1) resonance type interaction, (2) finite size effect
and (3) effect of magnetic field, which might be responsible for (nearly) perfect fluid nature
of RHIC/LHC matter.
}

\bigskip

%% 
%\end{abstract}

%\maketitle
\subsection{Introduction}
In 2002, researchers at Duke university discovered a super-cold lithium fluid having 
very small viscosity to entropy density ratio $\eta/s$ ($< 0.5$), close to its quantum lower 
bound, $\frac{1}{4\pi}$. Whereas, three years latter, Relativistic Heavy Ion
Collider (RHIC) experiment at BNL created a super-hot Quark Gluon Plasma
(QGP) with smallest $\eta/s$, almost equal to the lower bound. This nearly perfect
fluid nature of many body system at two extreme conditions (super-cold and
super-hot) has attracted the attention of large band of theoretical communities
from condense matter physics to nuclear physics to string theory~\cite{Schafer:2009dj}. 
RHIC data indicated a strongly interacting sQGP medium instead of a weakly interacting
gas, which is naturally expected from high temperature Quantum Chromo Dynamics (QCD)~\cite{Arnold:2000dr}, 
owing to the asymptotic freedom of QCD. To understand the
dynamical origin of low $\eta/s$ of RHIC matter, 
several microscopic calculations, based on effective QCD 
models~\cite{Marty:2013ita,Sasaki:2008um,Ghosh:2015mda,Deb:2016myz,Chakraborty:2010fr,Singha:2017jmq,Lang:2013lla,Cassing:2013iz,Ghosh:2019lmx} 
as well as hadronic models~\cite{FernandezFraile:2009mi,Ghosh:2014qba, Ghosh:2014ija, Ghosh:2015lba, Rahaman:2017sby,Kalikotay:2019fle,Ghosh:2018nqi,Ghosh:2019fpx,NoronhaHostler:2008ju,Kadam:2015xsa,Ghosh:2016clt,Ghosh:2016yvt}, have been done in
recent time. 
Present draft is intended to review briefly on the estimated values of $\eta/s$
and other transport coefficients, obtained from different microscopic calculations.
Based on investigation series, given in Refs.~\cite{Ghosh:2015mda,Singha:2017jmq,Ghosh:2014qba, Ghosh:2014ija, Ghosh:2015lba, Rahaman:2017sby,Ghosh:2018nqi,Ghosh:2019fpx,Ghosh:2019lmx},
three possible sources - (1) resonance type interaction, (2) finite size effect
and (3) effect of magnetic field are identified for getting low $\eta/s$
in RHIC/LHC matter. Addressing a brief framework of transport coefficient in
next section, Sec.~(\ref{sabya_sec:results}) has gone through the discussion on 
their estimated values, which are listed in tables and pointed out three
possible sources for getting low $\eta/s$. Lastly, a brief summary is made in Sec.~(\ref{sabya_sec:sum}).

\subsection{Brief framework of transport coefficient}
Let us start with a brief framework of transport coefficients like shear 
viscosity ($\eta$), bulk viscosity ($\zeta$) and electrical conductivity ($\sigma$).
The macroscopic definition of ideal part of energy-momentum tensor,
\beq
T^{\mu\nu}_0=-g^{\mu\nu} P + (\epsilon + P)u^\mu u^\nu~,
\eeq
which can be connected to its microscopic (kinetic theory) definition,
\beq
T^{\mu\nu}_0=\int\frac{d^3p}{(2\pi)^3}\frac{p^{\mu}p^{\nu}}{E} f~,
\eeq
where pressure $P$, energy density $\ep$, four velocity $u^\mu$ are macroscopic/fluid quantities
but energy $E=\sqrt{\vp^2+m^2}$, thermal distribution function $f$ are microscopic/particle quantities.
For boson/fermion, $f$ will be Bose-Einstein and Fermi-Dirac distribution function at temperature $T=1/\beta$. 
Now if the medium is slightly deviated from equilibrium distribution function $f$ to
$f+\delta f$, then deviation $\delta f$ will build a dissipative part of energy-momentum tensor
$T^{\mu\nu}_D$, whose microscopic expressions,
\beq
T^{\mu\nu}_D=\int\frac{d^3p}{(2\pi)^3}\frac{p^{\mu}p^{\nu}}{E} \delta f~,
\label{Tmn_micro}
\eeq
again can be connected with its macroscopic expression,
\beq
T^{\mu\nu}_D=\eta {\cal U}_\eta^{\mu\nu} + \zeta \Delta^{\mu\nu}\partial_\rho u^\rho~,
\label{Tmn_macro}
\eeq
where 
\bea
{\cal U}^{\mu\nu}_\eta&=&
\FB{D^\mu u^\nu + D^\nu u^\mu -\frac{2}{3}\Delta^{\mu\nu}\partial_\rho u^\rho}~,
\nn\\
\Delta^{\mu\nu} &=& g^{\mu\nu} -u^\mu u^\nu~,
D^\mu = \partial^\mu -u^\mu u^{\sigma}\partial_{\sigma}~.
%\nn\\
%{\cal U}_\zeta^{\mu\nu}&=&\Delta^{\mu\nu}\partial_\rho u^\rho~.
\eea
Similar to $T^{\mu\nu}_D$, dissipative part of electric current density $J^{\mu}_D$
will also have connection from macroscopic (Ohm's law) to microscopic expressions
\beq
\sigma^{\mu\nu}E_{\nu} = J^\mu_D = e\int\frac{d^3p}{(2\pi)^3}\frac{p^{\mu}}{E}\delta f~.
\label{J_E}
\eeq
Now, through relaxation time approximate (RTA) of Boltzmann equation, the $\delta f$
can be expressed in terms of ${\cal U}_\eta^{\mu\nu}$, $\del_\rho u^\rho$ and $E^\mu$
as
\bea
\delta f &=& \beta f(1\pm f)\frac{\tau_c}{E} \Big[ p_\mu p_\nu{\cal U}^{\mu\nu}_\eta
+ \Big\{\Big(\frac{1}{3}-c_s^2\Big)\vp^2 
\nn\\
&&-c_s^2\frac{\del}{\del \beta^2}\Big(\beta^2m^2\Big) \Big\}^2
\partial_\rho u^\rho +e p_\mu E^\mu\Big]
\label{df_RBE}
\eea
Implementing Eq.~(\ref{df_RBE}) in Eqs.~(\ref{Tmn_micro}), (\ref{Tmn_macro}), (\ref{J_E})
we get the final expressions of $\eta$, $\zeta$ and $\sigma$
for boson/fermion~\cite{Chakraborty:2010fr}:
% \bea
% \eta &=& \frac{\beta}{15} \int\frac{d^3\vp}{(2\pi)^3} \left(\frac{\vp^2}{E}\right)^2 \tau_c
% \: f(1\pm f)
% \\
% \zeta &=& \beta \int\frac{d^3\vp}{(2\pi)^3} \Big\{\Big(\frac{1}{3}-c_s^2\Big)\vp^2 
% -c_s^2\frac{\del}{\del \beta^2}\Big(\beta^2m^2\Big) \Big\}^2 \tau_c
% \: f(1\pm f)
% \\
% \sigma &=& \frac{\beta}{3} \int\frac{d^3\vp}{(2\pi)^3} \left(\frac{\vp}{E}\right)^2 \tau_c
% \: f(1 \pm f)~.
% \label{eta_el}
% \eea
%\begin{widetext}
\beq
\left(
\begin{array}{c}
\eta \\[0.2true cm]
\zeta \\[0.2true cm]
\sigma
\end{array}
\right) = \beta \int\frac{d^3\vp}{(2\pi)^3}\tau_c\: f(1\pm f)
\left(
\begin{array}{c}
\frac{1}{15}\left(\frac{\vp^2}{E}\right)^2 \\[0.2true cm]
\frac{1}{E^2}\Big\{\Big(\frac{1}{3}-c_s^2\Big)\vp^2 
-c_s^2\frac{\del}{\del \beta^2}\Big(\beta^2m^2\Big) \Big\}^2 \\[0.2true cm]
\frac{1}{3}\left(\frac{\vp}{E}\right)^2
\end{array}
\right)~.
\label{Tr_tc}
\eeq
%\end{widetext}
The degeneracy factor of medium constituents has to be multiplied in above equation and different
species has to be summed with appropriate care. It will depends on our dealing system (like
quark matter or pionic matter etc.) or model (effective quark or hadronic model or hadron resonance
gas model etc.).

\subsection{Results and Discussion} 
\label{sabya_sec:results}
\subsubsection{Shear viscosity of quark and hadronic matter}
In earlier section, we came to know the mathematical anatomy of transport coefficients expressions,
which can grossly be identified as
\bea
{\rm Transport~coefficient}&=&({\rm thermodynamical~ phase~space})
\nn\\
&&\times({\rm Relaxation~ Time})~,
\label{Tr_gross}
\eea
if we take momentum independent relaxation time or momentum averaged relaxation time.
The thermodynamical phase-space part of $\eta$ for massless boson or fermion is
$\frac{4\pi^2}{450}T^4$ or $\frac{7\pi^2}{900}T^4$ like thermodynamical quantity
- entropy density $s=\frac{4\pi^2}{90}T^3$ (boson) or $\frac{7\pi^2}{180}T^3$ (fermion).
Hence, the dimensionless quantity $\eta/s$ will be found as $\frac{\tau_c T}{5}$, which
is monotonically increasing function of $T$, if we consider a $T$-independent $\tau_c$.
Now, depending upon our dealing bosonic/fermionic system, different microscopic calculation
can get different $\tau_c(T)$, which will ultimately provide the temperature profile of $\eta/s$.

Arnold et al.~\cite{Arnold:2000dr} have well summarized $\eta/s$ calculations, based on 
the perturbative approach of finite temperature QCD with re-summed version, popularly known as 
hard thermal loop (HTL). By using the leading order results of HTL calculation~\cite{Arnold:2000dr} for quark matter and 
chiral perturbation theory (ChPT) for hadronic matter~\cite{Prakash:1993bt},
Ref.~\cite{Csernai:2006zz,Kapusta:2008vb} has interestingly shown
a possibility of valley type profile of $\eta/s(T)$ like helium, nitrogen, and water.
However, their order of magnitude ($\frac{10}{4\pi}$-$\frac{20}{4\pi}$) are quite far from the expectation of experimental
side~\cite{Adler:2003kt}, interpreted through macroscopic hydrodynamical simulation~\cite{Romatschke:2007mq}.
\begin{table}[tbh]
\begin{center}
\caption{Order of magnitude of $\eta/s$ from different model calculations (first column) with 
references at temperature range below (second column)
and above (third column) transition temperature $T_c$.}
%\begin{ruledtabular}
\begin{tabular}{l|cc}
	\hline
%\doubleRule
%& &  \\ 
Framework$^{\rm Reference}$ & $T ~\leq~T_c$ & $T ~\geq~T_c$  \\
%& &  \\
\hline
%& & \\
HTL~\cite{Arnold:2000dr} & - & $1.8$  \\
%& & \\
LQCD~\cite{Meyer:2007ic} & - & $0.1$  \\
%& & \\
NJL~\cite{Marty:2013ita} & $1$-$0.3$ & $0.3$-$0.08$  \\
%& & \\
NJL~\cite{Sasaki:2008um} & $1$-$0.5$ & $0.5$-$0.55$  \\
%& & \\
NJL~\cite{Ghosh:2015mda} & - & $0.5$-$0.12$  \\
%& & \\
NJL~\cite{Deb:2016myz} & $2$-$0.25$ & $0.25$-$0.5$  \\
%& & \\
LSM~\cite{Chakraborty:2010fr} & $0.87$-$0.55$ & $0.55$-$0.62$  \\
%& & \\
PQM~\cite{Singha:2017jmq} & $5$-$0.5$ & $0.3$-$0.08$  \\
%& & \\
URQMD~\cite{Demir:2008tr} & $1$ & -  \\
%& & \\
SMASH~\cite{Rose:2017bjz} & $1$ & -  \\
%& & \\
Unitarization~\cite{FernandezFraile:2009mi} & $0.8$-$0.3$ & -  \\
%& & \\
HFT~\cite{Ghosh:2014qba, Ghosh:2014ija, Ghosh:2015lba, Rahaman:2017sby} & $0.4$-$0.1$ & -  \\
%& & \\
HFT~\cite{Kalikotay:2019fle} & $0.8$-$0.25$ & -  \\
%& &  \\
HRG~\cite{Ghosh:2018nqi,Ghosh:2019fpx} & $0.13$-$0.28$ & -  \\
%& &  \\
\hline
\end{tabular}
%\end{ruledtabular}
\label{tab1}
\end{center}
\end{table}

Extracted values of $\eta/s$ by different hydro-groups are well sketched in Fig.~(4) of
Ref.~\cite{Jaiswal:2016hex}, from where a rough order of magnitude, $\eta/s=\frac{1}{4\pi}$-$\frac{5}{4\pi}$,
is expected for RHIC or LHC matter. Different alternatively models in quark sector like Nambu-Jona-Lasinio 
(NJL)~\cite{Marty:2013ita,Sasaki:2008um,Ghosh:2015mda,Deb:2016myz}, linear sigma model (LSM)~\cite{Chakraborty:2010fr}, 
polyakov-loop quark meson (PQM)~\cite{Singha:2017jmq} model have estimated $\eta/s$ of quark matter, while
$\eta/s$ of hadronic matter is estimated through different tools of hadronic phase like 
URQMD~\cite{Demir:2008tr}, SMASH~\cite{Rose:2017bjz} codes, Unitarization 
methodology~\cite{FernandezFraile:2009mi}, hadronic field theory (HFT)~\cite{Ghosh:2014qba, Ghosh:2014ija, Ghosh:2015lba, Rahaman:2017sby,Kalikotay:2019fle},
hadron resonance gas (HRG) model~\cite{Ghosh:2018nqi,Ghosh:2019fpx} etc. Their estimated values of $\eta/s$ of
hadronic and quark phases, located below and above transition temperature $T_c$, are
listed in Table~.(\ref{tab1}), which carry few selective works, whose results are quite close to
KSS bound. The present draft will zoom in the message of Refs.~\cite{Ghosh:2015mda,Singha:2017jmq,Ghosh:2014qba, Ghosh:2014ija, Ghosh:2015lba, Rahaman:2017sby,Ghosh:2018nqi,Ghosh:2019fpx,Ghosh:2019lmx},
indicating about three possible sources for which $\eta/s$ of
RHIC or LHC matter is appeared to be very low (near to KSS bound). They are discussed below.

{\bf (1). Resonance type interaction:}
\\
Among the references, listed in Table~(\ref{tab1}), Refs.~\cite{Ghosh:2015mda,Singha:2017jmq,Ghosh:2014qba, Ghosh:2014ija, Ghosh:2015lba, Rahaman:2017sby}
have gone through an effective quark-resonance~\cite{Ghosh:2015mda,Singha:2017jmq} and hadron-resonance~\cite{Ghosh:2014qba, Ghosh:2014ija, Ghosh:2015lba, Rahaman:2017sby} type 
interaction, which might be considered as one of the reason for low $\eta/s$ of quark and
hadronic matter. In 1994, it was Quack and Klevansky~\cite{Quack:1993ie}, who proposed about
the quark propagation with quark-meson loop correction in NJL model, which was implemented by
Refs.~\cite{Ghosh:2013cba,Lang:2013lla,Ghosh:2015mda} for viscosity calculations. Through this quark-pion and
quark-sigma loop calculations, quark relaxation time ($\tau_c$) is estimated from the imaginary part
of quark self-energy ($\Pi$) by using the connection $\tau_c\sim 1/{\rm Im}\Pi$. Along with the NJL model~\cite{Ghosh:2013cba,Lang:2013lla,Ghosh:2015mda}, PQM model~\cite{Singha:2017jmq} (through similar quark-meson loop calculations) 
also found very small $\tau_c$ and $\eta/s$, close to KSS bound but applicable for a narrow temperature
domain near transition temperature. Alternative way to calculate quark relaxation time by using
same quark-meson Lagrangian density~\cite{Quack:1993ie} has been adopted by Refs.~\cite{Sasaki:2008um,Marty:2013ita,Deb:2016myz}.
Similar to effective quark-resonance interaction, where $\pi$, $\sigma$ are appeared as resonances of 
quark matter, effective hadron-resonance interaction has been considered in Refs.~\cite{Ghosh:2014qba, Ghosh:2014ija, Ghosh:2015lba, Rahaman:2017sby,Kalikotay:2019fle},
where $\sigma$, $\rho$, $K^*$, $\phi$ mesons, $N^*$, $\Delta$, $\Delta^*$ baryons are appeared as resonances 
of $\pi$, $K$ and $N$ medium. Ref.~\cite{Ghosh:2014qba, Ghosh:2014ija, Ghosh:2015lba, Rahaman:2017sby} has obtained pion relaxation time from
$\pi\sigma$, $\pi\rho$ loops; kaon relaxation time from $KK^*$, $K\phi$ loops and nucleon relaxation
time from $\pi N^*$, $\pi\Delta$, $\pi\Delta^*$ loops. On the other hand Ref.~\cite{Kalikotay:2019fle} has
estimated those relaxation times via resonance-scattering type diagram. Unlike to standard ChPT calculation,
both HFT calculations~\cite{Ghosh:2014qba, Ghosh:2014ija, Ghosh:2015lba, Rahaman:2017sby,Kalikotay:2019fle} found very small $\eta/s$. Hence, the resonance
type interaction in quark and hadronic matter might be one of the responsible factor for getting
low $\eta/s$ in RHIC or LHC matter.

{\bf (2). Finite size effect:} 
\\
Another possible source is finite size effect of medium~\cite{Ghosh:2018nqi,Ghosh:2019fpx}. Owing to quantum effect of finite system size, thermodynamical phase
space of Eq.~\eqref{Tr_gross} can be reduced because lower limit of integration in Eq.~(\ref{Tr_tc})
can be transformed from $0$ to $\vp_{\rm min}=\pi/R$, where $R$ is system size.
On the other hand, relaxation time of hadrons can also face finite size effect by considering only those relaxation scales, which are lower than the system size. Ref~\cite{Ghosh:2018nqi,Ghosh:2019fpx} has shown elaborately how finite size of hadronic matter in HRG model can make impact on reducing the values of $\eta/s$. Finite size effect of $\eta/s$ in effective QCD model is also studied in Ref.~\cite{Saha:2017xjq}, which needs an extended investigation for finite size $\tau_c$ calculations, after which we can get a complete conclusive picture. 

{\bf (3). Effect of magnetic field:} 
\\
Another possibility for getting low $\eta/s$ is hinted from strong magnetic field,
which might be produced in the non-central heavy-ion collisions. Ref.~\cite{Ghosh:2019lmx}
have calculated shear viscosity of quark matter in presence of magnetic field, where
$\eta/s$ can be abruptly reduced because of lower effective relaxation time, build
by particle relaxation time and synchrotron frequency. However, further investigations
are necessary before getting the bold conclusion - magnetic field can be one of the source
for getting lower $\eta/s$ in RHIC/LHC matter.

\subsubsection{Bulk viscosity and electrical conductivity of quark and hadronic matter}
\begin{table}[!ht]
\begin{center}
 \caption{Same as Table~(\ref{tab1}) for $\zeta/s$.}
% \begin{ruledtabular}
 \begin{tabular}{l|cc}
 %& &  \\ 
 Framework$^{\rm Reference}$ & $T ~\leq~T_c$ & $T ~\geq~T_c$  \\
 %& &  \\
 \hline
 %& & \\
 LQCD~\cite{Meyer:2007dy} & - & $1$-$0$   \\
 %& & \\
 HTL~\cite{Arnold:2006fz} & - & $0.002$-$0.001$  \\
 %& & \\
 NJL~\cite{Marty:2013ita} & $0.9$-$0.02$ & $0.02$-$0.002$  \\
 %& & \\
 NJL~\cite{Sasaki:2008um} & $1.7$-$0.13$ & $0.13$-$0.005$  \\
 %& & \\
 NJL~\cite{Ghosh:2015mda} & - & $0.1$-$0.01$  \\
 %& & \\
 NJL~\cite{Deb:2016myz} & $0.61$-$0.11$ & $0.11$-$0.004$  \\
 %& & \\
 LSM~\cite{Chakraborty:2010fr} & $0.61$-$0.11$ & $0.11$-$0.004$  \\
 %& &  \\
 Unitarization~\cite{FernandezFraile:2009mi} & $0.04$-$0.027$ & -  \\
 %& & \\
 HRG~\cite{NoronhaHostler:2008ju} & $0.02$-$0.003$ & -  \\
 %& & \\
 HRG~\cite{Kadam:2015xsa} & $0.15$-$0.025$ & -  \\
 %& & \\
 HRG~\cite{Ghosh:2016clt} & $0.1$-$0.03$ & -  \\
 \end{tabular}
% \end{ruledtabular}
 \label{tab2}
 \end{center}
 \end{table}
 \begin{table}[!ht]
 \begin{center}
 \caption{Same as Table~(\ref{tab1}) for $\sigma/T$.}
% \begin{ruledtabular}
 \begin{tabular}{l|cc}
 %& &  \\ 
 & $m_\sigma$ & $\Gamma^0_\sigma$  \\
 %& &  \\
 \hline
 %& & \\
 LQCD~\cite{Gupta:2003zh} & - & $0.33$  \\
 %& & \\
 LQCD~\cite{Aarts:2014nba} & $0.002$ & $0.005$-$0.015$  \\
 %& & \\
 NJL~\cite{Marty:2013ita} & $0.02$-$0.015$ & $0.015$-$0.1$  \\
 %& & \\
 PHSD~\cite{Cassing:2013iz} & $0.1$-$0.02$ & $0.02$-$0.2$  \\
 %& & \\
 PQM~\cite{Singha:2017jmq} & $0.03$-$0.02$ & $0.01$  \\
 %& & \\
 Unitarization~\cite{FernandezFraile:2009mi} & $0.013$-$0.010$ & -  \\
 %& & \\
 HFT~\cite{Ghosh:2016yvt} & $0.004$-$0.001$ & -  \\
 %& &  \\
 \end{tabular}
% \end{ruledtabular}
 \label{tab3}
 \end{center}
 \end{table}
Similar to shear viscosity, other transport coefficients like bulk viscosity $\zeta$ and electrical
conductivity $\sigma$ are also rigorously investigated in recent times. List of references with
their estimated values of $\zeta/s$ and $\sigma/T$ are tabulated in Tables~(\ref{tab2}) and (\ref{tab3}).
Using same microscopic tools, which are able to estimate a very low $\eta/s$, one can found the order
of other transport coefficients, which are necessary  to know for complete dissipative hydrodynamical description. 
% By analyzing Eqs.~(\ref{Tr_tc}) and (\ref{Tr_gross}), one can notice the point - though
% all transport coefficients are proportional to the particle relaxation time but their thermodynamical
% phase-space are different. 
Based on the tabulated values, order of magnitude of transport 
coefficients of RHIC/LHC matter can be summarized by number: $\eta/s\approx 10^{0}$-$10^{-1}$,
$\zeta/s\approx 10^{0}$-$10^{-2}$, $\sigma/T\approx 10^{-1}$-$10^{-3}$.

\subsection{Summary}
\label{sabya_sec:sum}
To synchronize with the expectation from experimental side with macroscopic
hydrodynamical simulation, a long list of microscopic calculations are noticed in recent times.
Present draft has covered few selective microscopic model calculations,
which got a very low viscosity to entropy density ratio of quark/hadronic matter,
close to KSS bound. By providing a tabulated format of estimated values for transport
coefficients, present draft has attempt to make a catalogue on
microscopic calculations of transport coefficients.
Based on investigation series, given in Refs.~\cite{Ghosh:2015mda,Singha:2017jmq,Ghosh:2014qba, Ghosh:2014ija, Ghosh:2015lba, Rahaman:2017sby,Ghosh:2018nqi,Ghosh:2019fpx,Ghosh:2019lmx},
we have concluded that
three possible sources - (1) resonance type interaction, (2) finite size effect
and (3) effect of magnetic field, might be responsible for (nearly) perfect fluid nature
of RHIC/LHC matter.

%................................................................................

%% file: Arpan/arpan.tex
%%\documentclass[prd,superscriptaddress,nofootinbib,showpacs,preprint]{revtex4}
%%\documentclass[prc,twocolumn,nofootinbib]{revtex4}
%\documentclass[prc,nofootinbib]{revtex4}
%\usepackage{graphicx}
%\usepackage[usenames,dvipsnames]{color}
%\usepackage{amsmath,amssymb,bbold}
%% \usepackage[colorlinks]{hyperref}
%\usepackage{hyperref}
%\usepackage{comment}
%%\usepackage{refcheck}
%\usepackage{slashed}
%%\usepackage{txfonts}
%\usepackage{mathrsfs}
%\usepackage{bbm}
%\begin{document}

\section{Chiral transition in a chirally imbalanced plasma in the presence of magnetic field: a Wigner function approach}

%\author{Arpan Das$^{1}$}
%\email{arpan@prl.res.in}
%\author{Deepak Kumar$^{1,2}$}
%\email{deepakk@prl.res.in}
%\author{Hiranmaya Mishra$^{1}$}
%\email{hm@prl.res.in}

\textit{Arpan Das, Deepak Kumar, Hiranmaya Mishra}

\bigskip

%\affiliation{$^{1}$Theory Division, Physical Research Laboratory, 
%Navrangpura, Ahmedabad 380 009, India}
%\affiliation{$^2$ Indian Institute of Technology Gandhinagar,
%	Gandhinagar 382 355, Gujarat, India}

%\begin{abstract}
{\small
We discuss here the chiral transition and the associated chiral susceptibility for a chirally imbalanced plasma in the 
presence of a magnetic field ($B$) using the Wigner function approach within the framework of the 
Nambu-Jona-Lasinio model (NJL). As a regularization prescription, we use a medium separation regularization 
scheme (MSS) in the presence of magnetic field and chiral chemical potential ($\mu_5$)
to estimate the chiral condensate and chiral susceptibility.
We found that chiral transition temperature increases with the magnetic field,
while the transition temperature decreases with $\mu_5$. 
For a strong magnetic field, we find that the chiral transition temperature as well as susceptibility for up and down 
type quarks can be non degenerate.}
%\end{abstract}

\bigskip

%\pacs{}
%\maketitle

\subsection{Introduction}
\label{intro}
Relativistic heavy-ion collision experiments e.g. at RHIC and LHC, strongly indicate the
formation of deconfined quark-gluon plasma (QGP) phase of quantum chromodynamics (QCD) in the 
initial stages of heavy ion collision experiments as well as confined hadron phase in the subsequent evolution of QGP.
Two very important characteristic features of QCD are color confinement and spontaneous breaking of chiral symmetry. 
At vanishing temperature and/or density ground state of QCD does not have chirally symmetry.
The QCD vacuum undergoes a transition from a chiral symmetry broken phase to
a chiral symmetric phase, with an increase in temperature and/or baryon density.
Quark-antiquark scalar condensate is the order parameter of the chiral transition.

The study of fluctuations of conserved charges, e.g. net electric charge, baryon number, strangeness, etc.,
play an important role to explore the QCD phase diagram~\cite{Jeon:1999gr}.  
In this investigation, we study chiral transition and the associated chiral 
susceptibility using the Wigner function approach.
Wigner function is the quantum mechanical analog of
the classical distribution function. It encodes quantum corrections in the transport equation~\cite{Elze:1986qd,Elze:1986hq,DeGroot:1980dk}. 
The covariant Wigner function method for spin-1/2 fermions has been explored  
to study chiral magnetic effect (CME), dynamical 
generation of magnetic moment etc.~\cite{Weickgenannt:2019dks,Mao:2018jdo,Sheng:2018jwf,Sheng:2017lfu}.
Chiral susceptibility which is the measure of the response of the chiral condensate 
to the variation of the current quark mass, has been investigated earlier
using lattice QCD (LQCD)
simulations~\cite{Karsch:1994hm}, Nambu-Jona-Lasinio
(NJL) model~\cite{Zhuang:1994dw,Sasaki:2006ww} etc.

A nonvanishing magnetic field of the order of several $m_{\pi}^2$ is expected to be generated in noncentral relativistic heavy-ion collision experiments at RHIC and LHC
 \cite{Kharzeev:2007jp,Shovkovy:2012zn}. This apart   
 non-trivial topological field configurations of gluons and Adler-Bell-Jackiw (ABJ) anomaly can give rise to an asymmetry between the number of left and right chiral quarks, i.e. finite chiral chemical potential ($\mu_5$)~\cite{Kharzeev:2013ffa}.  
 In this work, we investigate the chiral  transition and chiral susceptibility in the presence of magnetic 
 field and chiral chemical potential ($\mu_5$) in quantum kinetic theory framework using Nambu Jona Lasinio (NJL) model~\cite{Buballa:2003qv}.
 It is important to mention that in Ref.\cite{Fukushima:2010fe} it was observed that the chiral transition temperature decreases with $\mu_5$.
 With a smooth cutoff for the three momentum
 it was observed that with increasing $\mu_5$ the chiral transition becomes a first-order transition 
 \cite{Chernodub:2011fr,Gatto:2011wc,Fukushima:2010fe}.
 Nonlocal NJL model analyzed in Ref.~\cite{Ruggieri:2016ejz} shows that the chiral transition temperature 
 increases with $\mu_5$ and the chiral transition is second order. However in Ref.~\cite{Yu:2015hym} it has been 
 observed that chiral
 transition temperature decreases with $\mu_5$ with a smooth cutoff and shows a first-order 
 transition at large $\mu_5$. On the other hand NJL model with``medium separation scheme'' (MSS) regularization,
 as investigated in Ref.~\cite{Farias:2016let} shows that the chiral transition temperature increases with $\mu_5$
 which is in accordance with some LQCD results~\cite{Braguta:2015owi,Braguta:2015zta}.
 In this work we use a medium separation scheme in the presence of magnetic field and $\mu_5$~\cite{Farias:2016let,Farias:2005cr,Avancini:2019ego}. We find that 
 chiral transition temperature decreases with $\mu_5$ as in Refs.~\cite{Fukushima:2010fe,Yu:2015hym}.

 We organize the paper in the following manner. 
 In Sec.\eqref{wignermag} we introduce the Winger 
 function in the presence of magnetic field as well as $\mu_5$ and calculate the chiral 
 condensate and chiral susceptibility for two flavor NJL model.
 In Sec.\eqref{arpan_results} we present the results and discussions.
 Finally, in Sec.\eqref{conclu} we conclude our results with an outlook to it.
 
 \subsection{Chiral condensate and chiral susceptibility in NJL model for non vanishing magnetic field and chiral chemical potential}
\label{wignermag}

Using the solutions of the Dirac equation in magnetic field and finite $\mu_5$ the Wigner function has been explicitly written down 
in Ref.\cite{Sheng:2017lfu}. Gauge invariant Wigner function in the presence of magnetic field as given in Ref.\cite{Sheng:2017lfu} is,
\begin{align}
 W_{\alpha\beta}(X,p)=\int\frac{d^4X^{\prime}}{(2\pi)^4} e^{(-ip_{\mu}X^{\prime\mu}-iq Byx^{\prime})}\bigg\langle\bar{\psi}_{\beta}\bigg(X+\frac{X^{\prime}}{2}\bigg)
 \otimes\psi_{\alpha}\bigg(X-\frac{X^{\prime}}{2}\bigg)\bigg\rangle,
 \label{equ32}
\end{align}
where a specific gauge choice of the external magnetic field is $A^{\mu}(X)=(0,-By,0,0)$.
$q$ is the electric charge of the particle. One can express the scalar condensate in terms of the Wigner
function as~\cite{Florkowski:1995ei},
\begin{align}
 \langle\bar{\psi}\psi\rangle = \int d^4p F(X,p), ~~~\text{where}, ~~F(X,p)= \text{Tr}~W(X,p).
 \label{equ16}
\end{align} 
For a system in global equilibrium with uniform temperature and 
chemical potential, as considered in this investigation,
Wigner function is independent of space time. Once the Wigner function is known it is straight forward 
to calculate the scalar condensate using Eq.\eqref{equ16}.
For two flavour NJL model with $u$ and $d$ quarks for non vanishing
magnetic field and chiral chemical potential as given by the following Lagrangian \cite{Buballa:2003qv,Dmitrasinovic:1996fi},
\begin{align}
 \mathcal{L}&=\bar{\psi}(i\slashed{D}-m+\mu_5\gamma^0\gamma^5)\psi+G_1\sum_{a=0}^3\left[(\bar{\psi}\tau^a\psi)^2+(\bar{\psi}i\gamma_5\tau^a\psi)^2\right]\nn\\
 &\hspace{1cm}+G_2\left[(\bar{\psi}\psi)^2-(\bar{\psi}\vec{\tau}\psi)^2-(\bar{\psi}i\gamma_5\psi)^2+(\bar{\psi}i\gamma_5\vec{\tau}\psi)^2\right],
 \label{equ53}
\end{align}
where $\psi$ is the $U(2)$ quark doublet $\psi=(\psi_u,\psi_d)^T$, 
the chiral condensate can be shown to be $\langle\bar{\psi}\psi\rangle^{\mu_5\neq0}_{B\neq0} =
 \sum_{f=u,d}\langle\bar{\psi}_f\psi_f\rangle^{\mu_5\neq0}_{B\neq0}$~\cite{Das:2019crc}, where 
\begin{align}
\langle \bar{\psi}_f \psi_f \rangle^{\mu_5\neq0}_{B\neq0} & = -\frac{N_c|q_f|B}{(2\pi)^2} \bigg[\int dp_z\, \frac{M_f}{E^{(0)}_{p_z,f}} 
\Big[ 1-f_{FD}(E^{(0)}_{p_z,f}-\mu)\, - 
f_{FD}(E^{(0)}_{p_z,f}+\mu)\Big]\,\nonumber\\
 & +  \sum_{n=1}^{\infty}\sum_s \int dp_z\,   \frac{M_f}{E^{(n)}_{p_z,s,f}}\Big[1-f_{FD}(E^{(n)}_{p_z,s,f}-\mu) 
 - \,f_{FD}(E^{(n)}_{p_z,s,f}+\mu)\Big]\bigg]\nonumber\\
 & = \langle \bar{\psi}_f \psi_f \rangle^{\mu_5\neq0}_{vac, B\neq0}+\langle \bar{\psi}_f 
 \psi_f \rangle^{\mu_5\neq0}_{med, B\neq0}.
 \label{equ57}
 \end{align}
 Here $N_c$ is the number of colors, $\mu$ is the quark chemical potential,
 $n$ denotes the Landau levels, $s=\pm$ denotes the spin states and $f_{FD}$ is the Fermi-Dirac distribution function.
The single particle energy of flavour $f$ can be expressed as,
$E_{p_z,f}^{(0)}= \sqrt{M^2_f+(p_z-\mu_5)^2}~~\text{for}~~n=0, ~~~~E_{p_z,s,f}^{(n)}=\sqrt{M^2_f+\left(\sqrt{p_z^2+2n|q_f|B}-s\mu_5\right)^2}~~\text{for}
~~n>0.$  The second term in Eq.~\eqref{equ53} is the four Fermi interaction.
$\tau^a, a=0,..3$ are the $U(2)$ generators in the flavour space. 
Third term is the t-Hooft interaction terms which introduces flavour mixing.
In the mean field approximation, the constituent quark masses for 
$u$ and $d$ quarks in terms of the chiral condensates can be given as, 
$M_u = m_u -4 G_1 \langle\bar{\psi}_u\psi_u\rangle-4 G_2 \langle\bar{\psi}_d\psi_d\rangle , 
~~M_d = m_d -4 G_1 \langle\bar{\psi}_d\psi_d\rangle-4 G_2 \langle\bar{\psi}_u\psi_u\rangle,$ respectively.
 $\langle \bar{\psi}_f \psi_f \rangle^{\mu_5\neq0}_{vac, B\neq0}$ which is the zero temperature and 
 zero quark chemical potential $(\mu)$ part,  contains 
 divergent integral which has been regularized using medium Separation Scheme (MSS)
 outlined in Ref.\cite{Duarte:2018kfd,Das:2019crc}.
 
 The chiral susceptibility measures the response of the chiral condensate to the infinitesimal
change of the current quark mass. Chiral susceptibility in two flavour NJL model can be defined as, 
$ \chi_c  =\frac{\partial\langle\bar{\psi}\psi\rangle}{\partial m} = \frac{\partial\langle\bar{\psi}_u\psi_u\rangle}{\partial m}
 +\frac{\partial\langle\bar{\psi}_d\psi_d\rangle}{\partial m}$.
To estimate chiral susceptibility we have to estimate
$\frac{\partial\langle\bar{\psi}_f\psi_f\rangle}{\partial M_{f}}$. Similar to chiral condensate,
$\frac{\partial\langle\bar{\psi}_f\psi_f\rangle}{\partial M_{f}}$ also has divergent integrals which can be 
regularized using the medium separation regularization scheme~\cite{Das:2019crc}.

\subsection{Results}
\label{arpan_results}
NJL model as described by Eq.~\eqref{equ53} has the following parameters, two couplings $G_1$, $G_2$, the three momentum cutoff 
$\Lambda$ and the current quark masses $m_u$ and $m_d$. 
To study the effects of flavour mixing, the couplings $G_1$ and $G_2$ are 
parametrized as $G_2=\alpha g$, $G_1=(1-\alpha)g$~\cite{Buballa:2003qv,Dmitrasinovic:1996fi}. The extent of flavour mixing 
is controlled by $\alpha$ and it can be argued to have a value $0\leq\alpha\leq0.5$~\cite{Boomsma:2009eh}. 
For phenomenological reason we take the parameters
$m_u=m_d = 6$ MeV, $\Lambda=590$MeV and  $g = 2.435/\Lambda^2$~\cite{Buballa:2003qv}. Next we show some of the important results of our study (for details see Ref.~\cite{Das:2019crc}).
\begin{figure}[!ht]
    \centering
    \begin{minipage}{.5\textwidth}
        \centering
        \includegraphics[width=1.1\linewidth]{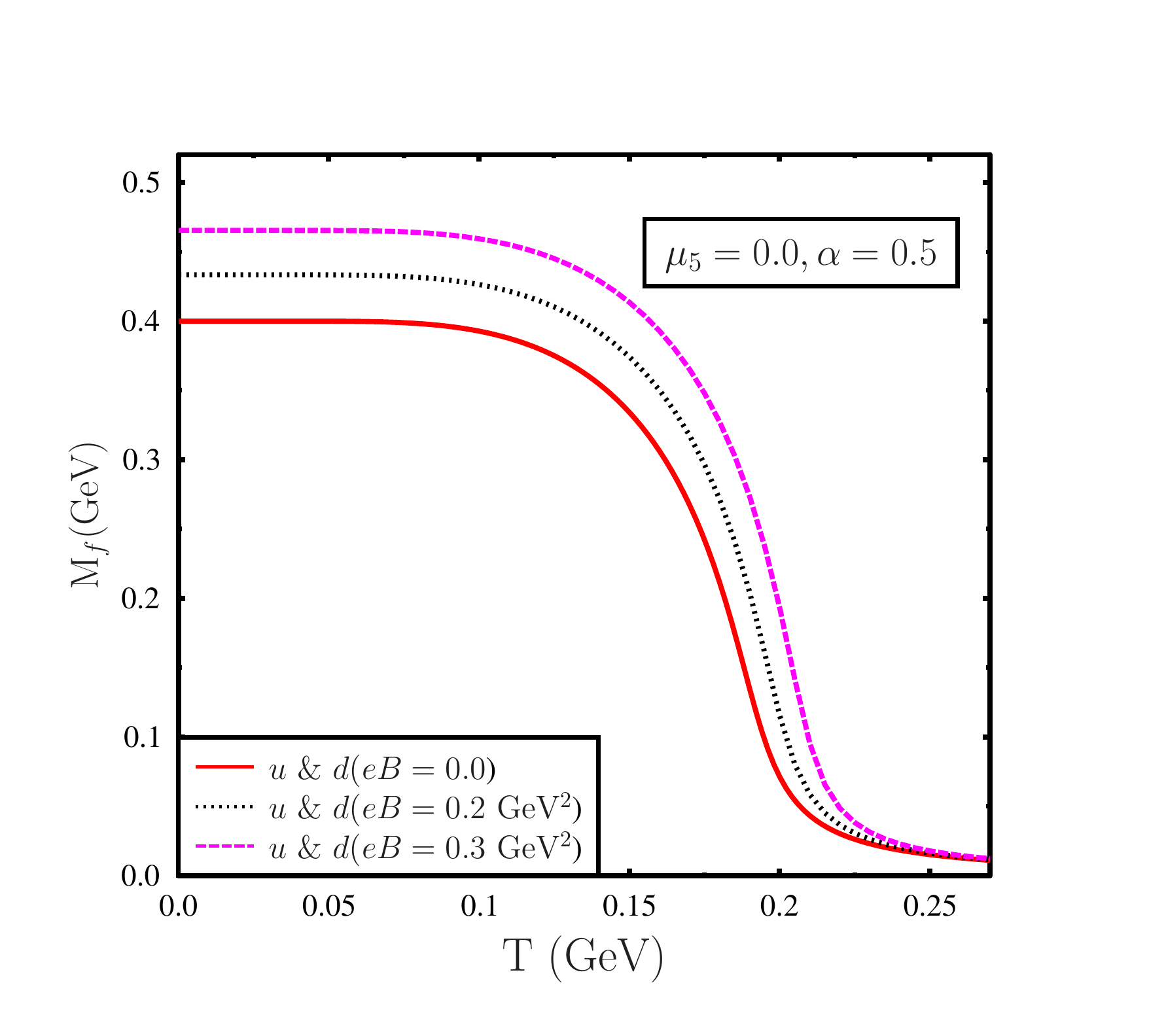}
    \end{minipage}%
    \begin{minipage}{0.5\textwidth}
        \centering
        \includegraphics[width=1.1\linewidth]{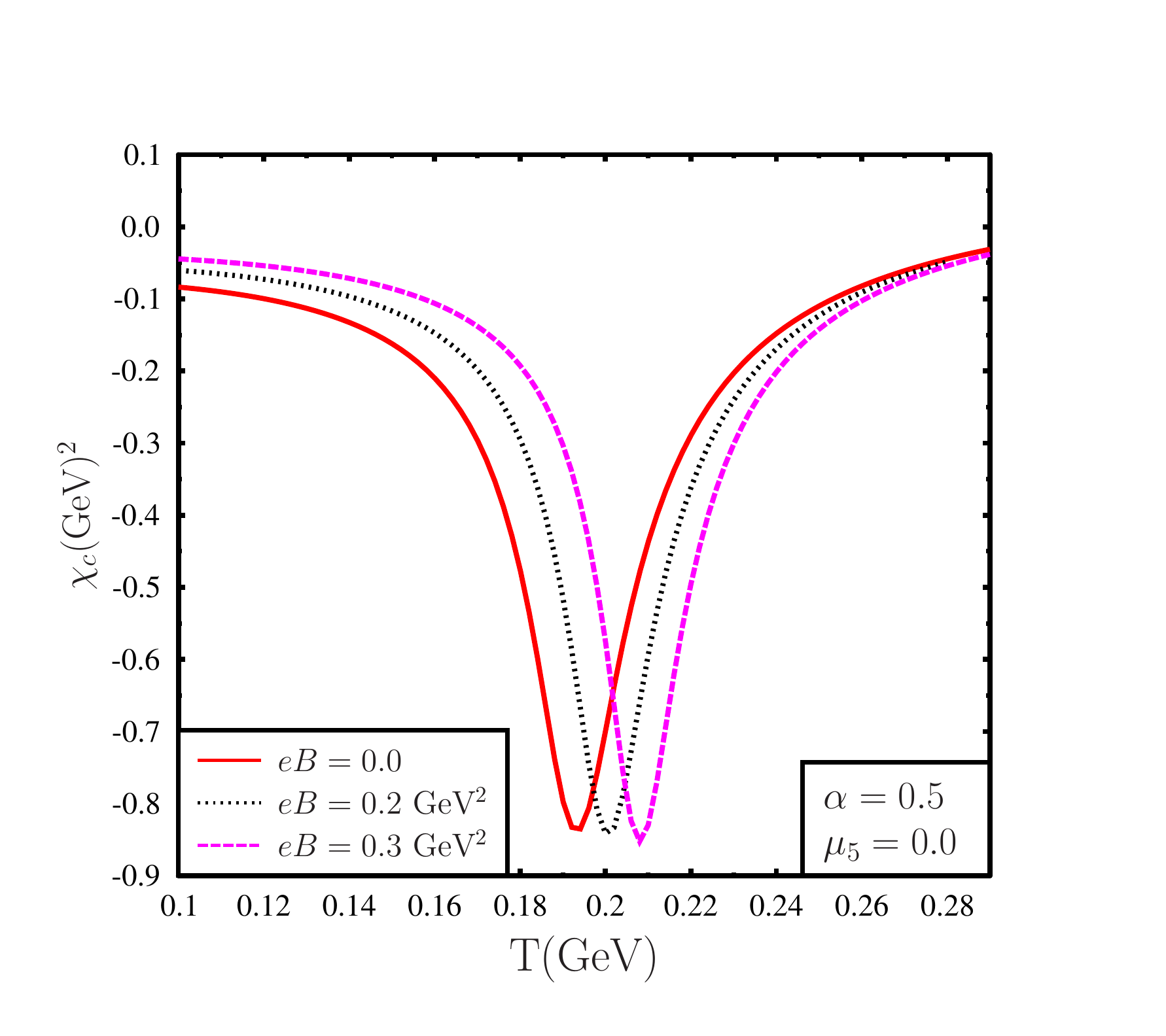}
    \end{minipage}
    \caption{Left plot: Variation of $M_u$ and $M_d$,
with temperature for $\mu_5=0$, but with different values of magnetic field for
$\alpha=0.5$. Right plot: Variation of 
$\chi_c$ with temperature ($T$) for $\mu_5=0$,
but with different values of magnetic field for $\alpha=0.5$~\cite{Das:2019crc}.}
    \label{arpan_fig4}
\end{figure}

In Fig.~\eqref{arpan_fig4} we show the variation of $M_u$ and $M_d$ and the chiral susceptibility ($\chi_c$), with temperature for $\mu_5=0$ and 
with different values of magnetic field for $\alpha=0.5$. Even in the presence of magnetic field $M_u=M_d$ for $\alpha = 0.5$. From the left and the right plot in Fig.~\eqref{arpan_fig4} it is clear that constituent quark mass and chiral transition temperature increases with increasing magnetic field.

\begin{figure}[!ht]
    \centering
    \begin{minipage}{.5\textwidth}
        \centering
        \includegraphics[width=1.1\linewidth]{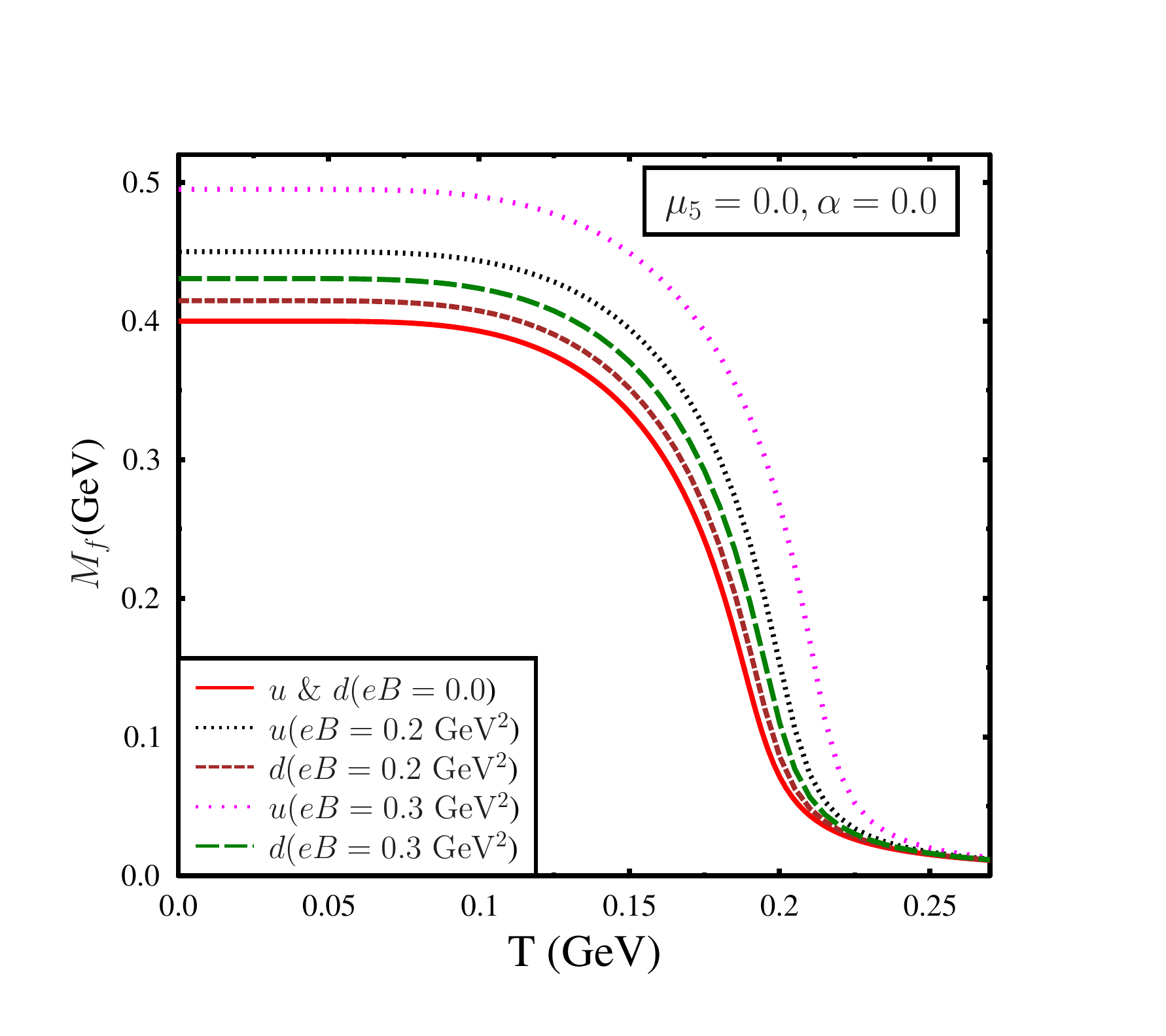}
    \end{minipage}%
    \begin{minipage}{0.5\textwidth}
        \centering
        \includegraphics[width=1.1\linewidth]{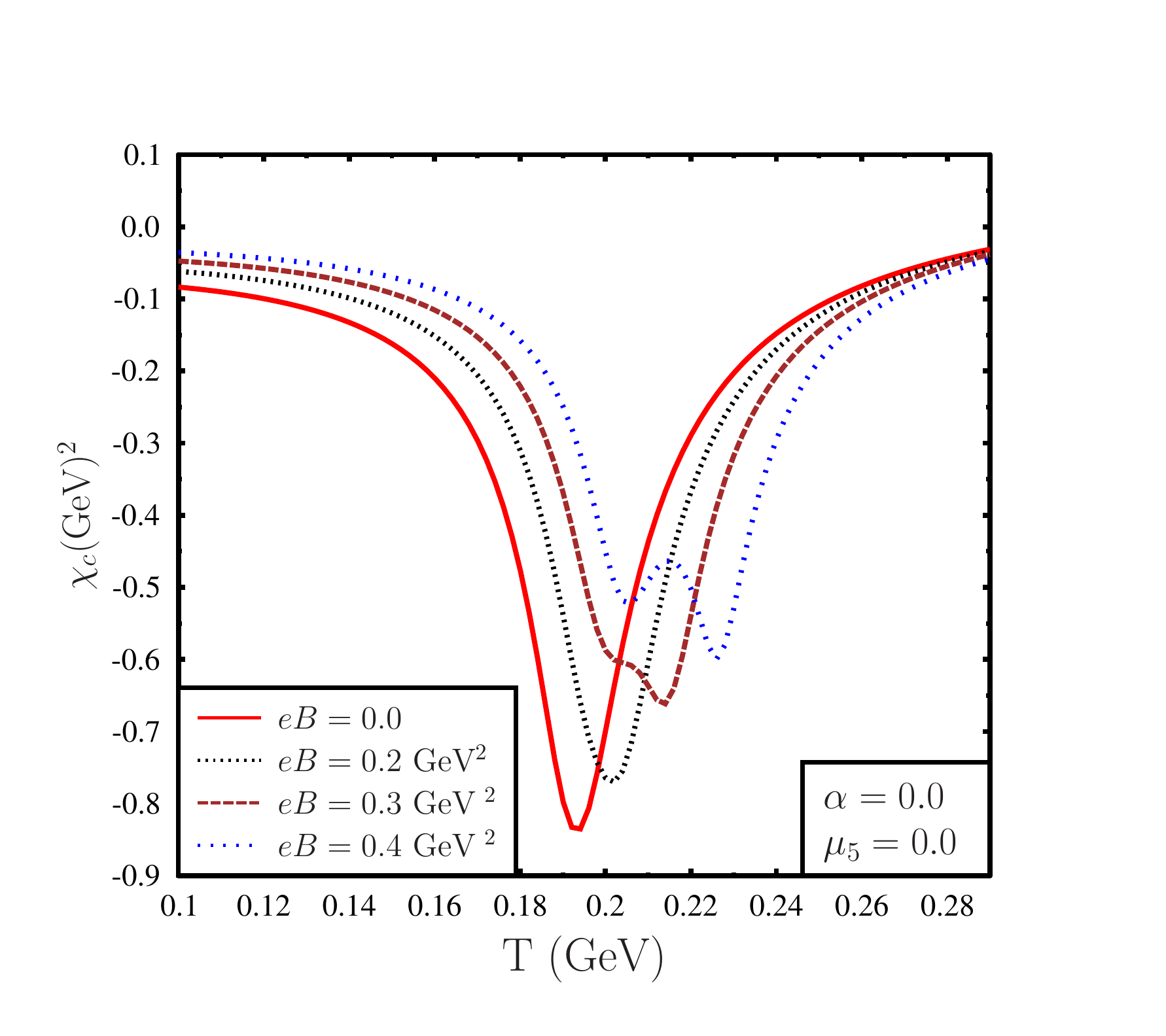}
    \end{minipage}
    \caption{Left plot: Variation of  $M_u$ and $M_d$,
with temperature for $\mu_5=0$, but with different values of magnetic field for $\alpha=0.0$.
Right plot: Variation of $\chi_c$ with temperature ($T$) for $\mu_5=0$,
but with different values of magnetic field for $\alpha=0.0$~\cite{Das:2019crc}.}
    \label{arpan_fig5}
\end{figure}

In Fig.~\eqref{arpan_fig5} we show the variation of $M_u$ and $M_d$ and 
the associated total chiral susceptibility ($\chi_c$),
with temperature for $\mu_5=0$ and with different values of magnetic field for $\alpha=0.0$. 
For $\alpha = 0.0$ there is no flavour mixing. From the 
left plot it is clear that at finite magnetic field $M_u\neq M_d$. For non vanishing magnetic field 
$u$ and $d$ quark condensates are different and for $\alpha=0.0$, $M_u$ is 
independent of $\langle\bar{\psi}_d\psi_d\rangle$. Similarly $M_d$ does not depend on  
$\langle\bar{\psi}_u\psi_u\rangle$  for $\alpha=0.0$. 
From the right plot in Fig.~\eqref{arpan_fig5} it is clear 
that chiral transition temperature increases with increasing magnetic field.
However unlike the case when $\alpha=0.5$, in this case susceptibility plot shows two distinct peaks for relatively large 
magnetic fields. These two peaks are associated with $u$ and $d$ quarks. In general for $\alpha\neq0.5$ chiral transition temperature associated with $u$ and $d$ type quarks are non degenerate.

\begin{figure}[!ht]
    \centering
    \begin{minipage}{.5\textwidth}
        \centering
        \includegraphics[width=1.1\linewidth]{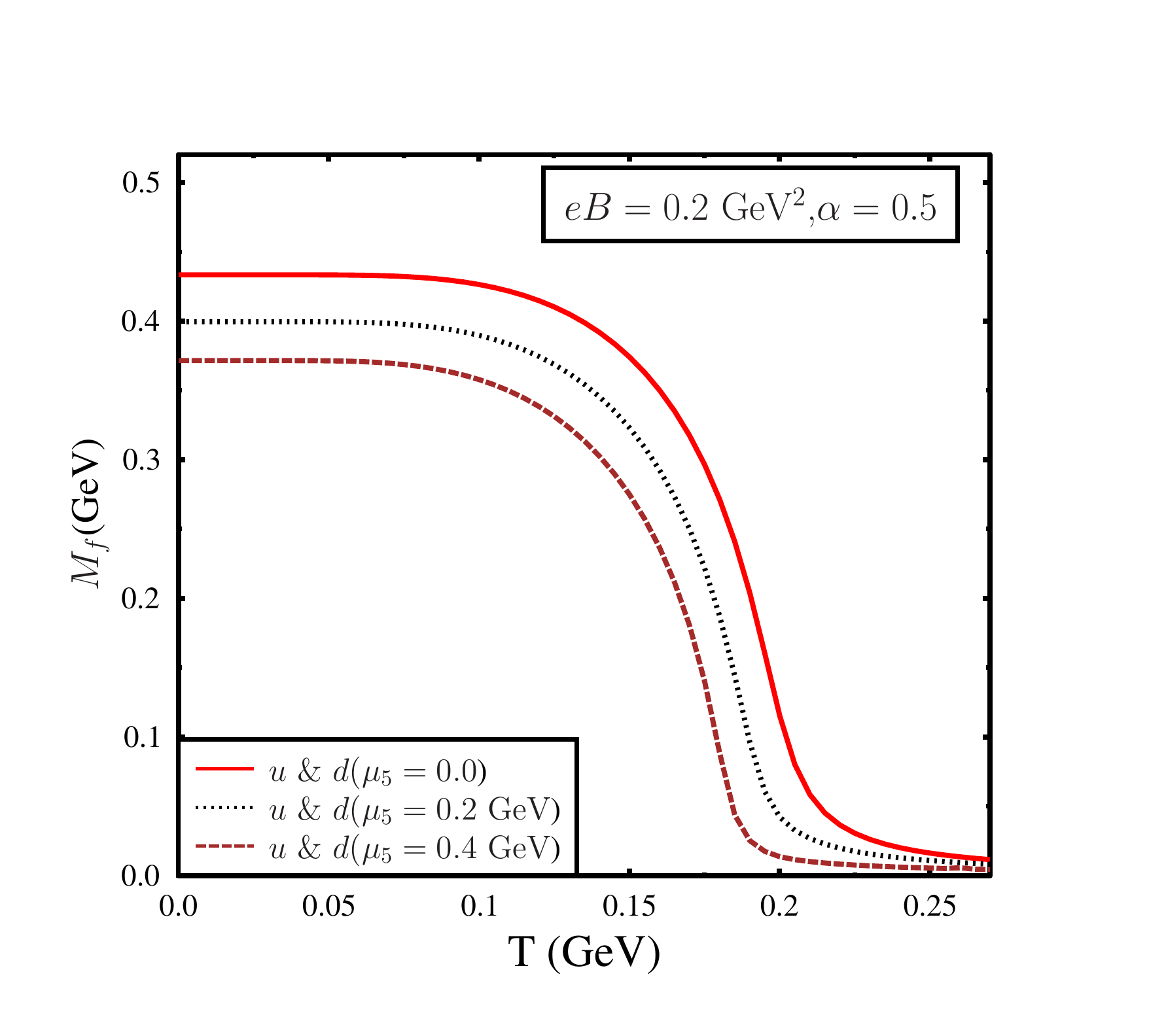}
    \end{minipage}%
    \begin{minipage}{0.5\textwidth}
        \centering
        \includegraphics[width=1.1\linewidth]{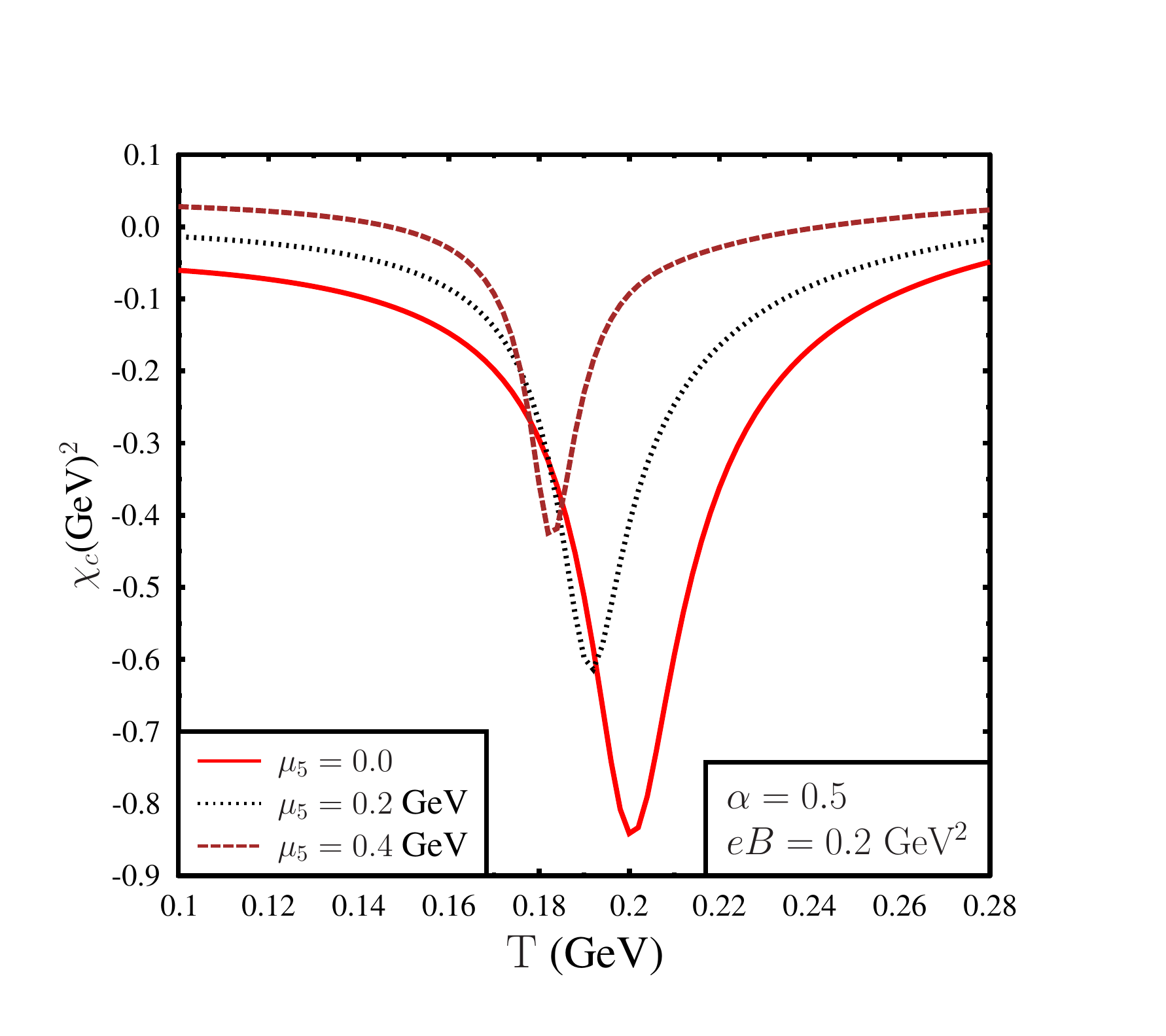}
    \end{minipage}
    \caption{Left plot: Variation of quark mass $M_u=M_d$, with temperature
    for finite $B$ and $\mu_5$. Right Plot: Variation of $\chi_c$ with temperature
    for finite $B$ and $\mu_5$ \cite{Das:2019crc}.}
    \label{arpan_fig7}
\end{figure}

 Finally in Fig.~\eqref{arpan_fig7} we show the variation of $M_u$ and $M_d$ and the associated
 susceptibilities $(\chi_c)$ with temperature for non vanishing magnetic field and 
 chiral chemical potential for $\alpha=0.5$. 
 From Fig.~\eqref{arpan_fig7} it is clear that with $\mu_5$ quark mass as well as transition temperature
 decreases.

 \subsection{Conclusion}
 \label{conclu}
 In this investigation we have studied chiral transition and the associated chiral susceptibility 
 for non vanishing magnetic field and $\mu_5$ using Wigner function approach within the framework of two flavour NJL model. We used 
 a medium separation regularization scheme to regulate divergent integral.
 With increasing $\mu_5$ constituent quark masses and the chiral transition temperature
 decreases. On the other hand with increasing magnetic field quark masses and chiral transition
 temperature increases. Further in the absence of maximal flavour mixing, i.e. $\alpha\neq0.5$, $u$ quark mass is larger than 
 $d$ quark mass for non vanishing magnetic field. Also chiral susceptibility shows two distinct peaks
 for high magnetic field associated with $u$ and $d$ quarks for $\alpha\neq0.5$.

%% file: Ranjita/ranjita.tex
\section{Electrical conductivity and Hall conductivity of hot and dense hadron 
gas in a magnetic field}

\textit{Arpan Das, Hiranmaya Mishra, Ranjita K. Mohapatra}

\bigskip

%\author{Arpan Das}
%\author{Hiranmaya Mishra}
%\author{Ranjita K. Mohapatra}
%\email{ranjita@iitb.ac.in}
%\affiliation{Theory Division, Physical Research Laboratory,
%Navrangpura, Ahmedabad 380009, India}
%\affiliation{Department of Physics, Indian Institute of Technology Bombay, Mumbai, 400076, India}
%
%\begin{abstract}
{\small
We estimate the electrical conductivity and the Hall conductivity of hot and dense hadron gas using the relaxation time approximation of 
the Boltzmann transport equation in the presence of electromagnetic field. We have investigated the temperature and the baryon chemical
potential dependence of these transport coefficients. We find that the electrical 
conductivity decreases in the presence of magnetic field. The Hall conductivity 
on the other hand shows a non monotonic behavior with respect to the dependence 
on magnetic field. We argue that for a pair plasma (particle-anti 
particle plasma) where $\mu_B=0$, Hall conductivity vanishes. Only for non vanishing  baryon chemical potential 
Hall conductivity has non zero value. We also estimate the electrical conductivity and the Hall conductivity as a function of the center
of mass energy along the freeze out curve.
}

\bigskip

%\end{abstract}
%
%\pacs{25.75.-q, 12.38.Mh}
%\maketitle

\subsection{Introduction}

\label{ranjita_intro}

  Transport coefficients of strongly interacting matter created in the relativistic 
heavy ion collision
experiments are of great importance for a comprehensive 
understanding of the hot and dense QCD (quantum chromodynamics) medium produced 
in these experiments. In the dissipative relativistic hydrodynamical model of the 
hot and dense medium, transport coefficients, e.g. 
shear and bulk viscosity etc plays an important role. In fact, it has been shown 
that a small shear viscosity to entropy ratio ($\eta/s$) is necessary to explain the 
flow data. The bulk viscosity $\zeta$, also plays a significant role  
in the dissipative hydrodynamics describing the QGP evolution. The bulk viscosity encodes the conformal measure $(\epsilon-3P)/T^4$ of the system and lattice QCD simulations shows a non monotonic
behaviour of both $\eta/s$ and $\zeta/s$  near the  critical temperature $T_c$.

In case of non central heavy ion collisions, due to the collision geometry, a large 
magnetic field is also expected to be produced. The magnitude of the produced magnetic field 
at the initial stages in these collisions are expected to be rather 
large, at least of the order of  several $m_{\pi}^2$. Since the strength
of the magnetic field is of hadronic scale, the effect of the magnetic field 
on the QCD medium can be significant. 

  In the present work, we investigate the electrical and the Hall conductivity of the hot 
and dense hadron gas produced in the subsequent evolution of QGP. 

\subsection{Boltzmann equation in relaxation time approximation}
\label{formalism}
%\pagebreak
The relativistic Boltzmann transport equation (RBTE) of a charged particle 
of single species in the presence of external electromagnetic field can 
be written as, 

\begin{align}
 p^{\mu}\partial_{\mu}f(x,p)+eF^{\mu\nu}p_{\nu}\frac{\partial f(x,p)}{\partial p^{\mu}} = \mathcal{C}[f],
 \label{equ1}
\end{align}

$\mathcal{C}[f]$ is the collision integral. In the relaxation time
approximation (RTA) the collision integral can be written as, 

\begin{align}
 \mathcal{C}[f]\simeq -\frac{p^{\mu}u_{\mu}}{\tau}(f-f_0)\equiv -\frac{p^{\mu}u_{\mu}}{\tau}\delta f ,
 \label{equ2}
\end{align}

Electric current is given by, 

\begin{align}
 j^i=e\int\frac{d^3p}{(2\pi)^3}v^i\delta f =\sigma^{ij}E_j=\sigma^{el}\delta^{ij}E_j+\sigma^H\epsilon^{ij}E_j,
\end{align}
where $\epsilon_{ij}$ is the anti symmetric $2\times2$ unity tensor, with $\epsilon_{12}=-\epsilon_{21}=1$. Then the electrical and the Hall 
conductivity can be identified as, 

\bea
 \sigma^{el}&=&\sum_i
 \frac{e^2_i\tau_i}{3T}\int\frac{d^3p}{(2\pi)^3}\frac{p^2}{\epsilon^2_i} \frac{1}{1+(\omega_{ci}\tau_i)^2}f_0,\\
 \label{equ19}
%%%%
 \sigma^{H}&= &\sum_i\frac{e^2_i\tau_i}{3T}\int\frac{d^3p}{(2\pi)^3}\frac{p^2}{\epsilon^2_i} \frac{\omega_{ci}\tau_i}{1+(\omega_{ci}\tau_i)^2}f_0,
 \label{equ20}
\eea

\subsection{Results and discussions}
\label{results}
We have considered an uniform radius of $r_h=0.5$ fm for all the mesons and baryons 
we have estimated the electrical conductivity and the Hall conductivity using 
Eq.\eqref{equ19} and Eq.~\eqref{equ20}.

\begin{figure}[!ht]
\begin{center}
\includegraphics[width=0.48\textwidth]{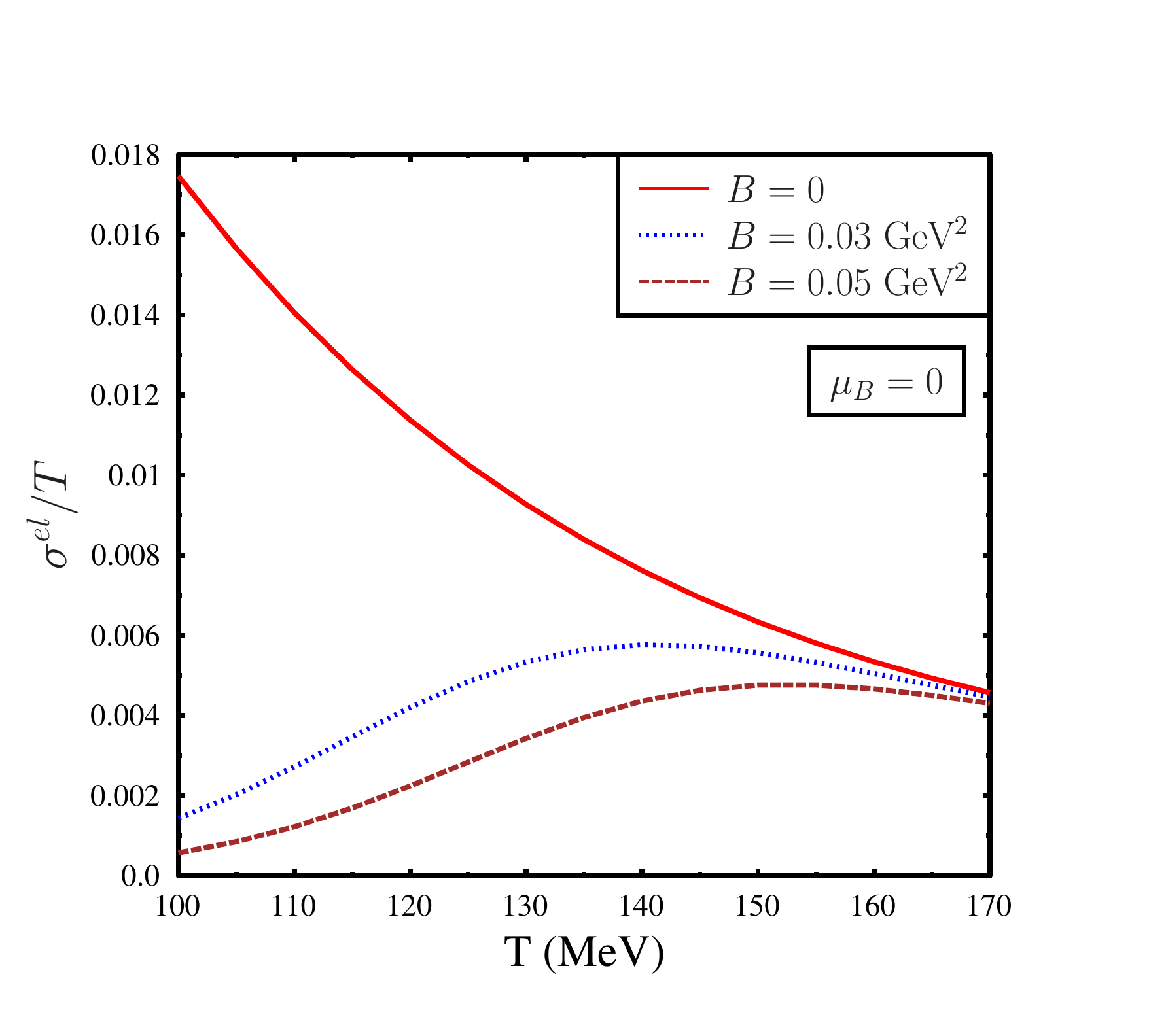}
\includegraphics[width=0.48\textwidth]{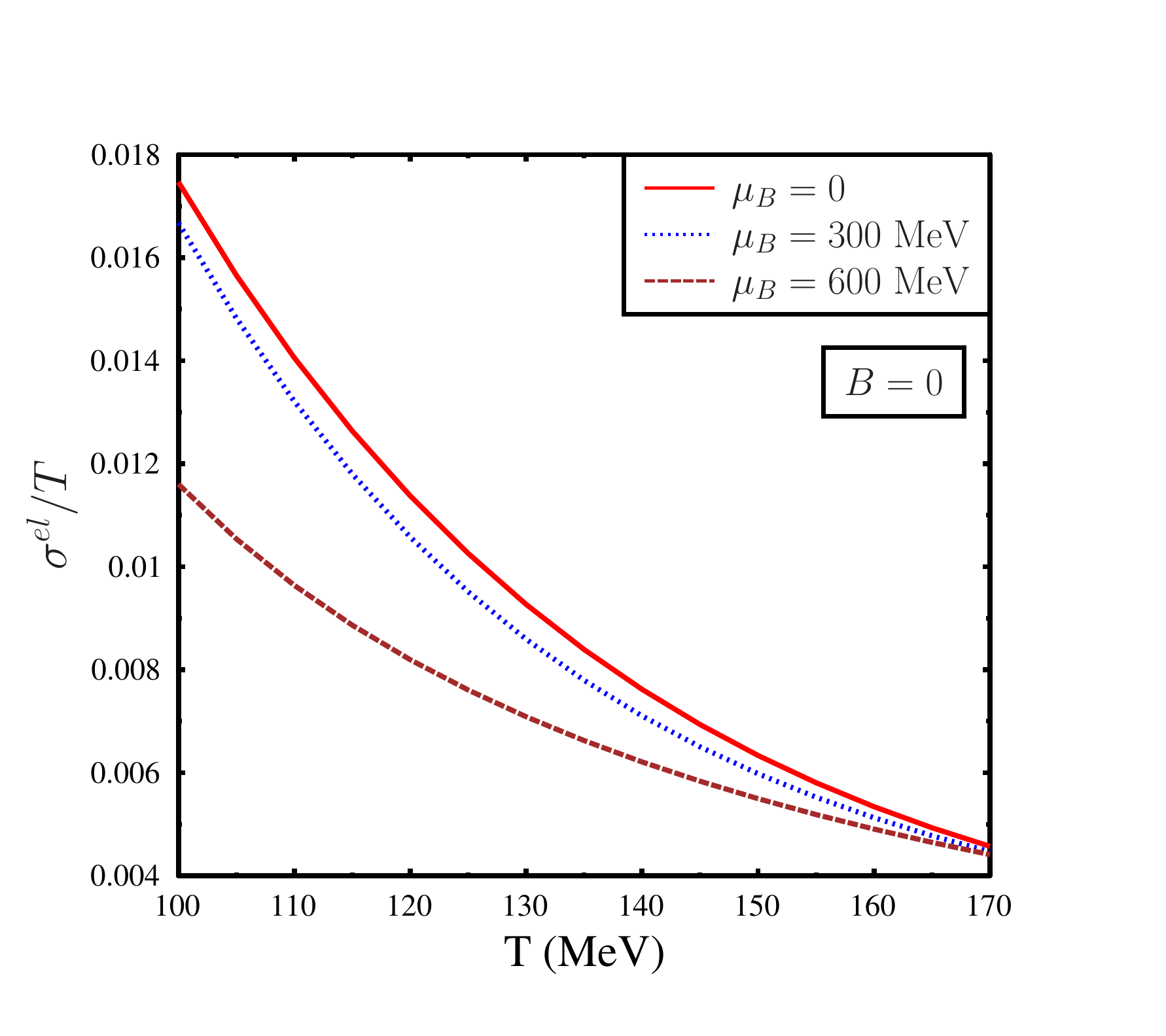}
\caption{Left plot: Variation of normalized electrical conductivity ($\sigma^{el}/T$) with
temperature ($T$) for different values of magnetic field ($B$) at
zero baryon chemical potential, Right Plot: Variation of $\sigma^{el}/T$ with 
temperature $T$ for different values of
baryon chemical potential $\mu_B$ at zero magnetic field.} 
\label{ranjita_fig1}
\end{center}
\end{figure}

It is clear from Fig.~\ref{ranjita_fig1} left plot, $\sigma^{el}/T$ decreases monotonically with 
temperature at $B=0$. This can be associated with the increase of randomness of the system with larger collision rate leading to smaller relaxation time. We point out here that the dominant contribution to the 
electrical conductivity arises from the charged pions due to the small mass of 
the pions as compared to that of other hadrons. Thus the monotonic decrease 
of $\sigma^{el}/T$ is due to the decrease of relaxation time of pions with increasing 
temperature.

For non vanishing magnetic field, the behaviour of $\sigma^{el} /T$ is very different 
as compared to $B=0$ counterpart. Firstly, it is observed that with increase in magnetic 
field strength the electrical conductivity decreases. This decrease in electrical 
conductivity with the magnetic field can be understood physically. At zero magnetic 
field, the electric current is along the direction of the electric field. However,
at finite magnetic field, charges also diffuse transverse to both electric and magnetic field, 
due to the Lorentz force, giving rise to a reduced current along the direction of electric field.
This is also reflected in the expression for electrical conductivity as in Eq.~\eqref{equ19}.

It is clear from Fig.~\eqref{ranjita_fig1} (right plot) that with increasing chemical potential ($\mu_B$) 
electrical conductivity decreases. For the range of $\mu_B$ considered here the 
contribution to the electrical conductivity from the charged hadrons is dominated by 
the charged pions similar to the case with vanishing chemical potential. At finite 
chemical potential the pion relaxation time decreases with $\mu_B$ due to scattering 
with the baryons, mostly from the nucleons. One would have naively expected 
the nucleon contribution to the electrical conductivity to 
increase with $\mu_B$, which will lead to an increase in the total electrical 
conductivity due to the $\mu_B$ dependent distribution function in the expression 
of electrical conductivity. However this increase of the baryonic contribution to 
the electrical conductivity is not enough to compensate the decreasing 
contribution arising from pions, at least for the chemical potential considered 
in the present investigation. This leads to a decrease of the total electrical 
conductivity with increase in baryon chemical potential at vanishing magnetic field.

\begin{figure}[!ht]
\begin{center}
\includegraphics[width=0.55\textwidth]{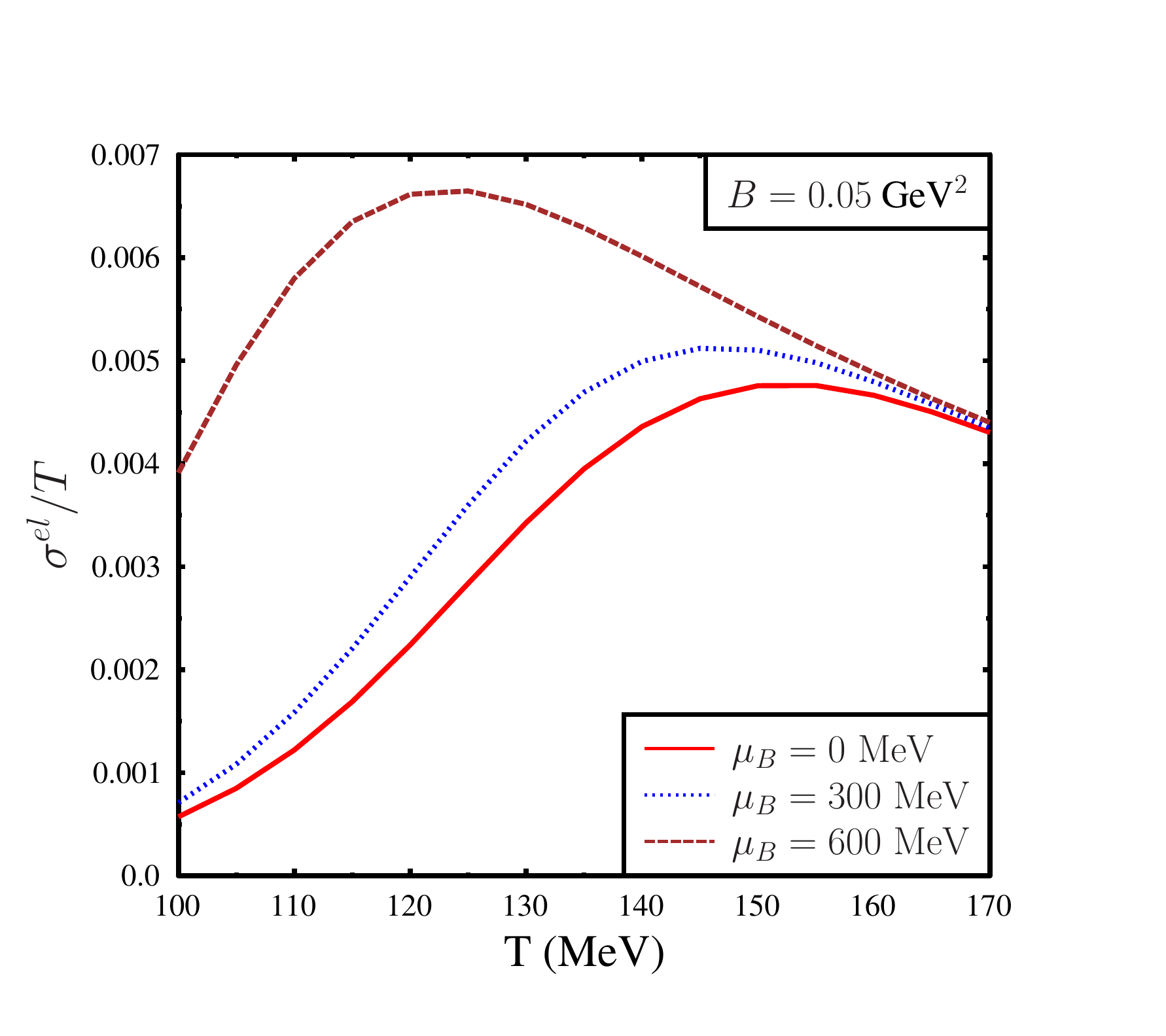}
\caption{Variation of normalized electrical conductivity $\sigma^{el}/T$ with 
temperature for different values of baryon chemical potential $\mu_B$ at $B=0.05$ GeV$^2$.} 
\label{ranjita_fig2}
\end{center}
\end{figure}

Next we discuss the variation $\sigma^{el}/T$ with temperature ($T$) in presence 
of magnetic field and for different values of baryon chemical potential ($\mu_B$). 
This is shown in Fig.~\eqref{ranjita_fig2}. Unlike the vanishing magnetic field case, it is 
seen that $\sigma^{el}/T$ increases with baryon chemical potential. This behaviour 
can be understood as follows. At finite magnetic field the contributions of the 
mesons to the electrical conductivity further decreases due to larger cyclotron 
frequency as compared to baryons, apart from the decrease in the relaxation time with 
increase in $\mu_B$.

\begin{figure}[tbh]
\begin{center}
\includegraphics[width=0.48\textwidth]{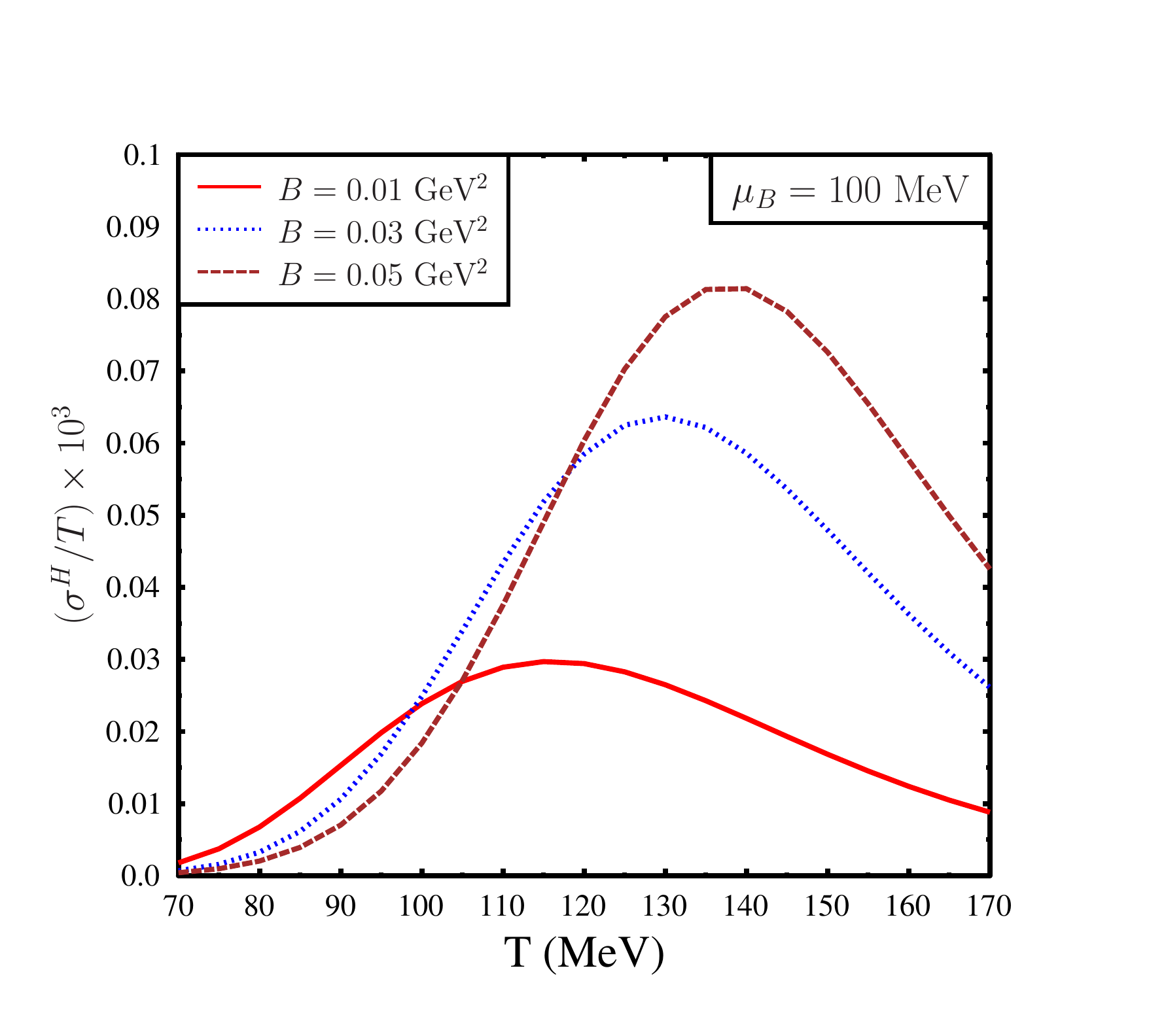}
\includegraphics[width=0.48\textwidth]{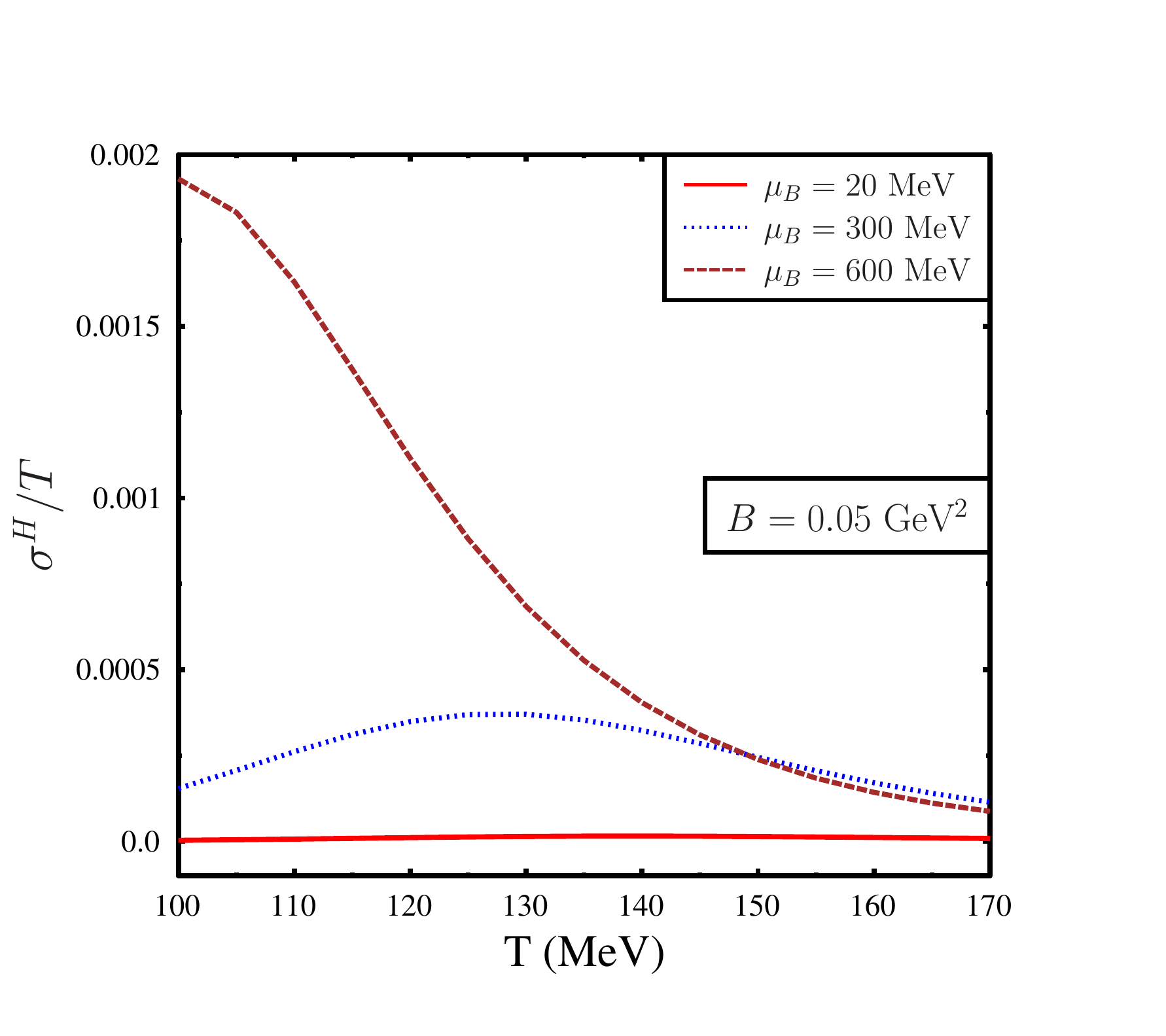}
\caption{Left plot: Variation of normalized Hall conductivity $\sigma^H/T$ with temperature 
for various values of magnetic field at $\mu_B=100$ MeV, Right Plot: Variation of 
normalized Hall conductivity $\sigma^H/T$ with temperature for various values of baryon 
chemical potential at $ B=0.05$GeV$^2$.}
\label{ranjita_fig3}
\end{center}
\end{figure}

Next, we discuss Hall conductivity in hadronic gas within HRG model. In 
Fig.\eqref{ranjita_fig3} (left plot), we
show the variation of Hall conductivity with temperature $(T)$ for different 
values of the magnetic field at finite baryon chemical potential $\mu_B=100$ MeV. Let us note that 
due to the opposite gyration of the particles and the antiparticles in a magnetic field, 
the mesonic contribution to the Hall conductivity gets exactly cancelled out. Hence, it is 
only the baryons which contribute to the Hall conductivity at finite baryon 
chemical potential. It may be observed in Fig.\eqref{ranjita_fig3} that for the small temperature 
the Hall conductivity decrease with increase in magnetic field, while for larger 
temperature the Hall conductivity increase with magnetic field. At low temperature 
since the relaxation time is smaller then the Hall conductivity the integrand 
$\sim \frac{1}{\omega_c\tau}$ ($\omega_c=\frac{eB}{\epsilon}$), which explains the 
suppression of Hall conductivity with increasing magnetic field.
On the other hand at large temperature with smaller relaxation time the 
integrand $\sim \omega_c\tau$ which explains the increase in the Hall conductivity with 
increasing magnetic field.
%%%%%%
In Fig.~\eqref{ranjita_fig3} (right plot) we plotted the variation of the normalized Hall conductivity 
$\sigma^H/T$ with temperature for different values of baryon chemical 
potential at $B=0.05$ GeV$^2$. As may be noted from this figure 
for smaller chemical potential the Hall conductivity is smaller. This is due to the fact that for finite Hall conductivity the imbalance between the number of particles and antiparticles is required.  With increase in baryon chemical potential, the number density of particles are significantly larger than that of antiparticles leading to a  non vanishing Hall current. Again the non monotonic behavior of normalized Hall conductivity with temperature for a specific value of the magnetic field is similar to Fig.~\eqref{ranjita_fig3}(left plot)

\begin{figure}[tbh]
\begin{center}
\includegraphics[width=0.48\textwidth]{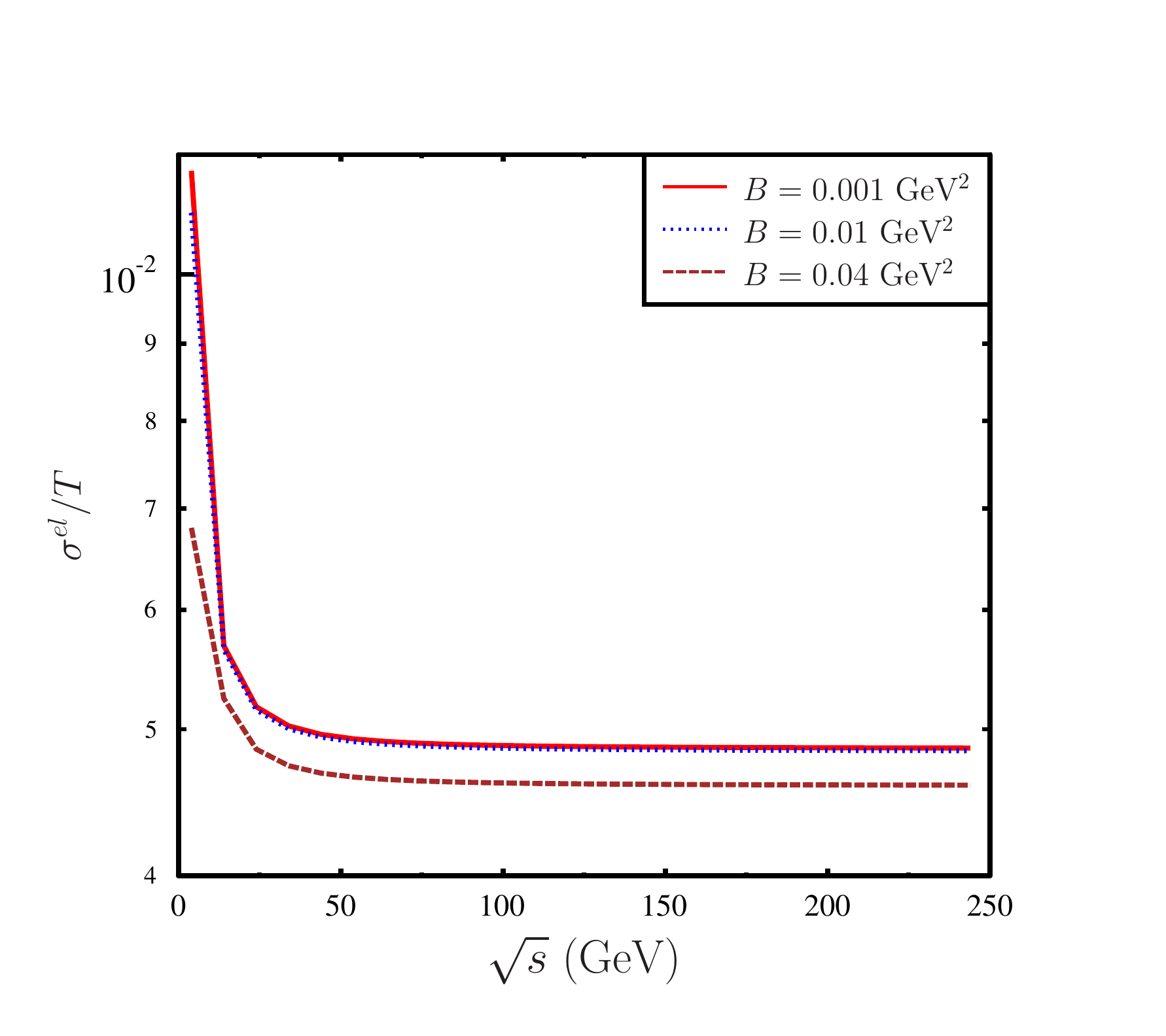}
\includegraphics[width=0.48\textwidth]{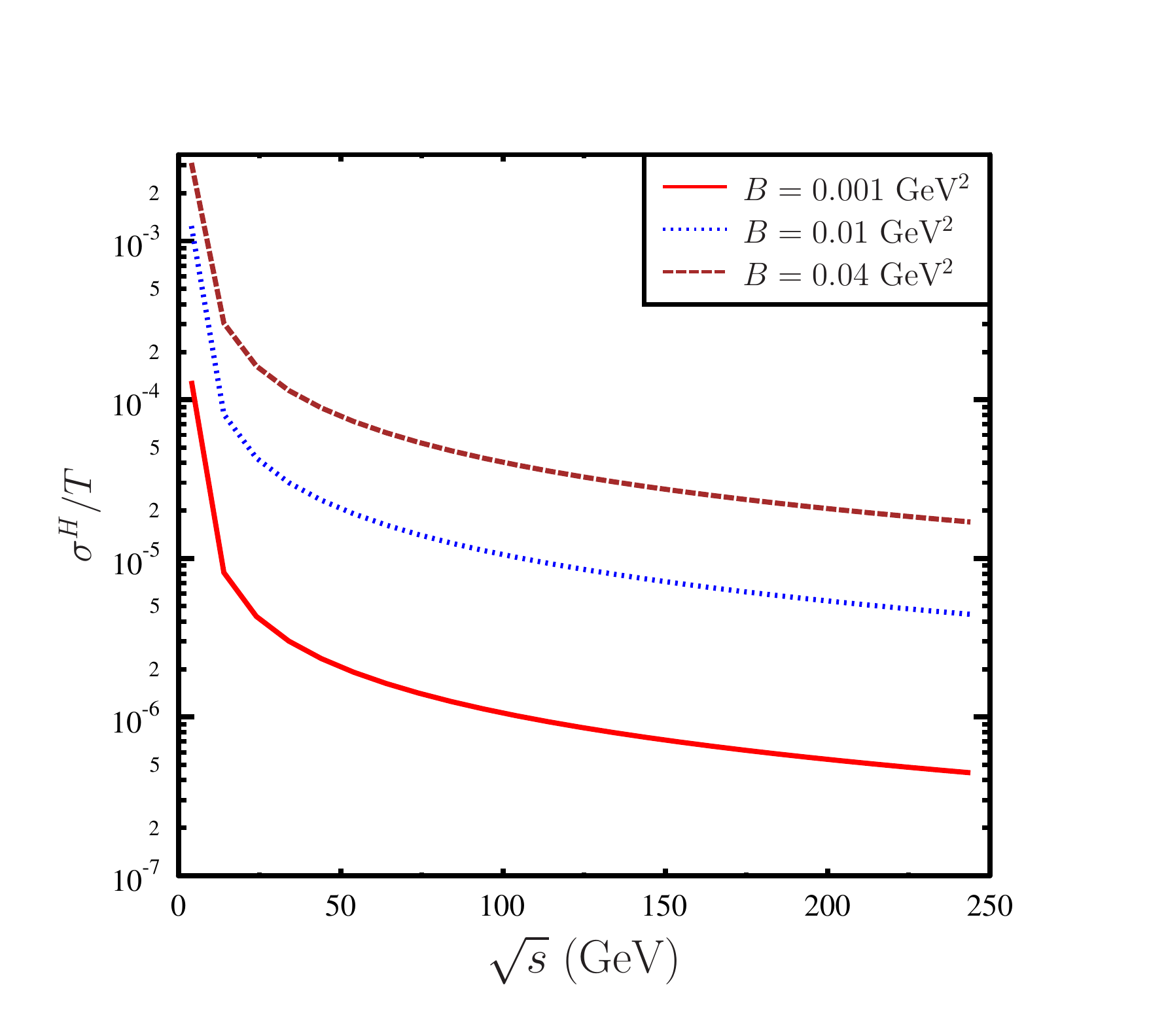}
\caption{Left plot: Variation of normalized electrical conductivity $\sigma^{el}/T$ with center of mass energy for difference values of magnetic fields, Right Plot: Variation of normalized Hall conductivity $\sigma^{H}/T$ with center of mass energy ($\sqrt{s}$) for different values of magnetic field. With increasing magnetic field $\sigma^{el}/T$ decreases and $\sigma^{H}/T$ increases.}
\label{ranjita_fig4}
\end{center}
\end{figure}

In Fig.~\eqref{ranjita_fig4} we have considered values of the magnetic field ranging 
from $B=0.001$ GeV$^2$ to $B=0.04$ GeV$^2$ for different center of mass energy. 
Let us note that for RHIC the center of mass energy 200 GeV and the estimated maximum 
magnetic field is of the order of $B=0.04$ GeV$^2$. For this value of the magnetic field 
and the collision energy, we get that the value of normalized Hall conductivity $\sigma^{H}/T$ 
for RHIC is of the order of $10^{-5}$ and the value of normalized electrical conductivity 
is of the order of $10^{-3}$. On the other hand for relatively low energy collisions, e.g., FAIR, 
the collision energy $E_{lab}~10 A GeV$ and the estimated maximum value of the magnetic field 
is of the order of $B=0.001$ GeV$^2$. For this value of the magnetic field and 
the collision energy relevant for FAIR, the value of normalized Hall conductivity 
is of the order of $10^{-4}$ and the normalized electrical conductivity is of the order of
$10^{-2}$.

\subsection{Summary and conclusions}
In this investigation, we have estimated the electrical ($\sigma^{el}$) and the 
Hall conductivity ($\sigma^{H}$) of the hot and dense hadron gas in the presence 
of an external magnetic field. 
We have not 
considered the Landau quantization of the charged particles as well as 
magnetic field dependent dispersion relation due to relatively smaller magnetic field. 

%\begin{thebibliography}{99}
%
%\end{thebibliography}
%
%\end{document}

%% file: Jayanta/jayanta.tex
\section{Effect of Dynamical Chiral Symmetry Breaking on the Electrical Resistivity of Magnetized Quark Matter}

\textit{Jayanta Dey, Sabyasachi Ghosh, Aritra Bandyopadhyay, Ricardo L. S. Farias, Gast\~ao Krein}

\bigskip

%
%\begin{abstract}

{\small
We studied the effects of dynamical chiral symmetry breaking (DCSB) on the resistivity of quark matter 
in presence of magnetic field. For massless quarks, we obtained the expected dissipation-less 
transverse Hall resistivity along with a longitudinal Drude's resistivity; while the former is 
independent of magnetic field, the latter is proportional to the magnetic field. At low temperatures 
and large magnetic fields, quarks become massive due to DCSB. We found that DCSB leads to a non-trivial 
temperature and field dependence for both longitudinal and transverse resistivity components.
}

\bigskip

%\end{abstract}
%
%\pacs{12.38.Mh, 21.65.Qr, 12.39.Ki}
%
%\keywords{Quark matter, Transport coefficients, Electrical conductivity, Dynamical chiral symmetry breaking}

%\maketitle

%%%%%%%%%%%%%%%%%%%%%%%%%%%%%%%%%%%%%%%%%%%%%%%%%%%%%%%%%%%%%%%%%%%%%%%%%%
%
\subsection{Introduction} 
Strong magnetic fields are expected to be produced in relativistic heavy-ion collision (HIC) 
experiments~\cite{Rafelski:1975rf, Kharzeev:2007jp}. Different physical properties of the quark matter created 
in these experiments can be affected by the magnetic field. In the present communication we present results
on the effect of dynamical chiral symmetry breaking (DCSB) on the resistivity of magnetized quark matter.
We first revisit the standard resistivity expression derived
by connecting the macroscopic Ohm's law and the microscopic Drude's approach. This leads to
the well known inverse relation between electrical resistivity and electrical conductivity. 
In the presence of a magnetic field, along with the normal, longitudinal resistivity, there appears
a Hall resistivity, transverse to the magnetic field. We study the temperature and 
magnetic field dependences of these resistivity components. We considered the case of massless quarks
and massive quarks, with the masses being temperature and magnetic field dependent. The masses are
generated through the mechanism of DCSB. We use the Nambu--Jona-Lasinio model
to obtain the quark masses.

%%%%%%%%%%%%%%%%%%%%%%%%%%%%%%%%%%%%%%%%%%%%%%%%%%%%%%%%%%%%%%%%%%%%%%%%%%
%
\subsection{Formalism}
\label{sec:formalism}
\subsubsection{Resistivity without magnetic field}
Let us assume an electric field applied along the x-axis, ${\vec E}={\hat x}E_x$; 
then, a current density ${\vec J_x}={\hat x}J_x$ along the same direction is generated.
The potential $V=\int E_xdx$ and current $I=\int J_x dydz$ follow the macroscopic 
Ohm's law, $V=RI$, where $R$ is the resistance of the medium. The vector form of 
Ohm's law can be written as
\bea
E_x{\hat x} = \rho_{xx} J_x {\hat x}
\hspace{0.5cm}{\rm or}\hspace{0.5cm}
\sigma_{xx} E_x{\hat x} = J_x {\hat x}~,
\label{E_J_macro}
\eea
where the resistivity, $\rho_{xx}=1/\sigma_{xx}$ ($\sigma_{xx}$ is conductivity), 
is a more appropriate dissipative quantity than~$R$. By this definition, the dimensions 
of $\rho$ and $R$ are related by $[\rho] = [R] \times {\rm Area}/{\rm Length}$.
A microscopic derivation of the resistivity can be obtained by using Drude's assumption that 
an external electric field $E_x$ accelerates a charge particle from rest to a finite momentum 
$m v_x$ within a relaxation time $\tau_c$. Therefore, a given quark with flavor~$f$ and
electric charge $e_f$ and mass $m_f$ experiences the force:
\be
e_f E_x=\frac{m_f v_x}{\tau_c}~,
\label{v_E}
\ee
and hence the current density can be expressed as
\bea
J_x  &=& e_f n_f v_x
= e_f n_f \left(\frac{e_f \tau_c E_x}{m_f} \right) =  \left[\frac{e_f^2n_f\tau_c}{m_f} \right]E_x~,
\label{E_J_micro}
\eea
where $n_f$ is the electric charge number density in the medium.
Comparing Eqs.~(\ref{E_J_macro}) and (\ref{E_J_micro}), one can get Drude's expression 
of conductivity or resistivity
\be
\sigma_{xx}=\frac{1}{\rho_{xx}}=\sigma^{NR}_D=\frac{e_f^2n_f\tau_c}{m_f}~.
\label{sig}
\ee
The NR in $\sigma^{NR}_D$ denotes the the nonrelativistic nature of this expression;
it can in principle be applied to quark matter with massive constituent quarks, where different flavor 
charges (e.g. $e_u=+\frac{2}{3}e$, $e_d=-\frac{1}{3}e$) with their spin, color and particle-anti-particle 
degeneracy factors have to be taken into account.  

A relativistic expression can be obtained using the energy-momentum relation $E=\sqrt{p^2 + m^2}$ and 
Boltzmann's equation for the quark distribution in medium:
\bea
\frac{\partial f}{\partial t} + \frac{\del x}{\del t} \frac{\del f}{\del x} 
+ e_f E_x \frac{\del f}{\del p_x} 
&=& \left(\frac{\partial f}{\partial t}\right)_{\rm coll}~.
\label{Boltz_B0}
\eea
Writing $f = f_0 + \delta f$, where $f_0$ is the equilibrium distribution $f_0=1/(e^{\beta E}+1)$ and
$\delta f$ the deviation of $f$ from $f_0$ within a time scale $\tau_c$, and using the relaxation-time 
approximation, in which $({\partial f}/{\partial t})_{\rm coll} = - \delta f/\tau_c$, Eq.~(\ref{Boltz_B0}) 
is solved by
\begin{equation}
e_f E_x \frac{p_x}{E}\beta f_0(1-f_0) 
= \frac{\delta f}{\tau_c}~.
\label{sol-B0}
\end{equation}
This leads to
\bea
J_x &=& e_f\int\frac{d^3p}{(2\pi)^3} \Big(\frac{p_x}{E}\Big) \delta f
\nn\\
&=& \left[e^2_f\int\frac{d^3p}{(2\pi)^3} \Big(\frac{p_x}{E}\Big)^2\tau_c\beta f_0(1-f_0)\right]\, E_x ~,
\label{J_E_micro_R}
\eea
and so
\bea
\sigma_{xx}&=&\sigma^R_D=1/\rho_{xx}
=e^2_f\int\frac{d^3p}{(2\pi)^3} \Big(\frac{p_x}{E}\Big)^2\tau_c\beta f_0(1-f_0)
\nn\\
&=&\frac{e^2_f}{3}\int\frac{d^3p}{(2\pi)^3} \Big(\frac{p}{E}\Big)^2\tau_c\beta f_0(1-f_0)~.
\eea

\subsubsection{Resistivity in the presence of a magnetic field}
In presence of a magnetic field along the z-axis, ${\vec B} = {\hat z} B$, the charge particle 
is subjected to a Lorentz force which in turn generates a current density perpendicular to 
${\vec B}$; the components $J_x$ and $J_y$ can be expressed in matrix form as~\cite{Tong:2016kpv}
\bea
\left(
\begin{array}{c}
J_x \\
J_y 
\end{array}
\right) &=&
\left(
\begin{array}{cc}
\sigma_{xx} & \sigma_{xy} \\
\sigma_{yx} & \sigma_{yy}
\end{array}
\right)
\left(
\begin{array}{c}
E_x \\
0 
\end{array}
\right)
\nn\\
\left(
\begin{array}{cc}
\rho_{xx} & \rho_{xy} \\
\rho_{yx} & \rho_{yy}
\end{array}
\right)
\left(
\begin{array}{c}
J_x \\
J_y 
\end{array}
\right) &=&
\left(
\begin{array}{c}
E_x \\
0 
\end{array}
\right)~,
\eea
where 
\bea
&&\sigma_{xx}=\sigma_{yy}=\sigma_D\frac{1}{1+(\tau_c/\tau_B)^2}
\nn\\
&&\sigma_{yx}=-\sigma_{xy}=\sigma_D\frac{\tau_c/\tau_B}{1+(\tau_c/\tau_B)^2}
\nn\\
&&\rho_{xx}=\rho_{yy}=\frac{1}{\sigma_D}
\nn\\
&&\rho_{xy}=-\rho_{yx}=\frac{\tau_c}{\tau_B \sigma_D}~,
\label{rhos}
\eea
where $\tau_B$ is another times scale along with collisional relaxation time $\tau_c$, with
$\tau_B=m_f/(eB)$ in a non-relativistic treatment or $\tau_B=E_{av}/(eB)$ in a relativistic 
treatment, where $E_{av}$ is the average value
\be
E_{av}= \frac{\int \frac{d^3p}{(2\pi)^3} E f_0}{\int \frac{d^3p}{(2\pi)^3} f_0}~.
\label{E_av}
\ee
In addition, in Eq.~(\ref{rhos}) we consider the Drude conductivities $\sigma_D=\sigma^{NR/R}_D$ 
for the non-relativistic or relativistic treatments.

\subsubsection{NJL model in presence of magnetic field}
Magnetized quark matter is described within a quasi-particle model within the Nambu-Jona-Lasinio (NJL) model 
framework. In the NJL model, the constituent quark mass $M = M(T,B)$, which is a function of the temperature~$T$ 
and magnetic field~$B$, is  obtained by solving the gap equation:
\bea
M = m - 2 G  \sum_{f=u,d}\langle \bar{\psi}_f\psi_f\rangle,
\label{Gap_B}
\eea
where $\langle \bar{\psi}_f\psi_f\rangle$ is the quark condensate, given by:
\bea
\langle \bar{\psi}_f\psi_f\rangle = \langle \bar{\psi}_f\psi_f\rangle^{vac} 
+ \langle \bar{\psi}_f\psi_f\rangle^B + \langle \bar{\psi}_f\psi_f\rangle^{T,B}~.
\label{gap}
\eea
The expressions for the different contributions can be found in Ref.~\cite{Farias:2014eca}.

%%%%%%%%%%%%%%%%%%%%%%%%%%%%%%%%%%%%%%%%%%%%%%%%%%%%%%%%%%%%%%%%%%%%%%%%%%
%
\subsection{Numerical results and discussion}
\label{sec:results}
Resistivity without magnetic field for massless fluid follows the simple
analytic expression
\be
\rho_{xx}=\frac{36}{e^2_f\tau_c T^2}~,
\label{rxx_m0}
\ee
which remains the same in the presence of magnetic field.
However, the Hall resistivity resistivity becomes a function
of the magnetic field, namely:
\bea
\rho_{xy}&=&\left(\frac{36}{e^2_f\tau_c T^2}\right)\left(\frac{\tau_c}{\tau_B}\right)
=\left(\frac{36}{e^2_f T^2}\right)\left(\frac{eB}{E_{av}}\right)~.
\label{rxy_m0}
\eea
While $\rho_{xy}\propto B$, $\rho_{xx}$ is independent of $B$.
The average energy, given in Eq.~(\ref{E_av}), will be $E_{av}=\frac{7\zeta(4)}{2\zeta(3)}T\propto T$.
Hence $\rho_{xx}\propto \frac{1}{\tau_cT^2}$ and $\rho_{xy}\propto \frac{1}{T^3}$, which indicates
the dissipation-free nature of Hall resistivity.

% %%%%%%%%%%%%%%%%%%%%%%%%%%%%%%%%%%%%%%%%%%%%%
\begin{figure}[!t]
\centering
\includegraphics[scale=0.35]{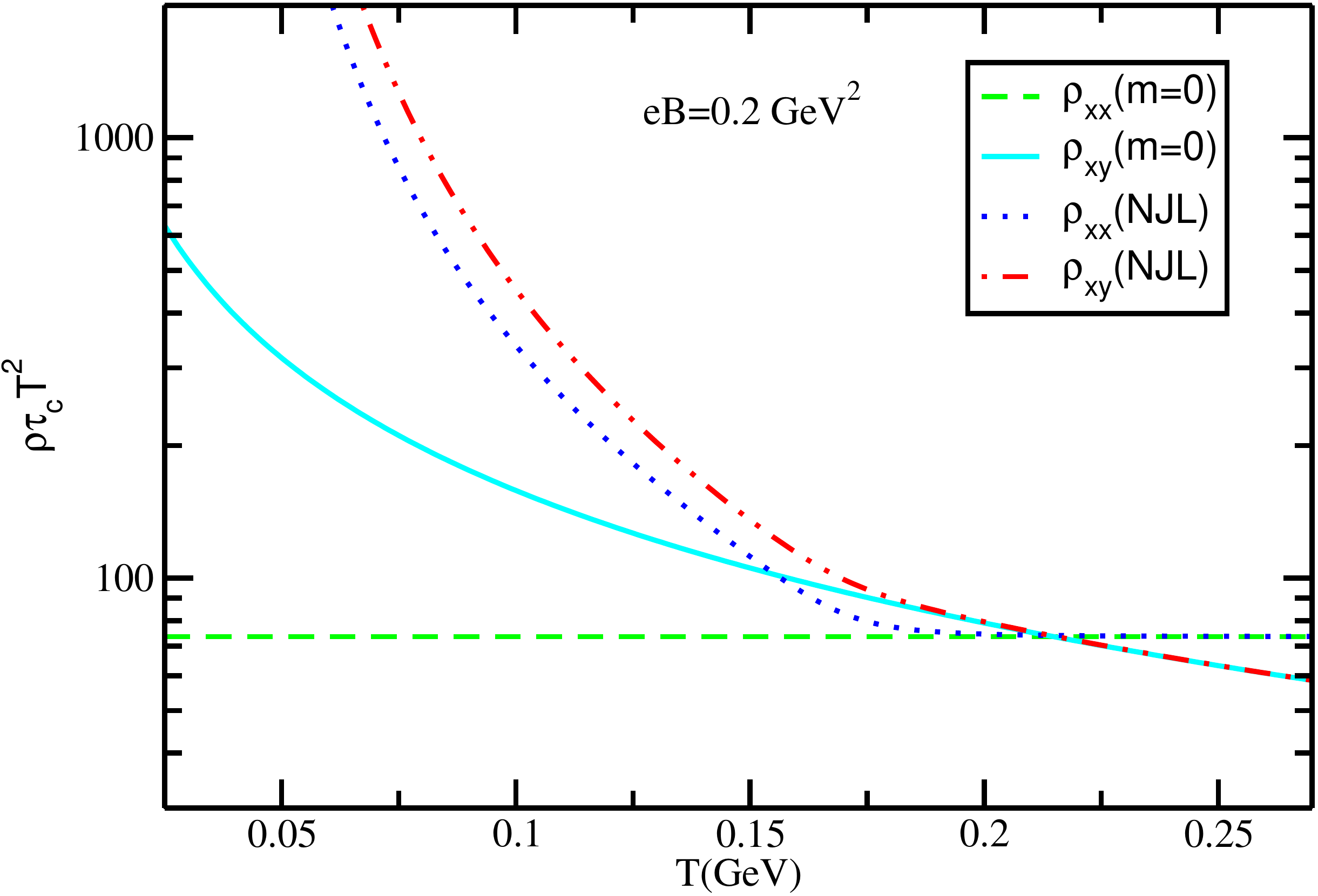}
\caption{Temperature dependence of normalized values of
$\rho_{xx}$, $\rho_{xy}$ for massless case and in NJL model.} 
\label{rho_T}
\end{figure}
% %%%%%%%%%%%%%%%%%%%%%%%%%%%%%%%%%%%%%%%%%%%%%

% %%%%%%%%%%%%%%%%%%%%%%%%%%%%%%%%%%%%%%%%%%%%%
\begin{figure}[!b]
\centering
\includegraphics[scale=0.35]{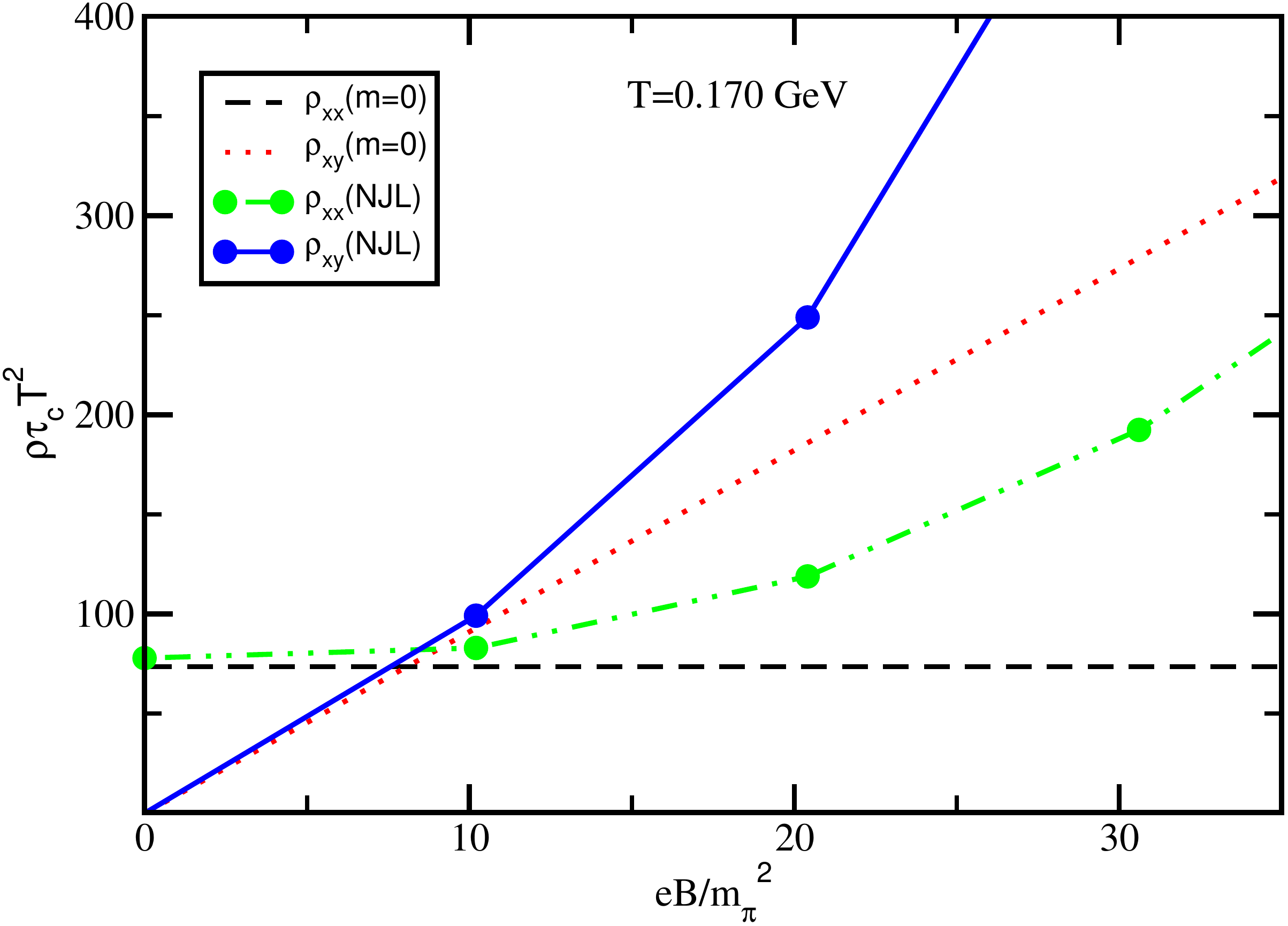}
\caption{Magnetic field dependence of normalized values of
$\rho_{xx}$, $\rho_{xy}$ for massless case and in NJL model.} 
\label{rho_B}
\end{figure}
% %%%%%%%%%%%%%%%%%%%%%%%%%%%%%%%%%%%%%%%%%%%%%

Fig.~(\ref{rho_T}) displays $u-$quark $\rho_{xx}$ and $\rho_{xy}$ resistivity components
(multiplied by $\tau_c T^2$) as a function of~$T$, for $eB=0.2$ GeV$^2$. For massless quarks 
(blue-solid and dashed-green curves), while the $xy$ component follows a $1/T$ dependence, 
the $xx$ component is independent of~$T$. For quarks with a mass $M(T,eB=0.2$ GeV$^2)$ 
obtained from the $NJL$ model (blue-dotted and red-dash-dotted curves), $\rho_{xx}$ and $\rho_{xy}$
have a similar $T$-dependence. For $T$ large, they approach the massless limits, as expected.
However, at low temperatures, when there is substantial  dynamical chiral symmetry breaking,
the resistivity components are enhanced.

Fig.~(\ref{rho_B}) displays the magnetic field dependence of $\rho_{xx}$ and $\rho_{xy}$ for $T=0.17$~GeV, 
again for $u$ quarks only. For massless quarks (black-dashed and red-dotted lines), while $\rho_{xx}$ is
$B-$independent, $\rho_{xy} \propto B$. This is well understandable from the massless relations, given 
in Eqs.~(\ref{rxx_m0}) and (\ref{rxy_m0}). Now when we use $T$ and $B$ dependent constituent quark mass, 
both $\rho_{xx}$ and $\rho_{xy}$ increase with~$B$. The deviations from the massless cases become more 
prominent at large values of $B$.

Figs.~(\ref{rho_T}) and (\ref{rho_B}) reveal the expected behavior of the normal ($\rho_{xx}$) and 
Hall ($\rho_{xy}$) resistivity components at low$-B$ and high$-T$ values. Whereas, the effect of a 
finite quark condensate, signaling dynamical chiral symmetry breaking, is clearly noticeable.  
%
%%%%%%%%%%%%%%%%%%%%%%%%%%%%%%%%%%%%%%%%%%%%%%%%
\subsection{Summary} 
\label{sec:summary}
In summary, we have studied the resistivity of quark matter in the presence of a magnetic field
along the z-direction. We obtained the normal Drude's resistivity along the x-axis, $\rho_{xx}$, and also
the Hall resistivity, $\rho_{xy}$, transverse to a magnetic field. We considered massless quarks and
$T-$ and $B-$dependent massive quarks, with the masses obtained with NJL model.
We have shown that the results for both components of the resistivity are drastically different 
in the massless and massive cases, particularly at low temperature and high magnetic field, 
for which dynamical chiral symmetry breaking is realized.

%\vspace{1.0cm}
%%
%%%%%%%%%%%%%%%%%%%%%%%%%%%%%%%%%%%%%%%%%%%%%%%%%%%%%%

%% file: Deeptak/deeptak.tex
\section{Non-conformal solution of viscous landau hydrodynamics}% Force line breaks with \\

\textit{Deeptak Biswas, Kishan Deka, Amaresh Jaiswal, Sutanu Roy}

\bigskip

%\begin{abstract}
%\leftskip1.0cm
%\rightskip1.0cm

{\small
We have solved viscous landau hydrodynamics for a non-conformal fluid with a constant speed of sound in the 1+1 dimension.  The analytic solution has been obtained considering relativistic Navier-Stokes form of the dissipative hydrodynamic equation. The non-conformal Landau flow has a better agreement with the experimental data than the conformal Landau flow solution with a fitted value of the speed of sound ($c_S^2$). 
}

\bigskip

%\end{abstract}\maketitle

\subsection{Introduction}
The phase structure of strongly interacting matter can be studied from the medium created in the collision of two highly relativistic nuclei~\cite{Lee:1974ma, Collins:1974ky, Itoh:1970uw}.  The created medium expands very fast due to the high-pressure gradient. This space-time evolution can be addressed using relativistic dissipative hydrodynamic simulations \cite{Romatschke:2007mq, Luzum:2008cw, Luzum:2009sb, Song:2010mg, Luzum:2010ag, Schenke:2011tv, Gale:2012rq, Bhalerao:2015iya, Jaiswal:2016hex}. In 1953, Landau first studied this hydrodynamical evolution for an ideal conformal fluid~\cite{Landau:1953gs} which gave rise to a gaussian rapidity distribution of produced particles which has better agreement \cite{Murray:2004gh, Bearden:2004yx, Murray:2007cy, Steinberg:2004wx, Steinberg:2007iv, Wong:2008ex} with $p_T$ integrated yield over whole rapidity range. In Landau's model, a fast longitudinal expansion is followed by a slower expansion in the transverse plane. Freezeout happens when the transverse displacement becomes larger than the initial transverse dimension. The final rapidity distribution of particles is therefore given by the rapidity distribution at the freeze-out time \cite{Murray:2004gh, Bearden:2004yx, Murray:2007cy, Steinberg:2004wx, Steinberg:2007iv, Wong:2008ex, Srivastava:1992xb, Srivastava:1992cg, Srivastava:1992gh, Mohanty:2003va, Hama:2004rr, Aguiar:2000hw, Pratt:2008jj, Bialas:2007iu, Csorgo:2006ax, Beuf:2008vd, Osada:2008cn}. Here we solve viscous landau hydrodynamics for a non-conformal equation of state in the Navier-stroke limit and employ the obtained solution to fit rapidity spectrum of observed pions in $\sqrt{s_{NN}}=$ 200, 17.3, 12.3, 8.76, 7.62, 6.27, 4.29, 3.83, 3.28 and 2.63 GeV collision energies. We find that the Landau flow with a non-conformal equation of state leads to a better agreement with the experimental data compared to the conformal Landau flow solution for a fitted value of $c_S^2$. The value of the squared speed of sound shows a monotonic decrease with decreasing collision energies. 
 
\subsection{Viscous Landau flow}
The energy-momentum tensor of a relativistic fluid in the Navier-Stokes limit is, \cite{Landau_book}
\begin{equation}\label{Tmunu}
T^{\mu \nu}=\epsilon\, u^\mu u^\nu - (P-\zeta\theta)\, \Delta^{\mu \nu} + 2\eta\sigma^{\mu\nu},
\end{equation}
where $\epsilon$ is the local energy density, $P$ is the thermodynamic pressure, $u^\mu$ is the fluid four-velocity and, $\eta$ and $\zeta$ are the coefficients of shear and bulk viscosity, respectively. Here, $\Delta^{\mu\nu}\equiv g^{\mu\nu}-u^\mu u^\nu$, $\theta\equiv\partial_\mu u^\mu$ and $\nabla^\mu\equiv\Delta^{\mu\nu}\partial_\nu$. The metric convention is $g^{\mu\nu}={\rm diag}(1,-1,-1,-1)$. Using these definitions, the shear tensor can be written as $\sigma^{\mu\nu}\equiv\frac{1}{2}\left(\nabla^\mu u^\nu + \nabla^\nu u^\mu \right)-\frac{1}{3}\Delta^{\mu\nu}\nabla_\alpha u^\alpha$. Here we have used the non-conformal equation of state, $P=c_s^2\epsilon$, where the speed of sound $c_s^2$ will be assumed to be constant for simplicity.

Following Ref.~\cite{Landau:1953gs}, the hydrodynamic equation for longitudinal expansion along $z$-direction can be written, $\partial_\mu T^{\mu\nu}=0$, leads to~\cite{Landau:1953gs, Wong:2008ex}
\begin{equation}\label{tz}
\frac{\partial T^{00}}{\partial t} + \frac{\partial T^{03}}{\partial z} = 0,\qquad
\frac{\partial T^{03}}{\partial t} + \frac{\partial T^{33}}{\partial z} = 0,
\end{equation}
here, $(t,x,y,z)\equiv(x^0,x^1,x^2,x^3)$. The non-zero velocity fields can be written in terms of longitudinal fluid rapidity, $y$ as  $u^0=\cosh{y}$ and $u^3=\sinh{y}$. Changing our set of co-ordinates following Ref.~\cite{Wong:2008ex} from $(t,z)$ to light-cone variables, $t_\pm\equiv t\pm z$ Eq.(\ref{tz}) become,
\begin{align}
\frac{\partial}{\partial t_+}\left[c_+\epsilon - \xi\,\nabla u \right]e^{2y} + \frac{\partial}{\partial t_-}\left[c_-\epsilon + \xi\,\nabla u \right] &= 0, \label{tptm1}\\
\frac{\partial}{\partial t_+}\left[c_-\epsilon + \xi\,\nabla u \right] + \frac{\partial}{\partial t_-}\left[c_+\epsilon - \xi\,\nabla u \right]e^{-2y} &= 0, \label{tptm2}
\end{align}
where $c_\pm\equiv1\pm c_s^2$ and $\nabla u\equiv\partial u^0/\partial t +\partial u^3/\partial z =e^y\,\partial y/\partial t_+- e^{-y}\,\partial y/\partial t_-$ is defined to simplify notations and $\xi\equiv\zeta+4\eta/3$.

For a non boost-invariant flow, the fluid rapidity $y$ can be related to space-time rapidity as, \cite{Landau:1953gs}
\begin{equation}\label{lnd_fl}
e^{2y}=f\,e^{2\eta_s}=f\,\frac{t_+}{t_-},
\end{equation}
where $f$ is a slowly varying function of $t_+$ and $t_-$ Ref.~\cite{Landau:1953gs, Wong:2008ex}.
Again we redefine our co-ordinate system in terms of, $y_\pm\equiv\ln\left(t_\pm/\Delta\right)$. First we solve for the ideal case with this velocity 
profile, i.e putting $\xi=0$ in Eq.[\ref{tptm1},\ref{tptm2}]. The evolution 
of energy density in case of ideal hydrodynamics is ,
\begin{equation}\label{sol_id}
\epsilon_{id}=\epsilon_0\exp\!\left[ -\frac{c_+^2}{4\,c_s^2}
\left( y_+ + y_-\right) + \frac{c_+c_-}{2\,c_s^2}\sqrt{y_+\, y_-}  \right]\!.
\end{equation}
Changing evolution variables to $y_\pm$, we see that Eq.~(\ref{tptm1})~$+$~Eq.~(\ref{tptm2}) leads to,
\begin{equation}\label{ypym1}
f\,\frac{\partial\epsilon}{\partial y_+} + \frac{\partial\epsilon}{\partial y_-} + \frac{1+f}{2}\left[ c_+\epsilon - \frac{\xi}{\Delta}\,e^{-(y_+ + y_-)/2} \right] = 0.
\end{equation}
Here we have assumed that the form of $f$ does not vary from that one obtains following Landau's prescription in the ideal case i.e $f=\sqrt{y_+/y_-}$. The evolution equations should have an even parity i.e invariant under $y_+\leftrightarrow y_-$ interchange due to symmetry of the colliding system which is there in Eq.~(\ref{ypym1}). Therefore the other combination Eq.~(\ref{tptm1})~$-$~Eq.~(\ref{tptm2}) has been neglected. Hence the solution of Eq.~(\ref{ypym1}) should lead to the evolution of energy density for viscous Landau flow in symmetric nucleus-nucleus collisions.

We assume the ratio $\xi/s$ to be a constant where $s$ is the entropy density. While this is a valid assumption in conformal case, it is not be strictly true for a non-conformal system. Therefore, for the case of constant $\xi/s$, one can write $\xi=\alpha\,\epsilon^{1/(1+c_s^2)}=\alpha\,\epsilon^{1/c_+}$, where $\alpha$ is a constant. Substituting in Eq.~\eqref{ypym1} and rearranging, we get 
\begin{equation}\label{ypym1_conf}
f\frac{\partial\epsilon}{\partial y_+} + \frac{\partial\epsilon}{\partial y_-} = \frac{1+f}{2}\!\left[ \frac{\alpha}{\Delta}\epsilon^{\frac{1}{c_+}}e^{-\frac{1}{2}(y_+ + y_-)} - c_+\epsilon \right]\!.
\end{equation}
Using method of characteristics, we get,
\begin{equation}\label{meth_char}
\frac{dy_+}{f} = \frac{dy_-}{1} = \frac{2\,d\epsilon}{(1+f)\!\left[ \frac{\alpha}{\Delta}\epsilon^{\frac{1}{c_+}}e^{-\frac{1}{2}(y_+ + y_-)} - c_+\epsilon \right]}.
\end{equation}
The above equations can be solved analytically to obtain the final form of $\epsilon$~\cite{Biswas:2019wtp},
\begin{equation}\label{sol_vis_fin}
\epsilon=\left[ g(\alpha)\,\epsilon_{id}^{c_s^2/c_+} -\frac{c_s^2\,\alpha}{c_+c_-\Delta}\,e^{-(y_+ + y_-)/2} \right]^{\frac{c_+}{c_s^2}},
\end{equation}
where $g(\alpha)$ is an arbitrary function of $\alpha$ such that $g(0)=1$. It is easy to see that the above form of energy density indeed satisfy Eq.~\eqref{ypym1}.  

\subsection{Rapidity Distribution}
In Landau's model a fast longitudinal expansion is followed by a slower expansion with constant acceleration in the transverse direction~\cite{Landau:1953gs, Wong:2008ex}. In this picture, the transverse expansion does not get any correction from viscosity so as the freeze-out time. Following Landau's freeze-out criteria~\cite{Landau:1953gs, Wong:2008ex} and considering the transverse expansion using the non-conformal equation of state the constant time freeze out time is given by, 
\begin{equation}\label{tfo}
t_{\rm FO} = a\,\sqrt{\frac{1+c_s^2}{c_s^2}}\,\cosh y.
\end{equation}
At the freeze-out hypersurface, $y$ takes the form $y_\pm=y_b'\pm y$, where $y_b'\equiv\frac{1}{2}\ln[c_+/(4c_s^2)] + y_b$ and $y_b\equiv\ln(\sqrt{s_{NN}}/m_p)$ is the beam rapidity and $m_p$ is mass of the proton~\cite{Wong:2008ex}. One can neglect the term proportional to $\alpha$ in Eq.~\eqref{sol_vis_fin} at freeze-out in the first approximation because this term is exponentially suppressed by large rapidity value at freeze-out. 

The ratio of entropy density to number density, $s/n$, is a conserved quantity in ideal evolution. As entropy density does not get any direct correction from dissipative term in the relativistic Navier-Stokes equation, i.e., $s\sim \epsilon^{1/c_+}$, $s/n$ is approximately conserved for viscous evolution. Neglecting the viscous correction to energy density the final expression for rapidity distribution turns out to be proportional to entropy density which is given by,
\begin{equation}\label{rap_sp}
\dfrac{dN}{dy} \sim \exp\!\left( \frac{c_-}{2c_s^2}\,\sqrt{{y_b'}^2 - y^2} \right).
\end{equation}
By setting $c_s^2=1/3$ the ideal rapidity spectrum for conformal system can be recovered \cite{Landau:1953gs, Wong:2008ex}. We note that $g(\alpha)$ contributes as an overall multiplicative factor and can be absorbed in the volume factor for calculation of the spectra. Therefore $g(\alpha)$ does not appear as an additional fitting parameter.

\subsection{Results and discussion}
%

%%%%%%%%%%%%%%%%%%%
\begin{figure}[!ht]
\centering
\includegraphics[width=7cm]{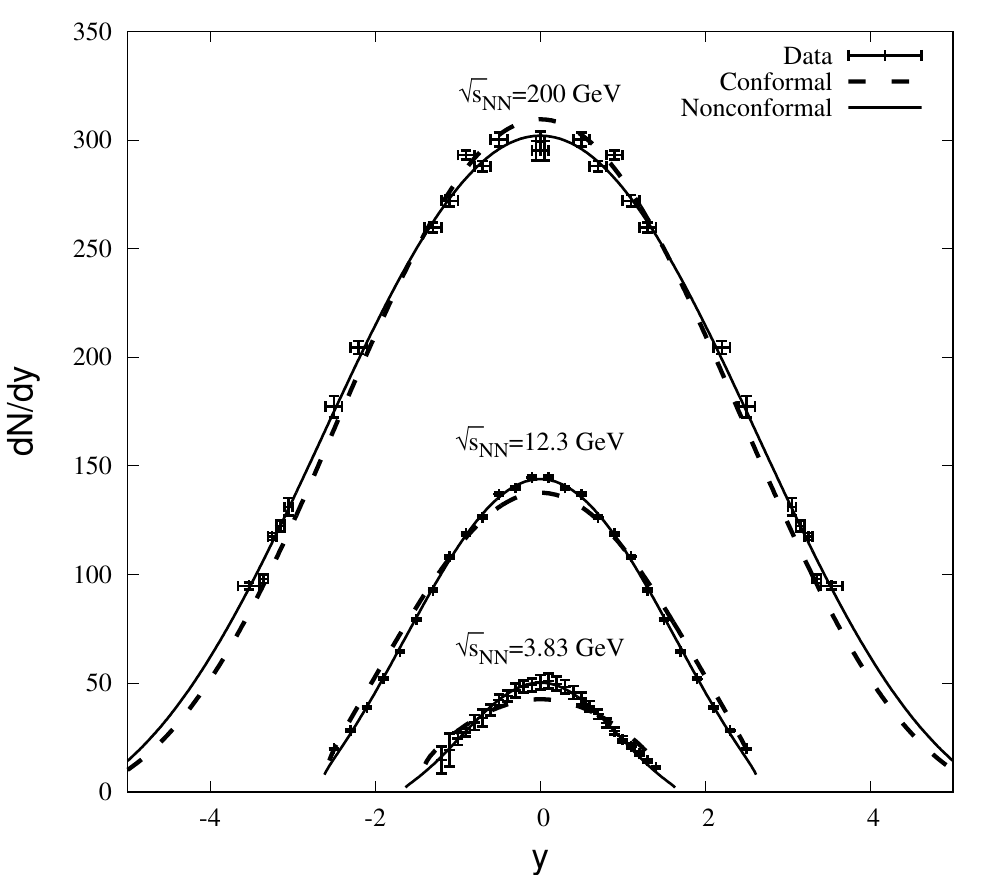}
\vspace*{-0.3cm} 
\caption{ Rapidity spectrum of pions fitted using Eq.~\eqref{rap_sp} ( solid curves) for three representative collision energies: $\sqrt{s_{NN}}=$ 200, 12.3 and 3.83 GeV. Also shown are the fit result using conformal solution of Landau hydrodynamics \cite{Wong:2008ex} (dashed curves) and the experimental results (with errorbars). Experimental data are from Refs.~\cite{Klay:2003zf, Alt:2007aa, Afanasiev:2002mx, Bearden:2004yx}. }
\label{spectra}
\end{figure}
%%%%%%%%%%%%%%%%%%%

In Fig.~\ref{spectra}, we show rapidity spectrum of pions fitted using Eq.~\eqref{rap_sp} for $\sqrt{s_{NN}}=$ 200, 12.3 and 3.83 GeV. The fitting is performed by keeping the overall normalization and $c_s^2$ as free parameters in  a minimization routine. We see that a better fit is obtained using the non-conformal solutions for these collision energies. We have also fitted the rapidity spectrum of pions for $\sqrt{s_{NN}}=$ 17.3, 8.76, 7.62, 6.27, 4.29, 3.28 and 2.63 GeV and found that there is an overall better fit with solutions from non-conformal equation of state. 

%%%%%%%%%%%%%%%%%%%
\begin{figure}[!ht]
\centering
\includegraphics[width=7cm]{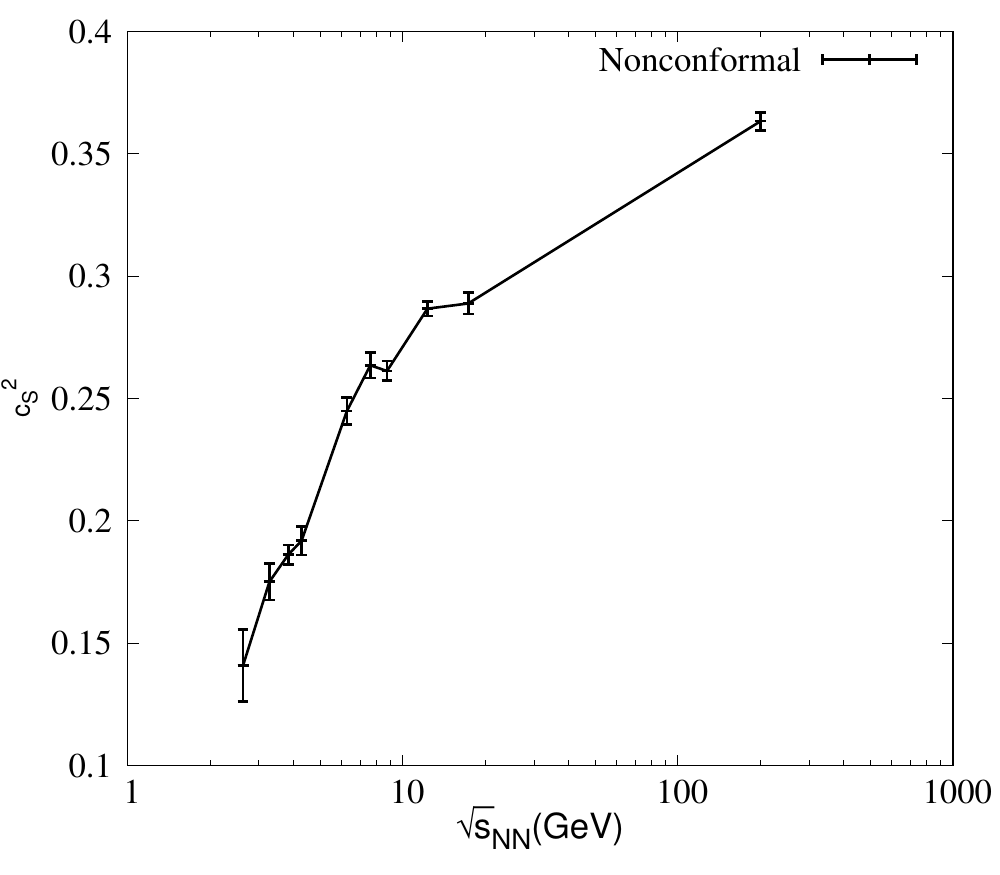}
\vspace*{-0.3cm} 
\caption {Squared speed of sound, extracted by fitting the pion rapidity spectra using the rapidity distribution obtained using non-conformal solution given in Eq.~\eqref{rap_sp}. The error bars corresponds to those obtained from chi-square fit on the fit parameters.}
\label{cs2}
\end{figure}
%%%%%%%%%%%%%%%%%%%

We note that in Fig.~\ref{spectra}, the conformal case is above the non-conformal one for small $y$ and below that at large $y$ for $\sqrt{s_{NN}}=$ 200 GeV. However as one moves to smaller energies an opposite trend is observed. This is due to the fact that an unconstrained fit at $\sqrt{s_{NN}}=$ 200 GeV leads to $c_s^2>1/3$ which is not reasonable. At such high energy, boost-invariance is a good symmetry and therefore Landau model with broken boost invariance is unable to reproduce the data with reasonable parameter values. In such cases, one should perform a constrained fit such that $c_s^2\leq1/3$ for all fitting range.

In Fig.~\ref{cs2}, we show a plot of squared speed of sound, extracted by fitting the pion rapidity spectra using Eq.~\eqref{rap_sp}, over various collision energies (red solid line). We see that at $\sqrt{s_{NN}}=$ 200 GeV, the fitted value of $c_s^2$ is slightly larger than $1/3$ which has also been observed in Refs.~\cite{Gazdzicki:2010iv, Gao:2015mha}. On the other hand, with lower collision energies, we find a monotonic decrease in the extracted value of $c_s^2$. For lower collision energies, rapidity spectra will provide a testing ground for determination of the correct value of $c_s^2$ and hence the equation of state.

For high-energy heavy ion collisions, such as those at RHIC and LHC, one expects the boost-invariance to be a good symmetry for evolution. This is the reason why Bjorken's boost invariant symmetry is extensively applied to model relativistic heavy ion collisions. However, as one goes to lower collisions energies, Bjorken symmetry is broken and one has to consider evolution which is dependent on space-time rapidity. Landau model provides an analytical framework to study the dynamics at low collision energies. Here, we have derived the evolution of the fireball with broken conformal symmetry as well as in presence of viscosity. Using the present analytical solution, one can directly extract the value of $c_s^2$ of QCD medium formed in heavy-ion collisions by analyzing the rapidity spectrum of produced particles. We claim that the rapidity spectra will be important for determination of the correct equation of state.

%% file: Sumana/sumana.tex
%\documentclass[twocolumn,amsmath,amssymb]{snp}
%\pagestyle{empty}
%\usepackage{graphicx}% Include figure files
%\usepackage{dcolumn}% Align table columns on decimal point
%\usepackage{bm}% bold math
%\topmargin 1.5 cm
%\textwidth14.5cm
%\textheight20cm
%\oddsidemargin0.7cm
%\columnsep0.2in
%\begin{document}

\section{Viscous corrections to the Coalescence model for hadron 
production in relativistic heavy-ion collisions}% Force line breaks with \\

\textit{Sumana Bhattacharyya, Amaresh Jaiswal}

\bigskip

{\small
We incorporate viscous corrections to the coalescence model for hadron production from a dissipative quark-gluon plasma. We use this viscous coalescence model to fit the spectra and elliptic flow of hadrons for 2.76 TeV Pb-Pb collisions at LHC. 
}

\bigskip

%\author{\large Sumana Bhattacharyya$^1$}
%\email{response2sumana91@gamil.com}
%\author{\large Amaresh Jaiswal$^2$}
%\email{a.jaiswal@niser.ac.in}
%%\email{abc@barc.gov.in}
%% \altaffiliation[Also at ]{Physics Department, XYZ University.}%Lines break automatically or can be forced with \\
%\affiliation{$^1$Center for Astroparticle Physics \& Space Science,
%Bose Institute, EN-80, Sector-5,
%Bidhan Nagar, Kolkata-700091, India}
%\affiliation{$^2$School of Physical Sciences
%National Institute of Science Education and Research
%Jatni, Odisha 752050, India}
%
%\maketitle

\subsection{Introduction}
Recombination models, along with fragmentation processes, have been used quite successfully to describe hadronization in heavy ion collisions. Coalescence model in heavy ion collision is mainly based on an instantaneous projection of thermalised quark states, those are close to each other both in space and in momentum space, onto hadron states \cite{Gupt:1983rq}. This model characterize numerous salient features of hadronization in heavy-ion collisions, including baryon enhancement \cite{Fries:2003vb} and the robust scaling of the elliptic flow with the number of valence quarks \cite{Kolb:2000sd}. It has been argued that the flow anisotropy originates in the partonic phase and it obeys a simple valence quark scaling for low transverse momentum, that naturally arises from a recombination model \cite{Voloshin:2002wa, Lin:2002rw}. However, it is assumed in this model that densely populated phase space distribution of partons do not change with hadronization, there are no dynamical thermal gluons in the medium and QCD plays a background part. Under these assumptions, temperature merely plays any part than scale the momentum. Quark numbers do not change with temperature. Here we modify the coalescence model to incorporate viscous corrections in the distribution function. We subsequently use this viscous coalescence model for hadron production from a dissipative quark-gluon plasma and to fit the spectra and elliptic flow of hadrons for 2.76 TeV Pb-Pb collisions at LHC.
 %No interactions are there. Constant volume, static model. Homogeneous description of recombining partons.

\subsection{Formalism}
In order to consider a boost invariant framework, it is easier to 
work in the Milne co-ordinate system where, 
\begin{equation}
\tau = \sqrt{t^2-z^2}, \quad
\eta_s = \tanh^{-1}(z/t),\quad
r = \sqrt{x^2+y^2},\quad
\varphi = {\rm atan2}(y,x).\\ %\label{MC4}
\end{equation}
The metric tensor for this co-ordinate system is $g_{\mu\nu}={\rm 
diag}(1,\,-\tau^2,\,-1,\,-r^2)$. 
Boost invariance and rotational 
invariance implies $u^\varphi=u^{\eta_s}=0$.  In this model, we 
further assumes that the particle freeze-out happens at a proper 
time $\tau_f$ having a constant temperature $T_f$ and uniform 
matter distribution, in the transverse plane. In summary, the hydrodynamic 
fields are parametrized as
\begin{equation}
T = T_f ,\quad
u^r = \gamma_T  \beta_T, \quad
u^\varphi = u^{\eta_s}= 0, \quad
u^\tau = \gamma_T, %\label{BW4}
\end{equation}
where $R$ is the transverse radius of the fireball at freeze-out, 
$\gamma_T=1/\sqrt{1-\beta_T^2}$ is the Lorentz factor in the 
transverse direction and $\beta_T$ is the transverse expansion 
velocity. \\
For central collisions, a power-law relation for transverse velocity 
flow profile leads to
\begin{equation}\label{betaT}
\beta_T = \beta_0\left( \frac{r}{R} \right)^m ,
\end{equation}
where $\beta_0$ is the maximum transverse velocity.
For non-central collisions, the transverse fluid velocity profile 
can be parametrized as
\begin{equation}\label{RFP}
\beta_T = \beta_0 \left(\frac{r}{R}\right)^m\left[1 + 2\sum_{n=1}^\infty
\hspace{-.03cm}\beta_ncos[n(\varphi-\psi_n)]\right],
\end{equation}
where $\beta_n$ are the strength of flow anisotropies in the 
transverse direction and $\psi_n$ are the angles between the $x$ 
axis and the major axis of the participant distribution.
However, in the present calculation we can only consider elliptic 
flow and treat $\beta_2$ as a parameter.

\subsection{Viscous correction}

The distribution function with viscous correction is written as $f=f_0+
\delta f$.
The equilibrium distribution function is given by
\begin{equation}\label{IDF}
f_0 = \frac{1}{\exp(u_\mu p^\mu/T) + a}~,
\end{equation}
where $a=+1$ for baryons and $a=-1$ mesons.

Approximating the shear stress tensor with its first-order 
relativistic Navier-Stokes expression, $\pi_{\alpha\beta}= 
2\eta\nabla_{\langle\alpha} u_{\beta\rangle}$, the expression for 
the Grad's 14-moment approximation reduces to \cite{H. Grad,Romatschke:2009im}
\begin{equation}\label{GradT}
\delta f^{(1)}_G = \frac{f_0\tilde f_0}{T^3}\left(\frac{\eta}{s}\right)
p^\alpha p^\beta \nabla_{\langle\alpha} u_{\beta\rangle},
\end{equation}
whereas that due to the Chapman-Enskog method leads to \cite{Jaiswal:2013vta, Jaiswal:2013npa}
\begin{equation}\label{CET}
\delta f^{(1)}_{CE} = \frac{5f_0\tilde f_0}{T^2(u\cdot p)}\left(\frac{\eta}
{s}\right)
p^\alpha p^\beta \nabla_{\langle\alpha} u_{\beta\rangle}.
\end{equation}
Here $\eta$ is the coefficient of shear viscosity, $s=(\epsilon+P)/T$
is the entropy density and the angular brackets denote traceless 
symmetric projection orthogonal to the fluid four-velocity.

For a particle at the space-time point $(\tau,\,\eta_s,\,r,\,\varphi)$ 
with the four momentum $p^\mu=(m_T\cosh y, 
\,p_T\cos\varphi_p,\,p_T\sin\varphi_p,\,m_T\sinh y)
$, we get
\begin{align}%\label{final}
p_\tau &= m_T\cosh(y-\eta_s), \quad
p_{\eta_s} = -\tau\, m_T\sinh(y-\eta_s), \\
p_r &= -p_T\cos(\varphi_p-\varphi), \quad
p_\varphi = -r\, p_T\sin(\varphi_p-\varphi).
\end{align}

Next step is to obtain $\nabla_{\langle\alpha} u_{\beta\rangle}$. We work in Milne co-ordinate system with the metric tensor $g_{\mu\nu}={\rm diag}(1,\,-\tau^2,\,-1,\,-r^2)$. Therefore, the inverse metric tensor is $g^{\mu\nu}={\rm 
diag}(1,\,-1/\tau^2,\,-1,\, -1/r^2)$, its determinant $g$ is 
$\sqrt{-g}=\tau r$ and the non-vanishing Christoffel symbols are 
$\Gamma^\tau_{\eta_s\eta_s}=\tau$, $\Gamma^{\eta_s}_{\tau\eta_s} 
=1/\tau$, $\Gamma^r_{\varphi\varphi}=-r$, and 
$\Gamma^\varphi_{r\varphi}=1/r$. Using the parametrization of the 
fluid velocity given in Eqs.~\eqref{RFP}, we get
\begin{equation}\label{identity1}
\Delta^{r\varphi} = 0, \quad
\Delta^{\varphi\varphi} = -\frac{1}{r^2}, \quad
\Delta^{rr} = - 1 - (u^r)^2, 
\end{equation}
where $\Delta^{\mu\nu}\equiv g^{\mu\nu}-u^\mu u^\nu$ is the 
projection operator orthogonal to the fluid four-velocity. For the 
derivatives of the velocity, we get
\begin{equation}\label{identity23}
\partial_r u^r = \frac{m\,u^r\,(u^\tau)^2}{r}, \quad
\partial_\varphi u^r = -2\,(u^\tau)^3\,\beta_0 \left(\frac{r}{R}\right)^{\!m}
\sum_{n=1}^{\infty}n\,\beta_n\sin[n(\varphi-\psi_n)].
\end{equation}

To fix the time derivatives of the fluid velocity, we assume that if 
the particles are freezing-out, they are free streaming, which means 
that $Du^\mu=0$. Here $D\equiv u^\mu d_\mu$ is the co-moving 
derivative and $d_\mu$ is the covariant derivative. With this 
prescription, we have
\begin{equation}\label{TD123}
\partial_\tau u^\varphi = 0, \quad
\partial_\tau u^r = -\beta_T\,\partial_r u^r = -m\,\frac{(u^r)^2\,(u^\tau)^2}{r\,u^\tau}, \quad
\partial_\tau u^\tau = \beta_T\,\partial_\tau u^r = -m\,\frac{(u^{r})^3}{r}
\end{equation}
where $\beta_T=u^r/u^\tau$ is the radial velocity in the transverse 
plane. The expansion scalar is
\begin{equation}\label{ES}
\frac{1}{\sqrt{-g}} \partial_{\mu} (\sqrt{-g} u^{\mu}) = 
\frac{u^{\tau}}{\tau} + \frac{u^{r}}{r} + \partial_\varphi u^\varphi + \partial_r u^r + \partial_\tau u^\tau
= \frac{u^{\tau}}{\tau} + (m+1) \frac{u^{r}}{r}.
\end{equation}
Assuming boost invariance, the spatial components of the viscous 
tensor are given by
\begin{align}
r \nabla^{\langle r} u^{\varphi\rangle} =&\, -\frac{r}{2} \partial_r u^{\varphi} - \frac{1}{2r}\partial_{\varphi} {u^{r}} 
- \frac{r}{2}u^r Du^\varphi - \frac{r}{2}u^\varphi Du^r 
- \frac{1}{3}r\Delta^{r\varphi}\frac{1}{\sqrt{-g}} \partial_{\mu}(\sqrt{-g} u^\mu ) \nonumber\\
=&~ \,(u^\tau)^3\,\frac{\beta_0}{r} \left(\frac{r}{R}\right)^{\!m}
\sum_{n=1}^{\infty}n\,\beta_n\sin[n(\varphi-\psi_n)]\,, \label{SRP}\\
r^2\nabla^{\langle\varphi}u^{\varphi\rangle} =&\, -\partial_\varphi u^{\varphi} - \frac{u^{r}}{r} - r^2 u^{\varphi} Du^{\varphi} 
-\frac{1}{3} r^2 \Delta^{\varphi\varphi} \frac{1}{\sqrt{-g}} \partial_{\mu} (\sqrt{-g} u^{\mu} ) 
=~ \frac{1}{3}\left[\frac{u^{\tau}}{\tau} + (m-2) \frac{u^{r}}{r}\right], \label{SPP}\\
\nabla^{\langle r}u^{r\rangle} =&\, - \partial_r u^{r} - u^{r} Du^{r} 
- \frac{1}{3} \Delta^{rr} \frac{1}{\sqrt{-g}} \partial_{\mu} (\sqrt{-g} u^{\mu} ) 
=~ \frac{(u^{\tau})^2}{3}\left[\frac{u^{\tau}}{\tau} + (1-2m) \frac{u^{r}}{r}\right], \label{SRR}
\end{align}
where we have used the fact that $(u^\tau)^2 = 1 + (u^r)^2$. Therefore,
\begin{align}
\tau^2\nabla^{\langle\eta_s}u^{\eta_s\rangle} =&\, - \frac{u^{\tau}}{\tau}  
+ \frac{1}{3} \frac{1}{\sqrt{-g}} \partial_{\mu} (\sqrt{-g} u^{\mu} ) 
=~ \frac{1}{3}\left[(m+1) \frac{u^{r}}{r} - 2\frac{u^{\tau}}{\tau} \right], \label{SEE}\\
\nabla^{\langle r}u^{\eta_s\rangle} =&~ \nabla^{\langle\varphi}u^{\eta_s\rangle} = 0  \label{SRESPE}\; .
\end{align} 

To obtain the temporal components of the viscous stress energy 
tensor, we use the Landau frame condition, 
$\nabla^{\langle\alpha}u^{\beta\rangle}u_{\beta}=0$.
\begin{align}
\nabla^{\langle\tau}u^{\tau\rangle}u_\tau + \nabla^{\langle\tau}u^{r\rangle}u_r &= 0 
~~\Rightarrow~~ \nabla^{\langle\tau}u^{\tau\rangle} = \beta_T\nabla^{\langle\tau}u^{r\rangle}, \label{STT}\\
\nabla^{\langle\eta_s}u^{\tau\rangle}u_\tau + \nabla^{\langle\eta_s}u^{r\rangle}u_r &= 0 
~~\Rightarrow~~ \nabla^{\langle\tau}u^{\eta_s\rangle} = 0, \label{STE}\\
\nabla^{\langle r}u^{\tau\rangle}u_\tau + \nabla^{\langle r}u^{r\rangle}u_r &= 0 
~~\Rightarrow~~ \nabla^{\langle\tau}u^{r\rangle} = \beta_T\nabla^{\langle r}u^{r\rangle}, \label{STR}\\
\nabla^{\langle\varphi}u^{\tau\rangle}u_\tau + \nabla^{\langle\varphi}u^{r\rangle}u_r &= 0 
~~\Rightarrow~~ \nabla^{\langle\tau}u^{\varphi\rangle} = \beta_T\nabla^{\langle r}u^{\varphi\rangle}. \label{STP}
\end{align}

Therefore, from Eqs. (\ref{STT}) and (\ref{STR}), we see that
\begin{equation}\label{STTF}
\nabla^{\langle\tau}u^{\tau\rangle} = \beta_T\nabla^{\langle\tau}u^{r\rangle} = \beta_T^2\nabla^{\langle r}u^{r\rangle} = \frac{(u^r)^2}{3}\left[\frac{u^{\tau}}{\tau} + (1-2m) \frac{u^{r}}{r}\right].
\end{equation}
Next, in order to verify our algebra, we confirm that the viscous 
stress tensor is traceless, i.e., $g_{\mu\nu}\nabla^{\langle\mu} 
u^{\nu\rangle}=0$. Using Eqs.~(\ref {SPP}),~(\ref{SRR}),~(\ref{SEE}) 
and~(\ref{STTF})
\begin{equation}\label{check}
g_{\mu\nu}\nabla^{\langle\mu}u^{\nu\rangle} = \nabla^{\langle\tau}u^{\tau\rangle} - 
\tau^2\nabla^{\langle\eta_s}u^{\eta_s\rangle} - \nabla^{\langle r}u^{r\rangle} - 
r^2\nabla^{\langle\varphi}u^{\varphi\rangle} = 0
\end{equation}

The components of viscous tensor are collected below for quick reference,
\begin{align*}
\nabla^{\langle\tau}u^{\tau\rangle} &= \frac{(u^r)^2}{3}\left[\frac{u^{\tau}}{\tau} + (1-2m) \frac{u^{r}}{r}\right],  \\
\nabla^{\langle\eta_s}u^{\eta_s\rangle} &= \frac{1}{3\tau^2}\left[(m+1) \frac{u^{r}}{r} - 2\frac{u^{\tau}}{\tau} \right],   \\
\nabla^{\langle r}u^{r\rangle} &= \frac{(u^{\tau})^2}{3}\left[\frac{u^{\tau}}{\tau} + (1-2m) \frac{u^{r}}{r}\right],  \\
\nabla^{\langle\varphi}u^{\varphi\rangle} &= \frac{1}{3r^2}\left[\frac{u^{\tau}}{\tau} + (m-2) \frac{u^{r}}{r}\right],  \\
\nabla^{\langle\tau}u^{r\rangle} &= \beta_T\nabla^{\langle r}u^{r\rangle}, \\
\nabla^{\langle r} u^{\varphi\rangle} &= (u^\tau)^3\,\frac{\beta_0}{r^2} \left(\frac{r}{R}\right)^{\!m} \sum_{n=1}^{\infty}n\,\beta_n\sin[n(\varphi-\psi_n)],  \\
\nabla^{\langle\tau}u^{\varphi\rangle} &= \beta_T\nabla^{\langle r}u^{\varphi\rangle}.
\end{align*}
In the above equations, $\tau$ is the freeze-out time which we will 
now denote as $\tau_f$. 

\subsection{Results and discussions}

The parameters that needs to be fit are: $T_f$, $\beta_0$, $m$, 
$\tau_f$, $R$, $\beta_n$ and $\eta/s$. Out of these, $T_f$, $\beta_0$
and $m$ are sensitive to the slope of the transverse momentum 
spectra. The magnitude of the spectra is sensitive to $\tau_f$ and 
$R$. However, since we are considering viscous corrections, $\tau_f$ 
might also effect the anisotropic flow $v_n$. 
The key parameters that are actually needed to fit $v_n$ are 
$\beta_n$ and $\eta/s$.

The freeze-out hyper-surface is~$d\Sigma_\mu=(\tau 
d\eta_s\,rdr\,d\varphi,\,0,\,0,\,0)$, and therefore the integration 
measure is given by
\begin{equation}
p^\mu d\Sigma_\mu = m_T\cosh(y-\eta_s)\,\tau d\eta_s\,rdr\,d\varphi.
\end{equation}
Momentum distribution of number density is given by,
\begin{align}\label{spectra_coal}
& \frac{d^{2}N}{d^{2}p_{T}dy} =\frac{1}{(2\pi)^{3}} \int_0^R r dr \int_0^{2\pi}d\varphi \int_{-\infty}^{\infty}\tau  d\eta_s m_T cosh(y-\eta_s) (f_{0}+\delta f).
\end{align}
For Coalescence Model distribution functions are modified as, $f(u,E,\vec{p})\to $
\begin{equation}
\begin{cases}
f\left(u,E,\frac{\overrightarrow{p}}{2}\right)   f\left(u,E,\frac{\overrightarrow{p}}{2}\right) \quad\hspace{1.55cm} \text{meson,}\\[0.2cm]
f\left(u,E,\frac{\overrightarrow{p}}{3}\right)   f\left(u,E,\frac{\overrightarrow{p}}{3}\right) f\left(u,E,\frac{\overrightarrow{p}}{3}\right) \hspace{0.1cm}\text{ baryon.}
\end{cases}
\end{equation}     
The anisotropy flow is given by
\begin{equation}
v_n(p_T)=\frac{\int_{-\pi}^{\pi} d\varphi_p\, cos[n(\varphi_p-\psi_n)] \frac{dN}{dyp_{T}dp_{T}d\varphi_p}}{\int_{-\pi}^{\pi} \frac{dN}{dyp_{T}dp_{T}d\varphi_p}},
\end{equation}
where $\psi_n$ is the event plane angle. To calculate elliptic flow parameter $v_2(p_T)$, we use \eqref{spectra_coal} in the above equation and perform the integral numerically.
   
\begin{figure}[!ht]
\begin{center}
\includegraphics[width=66mm]{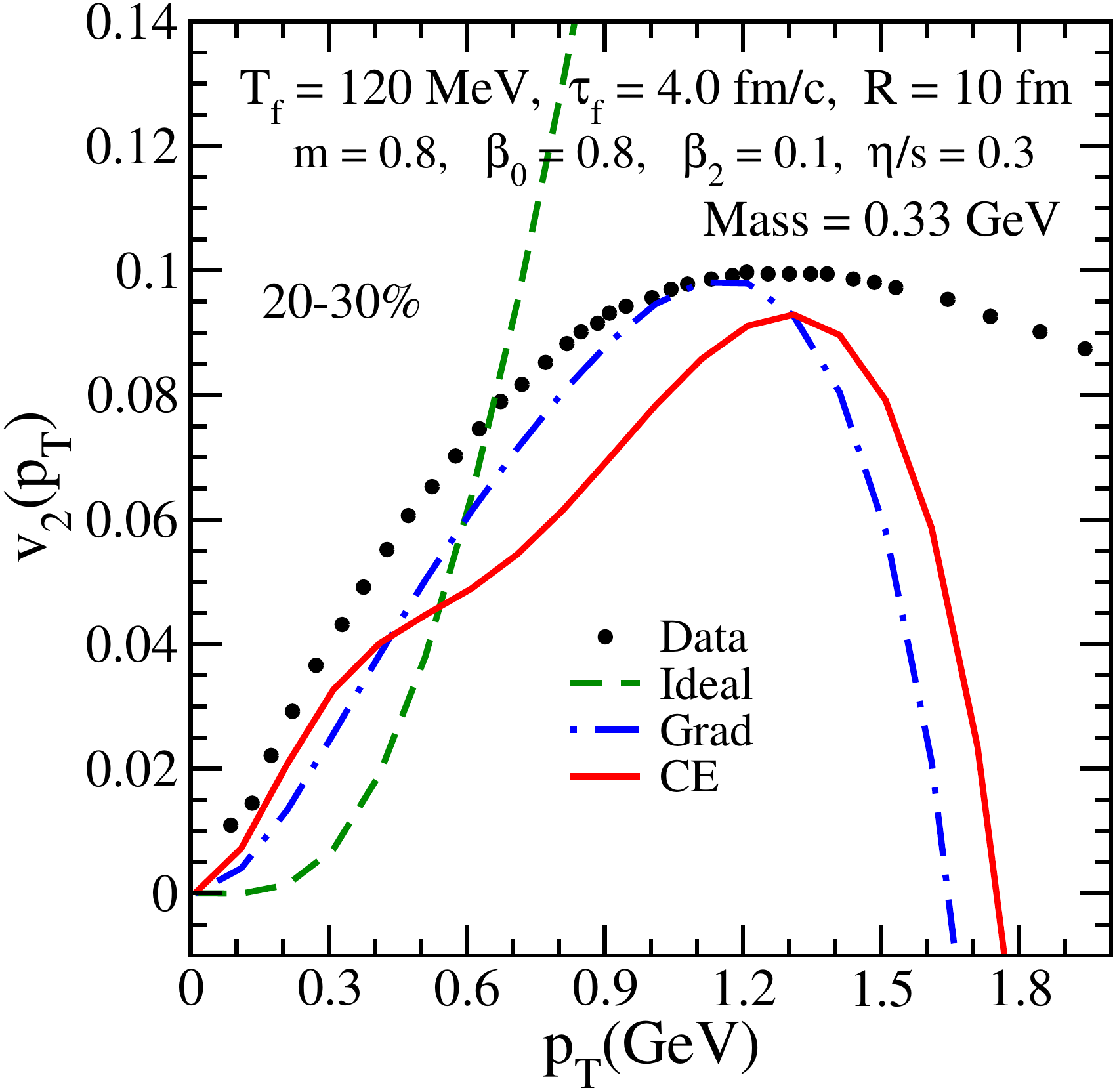}
\caption{\label{sumana_fig1} Elliptic flow $v_2$ as a function of the transverse momentum $p_T$ of pions at LHC for 20-30\% centrality.}
\end{center}
\end{figure}

Figure~\ref{sumana_fig1} shows our results obtained for elliptic flow as a function of transverse momentum at LHC for 20-30\% centrality class. We see that a reasonable agreement is obtained by including viscous correction to in the coalescence model. The fit parameters, we get from this analysis, are in reasonable agreement with literature~\cite{Gale:2012rq, Song:2009gc}. 

%\subsection*{Acknowledgments}
%SB was supported by CSIR. 

%\begin{thebibliography}{50}
%\bibitem{ref1} C. Gupt et al., Nuovo Cim. A \textbf{75}, 408 (1983).
%\bibitem{ref2}  R. J. Fries et al., Phys. Rev. Lett. 
%\textbf{90}, 202303 (2003).
%\bibitem{ref3} Peter F. Kolb et al., Phys. Rev. C \textbf{62}, 054909 (2000).
%\bibitem{ref4} S. A. Voloshin, Nucl. Phys. A \textbf{715}, 379 (2003).
%\bibitem{ref5} Zi-wei Lin et al., Phys. Rev. Lett. \textbf{89}, 202302 (2002).
%\bibitem{H. Grad} H. Grad, Comm. Pure Appl. Math. \textbf{2}, 331 (1949).
%\bibitem{Romatschk} P. Romatschke, Int. J. Mod. Phys. E \textbf{19}, 1 (2010).
%\bibitem{AJ} A. Jaiswal, Phys. Rev. C 87, 051901 (2013); Phys. Rev. C
%\textbf{88}, 021903 (2013).
%\end{thebibliography}
%
%\end{document}

%% file: Sreekanth/sreekanth.tex
%\documentclass[aps,prd,twocolumn,a4paper]{revtex4-1}
%\usepackage{ulem}
%\usepackage{color}
%\usepackage{graphicx}
%\usepackage{epsfig}
%\newcommand{\be}{\begin{equation}}
%\newcommand{\ee}{\end{equation}}
%\newcommand{\ba}{\begin{eqnarray}}
%\newcommand{\ea}{\end{eqnarray}}
%\newcommand{\nn}{\nonumber\\}
%%opening
%\begin{document}

\section{Shear viscosity and Chemical Equilibration in QGP} 

\textit{V. Sreekanth}

\bigskip

%\author{V. Sreekanth}
%\email{v$_$sreekanth@cb.amrita.edu}
%\affiliation{Department of Sciences, Amrita School of Engineering, Coimbatore, Amrita Vishwa Vidyapeethom, India}
%\date{\today}
% \begin{abstract}
% We study the effect of Chemical equilibration in shear viscosity induced cavitation scenarios 
% in the early stage of ultrarelativistic heavy-ion collisions at LHC energies. 
% 
% 
% \vspace{2mm}
% \noindent {\bf PACS}: 25.75.-q; 24.85.+p; 05.20.Dd; 12.38.Mh
% 
% \vspace{2mm}
% \noindent{\bf Keywords}: Quark-Gluon-Plasma
% \end{abstract}
%\maketitle

%\begin{abstract}
{\small
By using various temperature dependent $\eta/s$ prescriptions, we study the chemical equilibration of the hot quark gluon matter
created in the early stages of heavy-ion collisions using causal second order viscous hydrodynamics. Chemical equilibration is studied by introducing fugacity parameters in parton distribution functions. Solving the rate equations, energy and viscous evolution equations within scale invariant Bj\"orken hydrodynamics, we show that the equilibration gets delayed because of slower cooling rate of the fireball in presence of viscosity. Furthermore, we studied shear induced cavitation - negative pressure scenarios during expansion, in presence of chemical non-equilibrium. It has been observed that, cavitation sets in early times for the shear viscosity prescriptions used, invalidating the hydrodynamical modelling. 
}

\bigskip

\subsection{Introduction}
Properties of the quark-gluon plasma (QGP) formed in the early stages of relativistic heavy-ion collisions 
is under intense investigation. Hydrodynamical models explaining the expansion of the strongly coupled QGP 
has been met with great success. 
Several non-equilibrium effects are studied in this context \cite{Romatschke:2017ejr}. Particularly of our 
interest is that of chemical equilibration of the QGP produced. It is a matter of 
investigation whether the QGP produced achieves thermal, mechanical and chemical 
equilibration. Assuming thermal and mechanical equilibration, we study chemical equilibration of 
the fireball. In doing so we follow Ref. \cite{Biro:1993qt}, where the non-equilibrium measure is prescribed through 
the introduction of fugacities in the particle distribution functions. 
Essentially, one studies the evolution of quark (anti-quark) and gluon fugacities ($\lambda_i$) by means of 
relativistic fluid dynamical equations coupled to the rate equations; once the initial time, temperature and 
species number densities are provided.
\par
Shear viscosity of QGP formed in RHIC created huge interest in the scientific community due to its extreme small 
value (to be precise $\eta/s\sim1/4\pi$). 
In order to understand the equilibration dynamics of viscous quark-gluon matter, one need to use 
the causal dissipative hydrodynamics instead of familiar first order Navier-Stoke's equations to 
avoid acausal issues \cite{Muronga:2001zk,Romatschke:2009im,Jaiswal:2013npa, Jaiswal:2013vta, Jaiswal:2013fc, Florkowski:2015lra}. 
The study of evolution of chemically equilibrating QGP with constant shear viscosity
within causal second order viscous hydrodynamics 
has shown distinct properties with possible bearings on signals \cite{Bhatt:2009zg}. While considering higher energy 
collisions, temperature dependent shear viscosity prescriptions are considered, 
owing to the fact that in general physical systems show 
strong temperature dependence for $\eta/s$. On the other hand, 
certain temperature dependent shear viscosity prescriptions, 
results in negative pressure scenarios known as cavitation in the early stages of the fluid dynamical evolution itself 
\cite{Bhatt:2011kr}. 
Cavitation in such situations is problematic and even questions the validity of hydrodynamical modeling used 
and is known to affect the signals \cite{Rajagopal:2009yw,Bhatt:2010cy,Bhatt:2011kx}. 
We attempt to study 
these temperature dependent shear viscosity 
induced cavitation scenarios under chemical non-equilibrium for an expanding quark-gluon matter.   

\subsection{Model}
\par
We denote the single-particle distribution function of parton gas with momentum isotropy as
\cite{Biro:1993qt}:  
\begin{equation}\label{eq1}
 f_i \simeq \lambda_i \left(e^{\beta\cdot p}\pm 1 \right)^{-1},
\end{equation}
where $\beta\cdot p = \beta^\mu p_\mu=T^{-1} u^\mu p_\mu$, with $u^\mu$ being the four-velocity in the comoving frame. 
Note that the fugacities lie between the values zero and one, with latter denoting complete 
chemical equilibration.
Using the above definition of distribution functions, number ($n$) and energy ($\varepsilon$) 
densities and pressure ($P$) of the system with quarks ($q$), anti-quarks ($\bar q$) and gluons ($g$) can be 
calculated as \cite{Biro:1993qt} 
 \bea \label{EoS}
 n &=& (a_1 \lambda_g + b_1 [\lambda_q+\lambda_{\bar{q}}])\,T^3, \nonumber\\
 \varepsilon &=& 3P = (a_2 \lambda_g + b_2 [\lambda_q+\lambda_{\bar{q}}])\,T^4,
% %,\nn &=& 3P
 \eea
where $a_1=\frac{16\zeta(3)}{\pi^2},b_1=\frac{9\zeta(3)N_f}{2\pi^2}, a_2=\frac{8\pi^2}{15}, b_2=\frac{7\pi^2N_f}{40}$ 
with $N_f$ being the number of dynamical quark flavors. Note that for the baryonless QGP under 
consideration we have $\lambda_q=\lambda_{\bar{q}}$.

There are several formulations of second-order hydrodynamics and it is an active field of ongoing research \cite{Jaiswal:2016hex}. 
In this causal theory, shear stress $\pi^{\mu\nu}$ dynamically evolves with a characteristic relaxation time 
$\tau_{\pi}$. 
We use the equation for shear evolution and relaxation time $\tau_\pi=\frac{3\eta}{sT}$ from Ref. \cite{Shen:2011eg}. 
For a longitudinal boost invariant Bj\"orken flow, the energy equation and shear pressure 
$\Phi=\pi^{00}-\pi^{zz}$ evolution becomes \cite{Biro:1993qt,Bhatt:2009zg} 
\begin{align}
\frac{\dot{T}}{T}+\frac{1}{3\tau}+\frac{1}{4}
\frac{a_2\dot{\lambda}_g+2b_2\dot{\lambda}_q}
{a_2{\lambda}_g+2b_2{\lambda}_q} &= \frac{\Phi}{4\tau}\frac{1}{(a_2{\lambda}_g+2b_2{\lambda}_q)T^4}
  \, \label{eq:tempevol1} \\
\dot{\Phi}+\frac{\Phi}{\tau_\pi}
%
%&=&\frac{1}{2\beta_2} \nabla_{\langle\alpha} u_{\beta\rangle} \nonumber\\,
&= \frac{8}{27\tau}(a_2 \lambda_g + 2b_2 \lambda_q)\,T^4.
\, \label{eq:shearvisco4}
\end{align}
In all the equations, proper time derivative is denoted by the dot. 
There are several temperature dependent shear viscosity prescriptions available\cite{Bhatt:2011kr} and we 
use the one used in Ref.\cite{Shen:2011eg}: 
$\left(\eta/s\right)_1 = 0.2 + 0.3\,\frac{T-T_{chem}}{T_{chem}}$ (with $T_{chem}=0.165$ GeV)
which is known to 
result in cavitation, in the equilibrium case within one dimensional Bjorken flow, at 
early times itself \cite{Bhatt:2011kr}. 

Under chemical non-equilibrium, above set of equations have to be solved together with the 
parton density evolution equations prescribed through master equations. By considering 
the relevant reactions, $gg \longleftrightarrow ggg$ and $gg \longleftrightarrow q\bar q$, 
in the context of baryonless QGP, rate equations for the fugacities can be written in Bj\"orken flow 
as \cite{Biro:1993qt,Bhatt:2009zg} 
\begin{align}
\frac{\dot{\lambda_g}}{\lambda_g}+3\frac{\dot{T}}{T}+\frac{1}{\tau}
&=R_{gg\rightarrow ggg}\left(1-\lambda_g\right) - 2R_{g\rightarrow q}\left(1-\frac{\lambda^2_q} {\lambda^2_g}\right)
\,\label{eq:gluonfugacityevol2}\\
\frac{\dot{\lambda_q}}{\lambda_q}+3\frac{\dot{T}}{T}+\frac{1}{\tau}
&=R_{g\rightarrow q}\frac{a_1}{b_1}
\left( \frac{\lambda_g}{\lambda_q}-\frac{\lambda_q}{\lambda_g}
\right). \label{eq:quarkfugacityevol3} 
\end{align}
Here the rates are given as $R_{gg\rightarrow ggg} = 2.1\alpha^2_sT(2\lambda_g-\lambda^2_g)^{1/2}$ 
and $R_{g\rightarrow q} = 0.24N_f\alpha^2_s\lambda_gT\,ln(5.5/\lambda_g)$, with $\alpha_s$ being the strong coupling constant \cite{Biro:1993qt}. 

\begin{figure}
\begin{center}
\includegraphics[width=8.6cm]{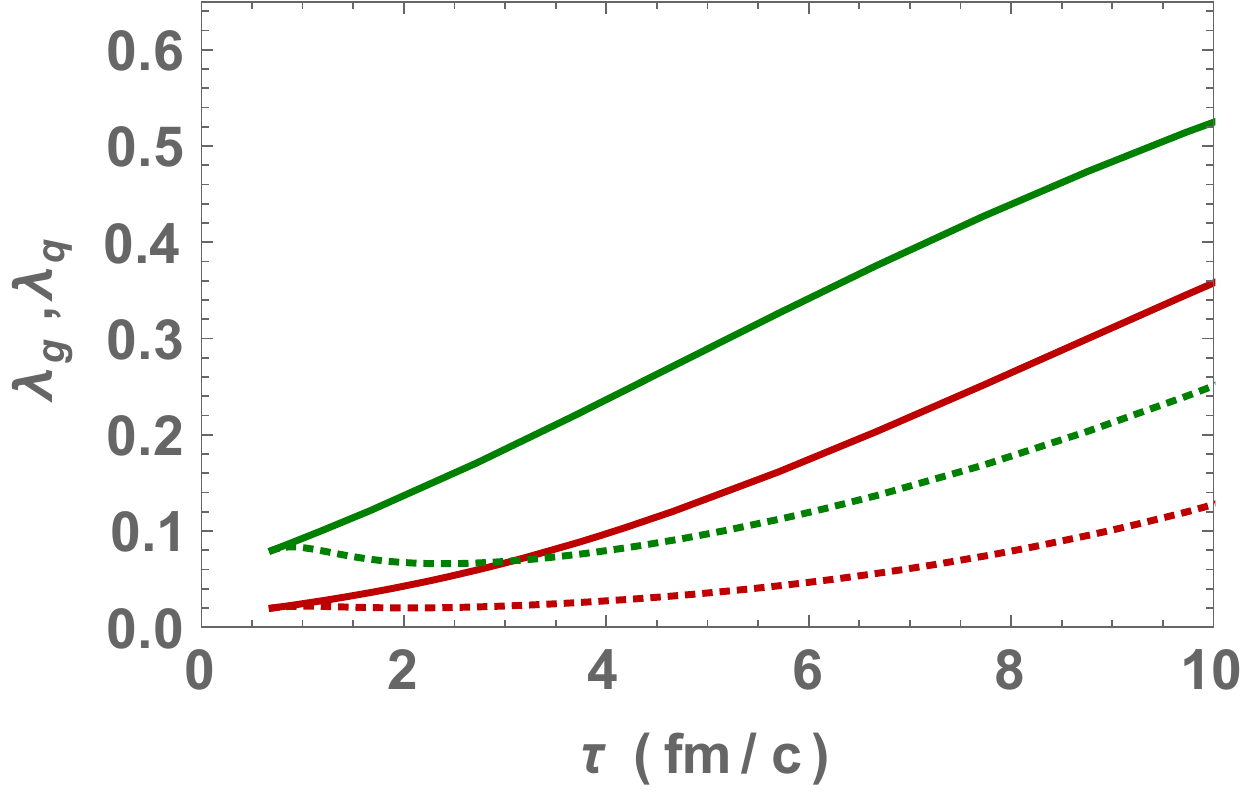}
\caption{Evolution of fugacities in presence of temperature dependent shear viscosity. Green (Red) line denotes 
gluon (quark) fugacity.  
Dotted lines denote the case with viscosity while thick lines correspond to zero viscosity.}\label{sreekanth_fig:1}
\end{center}
\end{figure}  

\begin{figure}
\begin{center}
\includegraphics[width=8.6cm]{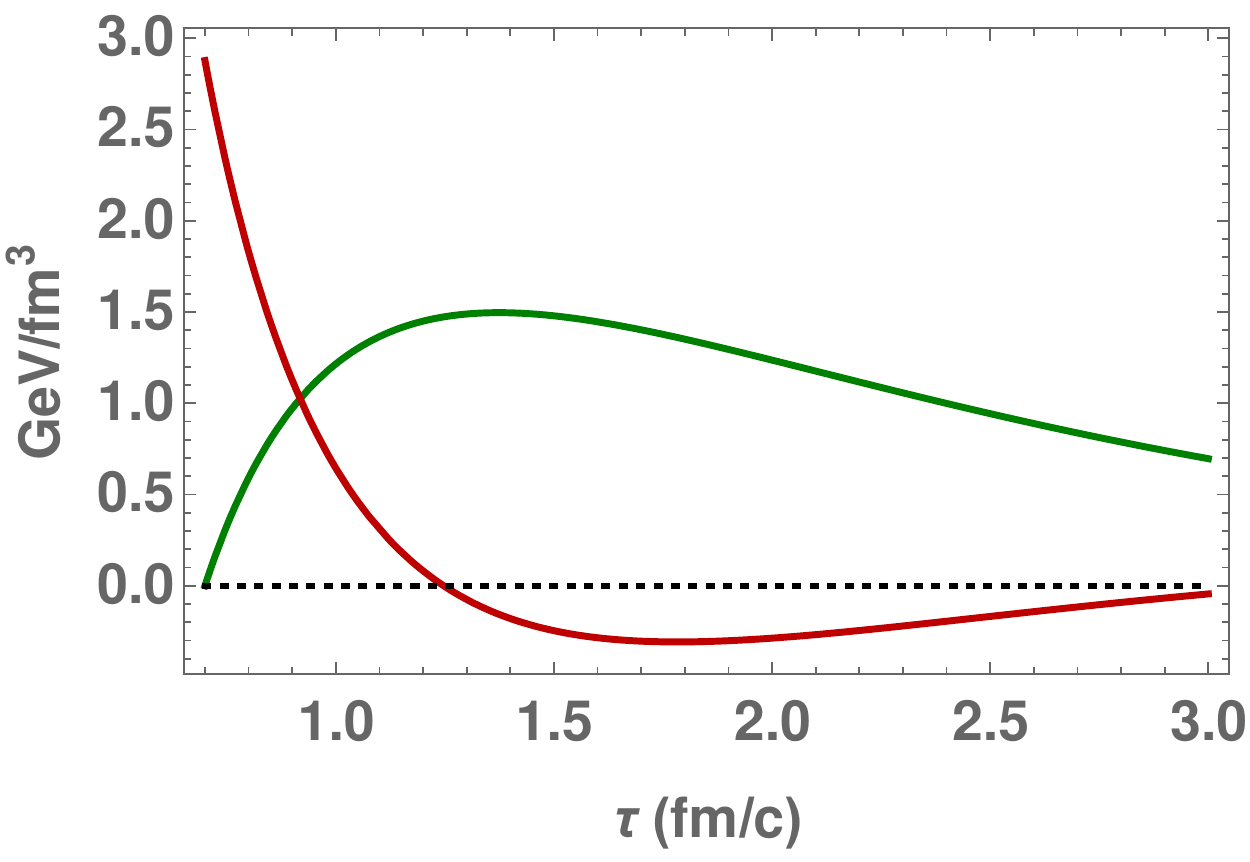}
\caption{Evolution of longitudinal pressure $P_z$ (red curve) 
and shear stress $\Phi$ (green curve). One can see $Pz=0$ at early times itself.}\label{sreekanth_fig:2}
\end{center}
\end{figure}  

Now Eqs.~\eqref{eq:tempevol1}-\eqref{eq:quarkfugacityevol3} describe the longitudinally expanding chemically equilibrating 
viscous hot QGP \cite{Bhatt:2009zg} and it can be solved numerically once 
all the initial conditions are given. While evolving the hydrodynamical code one 
need to make sure that the effective pressures- which has contributions from viscosities remain
positive. This effective pressure of the expanding fireball  
in the longitudinal direction is given by \cite{Bhatt:2011kr}
\begin{eqnarray}
P_{z} &=& P - \Phi,
\label{pressures}
\end{eqnarray}
which denotes the deviation from equilibrium pressure due to dissipative effects. 
The cavitation condition is then given as $P_z=0$, which in our case readily translates to 
\bea
%\frac{1}{3}(a_2 \lambda_g(\tau) + b_2 [\lambda_q (\tau)+\lambda_{\bar{q}} (\tau)])\,T(\tau)^4 -\Phi (\tau) &=& 0 \nn
\left[a_2 \lambda_g(\tau) +2 b_2 \lambda_q (\tau)\right]\,T(\tau)^4 -3\Phi (\tau) &=& 0.
\eea

\begin{figure}[!ht]
\begin{center}
\includegraphics[width=8.6cm]{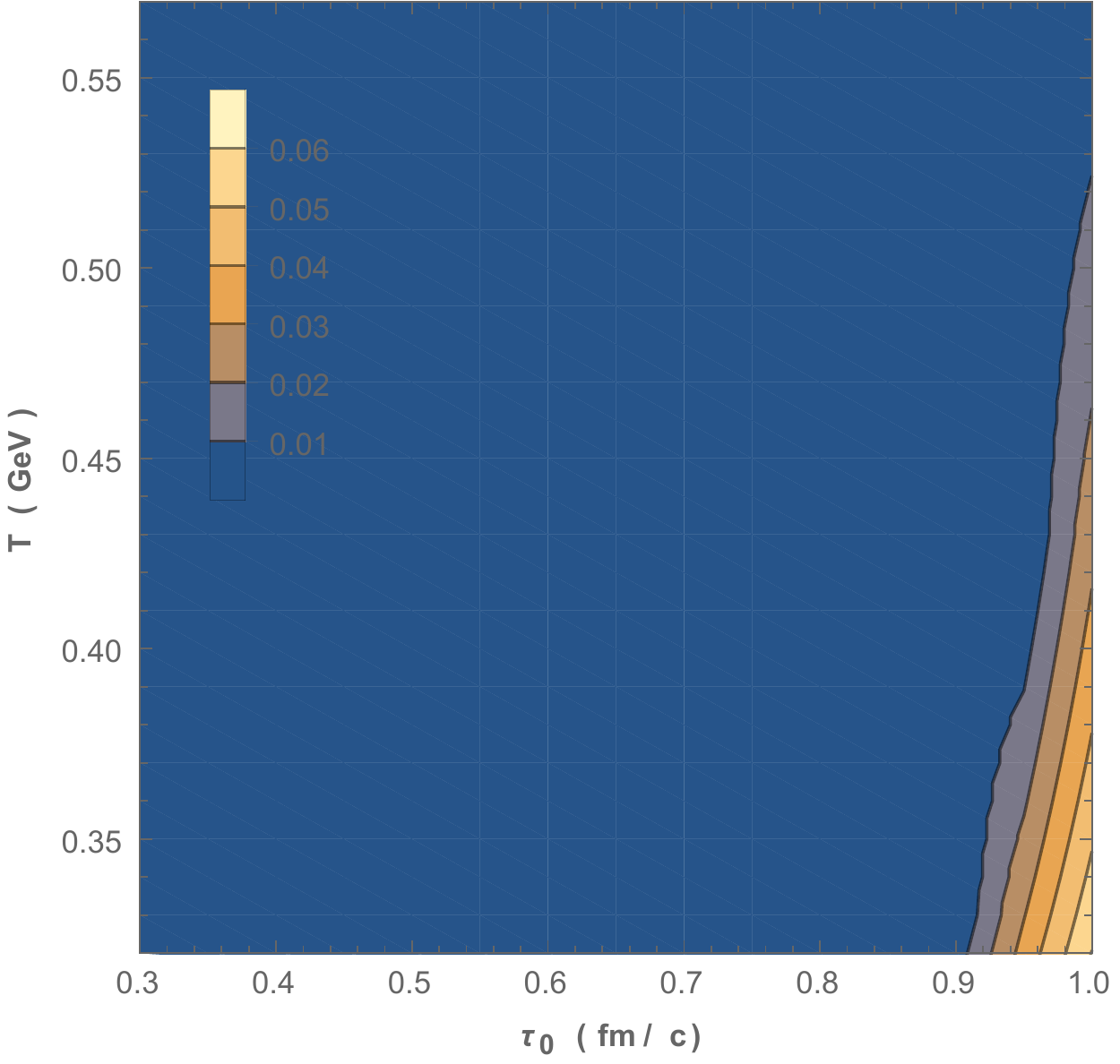}
\caption{Cavitation scenarios for various initial conditions. Darkest region denotes the existence of $P_z=0$.}\label{sreekanth_fig:3}
\end{center}
\end{figure}

\subsection{Results and discussions}
With the initial conditions relevant for LHC energies, we evolve the system, while monitoring the 
effective longitudinal pressure. The initial values taken are $\lambda_g^0=0.08$, $\lambda_q^0=0.02$, 
$T_0=0.570$ GeV and $\tau_0=0.7$ fm/c \cite{Biro:1993qt}. We evolve the system till its temperature drops to $T_c=.180$ GeV. 
Also we set minimum value of shear stress at the beginning: $\Phi(\tau_0)=0$. 

First we plot the evolution of quark and gluon fugacities in the case of temperature dependent $\eta/s$ 
in Figure [\ref{sreekanth_fig:1}]. It is observed that the 
equilibration process is delayed in the presence of viscosity. Now we plot the longitudinal pressure in Figure [\ref{sreekanth_fig:2}]. 
It can be seen that in the early times itself system reaches cavitation triggered by the peak in shear stress value. 
It need to be noted that such initial temperature and time had resulted in cavitations in 
chemically equilibrated case too \cite{Bhatt:2011kr}. Finally, we look into possible combinations of initial times and temperatures 
for which, under the considered chemical non-equilibrium scenario, cavitations occur. It is clear from Figure [\ref{sreekanth_fig:3}] 
that in order to avoid negative effective pressure scenarios one need to go for 
large initial time and relatively low initial temperatures. This might not be compatible with the 
usual early times and high temperatures associated with the chemical equilibration studies. It must also be noted that the occurrence of cavitation in early times will force us to look for alternate ways to understand the evolution of the system and freezeout. On the other hand, the successful statistical hadronization models predict a chemical freezeout in the vicinity of $T_c$, thus such an early time cavitation observed here, perhaps, point towards non-applicability of such high valued viscosity prescriptions used in the literature. 

\subsection{Summary and conclusions}
In conclusion, we have studied the effect of temperature dependent $\eta/s$ in the chemical equilibration of 
hot baryonless QGP produced in high energy heavy-ion collisions. It has observed that equilibration gets delayed 
because of slower cooling rate of the fireball in presence of viscosity. Further, it was seen that 
cavitations set in rather early time itself making the hydrodynamical evolution model into trouble. 
It is interesting to note that cavitation scenarios are not getting washed away with the introduction of 
chemical non-equilibration in the problem. One can also think of giving a bound to shear viscosity by assuming 
that the cavitations do not occur. This work is in progress and 
will be reported elsewhere \cite{Sreekanth:2019}.

%% file: Samapan/samapan.tex
\section{First Order Dissipative Hydrodynamics from an Effective Covariant Kinetic Theory}
\textit{Samapan Bhadury, Manu Kurian, Vinod Chandra, Amaresh Jaiswal}

\bigskip

{\small
Relativistic Dissipative Hydrodynamics has been used successfully as a tool to describe the space-time evolution of hot QCD matter created in high energy heavy ion collisions. We will describe how the hot QCD medium can be modeled using a quasiparticle  picture that is consistent with the equation of state of the system, estimated from Lattice QCD. In this model, we investigate evolution equation for the shear stress tensor, the bulk viscous pressure and the charge current under first order dissipative relativistic hydrodynamic. This shows some modification in the behavior of the transport coefficients.% The temperature evolution and pressure anisotropy for one-dimensional boost invariant Bjorken expansion is also studied.
}

\bigskip

%\keywords{Effective fugacity, Dissipative evolution, Quark chemical potential}
%\end{abstract}

%
\subsection{Introduction}

Collision of two heavy nuclei with ultra-relativistic velocities at RHIC and LHC, produces a hot QCD medium commonly known as Quark-Gluon-Plasma(QGP). The relativistic dissipative hydrodynamics serves as an efficient theoretical approach to describe the space-time evolution of the created QGP. In this contribution, we study the first order dissipative evolution equation of the QGP for a non-zero baryon chemical potential and quark mass following a recently proposed effective covariant kinetic theory~\cite{Mitra:2018akk}. We utilize the effective fugacity quasiparticle model (EQPM)~\cite{Chandra:2011en} to encode the effects of hot QCD equation of state (EoS) in terms of temperature dependent fugacity parameter. By employing the iterative Chapman-Enskog like expansion, we solve the relativistic transport equation in presence of an EQPM mean field term under the relaxation time approximation (RTA)~\cite{Anderson_Witting} . We study the ratios of the dissipative quantities with mean field correction at finite baryon chemical potential. %We investigate the viscous corrections to the temperature evolution and pressure anisotropy for the boost invariant Bjorken expansion.
\subsection{Effective covariant kinetic theory}
Under the RTA, the relativistic Boltzmann equation, which describes the change of momentum distribution function of each particle species $k$ is given by~\cite{Mitra:2018akk},

\begin{equation}\label{1}
\tilde{p}^{\mu}_k\,\partial_{\mu}f_k(x,\tilde{p}_k)+F_k^{\mu}\left(u\!\cdot\!\tilde{p}_k\right)\partial^{(p)}_{\mu} f_k = -\left(u\!\cdot\!\tilde{p}_k\right)\frac{\delta f_k}{\tau_R},
\end{equation}
where, $\tau_R$ is the thermal relaxation time and $u_\mu$ is the fluid velocity. The mean field force term, $F_k^{\mu}=-\partial_{\nu}(\delta\omega_k u^{\nu}u^{\mu})$ can be realized from the conservation laws~\cite{Mitra:2018akk}. The covariant form of EQPM distribution function for quarks, antiquarks and gluons at non-zero baryon chemical potential $\mu_q$ can be written as,
\begin{align}
f^0_q &=\frac{z_q \exp{[-\beta (u\!\cdot\! p_q - \mu_q)]}}{1 + z_q\exp{[-\beta (u\!\cdot\! p_q - \mu_q)]}}, \label{2}\\
f^0_{\bar{q}} &=\frac{z_{\bar{q}} \exp{[-\beta (u\!\cdot\! p_{\bar{q}} + \mu_q)]}}{1 + z_{\bar{q}}\exp{[-\beta (u\!\cdot\! p_{\bar{q}} + \mu_q)]}}, \label{3}\\
f^0_g &=\frac{z_g \exp{[-\beta\, u\!\cdot\! p_g]}}{1 - z_g\exp{[-\beta\, u\!\cdot\! p_g]}}, \label{2a}
\end{align}
where $z_{q}$ and $z_{g}$ are the temperature dependent effective fugacity parameter for quarks and gluons, respectively. It is to be noted, that the fugacity parameters $z_k$ that encode the thermal medium effects are same for quarks and antiquarks, $i.e.$ $z_q=z_{\bar{q}}$, in the present context. The dispersion relation relates the dressed (quasiparticle) four-momenta $\Tilde{p}_k^{\mu}$ and the bare particle four-momenta $p_k^{\mu}$ as, 
\begin{equation}\label{3a}
\tilde{p_k}^{\mu} = p_k^{\mu}+\delta\omega_k\, u^{\mu}, \qquad
\delta\omega_k= T^{2}\,\partial_{T} \ln(z_{k}),
\end{equation}
which implies, zeroth component of the four-momenta is given by $\tilde{p_k}^{0}\equiv\omega_{k}=E_k+\delta\omega_k$. 
We assume the system to be near local equilibrium i.e. $f_k=f^0_k+\delta f_k$ and solve the relativistic Boltzmann equation, employing an iterative Chapman-Enskog like expansion, where $\delta f_k/f^0_k \ll1$ and $\delta f_k$ have the forms,
\begin{align}
\!\!\delta f_q &= \tau_R\bigg[ \tilde{p}_q^\gamma\partial_\gamma \beta \!+\! \frac{\tilde{p}_q^\gamma}{u\!\cdot\!\tilde{p}_q} \!\Big( \beta\, \tilde{p}_q^\phi \partial_\gamma u_\phi \!-\! \partial_\gamma \alpha \Big) \!-\! \beta\theta\,\delta\omega_q \bigg]f_q\Tilde{f}_q, \label{4}\\
\!\!\delta f_{\bar{q}} &= \tau_R\bigg[ \tilde{p}_{\bar{q}}^\gamma\partial_\gamma \beta \!+\! \frac{\tilde{p}_{\bar{q}}^\gamma}{u\!\cdot\!\tilde{p}_{\bar{q}}} \!\Big( \beta\, \tilde{p}_{\bar{q}}^\phi \partial_\gamma u_\phi \!+\! \partial_\gamma \alpha \Big) \!-\! \beta\theta\,\delta\omega_{\bar{q}} \bigg] f_{\bar{q}} \tilde{f}_{\bar{q}}, \label{5}\\
\!\!\delta f_g &= \tau_R\bigg( \tilde{p}_g^\gamma\partial_\gamma \beta + \frac{\beta\, \tilde{p}_g^\gamma\, \tilde{p}_g^\phi}{u\!\cdot\!\tilde{p}_g}\partial_\gamma u_\phi -\beta\theta\,\delta\omega_g \bigg) f_g\tilde{f}_g. \label{6}
\end{align}
where, $\theta\equiv\partial_{\mu}u^{\mu}$ is the expansion scalar and $\alpha=\beta\mu_q$. The thermal relaxation time $\tau_{R}$ is assumed to be independent of particle four-momenta.

\subsection{Dissipative evolution equation}
\begin{figure*}[tbh]
\centering
\includegraphics[scale=0.34]{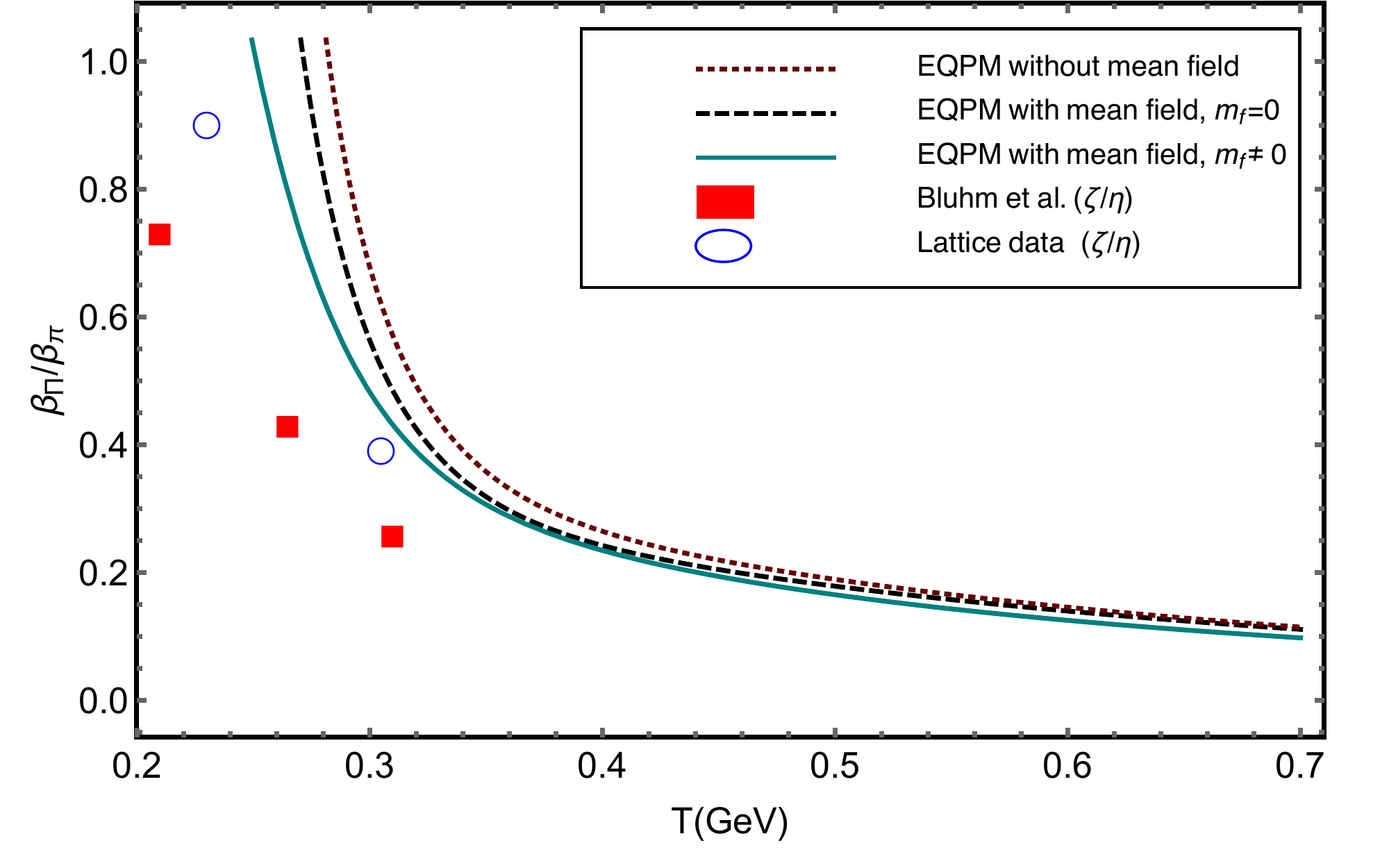}
\caption{Evolution of $(\beta_{\Pi}\!/\beta_{\pi})$ with temperature and comparison with results in~\cite{Bluhm:2011xu,Meyer:2007ic,Meyer:2007dy}. }
\label{f1}
\end{figure*}
Shear stress tensor is defined in terms of the non-equilibrium part of the distribution function $\delta f_k$ within EQPM as,
\begin{align}\label{7}
\pi^{\mu\nu}=&\sum_k g_k\Delta^{\mu\nu}_{\alpha\beta}\int{d\tilde{P}_k~ \tilde{p}_k^{\alpha}\,\tilde{p}_k^{\beta}\,\delta f_k}+\sum_k g_k\,\delta \omega_k\, \Delta^{\mu\nu}_{\alpha\beta}\int{d\tilde{P}_k~ \tilde{p}_k^{\alpha}\,\tilde{p}_k^{\beta}\,\frac{1}{E_k}\delta f_k},
\end{align}
where $g_k$ is the degeneracy factor and $d\tilde{P}_k\equiv\frac{d^3\mid\vec{\tilde{p}}_k\mid}{(2\pi)^3\omega_{k}}$ is the momentum integral factor. We use the two index projection operator $\Delta^{\mu\nu}\equiv g^{\mu\nu}-u^\mu u^\nu$  and a  four-index tensor $\Delta^{\mu\nu}_{\alpha\beta}\equiv\frac{1}{2}(\Delta^\mu_\alpha\Delta^\nu_\beta +\Delta^\mu_\beta\Delta^\nu_\alpha)-\frac{1}{3}\Delta^{\mu\nu}\Delta_{\alpha\beta}$ which is a traceless symmetric projection operator orthogonal to the fluid velocity. The bulk viscous pressure $\Pi$ and the particle diffusion current $n^{\mu}$ can be defined respectively as,
\begin{align}\label{8}
&\Pi=-\frac{1}{3}\!\sum_k\! g_k\Delta_{\alpha\beta}\int\! {d\tilde{P}_k\, \tilde{p}_k^{\alpha}\,\tilde{p}_k^{\beta}\,\delta f_k}-\frac{1}{3}\!\sum_k\! g_k\,\delta \omega_k\,\Delta_{\alpha\beta}\int\! {d\tilde{P}_k\, \tilde{p}_k^{\alpha}\,\tilde{p}_k^{\beta}\,\frac{1}{E_k}\delta f_k},\\
&n^{\mu}=g_q\Delta_{\alpha}^{\mu}\int{d\tilde{P}_q~\tilde{p}_q^{\alpha}\, (\delta f_q-\delta f_{\bar{q}})}-\delta\omega_q g_q\Delta_{\alpha}^{\mu}\int{d\tilde{P}_q~ \tilde{p}_q^{\alpha}\,\frac{1}{E_q}(\delta f_q-\delta f_{\bar{q}})}.
\end{align}
We replace $\delta f_k$ from Eqs.~(\ref{4})-(\ref{6}) and keep terms up to first order in gradients to obtain the Naiver-Stokes like equation as follows,
\begin{align}
&\pi^{\mu\nu} = 2\,\tau_R\,\beta_\pi\,\sigma^{\mu\nu},
&\Pi = -\tau_R\,\beta_\Pi\,\theta,
&&n^\mu = \tau_R\,\beta_n\,\nabla^\mu \alpha,\label{9}
\end{align}
with $\sigma^{\mu\nu}\equiv\Delta^{\mu\nu}_{\alpha\beta}\nabla^{\alpha}u^{\beta}$.
The dissipative coefficients $\beta_{\pi}$, $\beta_\Pi$ and $\beta_n$ are expressed in terms of thermodynamic integrals for massive and massless case  in the Ref.~\cite{Bhadury:2019xdf}. %We employ Bjorken's prescription to model the dissipative hydrodynamical evolution of the QGP~\cite{Bjorken:1982qr}. In terms of Milne coordinates,$(\tau,x,y,\eta_s)$, where $\tau=\sqrt{t^2-z^2}$, $\eta_s=\tanh^{-1}(z/t)$, $u^{\mu}=(1,0,0,0)$ and $g^{\mu\nu}=(1,-1,-1,-1/\tau^2)$, the energy density $\varepsilon$ evolution in the boost-invariant longitudinal expansion takes the form,
%
%\begin{equation}\label{10}
%\frac{d\varepsilon}{d\tau}=-\bigg(\frac{\varepsilon+P}{\tau}\bigg)+\bigg(\frac{\zeta+4\eta/3}{\tau^2}\bigg),
%\end{equation}
%
%with $\theta = 1/\tau,$ $\Pi=-\zeta/\tau$, $\pi^{\mu\nu}\sigma_{\mu\nu} = \Phi/\tau$, $\Phi=4\eta/3\tau$.
%
\subsection{Results and Discussions}
\begin{figure*}
\begin{center}
% \hspace{-.5 cm}
{\includegraphics[scale=0.343]{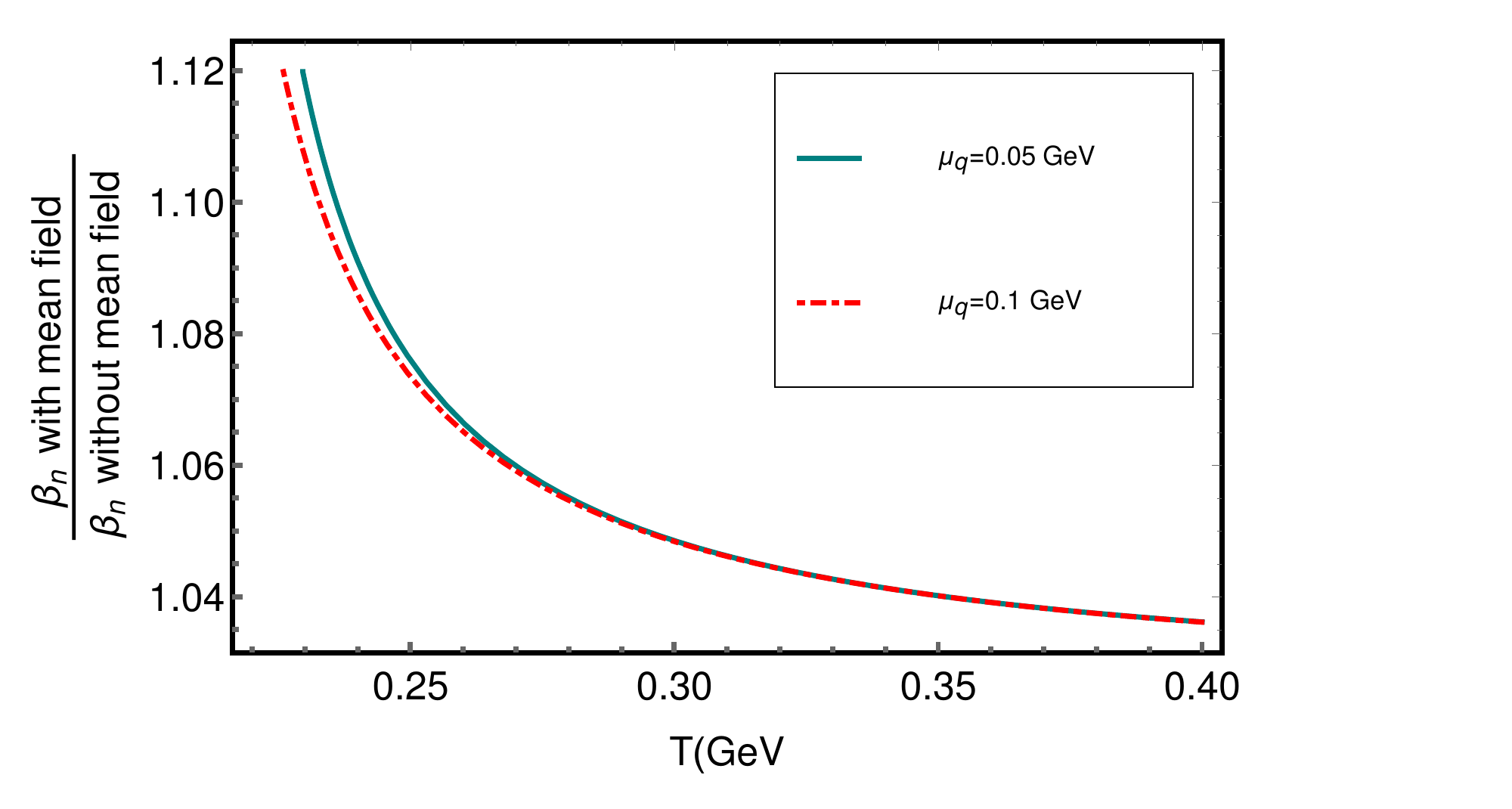}}
 \hspace{-1 cm}
{\includegraphics[scale=0.32]{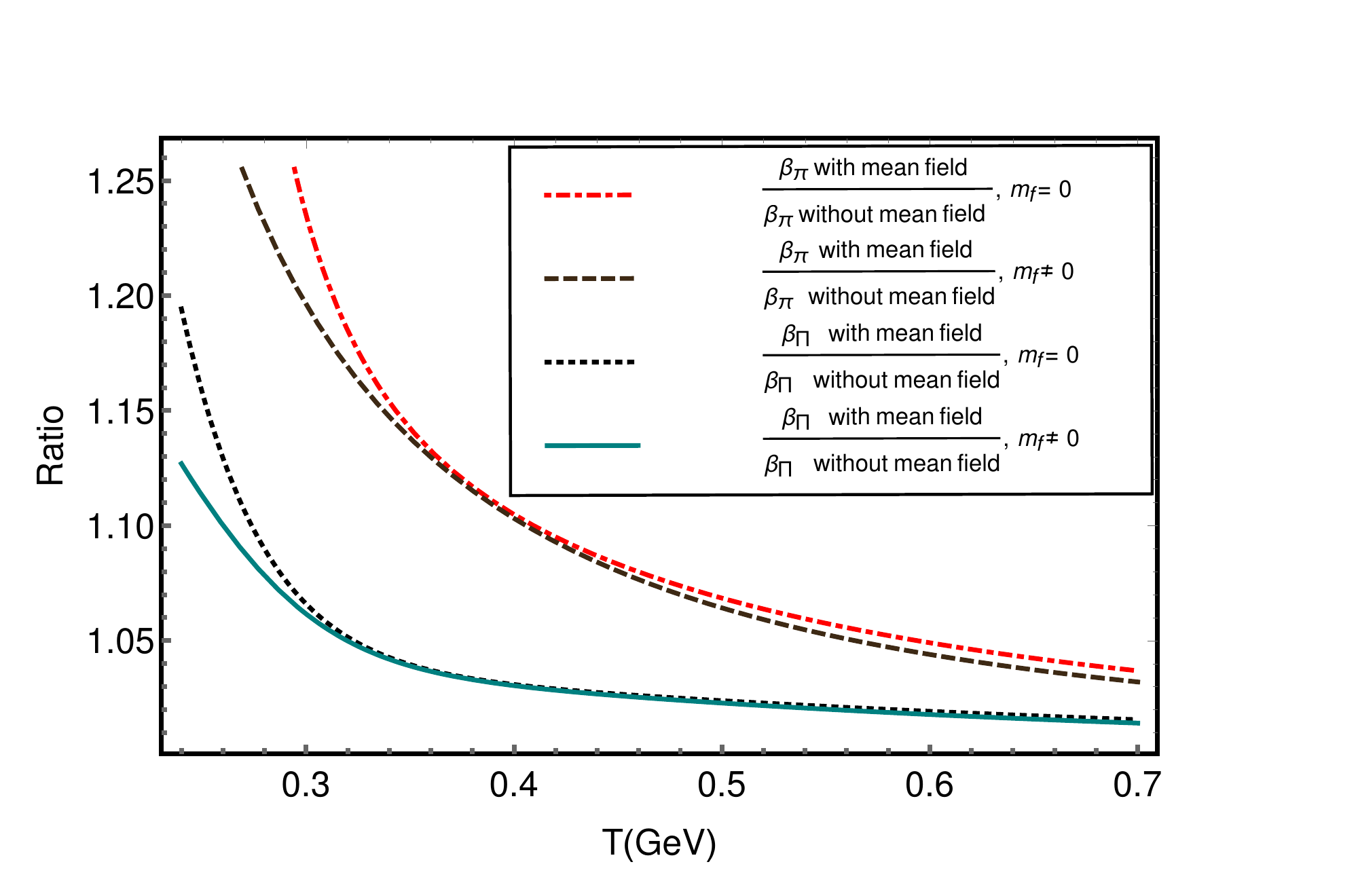}}
\caption{(Left panel) Temperature dependence of the mean field contribution to the particle
diffusion for different quark chemical potential. (Right panel) The effect of mean field contributions to the coefficients of bulk viscous pressure, shear tensor at $\mu = 0.1$ GeV with and without quark mass correction.}
\label{f2}
\end{center}
\end{figure*}   
The temperature dependence of the ratio of the coefficient of the bulk viscous tensor to that of the shear tensor $(\beta_{\Pi}/\beta_{\pi})$ at $\mu_q=0.1$ GeV is shown in Fig.~\ref{f1}. Under RTA, the ratio becomes $\beta_{\Pi}/\beta_{\pi}=\zeta/\eta$, where  $\zeta$ and $\eta$ are the bulk  and shear viscosities of medium. The ratio deceases with the increasing temperature. There are substantial affects due to quark mass correction and mean field corrections in the low temperature regime close to the transition temperature. 
In Fig.~\ref{f2} (left panel), the mean field effects to the first order coefficient of particle diffusion is shown for different quark chemical potential $\mu_q$. In the low temperature regimes, the effects of quark mass and chemical potential are visible whereas in the higher temperature regimes the mean field contributions are almost independent on $m_q$ and $\mu_q$. This may be attributed to the fact that in low temperature regime, when the temperature is of the same order of magnitude as quark mass and chemical potential, the effect of slight changes in these quantities become noticeable. On the other hand, at high temperature, there is a separation of scale the the effect due to changes in quark mass and chemical potential are not significant. The mean field correction to the transport parameters with binary, elastic collisions at $m_q = 0$ and $\mu_q = 0$ is described in Ref \cite{Mitra:2018akk} In Fig.~\ref{f2} (Right panel), the contributions to first order coefficients of the shear tensor and bulk viscous pressure due to mean field at quark chemical potential $\mu_q = 0.1$ GeV are depicted. Since the mean field corrections at high temperature regimes are negligible, the ratio asymptotically tends to unity. At this juncture, we note that the first order viscous hydrodynamics theory has issues with causality due to parabolic nature of the evolution equations. Deriving second-order causal hydrodynamic theory, within the present EQPM framework, is left for future work.

%% file: Avdhesh/avdhesh.tex
\section{Hydrodynamics with spin and its application to heavy-ion collisions} 
 
 %\thanks{Contribution to the Workshop on Dynamics of QCD Matter Proceedings}
 
\textit{Avdhesh Kumar} 

\bigskip   	
   	
%	\email{avdhesh.kumar@ifj.edu.pl} 
%	\affiliation{Institute of Nuclear Physics Polish Academy of Sciences, PL-31-342 Krak\'ow, Poland}
	
%\date{\today}

%\begin{abstract}
{\small
We briefly discuss, recently introduced equilibrium Wigner functions for spin 1/2 particles that are used in the semiclassical kinetic equations.  In the case of local thermodynamic equilibrium, we outline a procedure to formulate hydrodynamic framework for particles with spin 1/2 based on the semiclassical expansion of Wigner functions. For the case of a boost-invariant and transversely homogeneous expansion of the fireball produced in the heavy-ion collision we show that this formulation can be used to determine the space-time evolution of the spin polarization and physical observables related to the spin polarization.}
 %\end{abstract}
 
 \bigskip

%\pacs{24.10.Nz, 25.75.Ld, 25.75.-q}
	
%\keywords{Wigner functions, thermodynamic equilibrium, spin polarization}

%\maketitle

%\tableofcontents
\subsection{Introduction}
\label{sec:introduction}
%\smallskip

In the non-central heavy-ion collisions, nuclei colliding at ultra-relativistic energies carry a very large orbital angular momentum. After the collision a significant portion of this orbital angular momentum can be retained in the interaction region which can be further transformed from the initial purely orbital form into the spin part. The latter can be reflected in the spin polarization of the emitted particles.
Indded, recently global spin polarization of $\Lambda$ and $\bar\Lambda$ hyperons emitted from the fireball created in the non-central heavy ion collisions has been measured by the STAR collaboration ~\cite{STAR:2017ckg,Adam:2018ivw}. This result can be successfully explained by relativistic hydrodynamics (ideal or viscous)~\cite{Karpenko:2016jyx}. 

Hydrodynamic models that are used to describe the global spin polarization of $\Lambda$ and 
$\bar\Lambda$-hyperons make use of the fact that spin polarization effects are governed by thermal vorticity $ \varpi_{\mu \nu}$ which is defined by the expression $\varpi_{\mu \nu} = -\frac{1}{2} (\p_\mu \b_\nu-\p_\nu \beta_\mu)$, where $\beta_\mu$ is the ratio of the fluid flow vector $U_\mu$ and the local temperature $T$, {\it i.e.} $\beta_\mu = U_\mu/T$ ~\cite{Becattini:2016gvu,Karpenko:2016jyx,Becattini:2017gcx}. There remains, however, a puzzle known as sign problem {\it i.e.} the oscillations of the longitudinal polarization of $\Lambda$ as a function of the azimuthal angle observed by the STAR experiment has an opposite sign with respect to the results obtained using hydrodynamic calculations \cite{Niida:2018hfw}. 

On general thermodynamic grounds the spin polarization effects are expected to be governed by the tensor $ \omega_{\mu \nu} $ namely spin polarization tensor~\cite{Becattini:2018duy}
which can be independent of the thermal vorticity $\varpi_{\mu \nu}$.  This suggests a new hydrodynamic approach (known as hydrodynamics with spin), which allow the spin polarization tensor to be treated as an independent dynamical variable. Initial steps in this direction have been made in Refs.~\cite{Florkowski:2017ruc,Florkowski:2017dyn,Becattini:2018duy}, see also follow-up Refs.~\cite{Florkowski:2017njj,Florkowski:2018ual} and other related work~\cite{Weickgenannt:2019dks}. In this contribution, we briefly report on our recent works~\cite{Florkowski:2018ahw,Florkowski:2018fap, Florkowski:2019qdp} where we discuss Wigner function approach to formulate hydrodynamics with spin, for the case of the de Groot, van Leeuwen, and
van Weert (GLW) \cite{DeGroot:1980dk} formalism to study the space-time evolution of spin polarization and related physical observables for the boost invariant Bjorken flow. 
\subsection{Equilibrium Wigner functions}
\label{sec:eqwignerfunctions}
Our starting point are the relativistic
distribution functions $f^{\pm}_{rs}(x,p)$ for particles ($+$) and antiparticles ($-$) with spin 1/2 at local thermodynamical equilibrium as introduced in Ref.~\cite{Becattini:2013fla}
\beq
\fplusrsxp \!\!=\!\!
\frac{1}{2m} \ubarrp X^+ \usp, ~ \fminusrsxp\!\!=\!\!- \frac{1}{2m}\vbarsp X^- \vrp \nn %\\\label{eqd}
\eeq	
where $m$ is the (anti-)particle mass, $r$ and $s$ are the spin indices running from 1 to 2 and $\urp, $ and $\vrp$ are Dirac bispinors. The objects $X^{\pm}$ are the four-by-four matrices defined by the formula 
\beq
X^{\pm} =  \exp\left[\pm \xi(x) - \bmu(x) \pmu \pm \f{1}{2} \omega_{\mu\nu}\Sigma^{\mu\nu}\right] \nn
\eeq
where $\xi=\frac{\mu}{T}$, with $\mu$ and $T$ being the chemical potential and temperature. The quantity $\omnL$ is spin polarization tensor while $\SmunuU = (i/4) [\gamma^\mu,\gamma^\nu]$ is known as the Dirac spin operator. 

By assuming that the spin polarization tensor $\omnL$ satisfies the conditions, $\omnL \omnU \geq 0$ and $\omnL \omnUD = 0$~\cite{Florkowski:2017dyn}, where $\omnUD=\frac{1}{2}\epsilon_{\mu\nu\alpha\beta}\omega^{\alpha\beta}$ is the dual spin polarization tensor, we can introduce a new quantity $\zeta  = \f{1}{2} \sqrt{ \frac{1}{2} \omnL \omnU }$ which can be interpreted as the ratio of spin chemical potential $\Omega$ and temperature $T$~\cite{Florkowski:2017ruc}.

The equilibrium Wigner functions can be constructed by taking the above expressions for $\fplusrsxp$ and 
$\fminusrsxp$ as an input~\cite{DeGroot:1980dk} 
\beq
\Weqpxk &=& \frac{1}{2} \sum_{r,s=1}^2 \int dP\,
\delta^{(4)}(k-p) u^r(p) {\bar u}^s(p) f^+_{rs}(x,p),\nn \\
\Weqmxk &=&-\frac{1}{2} \sum_{r,s=1}^2 \int dP\,
\delta^{(4)}(k+p) v^s(p) {\bar v}^r(p) f^-_{rs}(x,p),\nn
\eeq	
where $dP = \frac{d^3p}{(2 \pi )^3 E_p}$ is the Lorentz invariant measure with $E_p = \sqrt{m^2 + \pv^2}$ being the on-mass-shell particle energy. Four momentum $k^\mu = (k^0, \boldsymbol{k})$ in the Wigner functions is not necessarily on the mass shell.

Being  4$\times$4 matrices that satisfy the relation $\Weqpmxk =\gamma_0 \Weqpmxk^\dagger \gamma_0$, equilibrium Wigner functions can always be expressed as combinations of the $16$ independent generators of the Clifford algebra
~\cite{Elze:1986qd,Vasak:1987um}
\beq
\Weqpmxk&=&\f{1}{4} \Big[ \Feqpmxk + i \gamma_5 \Peqpmxk + \gamma^\mu {\cal V}^\pm_{{\rm eq}, \mu}(x,k) \nn\\
&&+ \gamma_5 \gamma^\mu {\cal A}^\pm_{{\rm eq}, \mu}(x,k)
+ \SmnU {\cal S}^\pm_{{\rm eq},\mu \nu}(x,k) \Big]. \label{eq:equiwfn}
\eeq
Note that the coefficient functions appearing in the above decomposition can be obtained by contracting $\Weqpmxk$ with appropriate gamma matrices and then taking the trace~\cite{Florkowski:2018ahw}. The total Wigner function is the sum of the particle and antiparticle contributions $\Weqxk = \Weqpxk + \Weqmxk$. 
%
%%%%%%%%%%%%%%%%%%%%%%%%%%%%%%%%%%%%%%%%%%%%
\subsection{Semi-classical equation and formulation of hydrodynamics with spin} 
\label{sec:kinetic}
 A similar decomposition to \rf{eq:equiwfn} can be done  for any arbitrary Wigner function $\Wxk$. 
 % \beq
%\Wpmxk&=&\f{1}{4} \Big[ \Fpmxk + i \gamma_5 \Ppmxk + \gamma^\mu {\cal V}^\pm_{\mu}(x,k) \nn\\
%&&+ \gamma_5 \gamma^\mu {\cal A}^\pm_{\mu}(x,k)
%+ \SmnU {\cal S}^\pm_{\mu \nu}(x,k) \Big]. \label{eq:equiwfn}
%\eeq
%
 In absence of any mean fields $\Wxk$ satisfies the following equation \cite{Vasak:1987um}
\bel{eq:eqforW}
\left(\gamma_\mu K^\mu - m \right) {\cal W}(x,k) \!\!=\!\! C[{\cal W}(x,k)]; {K^\mu = k^\mu + \frac{i \hbar}{2} \,\p^\mu},
\eel
where $ C[{\cal W}(x,k)]$ is the collision term. In global or local equilibrium the 
collision term vanishes. In this situation, solution of above equation 
can be written in the form of a series in $\hbar$,
\beq
{\cal X} = {\cal X}^{(0)}  + \hbar {\cal X}^{(1)}  +  \hbar^2 {\cal X}^{(2)}   + \cdots;   {\cal X} \in \{{\cal F}, {\cal P}, {\cal V}_\mu,{\cal A}_\mu,  {\cal S}_{\nu\mu} \}\nn
\eeq
Keeping the zeroth and first order terms in $\hbar$ expansion the following equations for the coefficient functions ${\cal F}_{(0)}(x,k)$ and ${\cal A}^\nu_{(0)} (x,k)$ can be obtained,
\bel{eq:kineqs}
k^\mu \p_\mu {\cal F}_{(0)}(x,k) = 0,~ k^\mu \p_\mu \, {\cal A}^\nu_{(0)} (x,k) = 0, ~
k_\nu \,{\cal A}^\nu_{(0)} (x,k) =0.\nn
\eel
We note here that only two functions ${\cal F}^{(0)}$ and ${\cal A}^{(0)}_\mu$ are basic independent ones; others can be easily expressed by using these two. It can also shown easily that the algebraic structure  the zeroth-order equations obtained from the semi-classical expansion of the Wigner function is consistent with the equilibrium coefficient functions. 
Therefore, we can replace ${\cal X}^{(0)}$ by ${\cal X}_{\rm eq}$. In this way we can get the following Boltzmann-like kinetic equations for the equilibrium coefficient functions
\bel{eq:kineqFC1}
k^\mu \p_\mu {\cal F}_{\rm eq}(x,k) = 0,~ k^\mu \p_\mu \, {\cal A}^\nu_{\rm eq} (x,k) = 0,~ k_\nu \,{\cal A}^\nu_{\rm eq}(x,k) = 0.\nn
\eel
These equations represent the case of global equilibrium and are exactly fulfilled if $\partial_{\mu}\beta^{\nu}-\partial_{\nu}\beta^{\mu}=0$, $\xi=const.$ and spin polarization tensor $\omega_{\mu \nu}=const.$. The equation for $\beta_{\mu}$ field is known as the Killing equation its solution can be written as $\beta_{\mu}=b^{0}_{\mu}+\varpi_{\mu\nu}x^{\nu}$ with thermal vorticity $\varpi_{\mu \nu}$ being constant, Thus we see that both spin polarization tensor $\omega_{\mu \nu}$ and  thermal vorticity $\varpi_{\mu \nu}$ are constant but no conclusion can be drawn whether two are equal in global equilibrium. 

In the local equilibrium only certain moments of above kinetic equations can be set equal to zero; this point has been discussed in great detail in Ref.~\cite{Florkowski:2018ahw}.  It was shown in Ref.~\cite{Florkowski:2018ahw} that the following 
equations for the conservation laws for charge and energy-momentum and spin tensor can be obtained 
\bel{eq:conslaws}
\p_\mu N^\mu(x) = 0, ~~ \p_\mu T^{\mu\nu}_{\rm GLW}((x) = 0,  ~~\p_\lambda S^{\lambda , \mu \nu }_{\rm GLW}(x) = 0. \label{conslaw}
\eel
Exact expressions for charge current $N^\mu(x)$, energy-momentum $T^{\mu\nu}_{\rm GLW}$, and spin tensor $S^{\lambda , \mu \nu }_{\rm GLW}$ are given in Ref.~\cite{Florkowski:2018ahw}. 
As this formulation does not allow for arbitrary large values of the polarization tensor, we restrict ourself to the leading-order expressions in the $\omega_{\mu\nu}$~\cite{Florkowski:2018fap}. In this case the  
expressions for charge current $N^\mu(x)$ and energy momentum tensor $T^{\mu\nu}_{\rm GLW}(x)$ are given by, 
\bel{eq:Nmu}
N^\alpha = n U^\alpha, \quad T^{\a\b}_{\rm GLW} &=& (\varepsilon + P ) U^\a U^\b - P g^{\a\b}.
\eel
In \rf{eq:Nmu}, $n$ is the number density which is given by the expression,
\bel{nden}
n = 4 \, \sinh(\xi)\, \nU(T),
\eel
where the factor $4 \, \sinh(\xi) = 2 \left(e^\xi - e^{-\xi} \right)$ accounts for spin degeneracy and presence of both particles and antiparticles and $\nU(T) = \langle p\cdot U\rangle_0$ is the number density of spinless and neutral massive Boltzmann particles, with $\langle\, \cdots \rangle_0$ denoting a thermal average as defined in Ref.~\cite{Florkowski:2019qdp}.%
%\bel{thermal_av}
%\langle\, \cdots \rangle_0 \equiv \int{dP}  \,(\cdots) \,  e^{- \beta \cdot p}.
%\eel
%

In \rf{eq:Nmu}, $\varepsilon$ and $P$ are the energy density and pressure which are given by following expressions 
\bel{enden}
\varepsilon = 4 \, \cosh(\xi) \, \eU(T), P = 4 \, \cosh(\xi) \, \PU(T),
\eel
similar to $\nU(T)$, the auxiliary quantities $\eU(T)$ and $\PU(T)$ are expressed as $\eU(T) = \langle(p\cdot U)^2\rangle_0$ and $\PU(T) = -(1/3) \langle  p\cdot p - (p\cdot U)^2   \rangle_0$. 

In the leading order in $\omega_{\mu\nu}$ the expression for the spin tensor $S^{\alpha , \beta \gamma }_{\rm GLW}$ is given by 
\beq
S^{\alpha , \beta \gamma }_{\rm GLW}
&=&  {\ch(\xi)} \left( n_{(0)}(T) U^\alpha \omega^{\beta\gamma}  +  S^{\a, \b\g}_{\Delta\GLW} \right),
\label{eq:SGLW}
\eeq
In the above expreesion, the auxiliary tensor $S^{\a, \b\g}_{\Delta\GLW}$ is defined as~\cite{Florkowski:2017dyn}
\beq
S^{\a, \b\g}_{\Delta\GLW} 
&=&  {\cal A}_{(0)} \, U^\a U^\d U^{[\b} \omega^{\g]}_{\HP\d} \lab{SDeltaGLW} \\
&& \hspace{-0.5cm} + \, {\cal B}_{(0)} \, \Big( 
U^{[\b} \Delta^{\a\d} \omega^{\g]}_{\HP\d}
+ U^\a \Delta^{\d[\b} \omega^{\g]}_{\HP\d}
+ U^\d \Delta^{\a[\b} \omega^{\g]}_{\HP\d}\Big),
\nn
\eeq
where
\beq 
{\cal B}_{(0)} &=&-\frac{2}{\hat{m}^2}  \frac{\varepsilon_{(0)}(T)+P_{(0)}(T)}{T},~~{\cal A}_{(0)} = -3{\cal B}_{(0)} +2 n_{(0)}(T). \nn%=-\frac{2}{\hat{m}^2} s_{(0)}(T)\label{coefB}
\eeq
%
%%%%%%%%%%%%%%%%%%%%%%%%%%%%%%%%%%%%%
\subsection{Application of hydrodynamics with spin to heavy ion collsions}
%
%\subsection{Implementation of boost invariance}
We consider the case of transversely homogeneous and  boost-invariant expansion of the fireball produced in the heavy ion collisions. Such a case can be described by the following four boost invariant basis
\beq
U^\a &=& \frac{1}{\tau}\LR t,0,0,z \RR = \LR \ch(\eta), 0,0, \sh(\eta) \RR, \nn \\
X^\a &=& \LR 0, 1,0, 0 \RR,\nn\\
Y^\a &=& \LR 0, 0,1, 0 \RR, \nn\\
Z^\a &=& \frac{1}{\tau}\LR z,0,0,t \RR = \LR \sh(\eta), 0,0, \ch(\eta) \RR
\lab{BIbasis}
\eeq
where $\tau = \sqrt{t^2-z^2}$ is the longitudinal proper time, while $\eta =  \ln((t+z)/(t-z))/2$ is the space-time rapidity. The four-vector $U^\a$ is normalized to unity and a time like vector while four-vectors $X^\a$, $Y^\a$ and $Z^\a$ are space-like and orthogonal to $U^\a$ as well as to each other.

Using the basis \rfn{BIbasis}, the following representation of the spin polarization tensor $\omega_{\mu\nu}$ can be introduced,  (for details, see Refs.\cite{Florkowski:2019qdp}), 
\beq
\omega_{\mu\nu} &=& C_{\kappa Z} (Z_\mu U_\nu - Z_\nu U_\mu) 
 + C_{\kappa X} (X_\mu U_\nu - X_\nu U_\mu)  \nonumber \\
&& + C_{\kappa Y} (Y_\mu U_\nu - Y_\nu U_\mu)\label{eq:omegamunu} \\
&& + \, \epsilon_{\mu\nu\alpha\beta} U^\alpha (C_{\omega Z} Z^\beta + C_{\omega X} X^\beta + C_{\omega Y} Y^\beta). \nn
\eeq
Here we note that due to boost invariant notion the scalar coefficients ${C}'s$
are functions of the proper time $\tau$ only.
Knowing the boost boost-invariant decompostion of $\omega_{\mu\nu}$ a boost-invariant expression for the spin tensor $S^{\a , \b\g}_{\GLW}$ can also be obtained and finally the boost invariant form of conservation laws (given in  \rf{eq:conslaws}) can respectively be  written as follows,
\beq
\dot{n}+\frac{n}{\tau}=0.\lab{eq:charge},\\ 
\dot{\varepsilon}+\frac{(\varepsilon+P)}{\tau}=0.\lab{eq:en}
\eeq
%
%\begin{widetext}
\begin{equation}
\begin{bmatrix}
{\cal L}(\tau) & 0 & 0 & 0 & 0 & 0 \\
0 & {\cal L}(\tau) & 0 & 0 & 0 & 0 \\
0 & 0 & {\cal L}(\tau) & 0 & 0 & 0 \\
0 & 0 & 0 & {\cal P}(\tau) & 0 & 0 \\
0 & 0 & 0 & 0 & {\cal P}(\tau)  & 0 \\
0 & 0 & 0 & 0 & 0 & {\cal P}(\tau)\end{bmatrix}
\begin{bmatrix}
\Dot{C}_{\kappa X} \\
\Dot{C}_{\kappa Y} \\
\Dot{C}_{\kappa Z} \\
\Dot{C}_{\omega X} \\
\Dot{C}_{\omega Y} \\
\Dot{C}_{\omega Z} \end{bmatrix}=\begin{bmatrix}
{\cal{Q}}_1(\tau) & 0 & 0 & 0 & 0 & 0 \\
0 & {\cal{Q}}_1(\tau) & 0 & 0 & 0 & 0 \\
0 & 0 & {\cal{Q}}_2(\tau) & 0 & 0 & 0 \\
0 & 0 & 0 & {\cal{R}}_1(\tau) & 0 & 0 \\
0 & 0 & 0 & 0 & {\cal{R}}_1(\tau)  & 0 \\
0 & 0 & 0 & 0 & 0 & {\cal{R}}_2(\tau)\end{bmatrix}
\begin{bmatrix}
{C}_{\kappa X} \\
{C}_{\kappa Y} \\
{C}_{\kappa Z} \\
{C}_{\omega X} \\
{C}_{\omega Y} \\
{C}_{\omega Z} \end{bmatrix}, \label{cs}
\end{equation}
%\end{widetext}
%
where
${\cal L}(\tau)={\cal A}_1-\frac{1}{2}{\cal A}_2-{\cal A}_3$,
${\cal P}(\tau)={\cal A}_1$, \\
${\cal{Q}}_1(\tau)=-\left[\dot{{\cal L}}+\frac{1}{\tau}\left( {\cal L}+ \frac{1}{2}{\cal A}_3\right)\right]$,
${\cal{Q}}_2(\tau)=-\left(\dot{{\cal L}}+\frac{{\cal L}}{\tau}   \right)$,\\
${\cal{R}}_1(\tau)\!\!=\!\!\!-\left[\Dot{\cal P}+\frac{1}{\tau}\left({\cal P} -\frac{1}{2} {\cal A}_3 \right) \right]$, and
${\cal{R}}_2(\tau)=-\left(\Dot{{\cal P}} +\frac{{\cal P}}{\tau}\right)$
 with ${\cal A}_1$, ${\cal A}_2$ and ${\cal A}_3$ given by, 
${\cal A}_1 = {\cal C} \LR \nU -  {\cal B}_{(0)} \RR \label{A1}$,
${\cal A}_2 = {\cal C} \LR {\cal A}_{(0)} - 3{\cal B}_{(0)} \RR  \label{A2}$, and  
${\cal A}_3 =  {\cal C}\, {\cal B}_{(0)}$. 

One can see that all the $C$ coefficients evolve independently. Moreover, due to the rotational invariance in the transverse plane coefficients  ${C}_{\kappa X}$ and ${C}_{\kappa Y}$ (also ${C}_{\omega X}$ and ${C}_{\omega Y}$) obey the same differential equations.
The system of  Equations \rfn{eq:charge}, \rfn{eq:en} and \rfn{cs} can be easily solved numerically. We first solve Eqs.~\rfn{eq:charge} and \rfn{eq:en} to find the proper-time dependence of the temperature $T$ and chemical potential $\mu$. Once the  functions $T(\tau)$ and $\mu(\tau)$ are known, we can easily determine the functions ${\cal L}$, ${\cal P}$, ${\cal Q}$ and ${\cal R}$ in the \rfn{cs} and finally the proper-time dependence of $C$ coefficients . 

\begin{figure}[ht!]
\centering
\includegraphics[width=0.65\textwidth]{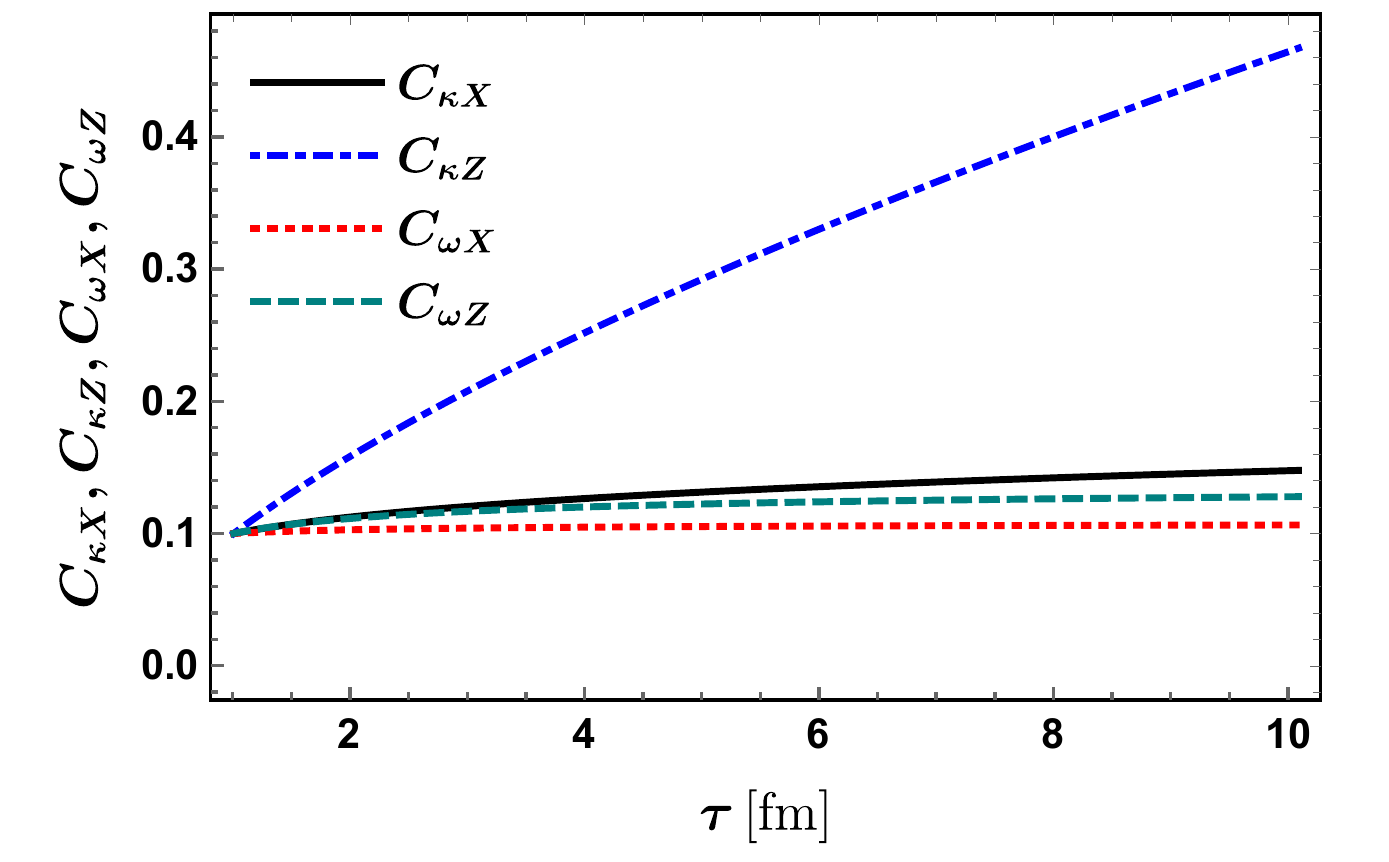}
\caption{Proper-time evolution of the coefficients $C_{\kappa X}$, $C_{\kappa Z}$, $C_{\omega X}$ and $C_{\omega Z}$. . }
\label{fig:c_coef}
\end{figure}

In order to study situations similar to experiments, we consider matter with the initial baryon chemical potential $\mu_0=800$~MeV and the initial temperature $T_0=155$~MeV. We take particle mass to be equal to the mass of $\Lambda$-hyperon, $m~=~1116$~MeV and continue the hydrodynamic evolution from  $\tau_0=1$~fm, till the final time $\tau_f=$~10~fm. In this scenario the proper-time dependence of the coefficients $C_{\kappa X}$, $C_{\kappa Z}$, $C_{\omega X}$ and $C_{\omega Z}$ is shown in Fig. \ref{fig:c_coef} for the same initial values (0.1) of all the $C$ coefficients. We have omitted $C_{\kappa Y}$ and $C_{\omega Y}$  because they fulfill the same equations as $C_{\kappa X}$ and $C_{\omega X}$). It can seen that  the coefficient $C_{\kappa Z}$ has the strongest time dependence as it increases by about 0.1 within 1 fm. 
%\bigskip
%%%%%%%%%%%%%%%%%%%%%%%%%%%%%%%%%%%%%%%%
\subsection{Physical observable}
In this section we demonstrate how hydrodynamics with spin can be used to obtain the information about the spin polarization of particles at freeze-out. The spin polarization of particles can be determined by the average Pauli-Luba\'nski vector $\langle\pi_{\mu}(p)\rangle$ in the rest frame of the particles. The average Pauli-Luba\'nski vector $\langle\pi_{\mu}(p)\rangle$ of particles with momentum $p$ emitted from a given the freeze-out hypersurface is obtained by the expression~\cite{Florkowski:2018ahw}
\beq
\langle\pi_{\mu}\rangle=\frac{E_p\frac{d\Pi _{\mu }(p)}{d^3 p}}{E_p\frac{d{\cal{N}}(p)}{d^3 p}}.
\eeq
where,
\beq
E_p\frac{d\Pi _{\mu }(p)}{d^3 p} &=& -\f{ \cosh(\xi)}{(2 \pi )^3 m}
\int
\Delta \Sigma _{\lambda } p^{\lambda } \,
e^{-\beta \cdot p} \,
\tilde{\omega }_{\mu \beta }p^{\beta }. \lab{PDPLV}\\
E_p\frac{d{\cal{N}}(p)}{d^3 p}&=&
\f{4 \cosh(\xi)}{(2 \pi )^3}
\int
\Delta \Sigma _{\lambda } p^{\lambda } 
\,
e^{-\beta \cdot p} \,.
\eeq
In the above expressions $\Delta \Sigma _{\lambda}$ is the element of freeze-out hypersurface and $p^{\lambda}$ is the particle four momentum. The above integrations can be carried out very easily by parametrizing $p^{\lambda}$ in terms of transverse mass $m_T$ and rapidity $y_p$ as;  
$p^\lambda = \left( m_T\ch(y_p),p_x,p_y,m_T\sh(y_p) \right)$ and assuming that freeze-out takes place at
a constant value of the proper time ($\Delta \Sigma _{\lambda }= U_{\lambda}dx dy\, \tau d\eta$). 
To compare the final result with experimental data we have to boost the average Pauli-Luba\'nski vector to the rest frame of the particles.
%By doing so we can obtain final result for spin polarization that can be compared with the experimental data.
 Note that since most of the experimental measurements of the spin polarization are done at midrapidity, therefore we can consider particles with $y_p = 0$. Moreover, since the mass of the $\Lambda$ is much larger then the values of temperature considered by us, we may take $m_T>>1$. In this case the spin polarization obtained in the particle rest frame can be cast to much simpler formula as follows
\beq
\langle \piv^* \rangle = -\frac{1}{4m} \left[
E_p \Cv_\omega - \pv \times \Cv_\kappa - \frac{\pv \cdot \Cv_\omega}{E_p + m} \pv
\right],
\eeq
where, $\pv = (p_x, p_y, 0)$. From the above formula one can see that for particles with small transverse momenta the polarization is directly determined by the coefficients $\Cv_\omega$ whose value can be obtained using hydrodynamic equations as discussed above.

%Moreover, since the coefficient functions $\Cv_\omega$ and $\Cv_\kappa$ depend on the freeze-out time in different way, see Fig.~\ref{fig:c_coef}, both the length and direction of the mean polarization three-vector $\langle \piv^* \rangle$ depend on the evolution time. This result may be interpreted also as a change of the polarization during the system expansion.

\subsection{Summary and conclusions}
In this contribution, using the relativistic
distribution functions for particles and antiparticles with spin 1/2 at local thermodynamical equilibrium as introduced in Ref. ~\cite{Becattini:2013fla} we have constructed the equilibrium Wigner functions.  Using the kinetic equation for Wigner function and its semiclassical expansion we have formulated hydrodynamics with spin. For the transversely homogeneous and boost invariant expansion of heavy ion collision fireball we have shown that hydrodynamics with spin can be used to determine the space-time evolution of the spin polarization tensor and finally spin polarization of the particles.

%% file: Sunil/sunil.tex
\section{Solutions and attractors of causal dissipative hydrodynamics for Bjorken flow}

\textit{Sunil Jaiswal, Chandrodoy Chattopadhyay, Amaresh Jaiswal, Subrata Pal, Ulrich Heinz}

\bigskip

{\small
Causal higher-order theories of relativistic viscous hydrodynamics in the limit of one-dimensional boost-invariant expansion is considered. Evolution equations for the inverse Reynolds number as a function of Knudsen number is obtained for three different choices of time dependence of the shear relaxation rate. It is shown that solutions of these equations exhibit attractor behavior. These dynamical attractors are characterized and uniquely determined by studying the analytical solutions at both small and large Knudsen numbers. 
}

\bigskip

%\end{abstract}

%\pacs{25.75.-q, 24.10.Nz, 47.75+f}

% 25.75.-q Relativistic heavy-ion collisions
% 24.10.Nz Hydrodynamic models
% 47.75.+f Relativistic fluid dynamics

%\maketitle

%%%%%%%%%%%%%%%%%%%%%%%%%%%%%%%%%%%%%%%%%%%%%%%%%
\subsection{Introduction}
\label{sunil_intro}
%%%%%%%%%%%%%%%%%%%%%%%%%%%%%%%%%%%%%%%%%%%%%%%%%

Relativistic viscous hydrodynamics has been very successful in explaining a wide range of collective phenomena observed in heavy-ion collisions.
Based on the paradigm that hydrodynamics requires local thermodynamic equilibrium to be applicable \cite{Florkowski:2017olj}, this successful hydrodynamic description led to the belief that these collisions create a nearly thermalized medium close to local thermal equilibrium \cite{Heinz:2001xi}. However, the advent of numerical dissipative relativistic fluid dynamics provides evidence of large deviations from local thermal equilibrium. This ``unreasonable effectiveness'' of hydrodynamics has generated much recent interest in the very foundations of fluid dynamics, culminating in the formulation of a new ``far-from-local-equilibrium fluid dynamics'' paradigm \cite{Romatschke:2017vte, Romatschke:2017ejr}. 

The simplest relativistic dissipative theory, relativistic Navier-Stokes (NS) theory, imposes instantaneous constitutive relations between the dissipative flows and their generating forces.
This approach was found to be plagued by acausality and intrinsic instability \cite{Hiscock:1985zz}.
The phenomenological second-order theory developed by M\"uller, Israel and Stewart (MIS) \cite{Muller:1967zza,Israel:1979wp} cures these problems by introducing a relaxation type equation for the dissipative flows and thus turning them into independent dynamical fields.
As discussed in \cite{Romatschke:2017ejr}, even the minimal causal  Maxwell-Cattaneo theory \cite{Maxwell:1867} resolves the causality issue, but introduces new ``non-hydrodynamic modes" that were absent in NS theory. These non-hydrodynamic modes are now known to play an important role in the approach to the regime of applicability of hydrodynamics, also known as the ``hydrodynamization'' process \cite{Florkowski:2017olj}. 
In the present study, we will focus on yet another feature that appears in causal theories of relativistic dissipative hydrodynamics, ``the hydrodynamic attractor" \cite{Heller:2015dha}.

%%%%%%%%%%%%%%%%%%%%%%%%%%%%%%%%%%%%%%%%%%%%%%%%%
\subsection{Attractor in ``minimal causal theory" }
\label{mct}
%%%%%%%%%%%%%%%%%%%%%%%%%%%%%%%%%%%%%%%%%%%%%%%%%

The energy-momentum tensor for a conformal system in the Landau frame has the form
\begin{align}\label{tmunu}
T^{\mu\nu} = \epsilon u^\mu u^\nu - P\Delta ^{\mu \nu} +  \pi^{\mu\nu},  
\end{align}
where $\epsilon$ and $P$ are the local energy density and pressure. Conformal symmetry implies an equation of state (EoS) $\epsilon=3P$ and zero bulk viscous pressure, $\Pi=0$. We define $\Delta^{\mu\nu} \equiv g^{\mu\nu}{-}u^{\mu}u^{\nu}$ which serves as a projection operator to the space orthogonal to $u^{\mu}$. Notations used: the metric convention used here is $g^{\mu\nu} = {\rm diag}(+\,-\,-\,-)$. We use $\sigma_{\mu\nu}\equiv \frac{1}{2} (\nabla_{\mu}u_{\nu}{+}\nabla_{\nu}u_{\mu}) - \frac{1}{3} \theta \Delta_{\mu\nu}$ for the velocity shear tensor, $\nabla^\alpha\equiv\Delta^{\mu\alpha} D_\mu$ for space-like derivative, $D_\mu$ for the covariant derivative and  $\theta\equiv D_\mu u^\mu$ for the expansion scalar.

The shear stress tensor, $\pi^{\mu\nu}$, is traceless and orthogonal to $u^{\mu}$.
The simplest form of $\pi^{\mu\nu}$ is the Navier-Stokes form, which is first order in velocity gradients, $\pi^{\mu\nu}_\mathrm{NS} = 2\eta\sigma^{\mu\nu}$ where $\eta$ is the shear viscosity coefficient. However, relativistic Navier-Stokes theory imposes instantaneous constitutive relations between the dissipative flows and their generating forces which results to superluminal signal propagation. The simplest way to restore causality is by introducing a dynamic relaxation-type equation for $\pi^{\mu \nu}$. This prescription, also known as the ``Maxwell-Cattaneo theory", requires that the dissipative forces relax to their Navier-Stokes values in some finite relaxation time, i.e.,
\begin{equation}
\label{max_cat}
\tau_\pi\dot\pi^{\langle\mu\nu\rangle} + \pi^{\mu\nu} = 2\eta\sigma^{\mu\nu},
\end{equation}
where $\tau_{\pi}$ is the shear relaxation time. We will now demonstrate the hydrodynamic attractor which appears in this minimal causal theory for Bjorken flow. We use the notation $\dot A\equiv u^\mu D_\mu A$ for the co-moving time derivative. Angular brackets around pairs of Lorentz indices indicate projection of the tensor onto its traceless and locally spatial part, e.g., $\dot{\pi}^{\langle\mu\nu\rangle} \equiv \Delta^{\mu\nu}_{\alpha\beta} \dot{\pi}^{\alpha\beta}$, where $\Delta^{\mu\nu}_{\alpha\beta} \equiv \frac{1}{2}(\Delta^\mu_\alpha\Delta^\nu_\beta{+}\Delta^\mu_\beta\Delta^\nu_\alpha) - \frac{1}{3}\Delta^{\mu\nu}\Delta_{\alpha\beta}$.

%------------------------------------------------

\paragraph*{Bjorken Flow$-$}
We will now simplify evolution equations for $\epsilon$ and $u^\mu$ obtained from energy-momentum conservation, $D_\mu T^{\mu\nu}=0$, and  Eq.~\eqref{max_cat} for Bjorken flow \cite{Bjorken:1982qr}. For transversally homogeneous and longitudinally boost-invariant systems expressed in Milne coordinates $x^\mu=(\tau,x,y,\eta_s)$ [with $\tau=\sqrt{t^2{-}z^2}$ and $\eta_s=\tanh^{-1}(z/t)$], the shear tensor is diagonal and space-like leaving only one independent component which we take to be the $\eta_s \eta_s$ component: $\pi^{xx}=\pi^{yy}= -\tau^2 \pi^{\eta_s \eta_s}/2 \equiv \pi/2$.

%------------------------------------------------

\paragraph*{Maxwell-Cattaneo Theory$-$}
Energy conservation and shear evolution equations for Maxwell-Cattaneo
\eqref{max_cat} reduce to a set of coupled ordinary differential equations (ODEs) in $\tau$:
\begin{align}
  \frac{d\epsilon}{d\tau} = -\frac{1}{\tau}\left(\frac{4}{3}\epsilon -\pi\right),
 \qquad
  \frac{d\pi}{d\tau} = - \frac{\pi}{\tau_\pi} + \frac{4}{3} \frac{\beta_\pi}{\tau} . 
  \label{mcbde}
\end{align}

Since $\beta_\pi\equiv\eta/\tau_\pi=4\epsilon/15$, Eqs.~\eqref{mcbde} are mutually coupled. The equation for shear stress can be decoupled by rewriting it in terms of dimensionless quantities, normalized shear stress (inverse Reynolds number) $\bar\pi=\pi/(\epsilon{+}P)=\pi/(4P)$, and rescaled time variable $\bar{\tau}\equiv\tau/\tau_\pi$ (which is the inverse Knudsen number for Bjorken flow).
Maxwell-Cattaneo evolution Eqs.~\eqref{mcbde} takes the form:
\begin{align}
\label{mcvt}
\frac{d \bar{\tau}}{d \tau} =& \left( \frac{ \bar{\pi} + 2 }{3} \right) \frac{\bar{\tau}}{ \tau},
\\ \label{mcdpibar}
 \left( \frac{ \bar{\pi} + 2 }{3} \right) \frac{d \bar{\pi}}{d \bar{\tau}} 
   =& - \bar{\pi} + \frac{1}{\bar{\tau}} \left( \frac{4}{15} + \frac{4}{3} \, \bar{\pi} - \frac{4}{3} \, \bar{\pi}^2 \right).
\end{align}
Here we also used that for a conformal system $\epsilon{\,\propto\,}T^4$ and $T\tau_{\pi} = 5 \bar{\eta}=\mathrm{const.}$ where  $\bar{\eta} \equiv \eta/s$ is the specific shear viscosity. 
Equation~\eqref{mcdpibar} is a first-order nonlinear ODE for the inverse Reynolds number that is completely decoupled from the evolution of the energy density.

%%%%%%%%%%%%%% Fig. 1 %%%%%%%%%%%
\begin{figure}[!ht]
 \begin{center}
  \includegraphics[width=7cm]{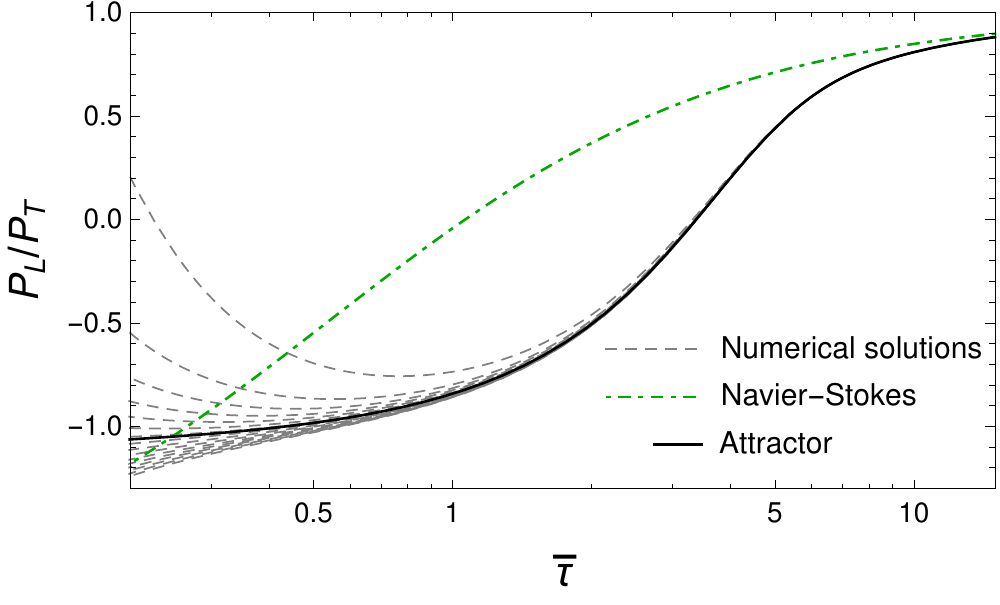}
 \end{center}
 %\vspace{-0.8cm}
 \caption{Gray dashed lines are numerical solutions of Eq.~\eqref{mcdpibar} for different initial conditions and solid black line represents the numerically determined attractor. Navier-Stokes is represented by dash-dotted green line.}
 \label{max_cat_att}
% \vspace{-0.4cm}
\end{figure}
%%%%%%%%%%%%%%%%%%%%%%%%%%%%%

We obtain the numerical attractor for Maxwell-Cattaneo from Eq.~\eqref{max_cat} following the prescription outlined in Ref.~\cite{Heller:2015dha}. Figure~\ref{max_cat_att} shows the evolution of pressure anisotropy, which is related to normalized shear through the equation $\frac{P_L}{P_T} = \frac{P-\pi}{P+\pi/2} = \frac{1-4\bar{\pi}}{1+2\bar{\pi}}$, for various initial conditions. One sees that numerical solutions for a broad range of initial conditions join the attractor at $\bar\tau \sim 2$, but that start to agree with the NS solution only for $\bar\tau \gtrsim 20$.

%%%%%%%%%%%%%%%%%%%%%%%%%%%%%%%%%%%%%%%%%%%%%%%%%
\subsection{Higher-order theories}
\label{hot}
%%%%%%%%%%%%%%%%%%%%%%%%%%%%%%%%%%%%%%%%%%%%%%%%%

We will now look at higher order hydrodynamic theories. For minimal causal conformally symmetric systems, one more term must be added in the evolution of shear stress to the Maxwell-Cattaneo theory:
\begin{equation}
\label{MIS}
   \tau_\pi\dot\pi^{\langle\mu\nu\rangle} + \pi^{\mu\nu} 
   = 2\eta\sigma^{\mu\nu} -\frac{4}{3}\tau_\pi\pi^{\mu\nu}\theta.
\end{equation}
This equation is a close variant \cite{Baier:2006um} of the one first derived by M\"uller, Israel and Stewart \cite{Israel:1979wp, Muller:1967zza}, and we will therefore refer to it as the ``MIS'' theory.

A systematic derivation of second-order (``transient'') relativistic fluid dynamics from  kinetic theory was performed in \cite{Denicol:2012cn}. 
For conformal systems and an RTA collision term, the result obtained in the 14-moment approximation differs from Eq.~(\ref{MIS}) by two additional terms:
\begin{equation}
\label{DNMR}
  \dot{\pi}^{\langle\mu\nu\rangle} \!+ \frac{\pi^{\mu\nu}}{\tau_\pi}\!= 
  2\beta_{\pi}\sigma^{\mu\nu}
  \!+2\pi_\gamma^{\langle\mu}\omega^{\nu\rangle\gamma}
  \!-\frac{10}{7}\pi_\gamma^{\langle\mu}\sigma^{\nu\rangle\gamma} 
  \!-\frac{4}{3}\pi^{\mu\nu}\theta.
\end{equation}
Here  $\omega^{\mu\nu}\equiv\frac{1}{2}(\nabla^\mu u^\nu{-}\nabla^\nu u^\mu)$ is the vorticity tensor. This ``DNMR'' theory \cite{Denicol:2012cn} can also be derived from a Chapman-Enskog like iterative solution of the RTA Boltzmann equation \cite{Jaiswal:2013npa}. 

Carrying the Chapman-Enskog expansion to one additional order, a third-order evolution equation for the shear stress was derived for the same system in \cite{Jaiswal:2013vta} which we will refer as the ``third-order" theory.

For Bjorken flow, the energy density and shear evolution equations for the above mentioned three theories can be brought into the following generic form:
\begin{align}
  \frac{d\epsilon}{d\tau} &= -\frac{1}{\tau}\left(\frac{4}{3}\epsilon -\pi\right), 
\label{bde1}\\
  \frac{d\pi}{d\tau} &= - \frac{\pi}{\tau_\pi} + \frac{1}{\tau}\left[\frac{4}{3}\beta_\pi 
  - \left( \lambda + \frac{4}{3} \right) \pi - \chi\frac{\pi^2}{\beta_\pi}\right]. 
  \label{bde2}
\end{align}
The coefficients $\beta_\pi$, $a$, $\lambda$, $\chi$, and $\gamma$ appearing in Eq.~(\ref{bde2}) above and in Eq.~(\ref{dpibar}) below are tabulated in Table~\ref{coeff} for these three theories.

%%%%%%%%%%%%%%%%%%%%%%%%%
\begin{table}[t]
 \begin{center}
  \begin{tabular}{|c|c|c|c|c|c|}
   \hline
   & $\beta_\pi$ & $a$ & $\lambda$ & $\chi$ & $\gamma$ \\
   \hline
   MIS & $4P/5$ & 4/15 & 0 & 0 & 4/3 \\
   \hline
   DNMR & $4P/5$ & 4/15 & 10/21 & 0 & 4/3\\
   \hline
   Third-order & $\,4P/5\,$ & \,4/15\, & \,10/21\, & \,72/245\, & \,412/147\, \\
   \hline
  \end{tabular}
  \caption{Coefficients for the causal viscous hydrodynamic evolution of the shear stress in Bjorken 
  	flow for the three theories studied in this work.}
  \label{coeff}
 \end{center}
 %\vspace{-0.8cm}
\end{table}
%%%%%%%%%%%%%%%%%%%%%%%%%

The shear evolution equation can be decoupled from the energy density evolution equation following similar procedure as mentioned in section \ref{mct}:
\begin{equation}
\label{dpibar}
   \left( \frac{ \bar{\pi} + 2 }{3} \right) \frac{d \bar{\pi}}{d \bar{\tau}} 
   = - \bar{\pi} + \frac{1}{\bar{\tau}} \left( a - \lambda \, \bar{\pi} - \gamma \, \bar{\pi}^2 \right),
\end{equation}
which has the same form as Eq.~\eqref{mcdpibar}. Note that Eqs.~\eqref{mcdpibar} and \eqref{dpibar} has the form of an Abel differential equation of the second kind for which, to the best of our knowledge, an analytical solution does not exist. The three hydrodynamic theories can be selected by choosing for $\lambda$ and $\gamma$ the appropriate combinations of constants given in Table~\ref{coeff}.

%%%%%%%%%%%%%%%%%%%%%%%%%%%%%%%%%%%%%%%%%%%%%%%%%
\subsection{Analytical solutions}
\label{analy_sol}
%%%%%%%%%%%%%%%%%%%%%%%%%%%%%%%%%%%%%%%%%%%%%%%%%

We will now derive analytical solutions for the evolution of $\bar\pi$ for Bjorken flow, at the expense of not being able to ensure the conformal relation $T\tau_\pi{\,=\,}$const. consistently with the evolution of the energy density. Instead, we find three separate classes of analytical solutions, corresponding to three different approximations of $\tau_\pi$ as a function of time. 

Starting from Eqs.~(\ref{bde1}),(\ref{bde2}), we decouple them as before by rewriting them in terms of the 
inverse Reynolds number $\bar\pi$ but without rescaling the proper time:
\begin{align}
   &\frac{1}{\epsilon {\tau}^{4/3}} \frac{d(\epsilon {\tau}^{4/3})}{d \tau} = \frac{4}{3}\frac{\bar{\pi}}{\tau}, 
\label{rbde1} \\
   &\frac{d\bar{\pi}}{d \tau } =  - \frac{\bar{\pi}}{{\tau}_{\pi}} + \frac{1}{\tau} 
      \left( a - \lambda \bar{\pi} - \gamma {\bar{\pi}^2} \right). 
\label{rbde2} 
\end{align} 
In the following, we find analytical solutions of Eq.~(\ref{rbde2}), using different approximations for the form of shear relaxation time $\tau_\pi$ \cite{Jaiswal:2019cju}.

%------------------------------------------------

\paragraph*{1. Constant relaxation time$-$}
In this approximation~\cite{Denicol:2017lxn}, the scaling of $\tau_\pi$ with temperature was ignored by simply setting it constant. This constitutes a rather drastic violation of conformal symmetry by introducing, in addition to the inverse temperature $1/T$, a second, independent length scale $\tau_\pi$. In the following two approximations, we will successively improve on this.

Introducing again the rescaled time $\bar\tau=\tau/\tau_\pi$, for constant $\tau_{\pi}$ Eq.~(\ref{rbde2}) turns directly into
\begin{equation}\label{RBED3}
   \frac{d\bar{\pi}}{d \bar{\tau} } = 
   -\bar{\pi}  + \frac{1}{\bar{\tau}} \left( a - \lambda \bar{\pi} - \gamma {\bar{\pi}^2} \right)
\end{equation}
which is similar to Eq.~(\ref{dpibar}) but without the nonlinearity on the left-hand side (l.h.s). As will be discussed in Section~\ref{analy_attr}, this difference has important consequences for the attractor solutions and Lyapunov exponents.

%------------------------------------------------

\paragraph*{2. Relaxation time from ideal hydrodynamics$-$}
A better approximation to Eq.~(\ref{dpibar}) can be obtained by setting $T\tau_\pi{\,=\,}$const. by approximating the time-dependence of $T$ at late times with the ideal fluid law
\begin{equation}
\label{Bj_evol}
   T_{\rm id}(\tau) = T_0\left(\frac{\tau_0}{\tau}\right)^{1/3},
\end{equation}
where $T_0$ is the temperature at initial time $\tau_0$. For $T\tau_\pi=5\bar\eta$ this yields $\tau_\pi(\tau) = b\,\tau^{1/3},$ with $b= \frac{5\bar\eta}{T_0\tau_0^{1/3}}.$
Using this to define the scaled time variable $\bar{\tau}\equiv\tau/\tau_\pi$,  Eq.~(\ref{rbde2}) turns into
\begin{equation}
\label{dpibar_id}
   \frac{2}{3}\frac{d\bar{\pi}}{d \bar{\tau} } =  -\bar{\pi}  
   + \frac{1}{\bar{\tau}} \left( a - \lambda \bar{\pi} - \gamma {\bar{\pi}^2} \right), 
\end{equation}
independent of $b$. This equation again misses the nonlinear term on the l.h.s. of Eq.~(\ref{dpibar}) and has the same structure as Eq.~(\ref{RBED3}). 

%------------------------------------------------

\paragraph*{3. Relaxation time from Navier-Stokes evolution$-$}
We can further improve our approximation by accounting for first-order gradient effects in the evolution of the temperature, by replacing the ideal fluid law (\ref{Bj_evol}) by the Navier-Stokes result \cite{Muronga:2001zk, Baier:2006um}
\begin{equation}
\label{Bj_evol_NS}
    T_{_{\rm NS}} = T_0\left(\frac{\tau_0}{\tau}\right)^{1/3}
    \left[1 + \frac{2\bar{\eta}}{3\tau_0T_0}\left\{ 1- \left(\frac{\tau_0}{\tau}\right)^{2/3} \right\} \right].
\end{equation}
Substituting this into $T\tau_\pi=5\bar\eta$ we find
\begin{equation}
\label{taupi_NS}
    \tau_\pi = \frac{\tau^{1/3}}{d - \frac{2}{15}\tau^{-2/3}}\, , \qquad 
    d\equiv\left( \frac{T_0\tau_0}{5\bar{\eta}} + \frac{2}{15} \right)\tau_0^{-2/3}.
\end{equation}
Using this relation, Eq.~(\ref{rbde2}) in terms of scaled time variable $\bar{\tau}\equiv\tau/\tau_\pi$:
\begin{equation}
\label{dpibar_NS}
    \left( \frac{a/\bar{\tau}+2}{3} \right)\frac{d\bar{\pi}}{d \bar{\tau} } 
    =  -\bar{\pi}  + \frac{1}{\bar{\tau}} \left( a - \lambda \bar{\pi} - \gamma {\bar{\pi}^2} \right),
\end{equation}
independent of the constant $d$. This shares with Eq.~(\ref{dpibar_id}) the same factor 2/3 on the l.h.s.. 

%------------------------------------------------

Equations~\eqref{RBED3},~\eqref{dpibar_id} and~\eqref{dpibar_NS} are first-order nonlinear ODE of Riccati type whose solutions can be given in the generic form \cite{Jaiswal:2019cju} 
\begin{align}
\label{generic_pibar}
   \bar{\pi}(w)=
   & \frac{(k{+}m{+}\frac{1}{2}) M_{k+1,m}(w) - \alpha W_{k+1,m}(w)}
             {\gamma |\Lambda| \left[ M_{k,m}(w)+\alpha W_{k,m}(w) \right]}.
 \end{align}               
The arguments and parameters appearing in the above equations are given in Table~\ref{T2}. Here $\epsilon_0$ is the initial energy density at time ${\bar{\tau}}_0$, and the constant $\alpha$ encodes the initial normalized shear stress $\bar\pi_0$. Note that $\alpha$ can only take values for which the energy density is positive-definite for $\bar{\tau}>0$. 
 %
 %%%%%%%%%%%%%%%%%%%%%%%%%
\begin{table}[t]
 \begin{center}
  \begin{tabular}{|c|c|c|c|c|}
   \hline
   $T(\tau)$ & $w$ & $\Lambda$ & $k$ & $m$ \\
   \hline
   \phantom{$\bigg|$} const.  \phantom{$\bigg|$}
   & $\bar{\tau}$ & $\, -1 \ $ & $-\frac{1}{2} \left(\lambda{+}1\right)$ 
   & $\frac{1}{2} \sqrt{4a\gamma{+}\lambda^2}$  \\
   \hline
   \phantom{$\bigg|$} ideal  \phantom{$\bigg|$}
   & $\frac{3}{2}\bar{\tau}$ &\, $-\frac{3}{2}$ & $-\frac{3 \lambda +2}{4}$ 
   & $\frac{3}{4} \sqrt{ 4 a \gamma{+}\lambda^2  }$  \\
   \hline
   \phantom{$\bigg|$} NS  \phantom{$\bigg|$}
   & $\frac{3}{2} \left( \bar{\tau}{+}\frac{a}{2} \right)$ & $\,-\frac{3}{2}$ 
   & $-\frac{ 6\lambda + 4 - 3 a}{8}$ &\, $\frac{3}{8} \sqrt{16a\gamma{+}a^2{-}4a\lambda{+}4 \lambda^2}$\,  
   \\
   \hline
  \end{tabular}
  \label{power_coeff}
 \end{center}
%\vspace*{-6mm}
\caption{Arguments and parameters of Eq.~\eqref{generic_pibar} for the analytic approximations studied.
	\label{T2}}
	%\vspace{-0.4cm}
\end{table}
%%%%%%%%%%%%%%%%%%%%%%%%%

%%%%%%%%%%%%%%%%%%%%%%%%%%%%%%%%%%%%%%%%%%%%%%%%%\\
\subsection{ Analytical attractors and  Lyapunov exponent} 
\label{analy_attr}
%%%%%%%%%%%%%%%%%%%%%%%%%%%%%%%%%%%%%%%%%%%%%%%%%

%%%%%%%%%%%%%% Fig. 2 %%%%%%%%%%%
\begin{figure}[!ht]
 \begin{center}
  \includegraphics[width=7cm]{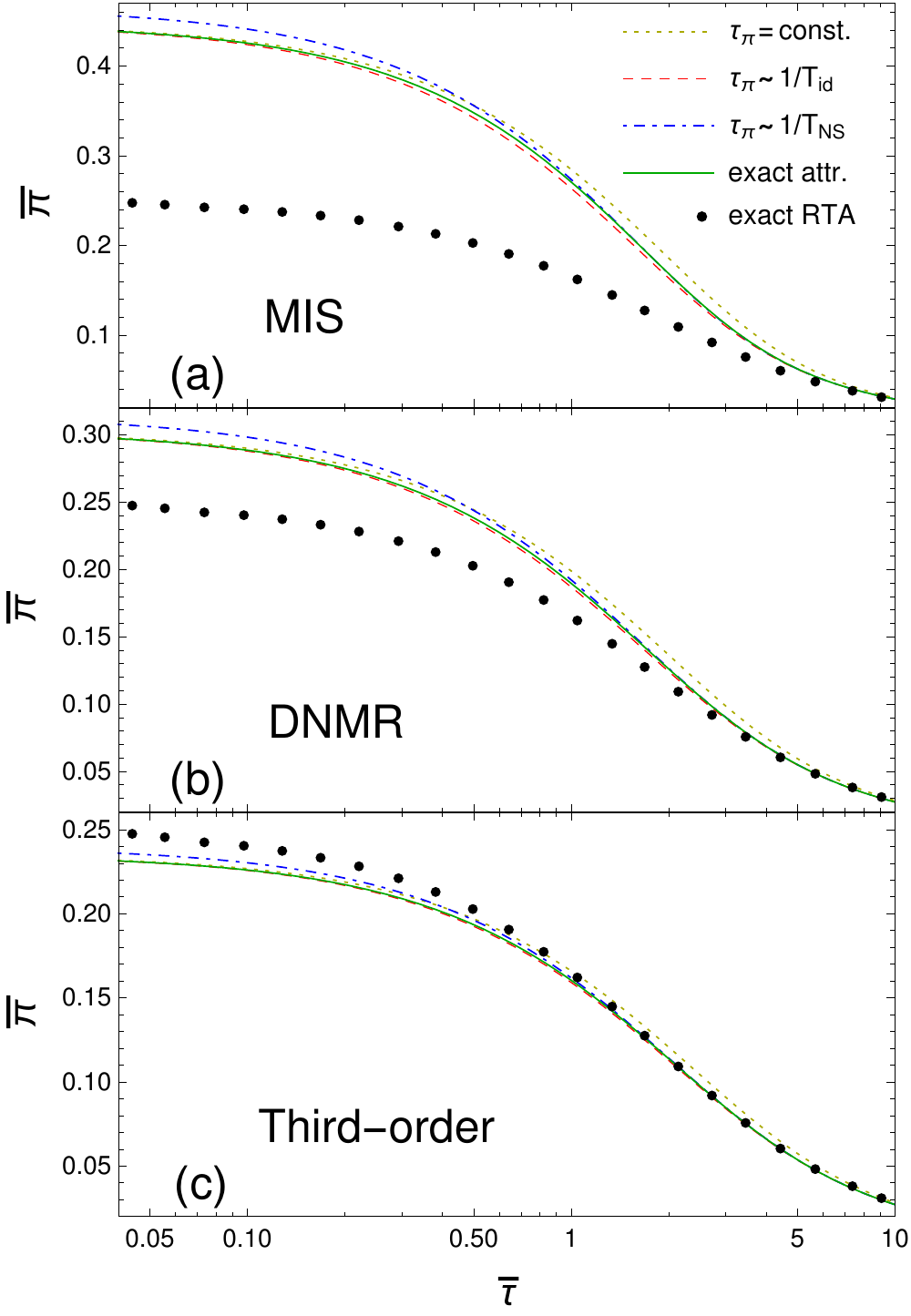}
 \end{center}
% \vspace{-0.8cm}
 \caption{Approximate analytical attractors for different theories
% the MIS (a), DNMR (b), and third-order (c) theories,
 	compared with their exact numerical attractors (solid green lines) and the exact analytical 	attractor for the RTA Boltzmann equation (black dots).}
 \label{att_all}
 %\vspace{-0.8cm}
\end{figure}
%%%%%%%%%%%%%%%%%%%%%%%%%%%%%

In this section, we determine the hydrodynamic attractors and obtain the Lyapunov exponent ($\Lambda$) from the analytic solutions.
We introduce the following procedure for identifying the hydrodynamic attractor  \cite{Chattopadhyay:2018pwe, Jaiswal:2019cju}: In terms of the parameter $\alpha$ encoding the initial condition for $\bar\pi$, we search for the value $\alpha_0$ at which the quantity
\begin{equation}
\label{psi}
   \psi(\alpha_0) \equiv \lim_{\bar{\tau} \to \bar{\tau}_0} 
   \frac{\partial\bar{\pi}}{\partial \alpha}\bigg{|}_{\alpha=\alpha_0}
\end{equation}
diverges at the scaled time $\bar{\tau}_0$ where the two fixed points of the evolution trajectories are located. Using this prescription, we obtain the attractor solutions from Eq.~(\ref{generic_pibar}) by setting the initial condition parameter $\alpha{\,=\,}0$:
\begin{align}
\label{gen_att}
   \bar{\pi}_\mathrm{attr}(w)=
   & \frac{k{+}m{+}\frac{1}{2}} {\gamma |\Lambda|}\, \frac{M_{k+1,m}(w)}{M_{k,m}(w)}.
 \end{align}
The attractors are shown in Fig.~\ref{att_all} for the three different hydrodynamic theories discussed in this paper (MIS (a), DNMR (b), and third-order (c)) and compared with the corresponding exact numerical attractors (obtained from Eq.~\eqref{dpibar} following the prescription outlined in Ref.~\cite{Heller:2015dha}) as well as with the attractor for the exact analytical solution of the RTA Boltzmann equation \cite{Strickland:2018ayk}.

To obtain the Lyapunov exponents ($\Lambda$) from the approximate analytic solutions, we use the formula  \cite{Jaiswal:2019cju}
	\begin{equation}
	\label{Lambda}
	    \Lambda = \lim_{\bar{\tau}\to\infty}\,\dfrac{\partial}{\partial\bar{\tau}} 
	              \left[ \ln\!\left( \dfrac{\partial\bar{\pi}}{\partial\alpha} \right)\right].
	\end{equation}
We find that the constant temperature approximation corresponds to a Lyapunov exponent of $\Lambda = -1$. This obviously differs from $\Lambda{\,=\,}-\frac{3}{2}$ \cite{Behtash:2018moe,Jaiswal:2019cju} for the conformally invariant theories described by Eq.~(\ref{dpibar}); the difference is a direct consequence of the breaking of conformal symmetry by setting $\tau_\pi$ constant instead of $\propto 1/T$. 
 However, we recover the same Lyapunov exponent $\Lambda= -\frac{3}{2}$ as for the conformally invariant theories from the approximate solutions obtained when temperature dependence is approximated from ideal and Navier-Stokes evolution.

While the ODE describing the evolution of the inverse Reynolds number for Bjorken flow can be solved numerically, the analytic approximations studied here are surprisingly accurate, and they yield valuable insights into the details of initial state memory loss \cite{Kurkela:2019set,Jaiswal:2019cju} and the approach to attractor dynamics in Bjorken flow.

%
%\bibliography{reference}
%
%\end{document}

%% file: Deependra/deependra.tex
\section{Non-perturbative dynamics of QCD and its phase structure: An overview}

\textit{Deependra Singh Rawat, H. C. Chandola, Dinesh Yadav, H. C. Pandey}

\bigskip

%\begin{abstract}
{\small
Keeping in view the dominance of non-perturbative phenomena in low energy regime of QCD, an infrared effective dual QCD based on topologically viable homogeneous fibre bundle approach, has been analysed for exploring the dynamics of quark confinement in its dynamically broken phase which has been shown to lead an unique multi-flux tube configuration and a typical glueball spectrum. The dynamics of confinement-deconfinement phase transition has been discussed by computing the critical parameters of phase transition and their possible implications to QGP formation and QCD phase structure has also been discussed. The intimate connection of chromoelectric field with the color confining features has been discussed to establish the validity of present dual QCD model in the infrared sector of QCD. 
}
%\end{abstract}
\bigskip

%neutrino interactions with, 13.15.+g
%neutrino-nucleus, 25.30.Pt
%Partially conserved axial-vector currents, 11.40.Ha

%\pacs{11.30.Jw, 14.80.Hv, 12.38.Aw, 11.15.Ex}


%\maketitle

%\tableofcontents

\subsection{Introduction}
  QCD is the leading theoretical formulation of the strong interactions~\cite{Gross:1973ju,Politzer:1973fx} which turns out to lead the outstanding description of the dynamics of quarks and gluons inside the hadrons. The phenomena associated with the high energy regimes of QCD where quarks and gluons are weakly interacting on account of asymptotic freedom \cite{Gross:1973ju} are rather well described by employing the perturbative mathematical techniques. However, its low energy limit where color isocharges becomes strongly coupled is still lacks of precise understanding due to the appearance of several non-pertervative peculiarities like confinement, chirally asymmetric behavior and dielectric nature of QCD vacuum, the hadron mass spectrum etc. In particular, the color confinement which is typically characterized by the absence of the asymptotic states of colored particles is one of the outstanding conjecture that direly needed a fundamental explanation using first principle in the domain of hadron physics. To sort out the puzzling confining structure of QCD vacuum, the first proposal was made by Nambu~\cite{Nambu:1974zg} and others, which asserts that the magnetic condensation could provide the confinement of color electric flux carried by the quarks through the dual Meissner effect. The physical realization of the magnetic condensation of QCD vacuum or introduction of QCD monopoles in the theory requires an analytical field theoretical investigation that may essentially incorporates the topological properties of the associated gauge group. In this direction, 't Hooft's~\cite{tHooft:1981bkw} proposal of Abelian dominance made a remarkable significance where fixing the gauge degrees of freedom reduces the QCD to the Abelian theory with the appearance of color magnetic monopoles as a topological excitations of the theory. The effective interaction of these colored monopoles brings the QCD vacuum in the state of color superconductivity and develops the confining features in the theory. Despite of its ordered mathematical description, the Abelian projection technique suffers from the serious problem of gauge dependency to project out the Abelian dominance which is in the violation of the fact that all the natural process must be gauge invariant. Keeping in view the gauge independent confining structure of QCD vacuum we have recently \cite{Pandey:2000bt,Chandola:2009zz,Rawat:2018mxg,Chandola:2019xwo} analyzed a dual version of color gauge theory by imposing an additional magnetic isometry as an effective theory of the nonperturbative QCD. In such a dual formulation, the flux tube solution may be interpreted as the excitation corresponding to the topological degrees of freedom and develops two characteristic mass scales (vector and scalar mass mode).
  
In the present study, we further focus on to discuss the gauge invariant color confining structure of dual QCD vacuum and its implications in the study of QCD phase transition under the extreme conditions of temperature and density.

\subsection{Color confinement in dual QCD}
The magnetic symmetry~\cite{Cho:1979nv,Cho:1980nx,Cho:2012pq} defined as an additional isometry of the internal fiber space is introduced with the non-Abelian formulation of a gauge theory in the ($4+n$) multidimensional unified space. For an arbitrary gauge group (G), the associated magnetic symmetry structure is introduced by Killing vector fields ($\hat{m}$). The Killing condition along the magnetic vector may be imposed by insisting that the gauge potential (${\bf W_\mu}$) must satisfy the gauge covariant magnetic symmetry condition~\cite{Cho:1979nv,Cho:1980nx} that keeps intact the gauge symmetry and for G $\equiv$ SU(2) is given by, $D_{\mu}\hat{m} = (\partial_{\mu}+g{\bf{W_{\mu}}}\times)\hat{m}=0$ (where $\hat{m}$ belongs to the adjoint representation of gauge group) leading to the decomposition of the gauge potential (${\bf W_\mu}$) in the following form,
\begin{equation}
{\bf{W_{\mu}}}=A_{\mu}\hat{m}-g^{-1}(\hat{m}\times\partial_{\mu}\hat{m})
\label{equation_one}
\end{equation}
where $\hat{m}.{\bf W_\mu}\equiv A_\mu$ is the color electric potential unrestricted by magnetic symmetry and is Abelian in nature. The second term retains the topological characteristics resulting from the imposition of magnetic symmetry. The isolated singularities of the multiplet $\hat{m}$ may then be viewed to define the homotopy of the mapping $\Pi_{2}(S^{2})$ on $\hat{m}:S_{R}^{2}\rightarrow S^{2}=SU(2)/U(1)$ which ensures the appearance of chromomagnetic monopoles in the theory. It clearly shows that the imposition of magnetic symmetry on the gauge potential brings the topological structure into the dynamics explicitly. The duality between color isocharges and topological charges becomes more evident when the gauge fields and the associated gauge potential Eq.\ref{equation_one} are expressed in terms of magnetic gauge (or Dirac gauge) obtained by rotating $\hat{m}$ to a prefixed space-time independent direction in isospace using a gauge transformation (U) as $\hat{m} \longrightarrow \hat{\zeta}_{3} = (0,0,1)^{T}$ which leads to the field decomposition as, ${\bf{G_{\mu\nu}}} = {\bf{W_{\nu,\mu}}} - {\bf{W_{\mu,\nu}}} + g{\bf{W_{\mu}}} \times {\bf{W_{\nu}}} \equiv (F_{\mu\nu}+B_{\mu\nu}^{(d)})\hat{m}$
with $F_{\mu\nu}=A_{\nu,\mu}-A_{\mu\nu}$ and $B_{\mu\nu}^{(d)}=B_{\nu,\mu}-B_{\mu,\nu} = g^{-1}\hat{m}.(\partial_{\mu}\hat{m}\times\partial_{\nu}\hat{m})$. The dynamics of the resulting dual QCD vacuum and its implications on color confinement, then follows from the SU(2) Lagrangian with a quark doublet source $\psi(x)$, as given by
\begin{equation}
\mathcal{L}=-\frac{1}{4}G_{\mu\nu}^{2}+\bar{\psi}(x)i\gamma^{\mu}D_{\mu}\psi(x)-m_{0}\bar{\psi}(x)\psi(x).
\label{equation_two}
\end{equation}
However, in order to avoid the problems of the singular behavior of the potential associated with monopoles and its point-like source we use the electric gauge in which the magnetic potential becomes regular dual magnetic potential $B_{\mu}^{(d)}$ and coupled with its point like magnetic source represented by a complex scalar field $\phi(x)$. Taking these consideration into the account, the modified form of the dual QCD Lagrangian (Eq.\ref{equation_two}) in quenched approximation is given below,
\begin{equation}
\mathcal{L}_{m}^{(d)}=-\frac{1}{4}B_{\mu\nu}^{2}+\vert[\partial_{\mu}+i\frac{4\pi}{g}B_{\mu}^{(d)}]\phi \vert^{2}- 3\lambda\alpha_{s}^{-2}(\phi^{\ast}\phi-\phi_{0}^{2})^{2}.
\label{equation_three}
\end{equation}
The quadratic effective renormalized potential is appropriate for inducing the dynamical breaking of magnetic symmetry that forces the magnetic condensation resulting in dual Meisner effect with the QCD vacuum in a state of magnetic supercondutor which, with the formation of flux tubes confine the color isocharges.

The nature of magnetically condensed vacuum and the associated flux tube structure may be analysed using the field equations resulting from the Lagrangian (Eq.\ref{equation_three}) into the following form,
\begin{equation}
\mathcal{D}^{\mu}\mathcal{D}_{\mu}\phi + 6\lambda\alpha_{s}^{-2}(\phi^{\ast}\phi-\phi_{0}^{2})\phi=0\nonumber
\end{equation}
\begin{equation}
\partial^{\nu}B_{\mu\nu}-i\frac{4\pi}{g}(\phi^{\ast}\buildrel \leftrightarrow \over\partial_{\mu}\phi)-8\pi\alpha_{s}^{-1}B_{\mu}^{(d)}\phi\phi^{\ast}=0.
\label{equation_four}
\end{equation}
with $\mathcal{D}_{\mu}=\partial_{\mu}+i4\pi g^{-1}B_{\mu}^{(d)}$. The close agreement of these field equations with the Neilsen and Olesen~\cite{Nielsen:1973cs} interpretation of vortex like solutions indicates the flux-tube like configurations inside the QCD vacuum. It leads to the possibility of the existence of the monopole pairs inside the superconducting vacuum in the form of thin flux tubes that may be responsible for the confinement of any colored fluxes. Under cylindrical symmetry $(\rho,\varphi,z)$ and the field ansatz $B_{\varphi}^{(d)}(x)=B(\rho),~~~B_{0}^{(d)}=B_{\rho}^{(d)}=B_{z}^{(d)}=0$~ and~ $\phi(x)=\exp(in\varphi)\chi(\rho)~~(n=0,\pm1,\pm2,--)$, the field equations given by Eq. \ref{equation_four} are transformed to the following form,
\begin{equation}
\frac{1}{\rho}\frac{d}{d\rho}\biggl(\rho\frac{d\chi}{d\rho}\biggr)-\biggl[\biggl(\frac{n}{\rho}+(4\pi\alpha_{s}^{-1})^\frac{1}{2}B(\rho)\biggr)^{2}-6\lambda\alpha_{s}^{-2}(\chi^{2}-\phi_{0}^{2})\biggr]\chi(\rho) = 0,\nonumber
\end{equation}
\begin{equation}
\frac{d}{d\rho}\biggl[\rho^{-1}\frac{d}{d\rho}(\rho B(\rho))\biggr]+8\pi g^{-1}\biggl(\frac{n}{\rho}-4\pi g^{-1}B(\rho)\biggr)\chi^{2}(\rho)=0.
\label{equation_five}
\end{equation}
Further, with these considerations, the form of the color electric field in the z-direction is given by,
\begin{equation}
E_{m}(\rho)=-\frac{1}{\rho}\frac{d}{d\rho}(\rho B(\rho))
\label{equation_five1}
\end{equation}
The equations (\ref{equation_five}) are desired field equations that governs the dynamics of dual QCD vacuum and coincides exactly with those of the ordinary single vortex solution in cylindrical framework. The highly non-linear coupled structure of these differential equations doesn't allow us to go through their exact solutions. Hence, their asymptotic solution using the boundary conditions $\phi\rightarrow\phi_{0}$ as $\rho\rightarrow\infty$ leading to the appropriate asymptotic solution for the dual gauge potential that ensures the formation of color flux tubes, is given by,
\begin{equation}
B(\rho)=-ng(4\pi\rho)^{-1}[1+F(\rho)]~and~F(\rho) {\buildrel\rho\rightarrow\infty\over\longrightarrow} C\rho^{\frac{1}{2}}exp(-m_{B}\rho),
\label{equation_six}
\end{equation}
where C is a constant and $m_{B}(=4\pi g^{-1}\sqrt{2}\phi_{0})$ is the glueball mass. In dual QCD scenario, the dynamical breaking of magnetic symmetry sets two characteristic mass scales, The vector mass mode $m_{B}$ determines the magnitude of dual Meissner effact whereas the scalar mass mode $m_{\phi}$ tells the rate of magnetic condensation. The ratio of these two mass scales reflects the nature of dual QCD vacuum in terms of dual Ginzberg-Landau parameter\cite{Pandey:2000bt,Chandola:2009zz,Rawat:2018mxg} which in the relatively weak coupling limit (for $\alpha_{s} = 0.12~and~0.24$) exhibit the type-II superconducting behavior ($K^{d}_{QCD}$ $>$ 1) with multi-flux tube configurations. Thus, the gauge invariant field decomposition formulation (based on magnetic symmetry) by utilizing its topological structure gives a viable explanation of the mysterious confining behavior of QCD at the fundamental level.

\subsection{Quark-hadron phase transition in dual QCD}
The color confining features of QCD vacuum as a result of the formation of color flux tubes can be visualized more effectively by evaluating the energy per unit length of the flux tube structure in cylindrical system \cite{Chandola:2019xwo} as,
\begin{eqnarray}
k &=& 2\pi\int_{0}^{\infty}\rho\,d\rho\biggl[\frac{n^{2}g^{2}\rho^{-2}}{32\pi^{2}}F'^{2}+\frac{n^{2}}{\rho^{2}}F^{2}(\rho)\chi^{2}(\rho)  
+ \chi'^{2}+\frac{48\pi^{2}}{g^{4}}(\chi^{2}(\rho)-\phi_{0}^{2})^{2}\biggr]~.
\label{equation_seven}
\end{eqnarray}
It, in turn, plays an important role in the phase structure of QCD vacuum, if we take the multi-flux tube system on a $S^{2}$-sphere with periodically distributed flux tubes and introduce a new variables $R$ on $S^{2}$ and express it as, $\rho=R\sin\theta$. As a result, a number of flux tubes considered inside a hadronic sphere of radius $R$ pass through the two poles of the hadronic sphere. Under such prescription, the flux tube solution governed by Eq.(\ref{equation_six}) corresponds to the case of large $R$ limit $(R\rightarrow\infty)$ such that $R\gg\rho$ and $\theta\rightarrow 0$. With these considerations, the finite energy expression given by above Eq.(\ref{equation_seven}) may be re-expressed as,
\begin{equation}
k=\varepsilon_{C}+\varepsilon_{D}+\varepsilon_{0}~~with~~\varepsilon_{C}=k_{C}R^{2},~~\varepsilon_{D}=k_{D}R^{-2},~~\varepsilon_{0}=k_{0}
\label{equation_eight}
\end{equation}
where the functions $k_{C}$,~$k_{D}$ and $k_{0}$ are given by,
\begin{equation}
k_{C}=\frac{96\pi^{3}}{g^{4}}\int_{0}^{\pi}[\chi^{2}(\theta)-\phi_{0}^{2}]^{2}\sin\theta~d\theta,
\label{equation_nine}
\end{equation}
\begin{equation}
  k_{D}=\frac{n^{2}g^{2}}{16\pi^{2}}\int_{0}^{\pi}\frac{1}{\sin\theta}\biggl(\frac{\partial F}{\partial\theta}\biggr)^{2}d\theta,
  \label{equation_ten}
\end{equation}
\begin{equation}
k_{0}=2\pi\int_{0}^{\pi}\biggl[\frac{n^{2}F^{2}(\theta)\chi^{2}(\theta)}{\sin\theta}+\sin\theta\biggl(\frac{\partial\chi}{\partial\theta}\biggr)^{2}\biggr]d\theta.
\label{equation_eleven}
\end{equation}
The energy expression (Eq.\ref{equation_eight}) provides an straightforward description of the behavior of QCD vacuum at different energy scales. At large distance scale, the first term ($\varepsilon_{C}$) in equation (\ref{equation_eight}) dominates which increases at increasing hadronic distances and gets minimized when the monopole field acquire its vacuum expectation value ($\phi_{0}$) which incidentally acts as an order parameter and will dynamically break the magnetic symmetry of the system. The resulting magnetic condensation then forces the color electric field to transform in to the form of the thin flux tubes extending from $\theta$ = 0 to $\theta$ = $\pi$ and the QCD vacuum is ultimately pushed to the confining phase. For the computation of associated critical parameters, we proceed by evaluating the functions associated with the expression (Eq.~\ref{equation_eight}) in the following way\cite{Chandola:2019xwo}, 
\begin{equation}
k_{C}=\frac{6\pi}{\alpha_{s}^{2}}\int_{0}^{\pi}\left[\chi^{2}(\theta)-\phi_{0}^{2}\right]^{2}\sin\theta~d\theta~~~~\Rightarrow~~~~k_{C}=\frac{3m_{B}^{4}}{16\pi},
\label{equation_twelve}
\end{equation}
which shows the direct dependency of confining energy on the vector glueball mass of the magnetically condensed vacuum. On the other hand, the dominating component of energy expression in short distance limit ($\varepsilon_{D}$), may also be evaluated in the following form:
\begin{equation}
k_{D} = \pi~R^{4}\int_{0}^{\pi}E_{m}^{2}(\theta)\sin\theta~d\theta~~~\text{where}~~
E_{m}(\theta)=\frac{ng}{4\pi~R^{2}\sin\theta}\frac{\partial F}{\partial\theta}.
\label{equation_thirteen}
\end{equation}
%%%%%%%%
Using Eq.(\ref{equation_six}) and Eq.(\ref{equation_ten}) alongwith the flux quantization condition given by\\
\begin{equation}
\int\rho~E_{m}(\rho)d\rho=\frac{ng}{4\pi}
\label{equation_fourteen}
\end{equation}
then leads to,
\begin{equation}
k_{D}=\frac{n^{2}g^{2}}{8\pi}=\frac{1}{2}n^{2}\alpha_{s}.
\label{equation_fifteen}
\end{equation}
For the confinement-deconfinement phase transition, we have $\varepsilon_{D}/\varepsilon_{C}=1$ and $R=R_{c}$ which leads to the critical radius and density of phase transition in the following form, 
\begin{equation}
R_{c}=\left(\frac{2}{3}n^{2}g^{2}\right)^{\frac{1}{4}}m_{B}^{-1}~\text{and}~d_{c}=\frac{1}{2\pi R_{c}^{2}}=\left(\frac{8}{3}\pi^{2}n^{2}g^{2}\right)^{-\frac{1}{2}}m_{B}^{2}.
\label{equation_sixteen}
\end{equation}
The Eq.~\eqref{equation_sixteen} exhibit that the critical radius and critical density of phase transition are clearly expressible in terms of free parameters of the QCD vacuum. In view of the running nature of QCD coupling constant, we can estimate these critical factors associated with the QCD vacuum in its infrared sector using the numerical estimations of glueball masses~\cite{Cho:1979nv,Cho:1980nx}. For instance, for the optimal value of ($\alpha_{s}$) as $\alpha_{s}\equiv 0.12$ with the glueball masses $m_{B}=2.102 GeV$ and $m_{\phi}=4.205\,$GeV, equations (17), lead to $R_{c}=0.094~fm~~~\text{and}~~~d_{c}=18.003~fm^{-2}.$.

The deconfinement phase transition in the dual QCD vacuum therefore expected to appears around the above-mentioned critical values for a typical coupling of $\alpha_{s}=0.12$ in the near infrared region of QCD. In this case, for $R_{c} \equiv$ 0.094 fm, the corresponding flux tube number density acquire its critical value of 18.003 $fm^{-2}$ and the first part of the energy expression (Eq.\ref{equation_eight}) dominates which demonstrates the confinement of color isocharges in the low energy-momentum scale of QCD vacuum. However, below $R_{c}=0.094$ fm, the quarks and gluons appear as free states and the system shifted towards electrically dominated deconfined phase and mathematically govern by the second part of the expression (Eq.\ref{equation_eight}). Consequently there is the possibility of sharp increase in the flux tube number density tube system, the flux tube annihilation may takes place which then leads to the generation of dynamical quarks and gluons. The gluon self-interactions are then expected to play a major role in the thermalization of QCD system and create an intermediatery state of quark-gluon plasma (QGP). As a result of such flux tube melting in the high momentum transfer sector of QCD vacuum, the system is expected to evolve with an intermediatory QGP phase. 

Furthermore, the general form of color-electric field may be evaluated by using equation (Eq.\ref{equation_five1}) for the case of multi-flux tube system on the $S^{2}$-sphere, is obtained as,
\begin{equation}
E_{m}(\theta)= \tilde{E}_{m}(\theta) \exp(-Rm_{B}\sin\theta),
\label{equation_seventeen}
\end{equation}
\begin{figure}[tbh]
\centering
\begin{tabular}{cc}
\includegraphics[width=65mm]{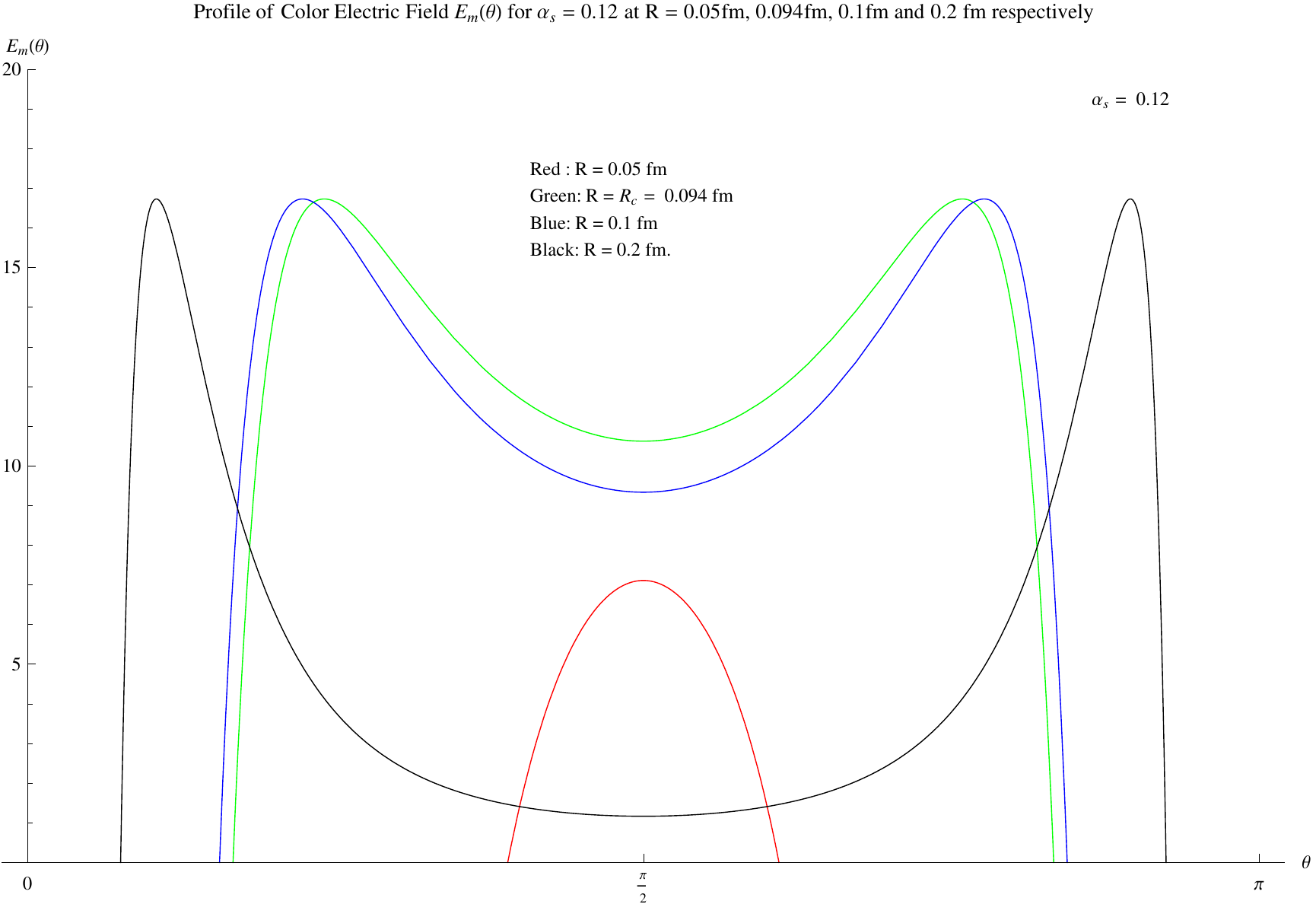}\\
\end{tabular}
\caption{(Color online) Profiles of color electric field $E_{m}(\theta)$ for $\alpha_{s}=0.12$ as a function of $\theta$.}
\label{Figure_EField}
\end{figure} 
where, $\tilde{E}_{m}(\theta)=\frac{nC\alpha_{s}^{1/2}}{4\pi^{1/2}R^{3/2}\sin^{3/2}\theta}\left(1-2Rm_{B} \sin\theta\right)$. The profile of such color electric field as a function of the polar angle $\theta$ for different values of radius ($R$) at $\alpha_{s}=0.12$ in the infrared sector of QCD has been presented by a $(2-d)$ graphics given by Fig.~(\ref{Figure_EField}). It clearly shows that, in the infrared sector of QCD, for a large sphere enclosing the flux tubes, the color electric flux gets localized or spread around the poles ($\theta$ = 0~and~$\pi$) while its gets uniformly distributed for the small sphere case and acquires a constant value at the critical radius $R_{c}$ as given by Eq.(\ref{equation_sixteen}).

\subsection{Summary and conclusions}
In the present paper, we have studied the color confining structure of dual QCD vacuum in the SU(2) pure gauge theory at zero temperature. In the present scenario for the non perturbative regime of QCD the flux tube configuration in dual QCD vacuum has been analyzed by breaking the magnetic symmetry of the system in a dynamical way. Its implication, in turn, are extended to discuss the dynamics of quark-hadron phase transition by computing the associated critical parameters of phase transition at zero temperature. Further, the study of chromoelectric field profiles at different length scale on $S^{2}$- sphere, has been shown to lead that at large distance scale, the color electric flux is localized towards the pole ($\theta=o,\pi$) where QCD monopole gets condensed while at small distance scale its uniform distribution takes place and at $R=R_{c}$ the whole flux is shifted at the central region where QCD monopole density vanishes. Such variation of color electric field with the length scale also indicates the possibility of the formation of QGP before acquiring the fully deconfined state in the high energy regime. It may, therefore, be concluded that below the $R_{c}$, the topological configurations appearing within the QCD vacuum and their condensation at large distance scales can produce confining features in the system. The chromoelectric field profiles in the thermal environment \cite{Chandola:2019xwo} indicates the decreasing amplitude of the field as the temperature is increased towards and above the deconfinement temperature and shows good agreement with Monte Carlo simulation studies~\cite{Cea:2015wjd}.

The present gauge independent dual QCD formulation with its essential validity in the non perturbative regime of QCD supported by recent lattice QCD studies~\cite{Cho:2012pq,Cea:2015wjd}, may be useful to investigate the dynamical problem like plasma oscillation and stability of flux tube configuration in a viable dual QCD formulation.

%\subsection{Acknowledgements}
%{\small The author (DSR) is thankful to UGC, New Delhi for the financial assistance. DSR is also thankful to the organizing committee of DQCD-2019 for their invitation and hospitality.}

%\clearpage
%%%%%%%%%%%%%%%%%%%%%%%%%%%%%%%%%%%%%%%%%%%%%%%%%%%%%%%%%%%%%%%%%%%
%\noindent
%\begin{thebibliography}{100}
%\medskip
%\bibitem{Gross} D. Gross, F. Wilczek: Phys. Rev. {\bf{D8}}, 3497 (1973).
%\bibitem{Polit} H.D. Politzer: Phys. Rev. Lett. {\bf{26}}, 1346 (1973).
%\bibitem{Nambu} Y. Nambu: Phys. Rev. {\bf{D10}}, 4262 (1974).
%\bibitem{hooft} G t' Hooft: Nucl. Phys. {\bf{B190}}, 455 (1981).
%\bibitem{hcc1} H.C. Panndey and H.C. Chandola: Phys. Lett. {\bf{B476}}, 193 (2000); H.C. Chandola and D. Yadav: Nucl. Phys. {\bf{A829}}, 151 (2009).
%D.S. Rawat, H.C. Chandola et. al.: Springer Proc. Phys. {\bf{203}}, 625 (2018).
%\bibitem{dsr} H.C. Chandola, D.S. Rawat et. al.: arXiv: 190411714v1[hep-th] 2019.
%\bibitem{cho1} Y.M. Cho: Phys. Rev. {\bf{D21}}, 1080 (1980); ibid {\bf{D23}}, 2415 (1981).
%\bibitem{cho3} Y.M. Cho et. al.: Phys. Rev. {\bf{D87}}, 085025 (2013).
%\bibitem{NO} H. Neilsen and B. Olesen: Nucl. Phys. {\bf{B61}}, 45 (1973).
%\bibitem{PC} P. Cea et. al.: JHEP {\bf{06}}, 033 (2016).
%
%
%
%
%\end{thebibliography}
%
%
%\end{document}
%%
% ****** End of file apssamp.tex ******

%% file: acknow.tex
\section*{Acknowledgments}

This writeup is a compilation of the contributions as presented in the `Workshop on Dynamics of QCD Matter' held from 15th to 17th August 2019 in NISER Bhubaneswar, India. The aim of this workshop was to enhance the direct exchange of scientific information among the younger members of the Relativistic Heavy Ion community in India, both from the experiments and theory. The focus of the discussions was on the fundamental understanding of strongly-interacting matter at extreme conditions, as formed in ultra-relativistic nucleus-nucleus collisions, as well as on emergent QCD phenomena in high-multiplicity proton-proton and proton-nucleus collisions.

Organizers of the Workshop on Dynamics of QCD Matter (2019), Amaresh Jaiswal, Najmul Haque, Victor Roy, Ranbir Singh, Varchaswi Kashyap and Sudipan De, would like to thank NISER, particularly its head Prof. Sudhakar Panda. Special thanks to Professor Ashok Mahapatra, the chairperson of the School of Physical Sciences, for the encouragement and support. Last but not the least, we thank Department of Atomic Energy, Govt. of India, for providing the financial support.

Lokesh Kumar acknowledges discussions with Bedangadas Mohanty and Natasha Sharma. The support from the SERB Grant No. ECR/2016/000109 is acknowledged and would like to thank the ALICE collaboration.

Victor Roy is supported by the DST-INSPIRE Faculty research grant, India.

Deeptak Biswas, Ritesh Ghosh and Deependra Singh Rawat are thankful to UGC, New Delhi for the financial assistance. 

Sabyasachi Ghosh thanks his collaborators Sourav Sarkar, Santosh K. Das, Jane Alam, Gastao Krein, Bedangadas Mohanty, Hiranmoy Mishra, Sandeep Chatterjee, Victor Roy, Sukanya Mitra, Anirban Lahiri, Sarbani Majumder, Rajarshi Ray, Sanjay K. Ghosh, whose direct/indirect impact is linked with the investigation series, which is briefly reviewed in present contribution. 

Sabyasachi Ghosh and Jayanta Dey acknowledge IIT-Bhilai, funded by Ministry of Human Resource Development (MHRD), Government of India. Ralf Rapp has been supported by the U.S. National Science Foundation under grant no. PHY-1913286. Work partially supported by Conselho Nacional de Desenvolvimento Cient\'{\i}fico e Tecnol\'ogico - CNPq, Grants. no. 304758/2017-5 (R.L.S.F), 305894/2009-9 (Gast\~ao Krein), and 464898/2014-5(G.K) (INCT F\'{\i}sica Nuclear e Aplicac\~oes), and Fundac\~ao de Amparo \`a Pesquisa do Estado do  Rio Grande do Sul - FAPERGS, Grant No. 19/2551-0000690-0 (Ricardo L. S. Farias), and Fundac\~ao de Amparo \`a Pesquisa do Estado de S\~ao Paulo - FAPESP, Grant No. 2013/01907-0 (Gast\~ao Krein) and Coordenac\~ao de Aperfeicoamento de Pessoal de N\'ivel Superior (CAPES) (A.B.) - Brasil (CAPES)- Finance Code 001. 

Deeptak Biswas acknowledges kind hospitality of NISER. Amaresh Jaiswal is supported in part by the DST-INSPIRE faculty award under Grant No. DST/INSPIRE/04/2017/000038. Sutanu Roy is supported in part by the SERB Early Career Research Award under Grant No. ECR/2017/001354.

Sumana Bhattacharyya was supported by Council of Scientific and Industrial Research (CSIR), India. 

Vinod Chandra would like to acknowledge SERB for the Early Career Research Award (ECRA/2016), and DST, Govt. of India for INSPIRE-Faculty Fellowship (IFA-13/PH-55).

Avdhesh Kumar thank Wojciech Florkowski and Radoslaw Ryblewski for a very fruitful collaboration. This research was supported in part by the Polish National Science Center Grant No. 2016/23/B/ST2/00717.

Arpan Das, Ritesh Ghosh, Sabyasachi Ghosh, Lokesh Kumar and Deependra Singh Rawat acknowledges the organizers of ``Workshop on Dynamics of QCD matter 2019" for the invitation.

%% file: DQM_ijmpe.bbl
\begin{thebibliography}{999}
%%%%%%%%Lokesh begins%%%%%%%%

%\cite{Acharya:2018hhy}
\bibitem{Acharya:2018hhy}
S.~Acharya \textit{et al.} [ALICE],
%``Centrality and pseudorapidity dependence of the charged-particle multiplicity density in Xe–Xe collisions at $\sqrt{s_{\rm NN}}$ =5.44TeV,''
Phys. Lett. B \textbf{790}, 35-48 (2019)
%doi:10.1016/j.physletb.2018.12.048
[arXiv:1805.04432 [nucl-ex]].
  
%\cite{Loizides:2016djv}
\bibitem{Loizides:2016djv}
C.~Loizides,
%``Glauber modeling of high-energy nuclear collisions at the subnucleon level,''
Phys. Rev. C \textbf{94}, 024914 (2016)
%doi:10.1103/PhysRevC.94.024914
[arXiv:1603.07375 [nucl-ex]].
  
%\cite{Knospe:2018mek}
\bibitem{Knospe:2018mek}
A.~G.~Knospe [ALICE],
%``New results on soft particle production in heavy-ion collisions with ALICE,''
PoS \textbf{LHCP2018}, 217 (2018).
%doi:10.22323/1.321.0217
  
%\cite{Abelev:2013vea}
\bibitem{Abelev:2013vea}
B.~Abelev \textit{et al.} [ALICE],
%``Centrality dependence of $\pi$, K, p production in Pb-Pb collisions at $\sqrt{s_{NN}}$ = 2.76 TeV,''
Phys. Rev. C \textbf{88}, 044910 (2013)
%doi:10.1103/PhysRevC.88.044910
[arXiv:1303.0737 [hep-ex]].

%\cite{Adam:2016dau}
\bibitem{Adam:2016dau}
J.~Adam \textit{et al.} [ALICE],
%``Multiplicity dependence of charged pion, kaon, and (anti)proton production at large transverse momentum in p-Pb collisions at $\mathbf{\sqrt{{\textit s}_{\rm NN}}}$ = 5.02 TeV,''
Phys. Lett. B \textbf{760}, 720-735 (2016)
%doi:10.1016/j.physletb.2016.07.050
[arXiv:1601.03658 [nucl-ex]].

%\cite{ALICE:2017jyt}
\bibitem{ALICE:2017jyt}
J.~Adam \textit{et al.} [ALICE],
%``Enhanced production of multi-strange hadrons in high-multiplicity proton-proton collisions,''
Nature Phys. \textbf{13}, 535-539 (2017)
%doi:10.1038/nphys4111
[arXiv:1606.07424 [nucl-ex]].

%\cite{Adam:2015vsf}
\bibitem{Adam:2015vsf}
J.~Adam \textit{et al.} [ALICE],
%``Multi-strange baryon production in p-Pb collisions at $\sqrt{s_\mathbf{NN}}=5.02$ TeV,''
Phys. Lett. B \textbf{758}, 389-401 (2016)
%doi:10.1016/j.physletb.2016.05.027
[arXiv:1512.07227 [nucl-ex]].

%\cite{Abelev:2013haa}
\bibitem{Abelev:2013haa}
B.~B.~Abelev \textit{et al.} [ALICE],
%``Multiplicity Dependence of Pion, Kaon, Proton and Lambda Production in p-Pb Collisions at $\sqrt{s_{NN}}$ = 5.02 TeV,''
Phys. Lett. B \textbf{728}, 25-38 (2014)
%doi:10.1016/j.physletb.2013.11.020
[arXiv:1307.6796 [nucl-ex]]. 

%\cite{Koch:1986ud}
\bibitem{Koch:1986ud}
P.~Koch, B.~Muller and J.~Rafelski,
%``Strangeness in Relativistic Heavy Ion Collisions,''
Phys. Rept. \textbf{142}, 167-262 (1986).
%doi:10.1016/0370-1573(86)90096-7

%\cite{Acharya:2018orn}
\bibitem{Acharya:2018orn}
S.~Acharya \textit{et al.} [ALICE],
%``Multiplicity dependence of light-flavor hadron production in pp collisions at $\sqrt{s}$ = 7 TeV,''
Phys. Rev. C \textbf{99}, 024906 (2019)
%doi:10.1103/PhysRevC.99.024906
[arXiv:1807.11321 [nucl-ex]].

%\cite{Sharma:2018owb}
\bibitem{Sharma:2018owb}
N.~Sharma, J.~Cleymans and L.~Kumar,
%``Thermal model description of p–Pb collisions at $\sqrt{s_{NN}} = 5.02$  TeV,''
Eur. Phys. J. C \textbf{78}, no.4, 288 (2018)
%doi:10.1140/epjc/s10052-018-5767-3
[arXiv:1802.07972 [hep-ph]].

%\cite{Knospe:2015nva}
\bibitem{Knospe:2015nva}
A.~G.~Knospe, C.~Markert, K.~Werner, J.~Steinheimer and M.~Bleicher,
%``Hadronic resonance production and interaction in partonic and hadronic matter in the EPOS3 model with and without the hadronic afterburner UrQMD,''
Phys. Rev. C \textbf{93}, 014911 (2016)
%doi:10.1103/PhysRevC.93.014911
[arXiv:1509.07895 [nucl-th]].
  
%\cite{Becattini:2007sr}
\bibitem{Becattini:2007sr}
F.~Becattini, F.~Piccinini and J.~Rizzo,
%``Angular momentum conservation in heavy ion collisions at very high energy,''
Phys. Rev. C \textbf{77}, 024906 (2008)
%doi:10.1103/PhysRevC.77.024906
[arXiv:0711.1253 [nucl-th]].  
  
%\cite{Kharzeev:2007jp}
\bibitem{Kharzeev:2007jp}
D.~E.~Kharzeev, L.~D.~McLerran and H.~J.~Warringa,
%``The Effects of topological charge change in heavy ion collisions: 'Event by event P and CP violation',''
Nucl. Phys. A \textbf{803}, 227-253 (2008)
%doi:10.1016/j.nuclphysa.2008.02.298
[arXiv:0711.0950 [hep-ph]].

%\cite{Liang:2004xn}
\bibitem{Liang:2004xn}
Z.~T.~Liang and X.~N.~Wang,
%``Spin alignment of vector mesons in non-central A+A collisions,''
Phys. Lett. B \textbf{629}, 20-26 (2005)
%doi:10.1016/j.physletb.2005.09.060
[arXiv:nucl-th/0411101 [nucl-th]].

%\cite{Yang:2017sdk}
\bibitem{Yang:2017sdk}
Y.~G.~Yang, R.~H.~Fang, Q.~Wang and X.~N.~Wang,
%``Quark coalescence model for polarized vector mesons and baryons,''
Phys. Rev. C \textbf{97}, 034917 (2018)
%doi:10.1103/PhysRevC.97.034917
[arXiv:1711.06008 [nucl-th]].

%\cite{Fano:1957zz}
\bibitem{Fano:1957zz}
U.~Fano,
%``Description of States in Quantum Mechanics by Density Matrix and Operator Techniques,''
Rev. Mod. Phys. \textbf{29}, 74-93 (1957).
%doi:10.1103/RevModPhys.29.74

%%%%%%%%Lokesh ends%%%%%%%%



%%%%%%%Aritra Start%%%%%%%%%%%

%\cite{Skokov:2009qp}
\bibitem{Skokov:2009qp}
V.~Skokov, A.~Y.~Illarionov and V.~Toneev,
%``Estimate of the magnetic field strength in heavy-ion collisions,''
Int. J. Mod. Phys. A \textbf{24}, 5925-5932 (2009)
%doi:10.1142/S0217751X09047570
[arXiv:0907.1396 [nucl-th]].

%\cite{Duncan:1992hi}
\bibitem{Duncan:1992hi}
R.~C.~Duncan and C.~Thompson,
%``Formation of very strongly magnetized neutron stars - implications for gamma-ray bursts,''
Astrophys. J. Lett. \textbf{392}, L9 (1992).
%doi:10.1086/186413

%\cite{Alexandre:2000jc}
\bibitem{Alexandre:2000jc}
J.~Alexandre,
%``Vacuum polarization in thermal QED with an external magnetic field,''
Phys. Rev. D \textbf{63}, 073010 (2001)
%doi:10.1103/PhysRevD.63.073010
[arXiv:hep-th/0009204 [hep-th]].

%\cite{Tsai:1974ap}
\bibitem{Tsai:1974ap}
W.~y.~Tsai,
%``Vacuum Polarization in Homogeneous Magnetic Fields,''
Phys. Rev. D \textbf{10}, 2699 (1974).
%doi:10.1103/PhysRevD.10.2699

%\cite{Das:2019ehv}
\bibitem{Das:2019ehv}
A.~Das and N.~Haque,
%``Neutral pion mass in the linear sigma model coupled to quarks at arbitrary magnetic field,''
Phys. Rev. D \textbf{101}, 074033 (2020)
%doi:10.1103/PhysRevD.101.074033
[arXiv:1908.10323 [hep-ph]].

%\cite{Ayala:2018zat}
\bibitem{Ayala:2018zat}
A.~Ayala, R.~L.~S.~Farias, S.~Hernández-Ortiz, L.~A.~Hernández, D.~M.~Paret and R.~Zamora,
%``Magnetic field-dependence of the neutral pion mass in the linear sigma model coupled to quarks: The weak field case,''
Phys. Rev. D \textbf{98}, 114008 (2018)
%doi:10.1103/PhysRevD.98.114008
[arXiv:1809.08312 [hep-ph]].

%\cite{Bali:2017ian}
\bibitem{Bali:2017ian}
G.~S.~Bali, B.~B.~Brandt, G.~Endr\"odi and B.~Gl\"a\ss le,
%``Meson masses in electromagnetic fields with Wilson fermions,''
Phys. Rev. D \textbf{97}, 034505 (2018)
%doi:10.1103/PhysRevD.97.034505
[arXiv:1707.05600 [hep-lat]].

%%%%%%%Aritra end%%%%%%%%%%%

%%%%%%%%%%Ashutosh start%%%%%%%%%%%%
\bibitem{Gavai:2004sd} 
R.~V.~Gavai and S.~Gupta,
%``The Critical end point of QCD,''
Phys.\ Rev.\ D {\bf 71}, 114014 (2005)
%doi:10.1103/PhysRevD.71.114014
%[hep-lat/0412035].
\bibitem{Niemi:2012aj} 
H.~Niemi, G.~S.~Denicol, H.~Holopainen and P.~Huovinen,
%``Event-by-event distributions of azimuthal asymmetries in ultrarelativistic heavy-ion collisions,''
Phys.\ Rev.\ C {\bf 87}, no. 5, 054901 (2013)
\bibitem{Dash:2019fsm} 
A.~Dash and V.~Roy,
%``Flow correlation as a measure of phase transition: results from a new hydrodynamic code,''
arXiv:1908.05292 [hep-ph].
%\cite{Huovinen:2009yb}

%%%%%%%%%%%Ashutosh end%%%%%%%%%%%

%%%%%%%%Mariyah begins%%%%%%%%

%\cite{Bacchetta:2016ccz}
\bibitem{Bacchetta:2016ccz} 
  A.~Bacchetta,
  %``Where do we stand with a 3-D picture of the proton?,''
  Eur.\ Phys.\ J.\ A {\bf 52}, no. 6, 163 (2016).
  %doi:10.1140/epja/i2016-16163-5.
  %%CITATION = doi:10.1140/epja/i2016-16163-5;%%

 %\cite{Gribov:1984tu}
\bibitem{Gribov:1984tu} 
  L.~V.~Gribov, E.~M.~Levin and M.~G.~Ryskin,
  %``Semihard Processes in QCD,''
  Phys.\ Rept.\  {\bf 100}, 1 (1983).
  %doi:10.1016/0370-1573(83)90022-4
  %%CITATION = doi:10.1016/0370-1573(83)90022-4;%%
  
  %\cite{Dominguez:2011wm}
\bibitem{Dominguez:2011wm} 
  F.~Dominguez, C.~Marquet, B.~W.~Xiao and F.~Yuan,
  %``Universality of Unintegrated Gluon Distributions at small x,''
  Phys.\ Rev.\ D {\bf 83}, 105005 (2011)
  %doi:10.1103/PhysRevD.83.105005
  [arXiv:1101.0715 [hep-ph]].
  %%CITATION = doi:10.1103/PhysRevD.83.105005;%%
  
%\cite{Buffing:2013eka}
\bibitem{Buffing:2013eka} 
  M.~G.~A.~Buffing, P.~J.~Mulders and A.~Mukherjee,
  %`Universality of Quark and Gluon TMD Correlators,''
  Int.\ J.\ Mod.\ Phys.\ Conf.\ Ser.\  {\bf 25}, 1460003 (2014)
  %doi:10.1142/S2010194514600039
  [arXiv:1309.2472 [hep-ph]].
  %%CITATION = doi:10.1142/S2010194514600039;%%

%\cite{Hatta:2016dxp}
\bibitem{Hatta:2016dxp} 
  Y.~Hatta, B.~W.~Xiao and F.~Yuan,
  %``Probing the Small- x Gluon Tomography in Correlated Hard Diffractive Dijet Production in Deep Inelastic Scattering,''
  Phys.\ Rev.\ Lett.\  {\bf 116}, no. 20, 202301 (2016)
  %doi:10.1103/PhysRevLett.116.202301
  [arXiv:1601.01585 [hep-ph]].
  %%CITATION = doi:10.1103/PhysRevLett.116.202301;%%

%\cite{Xiao:2017yya}
\bibitem{Xiao:2017yya} 
  B.~W.~Xiao, F.~Yuan and J.~Zhou,
  %``Transverse Momentum Dependent Parton Distributions at Small-x,''
  Nucl.\ Phys.\ B {\bf 921}, 104 (2017)
  %doi:10.1016/j.nuclphysb.2017.05.012
  [arXiv:1703.06163 [hep-ph]].
  %%CITATION = doi:10.1016/j.nuclphysb.2017.05.012;%%
  
  %\cite{Abir:2018hvk}
\bibitem{Abir:2018hvk} 
  M.~Siddiqah, N.~Vasim, K.~Banu, R.~Abir and T.~Bhattacharyya,
  %``Unintegrated dipole gluon distribution at small transverse momentum,''
  Phys.\ Rev.\ D {\bf 97}, 054009 (2018)
  %doi:10.1103/PhysRevD.97.054009
  [arXiv:1801.01637 [hep-ph]].
  %%CITATION = doi:10.1103/PhysRevD.97.054009;%%
  
%\cite{Levin:1999mw}
\bibitem{Levin:1999mw} 
  E.~Levin and K.~Tuchin,
  %``Solution to the evolution equation for high parton density QCD,''
  Nucl.\ Phys.\ B {\bf 573}, 833 (2000)
  [hep-ph/9908317].
  %%CITATION = HEP-PH/9908317;%%

  %\cite{Levin:2000mv}
\bibitem{Levin:2000mv} 
  E.~Levin and K.~Tuchin,
  %``New scaling at high-energy DIS,''
  Nucl.\ Phys.\ A {\bf 691}, 779 (2001)
  [hep-ph/0012167].
  %%CITATION = HEP-PH/0012167;%%  

%%%%%%%%Mariyah ends%%%%%%%%



%%%%%%%%Pallavi begins%%%%%%%%


%\cite{Prakash:1993bt}
\bibitem{Prakash:1993bt}
M.~Prakash, M.~Prakash, R.~Venugopalan and G.~Welke,
%``Nonequilibrium properties of hadronic mixtures,''
Phys. Rept. \textbf{227}, 321-366 (1993).
%doi:10.1016/0370-1573(93)90092-R

%\cite{Romatschke:2014gna}
\bibitem{Romatschke:2014gna}
P.~Romatschke and S.~Pratt,
%``Extracting the shear viscosity of a high temperature hadron gas,''
[arXiv:1409.0010 [nucl-th]].
	
%%%%%%%%Pallavi ends%%%%%%%%


%%%%%%%%Sarthak begins%%%%%%%%

\bibitem{Djouadi:2005gi} 
A.~Djouadi,
%``The Anatomy of electro-weak symmetry breaking. I: The Higgs boson in the standard model,''
Phys.\ Rept.\  {\bf 457}, 1 (2008)
%doi:10.1016/j.physrep.2007.10.004
[hep-ph/0503172].

\bibitem{dEnterria:2018bqi} 
D.~d'Enterria and C.~Loizides,
%``Final-state interactions of the Higgs boson in quark-gluon matter,''
arXiv:1809.06832 [hep-ph].
 
\bibitem{Ghiglieri:2019lzz} 
J.~Ghiglieri and U.~A.~Wiedemann,
%``Thermal width of the Higgs boson in hot QCD matter,''
Phys.\ Rev.\ D {\bf 99}, no. 5, 054002 (2019)
% doi:10.1103/PhysRevD.99.054002
[arXiv:1901.04503 [hep-ph]].
 
\bibitem{CaronHuot:2009ns} 
S.~Caron-Huot,
%``Asymptotics of thermal spectral functions,''
Phys.\ Rev.\ D {\bf 79}, 125009 (2009)
%doi:10.1103/PhysRevD.79.125009
[arXiv:0903.3958 [hep-ph]].

\bibitem{Wilson:1969zs} 
K.~G.~Wilson,
%``Nonlagrangian models of current algebra,''
Phys.\ Rev.\  {\bf 179}, 1499 (1969).
% doi:10.1103/PhysRev.179.1499
 
\bibitem{Wilson:1972ee} 
K.~G.~Wilson and W.~Zimmermann,
%``Operator product expansions and composite field operators in the general framework of quantum field theory,''
Commun.\ Math.\ Phys.\  {\bf 24}, 87 (1972).
% doi:10.1007/BF01878448
  
\bibitem{Moore:2008ws} 
G.~D.~Moore and O.~Saremi,
%``Bulk viscosity and spectral functions in QCD,''
JHEP {\bf 0809}, 015 (2008)
%doi:10.1088/1126-6708/2008/09/015

\bibitem{Laine:2010tc} 
M.~Laine, M.~Vepsalainen and A.~Vuorinen,
%``Ultraviolet asymptotics of scalar and pseudoscalar correlators in hot Yang-Mills theory,''
JHEP {\bf 1010}, 010 (2010)
% doi:10.1007/JHEP10(2010)010
[arXiv:1008.3263 [hep-ph]].
 
%%%%%%%%Sarthak ends%%%%%%%%


%%%%%%%%Ritesh begins%%%%%%%%

%\cite{Fukushima:2012vr}
\bibitem{Fukushima:2012vr}
K.~Fukushima,
%``Views of the Chiral Magnetic Effect,''
Lect. Notes Phys. \textbf{871}, 241-259 (2013)
%doi:10.1007/978-3-642-37305-3_9
[arXiv:1209.5064 [hep-ph]].	
		
%\cite{McLerran:2013hla}
\bibitem{McLerran:2013hla}
L.~McLerran and V.~Skokov,
%``Comments About the Electromagnetic Field in Heavy-Ion Collisions,''
Nucl. Phys. A \textbf{929}, 184-190 (2014)
%doi:10.1016/j.nuclphysa.2014.05.008
[arXiv:1305.0774 [hep-ph]].
	
%\cite{Bandyopadhyay:2016fyd}
\bibitem{Bandyopadhyay:2016fyd}
A.~Bandyopadhyay, C.~A.~Islam and M.~G.~Mustafa,
%``Electromagnetic spectral properties and Debye screening of a strongly magnetized hot medium,''
Phys. Rev. D \textbf{94}, no.11, 114034 (2016)
%doi:10.1103/PhysRevD.94.114034
[arXiv:1602.06769 [hep-ph]].
	
%\cite{Karmakar:2019tdp}
\bibitem{Karmakar:2019tdp}
B.~Karmakar, R.~Ghosh, A.~Bandyopadhyay, N.~Haque and M.~G.~Mustafa,
%``Anisotropic pressure of deconfined QCD matter in presence of strong magnetic field within one-loop approximation,''
Phys. Rev. D \textbf{99}, no.9, 094002 (2019)
%doi:10.1103/PhysRevD.99.094002
[arXiv:1902.02607 [hep-ph]].
	
%\cite{Das:2017vfh}
\bibitem{Das:2017vfh}
A.~Das, A.~Bandyopadhyay, P.~K.~Roy and M.~G.~Mustafa,
%``General structure of fermion two-point function and its spectral representation in a hot magnetized medium,''
Phys. Rev. D \textbf{97}, no.3, 034024 (2018)
%doi:10.1103/PhysRevD.97.034024
[arXiv:1709.08365 [hep-ph]].
		
%\cite{Karmakar:2018aig}
\bibitem{Karmakar:2018aig}
B.~Karmakar, A.~Bandyopadhyay, N.~Haque and M.~G.~Mustafa,
%``General structure of gauge boson propagator and its spectra in a hot magnetized medium,''
Eur. Phys. J. C \textbf{79}, no.8, 658 (2019)
%doi:10.1140/epjc/s10052-019-7154-0
[arXiv:1804.11336 [hep-ph]].
	
%\cite{PerezMartinez:2007kw}
\bibitem{PerezMartinez:2007kw}
A.~Perez Martinez, H.~Perez Rojas and H.~Mosquera Cuesta,
%``Anisotropic Pressures in Very Dense Magnetized Matter,''
Int. J. Mod. Phys. D \textbf{17}, 2107-2123 (2008)
%doi:10.1142/S0218271808013741
[arXiv:0711.0975 [astro-ph]].
	
%%%%%%%%Ritesh ends%%%%%%%%


%%%%%%%%Sabya begins%%%%%%%%

%\cite{Schafer:2009dj}
\bibitem{Schafer:2009dj}
T.~Sch\"afer and D.~Teaney,
%``Nearly Perfect Fluidity: From Cold Atomic Gases to Hot Quark Gluon Plasmas,''
Rept. Prog. Phys. \textbf{72}, 126001 (2009)
%doi:10.1088/0034-4885/72/12/126001
[arXiv:0904.3107 [hep-ph]].

%\cite{Arnold:2000dr}
\bibitem{Arnold:2000dr}
P.~B.~Arnold, G.~D.~Moore and L.~G.~Yaffe,
%``Transport coefficients in high temperature gauge theories. 1. Leading log results,''
JHEP \textbf{11}, 001 (2000)
%doi:10.1088/1126-6708/2000/11/001
[arXiv:hep-ph/0010177 [hep-ph]].

%\cite{Marty:2013ita}
\bibitem{Marty:2013ita}
R.~Marty, E.~Bratkovskaya, W.~Cassing, J.~Aichelin and H.~Berrehrah,
%``Transport coefficients from the Nambu-Jona-Lasinio model for $SU(3)_f$,''
Phys. Rev. C \textbf{88}, 045204 (2013)
%doi:10.1103/PhysRevC.88.045204
[arXiv:1305.7180 [hep-ph]].

%\cite{Sasaki:2008um}
\bibitem{Sasaki:2008um}
C.~Sasaki and K.~Redlich,
%``Transport coefficients near chiral phase transition,''
Nucl. Phys. A \textbf{832}, 62-75 (2010)
%doi:10.1016/j.nuclphysa.2009.11.005
[arXiv:0811.4708 [hep-ph]].

%\cite{Ghosh:2015mda}
\bibitem{Ghosh:2015mda}
S.~Ghosh, T.~C.~Peixoto, V.~Roy, F.~E.~Serna and G.~Krein,
%``Shear and bulk viscosities of quark matter from quark-meson fluctuations in the Nambu–Jona-Lasinio model,''
Phys. Rev. C \textbf{93}, no.4, 045205 (2016)
%doi:10.1103/PhysRevC.93.045205
[arXiv:1507.08798 [nucl-th]].

%\cite{Deb:2016myz}
\bibitem{Deb:2016myz}
P.~Deb, G.~P.~Kadam and H.~Mishra,
%``Estimating transport coefficients in hot and dense quark matter,''
Phys. Rev. D \textbf{94}, no.9, 094002 (2016)
%doi:10.1103/PhysRevD.94.094002
[arXiv:1603.01952 [hep-ph]].

%\cite{Chakraborty:2010fr}
\bibitem{Chakraborty:2010fr}
P.~Chakraborty and J.~I.~Kapusta,
%``Quasi-Particle Theory of Shear and Bulk Viscosities of Hadronic Matter,''
Phys. Rev. C \textbf{83}, 014906 (2011)
%doi:10.1103/PhysRevC.83.014906
[arXiv:1006.0257 [nucl-th]].

%\cite{Singha:2017jmq}
\bibitem{Singha:2017jmq}
P.~Singha, A.~Abhishek, G.~Kadam, S.~Ghosh and H.~Mishra,
%``Calculations of shear, bulk viscosities and electrical conductivity in the Polyakov-quark–meson model,''
J. Phys. G \textbf{46}, no.1, 015201 (2019)
%doi:10.1088/1361-6471/aaf256
[arXiv:1705.03084 [nucl-th]].

%\cite{Lang:2013lla}
\bibitem{Lang:2013lla}
R.~Lang and W.~Weise,
%``Shear viscosity from Kubo formalism: NJL model study,''
Eur. Phys. J. A \textbf{50}, 63 (2014)
%doi:10.1140/epja/i2014-14063-4
[arXiv:1311.4628 [hep-ph]].

%\cite{Cassing:2013iz}
\bibitem{Cassing:2013iz}
W.~Cassing, O.~Linnyk, T.~Steinert and V.~Ozvenchuk,
%``Electrical Conductivity of Hot QCD Matter,''
Phys. Rev. Lett. \textbf{110}, no.18, 182301 (2013)
%doi:10.1103/PhysRevLett.110.182301
[arXiv:1302.0906 [hep-ph]].

%\cite{Ghosh:2019lmx}
\bibitem{Ghosh:2019lmx}
S.~Ghosh, B.~Chatterjee, P.~Mohanty, A.~Mukharjee and H.~Mishra,
%``Impact of magnetic field on shear viscosity of quark matter in Nambu–Jona-Lasinio model,''
Phys. Rev. D \textbf{100}, no.3, 034024 (2019)
%doi:10.1103/PhysRevD.100.034024
[arXiv:1804.00812 [hep-ph]].

%\cite{FernandezFraile:2009mi}
\bibitem{FernandezFraile:2009mi}
D.~Fernandez-Fraile and A.~Gomez Nicola,
%``Transport coefficients and resonances for a meson gas in Chiral Perturbation Theory,''
%doi:10.1140/epjc/s10052-009-0935-0
[arXiv:0902.4829 [hep-ph]].

%\cite{Ghosh:2014qba}
\bibitem{Ghosh:2014qba}
S.~Ghosh, G.~Krein and S.~Sarkar,
%``Shear viscosity of a pion gas resulting from $\rho\pi\pi$ and $\sigma\pi\pi$ interactions,''
Phys. Rev. C \textbf{89}, no.4, 045201 (2014)
%doi:10.1103/PhysRevC.89.045201
[arXiv:1401.5392 [nucl-th]].

%\cite{Ghosh:2014ija}
\bibitem{Ghosh:2014ija}
S.~Ghosh,
%``Nucleon thermal width owing to pion-baryon loops and its contributions to shear viscosity,''
Phys. Rev. C \textbf{90}, no.2, 025202 (2014)
%doi:10.1103/PhysRevC.90.025202
[arXiv:1503.06927 [nucl-th]].

%\cite{Ghosh:2015lba}
\bibitem{Ghosh:2015lba}
S.~Ghosh,
%``Shear viscosity of pionic and nucleonic components from their different possible mesonic and baryonic thermal fluctuations,''
Braz. J. Phys. \textbf{45}, no.6, 687-698 (2015)
%doi:10.1007/s13538-015-0352-9
[arXiv:1507.01705 [nucl-th]].

%\cite{Rahaman:2017sby}
\bibitem{Rahaman:2017sby}
M.~Rahaman, S.~Ghosh, S.~Ghosh, S.~Sarkar and J.~e.~Alam,
%``Contribution of a kaon component in the viscosity and conductivity of a hadronic medium,''
Phys. Rev. C \textbf{97}, no.3, 035201 (2018)
%doi:10.1103/PhysRevC.97.035201
[arXiv:1708.08300 [nucl-th]].

%\cite{Kalikotay:2019fle}
\bibitem{Kalikotay:2019fle}
P.~Kalikotay, N.~Chaudhuri, S.~Ghosh, U.~Gangopadhyaya and S.~Sarkar,
%``Viscous coefficients and thermal conductivity of a $\pi K N$ gas mixture in the medium,''
Eur. Phys. J. A \textbf{56}, no.3, 79 (2020)
%doi:10.1140/epja/s10050-020-00074-3
[arXiv:1908.02933 [nucl-th]].

%\cite{Ghosh:2018nqi}
\bibitem{Ghosh:2018nqi}
S.~Ghosh, S.~Ghosh and S.~Bhattacharyya,
%``Phenomenological bound on the viscosity of the hadron resonance gas,''
Phys. Rev. C \textbf{98}, no.4, 045202 (2018)
%doi:10.1103/PhysRevC.98.045202
[arXiv:1807.03188 [hep-ph]].

%\cite{Ghosh:2019fpx}
\bibitem{Ghosh:2019fpx}
S.~Ghosh, S.~Samanta, S.~Ghosh and H.~Mishra,
%``Viscosity calculations from Hadron Resonance Gas model: Finite size effect,''
Int. J. Mod. Phys. E \textbf{28}, no.09, 1950036 (2019)
%doi:10.1142/S0218301319500368
[arXiv:1906.06029 [nucl-th]].

%\cite{NoronhaHostler:2008ju}
\bibitem{NoronhaHostler:2008ju}
J.~Noronha-Hostler, J.~Noronha and C.~Greiner,
%``Transport Coefficients of Hadronic Matter near T(c),''
Phys. Rev. Lett. \textbf{103}, 172302 (2009)
%doi:10.1103/PhysRevLett.103.172302
[arXiv:0811.1571 [nucl-th]].

%\cite{Kadam:2015xsa}
\bibitem{Kadam:2015xsa}
G.~P.~Kadam and H.~Mishra,
%``Dissipative properties of hot and dense hadronic matter in an excluded-volume hadron resonance gas model,''
Phys. Rev. C \textbf{92}, no.3, 035203 (2015)
%doi:10.1103/PhysRevC.92.035203
[arXiv:1506.04613 [hep-ph]].

%\cite{Ghosh:2016clt}
\bibitem{Ghosh:2016clt}
S.~Ghosh, S.~Chatterjee and B.~Mohanty,
%``Bulk viscosity for pion and nucleon thermal fluctuation in the hadron resonance gas model,''
Phys. Rev. C \textbf{94}, no.4, 045208 (2016)
%doi:10.1103/PhysRevC.94.045208
[arXiv:1607.04779 [nucl-th]].

%\cite{Ghosh:2016yvt}
\bibitem{Ghosh:2016yvt}
S.~Ghosh,
%``Electrical conductivity of hadronic matter from different possible mesonic and baryonic loops,''
Phys. Rev. D \textbf{95}, no.3, 036018 (2017)
%doi:10.1103/PhysRevD.95.036018
[arXiv:1607.01340 [nucl-th]].

%\cite{Csernai:2006zz}
\bibitem{Csernai:2006zz}
L.~P.~Csernai, J.~I.~Kapusta and L.~D.~McLerran,
%``On the Strongly-Interacting Low-Viscosity Matter Created in Relativistic Nuclear Collisions,''
Phys. Rev. Lett. \textbf{97}, 152303 (2006)
%doi:10.1103/PhysRevLett.97.152303
[arXiv:nucl-th/0604032 [nucl-th]].

%\cite{Kapusta:2008vb}
\bibitem{Kapusta:2008vb}
J.~I.~Kapusta,
%``Viscous Properties of Strongly Interacting Matter at High Temperature,''
Landolt-Bornstein \textbf{23}, 563-580 (2010)
%doi:10.1007/978-3-642-01539-7_18
[arXiv:0809.3746 [nucl-th]].

%\cite{Adler:2003kt}
\bibitem{Adler:2003kt}
S.~S.~Adler \textit{et al.} [PHENIX],
%``Elliptic flow of identified hadrons in Au+Au collisions at s(NN)**(1/2) = 200-GeV,''
Phys. Rev. Lett. \textbf{91}, 182301 (2003)
%doi:10.1103/PhysRevLett.91.182301
[arXiv:nucl-ex/0305013 [nucl-ex]].

%\cite{Romatschke:2007mq}
\bibitem{Romatschke:2007mq}
P.~Romatschke and U.~Romatschke,
%``Viscosity Information from Relativistic Nuclear Collisions: How Perfect is the Fluid Observed at RHIC?,''
Phys. Rev. Lett. \textbf{99}, 172301 (2007)
%doi:10.1103/PhysRevLett.99.172301
[arXiv:0706.1522 [nucl-th]].

%\cite{Meyer:2007ic}
\bibitem{Meyer:2007ic}
H.~B.~Meyer,
%``A Calculation of the shear viscosity in SU(3) gluodynamics,''
Phys. Rev. D \textbf{76}, 101701 (2007)
%doi:10.1103/PhysRevD.76.101701
[arXiv:0704.1801 [hep-lat]].

%\cite{Demir:2008tr}
\bibitem{Demir:2008tr}
N.~Demir and S.~A.~Bass,
%``Shear-Viscosity to Entropy-Density Ratio of a Relativistic Hadron Gas,''
Phys. Rev. Lett. \textbf{102}, 172302 (2009)
%doi:10.1103/PhysRevLett.102.172302
[arXiv:0812.2422 [nucl-th]].

%\cite{Rose:2017bjz}
\bibitem{Rose:2017bjz}
J.~B.~Rose, J.~M.~Torres-Rincon, A.~Sch\"afer, D.~R.~Oliinychenko and H.~Petersen,
%``Shear viscosity of a hadron gas and influence of resonance lifetimes on relaxation time,''
Phys. Rev. C \textbf{97}, no.5, 055204 (2018)
%doi:10.1103/PhysRevC.97.055204
[arXiv:1709.03826 [nucl-th]].

%\cite{Jaiswal:2016hex}
\bibitem{Jaiswal:2016hex}
A.~Jaiswal and V.~Roy,
%``Relativistic hydrodynamics in heavy-ion collisions: general aspects and recent developments,''
Adv. High Energy Phys. \textbf{2016}, 9623034 (2016)
%doi:10.1155/2016/9623034
[arXiv:1605.08694 [nucl-th]].

%\cite{Quack:1993ie}
\bibitem{Quack:1993ie}
E.~Quack and S.~P.~Klevansky,
%``Effective 1/N(c) expansion in the NJL model,''
Phys. Rev. C \textbf{49}, 3283-3288 (1994)
%doi:10.1103/PhysRevC.49.3283

%\cite{Ghosh:2013cba}
\bibitem{Ghosh:2013cba}
S.~Ghosh, A.~Lahiri, S.~Majumder, R.~Ray and S.~K.~Ghosh,
%``Shear viscosity due to Landau damping from the quark-pion interaction,''
Phys. Rev. C \textbf{88}, no.6, 068201 (2013)
%doi:10.1103/PhysRevC.88.068201
[arXiv:1311.4070 [nucl-th]].

%\cite{Saha:2017xjq}
\bibitem{Saha:2017xjq}
K.~Saha, S.~Ghosh, S.~Upadhaya and S.~Maity,
%``Transport coefficients in a finite volume Polyakov–Nambu–Jona-Lasinio model,''
Phys. Rev. D \textbf{97}, no.11, 116020 (2018)
%doi:10.1103/PhysRevD.97.116020
[arXiv:1711.10169 [nucl-th]].

%\cite{Meyer:2007dy}
\bibitem{Meyer:2007dy}
H.~B.~Meyer,
%``A Calculation of the bulk viscosity in SU(3) gluodynamics,''
Phys. Rev. Lett. \textbf{100}, 162001 (2008)
%doi:10.1103/PhysRevLett.100.162001
[arXiv:0710.3717 [hep-lat]].

%\cite{Arnold:2006fz}
\bibitem{Arnold:2006fz}
P.~B.~Arnold, C.~Dogan and G.~D.~Moore,
%``The Bulk Viscosity of High-Temperature QCD,''
Phys. Rev. D \textbf{74}, 085021 (2006)
%doi:10.1103/PhysRevD.74.085021
[arXiv:hep-ph/0608012 [hep-ph]].

%\cite{Gupta:2003zh}
\bibitem{Gupta:2003zh}
S.~Gupta,
%``The Electrical conductivity and soft photon emissivity of the QCD plasma,''
Phys. Lett. B \textbf{597}, 57-62 (2004)
%doi:10.1016/j.physletb.2004.05.079
[arXiv:hep-lat/0301006 [hep-lat]].

%\cite{Aarts:2014nba}
\bibitem{Aarts:2014nba}
G.~Aarts, C.~Allton, A.~Amato, P.~Giudice, S.~Hands and J.~I.~Skullerud,
%``Electrical conductivity and charge diffusion in thermal QCD from the lattice,''
JHEP \textbf{02}, 186 (2015)
%doi:10.1007/JHEP02(2015)186
[arXiv:1412.6411 [hep-lat]].

%%%%%%%%Sabya ends%%%%%%%%

%%%%%%%Arpan begins%%%%%%%%%%%

%\cite{Jeon:1999gr}
\bibitem{Jeon:1999gr}
S.~Jeon and V.~Koch,
%``Fluctuations of particle ratios and the abundance of hadronic resonances,''
Phys. Rev. Lett. \textbf{83}, 5435-5438 (1999)
%doi:10.1103/PhysRevLett.83.5435
[arXiv:nucl-th/9906074 [nucl-th]].

%\cite{Elze:1986qd}
\bibitem{Elze:1986qd}
H.~T.~Elze, M.~Gyulassy and D.~Vasak,
%``Transport Equations for the {QCD} Quark Wigner Operator,''
Nucl. Phys. B \textbf{276}, 706-728 (1986)
%doi:10.1016/0550-3213(86)90072-6

%\cite{Elze:1986hq}
\bibitem{Elze:1986hq}
H.~T.~Elze, M.~Gyulassy and D.~Vasak,
%``Transport Equations for the {QCD} Gluon Wigner Operator,''
Phys. Lett. B \textbf{177}, 402-408 (1986)
%doi:10.1016/0370-2693(86)90778-1

%\cite{DeGroot:1980dk}
\bibitem{DeGroot:1980dk}
S.~R.~De Groot, W.~A.~Van Leeuwen and C.~G.~Van Weert,
``Relativistic Kinetic Theory. Principles and Applications,''
Amsterdam, Netherlands: North-holland ( 1980).

%\cite{Weickgenannt:2019dks}
\bibitem{Weickgenannt:2019dks}
N.~Weickgenannt, X.~L.~Sheng, E.~Speranza, Q.~Wang and D.~H.~Rischke,
%``Kinetic theory for massive spin-1/2 particles from the Wigner-function formalism,''
Phys. Rev. D \textbf{100}, no.5, 056018 (2019)
%doi:10.1103/PhysRevD.100.056018
[arXiv:1902.06513 [hep-ph]].

%\cite{Mao:2018jdo}
\bibitem{Mao:2018jdo}
S.~Mao and D.~H.~Rischke,
%``Dynamically generated magnetic moment in the Wigner-function formalism,''
Phys. Lett. B \textbf{792}, 149-155 (2019)
%doi:10.1016/j.physletb.2019.03.034
[arXiv:1812.06684 [hep-th]].

%\cite{Sheng:2018jwf}
\bibitem{Sheng:2018jwf}
X.~L.~Sheng, R.~H.~Fang, Q.~Wang and D.~H.~Rischke,
%``Wigner function and pair production in parallel electric and magnetic fields,''
Phys. Rev. D \textbf{99}, no.5, 056004 (2019)
%doi:10.1103/PhysRevD.99.056004
[arXiv:1812.01146 [hep-ph]].

%\cite{Sheng:2017lfu}
\bibitem{Sheng:2017lfu}
X.~l.~Sheng, D.~H.~Rischke, D.~Vasak and Q.~Wang,
%``Wigner functions for fermions in strong magnetic fields,''
Eur. Phys. J. A \textbf{54}, no.2, 21 (2018)
%doi:10.1140/epja/i2018-12414-9
[arXiv:1707.01388 [hep-ph]].

%\cite{Karsch:1994hm}
\bibitem{Karsch:1994hm}
F.~Karsch and E.~Laermann,
%``Susceptibilities, the specific heat and a cumulant in two flavor QCD,''
Phys. Rev. D \textbf{50}, 6954-6962 (1994)
%doi:10.1103/PhysRevD.50.6954
[arXiv:hep-lat/9406008 [hep-lat]].

%\cite{Zhuang:1994dw}
\bibitem{Zhuang:1994dw}
P.~Zhuang, J.~Hufner and S.~P.~Klevansky,
%``Thermodynamics of a quark - meson plasma in the Nambu-Jona-Lasinio model,''
Nucl. Phys. A \textbf{576}, 525-552 (1994)
%doi:10.1016/0375-9474(94)90743-9

%\cite{Sasaki:2006ww}
\bibitem{Sasaki:2006ww}
C.~Sasaki, B.~Friman and K.~Redlich,
%``Susceptibilities and the Phase Structure of a Chiral Model with Polyakov Loops,''
Phys. Rev. D \textbf{75}, 074013 (2007)
%doi:10.1103/PhysRevD.75.074013
[arXiv:hep-ph/0611147 [hep-ph]].

%\cite{Shovkovy:2012zn}
\bibitem{Shovkovy:2012zn}
I.~A.~Shovkovy,
%``Magnetic Catalysis: A Review,''
Lect. Notes Phys. \textbf{871}, 13-49 (2013)
%doi:10.1007/978-3-642-37305-3_2
[arXiv:1207.5081 [hep-ph]].

%\cite{Kharzeev:2013ffa}
\bibitem{Kharzeev:2013ffa}
D.~E.~Kharzeev,
%``The Chiral Magnetic Effect and Anomaly-Induced Transport,''
Prog. Part. Nucl. Phys. \textbf{75}, 133-151 (2014)
%doi:10.1016/j.ppnp.2014.01.002
[arXiv:1312.3348 [hep-ph]].

%\cite{Buballa:2003qv}
\bibitem{Buballa:2003qv}
M.~Buballa,
%``NJL model analysis of quark matter at large density,''
Phys. Rept. \textbf{407}, 205-376 (2005)
%doi:10.1016/j.physrep.2004.11.004
[arXiv:hep-ph/0402234 [hep-ph]].

%\cite{Fukushima:2010fe}
\bibitem{Fukushima:2010fe}
K.~Fukushima, M.~Ruggieri and R.~Gatto,
%``Chiral magnetic effect in the PNJL model,''
Phys. Rev. D \textbf{81}, 114031 (2010)
%doi:10.1103/PhysRevD.81.114031
[arXiv:1003.0047 [hep-ph]].

%\cite{Chernodub:2011fr}
\bibitem{Chernodub:2011fr}
M.~N.~Chernodub and A.~S.~Nedelin,
%``Phase diagram of chirally imbalanced QCD matter,''
Phys. Rev. D \textbf{83}, 105008 (2011)
%doi:10.1103/PhysRevD.83.105008
[arXiv:1102.0188 [hep-ph]].

%\cite{Gatto:2011wc}
\bibitem{Gatto:2011wc}
R.~Gatto and M.~Ruggieri,
%``Hot Quark Matter with an Axial Chemical Potential,''
Phys. Rev. D \textbf{85}, 054013 (2012)
%doi:10.1103/PhysRevD.85.054013
[arXiv:1110.4904 [hep-ph]].

%\cite{Ruggieri:2016ejz}
\bibitem{Ruggieri:2016ejz}
M.~Ruggieri and G.~X.~Peng,
%``Critical Temperature of Chiral Symmetry Restoration for Quark Matter with a Chiral Chemical Potential,''
J. Phys. G \textbf{43}, no.12, 125101 (2016)
%doi:10.1088/0954-3899/43/12/125101
[arXiv:1602.05250 [hep-ph]].

%\cite{Yu:2015hym}
\bibitem{Yu:2015hym}
L.~Yu, H.~Liu and M.~Huang,
%``Effect of the chiral chemical potential on the chiral phase transition in the NJL model with different regularization schemes,''
Phys. Rev. D \textbf{94}, no.1, 014026 (2016)
%doi:10.1103/PhysRevD.94.014026
[arXiv:1511.03073 [hep-ph]].

%\cite{Farias:2016let}
\bibitem{Farias:2016let}
R.~L.~S.~Farias, D.~C.~Duarte, G.~Krein and R.~O.~Ramos,
%``Thermodynamics of quark matter with a chiral imbalance,''
Phys. Rev. D \textbf{94}, no.7, 074011 (2016)
%doi:10.1103/PhysRevD.94.074011
[arXiv:1604.04518 [hep-ph]].

%\cite{Braguta:2015owi}
\bibitem{Braguta:2015owi}
V.~V.~Braguta, E.~M.~Ilgenfritz, A.~Y.~Kotov, B.~Petersson and S.~A.~Skinderev,
%``Study of QCD Phase Diagram with Non-Zero Chiral Chemical Potential,''
Phys. Rev. D \textbf{93}, no.3, 034509 (2016)
%doi:10.1103/PhysRevD.93.034509
[arXiv:1512.05873 [hep-lat]].

%\cite{Braguta:2015zta}
\bibitem{Braguta:2015zta}
V.~V.~Braguta, V.~A.~Goy, E.~M.~Ilgenfritz, A.~Y.~Kotov, A.~V.~Molochkov, M.~Muller-Preussker and B.~Petersson,
%``Two-Color QCD with Non-zero Chiral Chemical Potential,''
JHEP \textbf{06}, 094 (2015)
%doi:10.1007/JHEP06(2015)094
[arXiv:1503.06670 [hep-lat]].

%\cite{Farias:2005cr}
\bibitem{Farias:2005cr}
R.~L.~S.~Farias, G.~Dallabona, G.~Krein and O.~A.~Battistel,
%``Cutoff-independent regularization of four-fermion interactions for color superconductivity,''
Phys. Rev. C \textbf{73}, 018201 (2006)
%doi:10.1103/PhysRevC.73.018201
[arXiv:hep-ph/0510145 [hep-ph]].

%\cite{Avancini:2019ego}
\bibitem{Avancini:2019ego}
S.~S.~Avancini, A.~Bandyopadhyay, D.~C.~Duarte and R.~L.~S.~Farias,
%``Cold QCD at finite isospin density: confronting effective models with recent lattice data,''
Phys. Rev. D \textbf{100}, no.11, 116002 (2019)
%doi:10.1103/PhysRevD.100.116002
[arXiv:1907.09880 [hep-ph]].

%\cite{Florkowski:1995ei}
\bibitem{Florkowski:1995ei}
W.~Florkowski, J.~Hufner, S.~P.~Klevansky and L.~Neise,
%``Chirally invariant transport equations for quark matter,''
Annals Phys. \textbf{245}, 445-463 (1996)
%doi:10.1006/aphy.1996.0016
[arXiv:hep-ph/9505407 [hep-ph]].

%\cite{Dmitrasinovic:1996fi}
\bibitem{Dmitrasinovic:1996fi}
V.~Dmitrasinovic,
%``U-A(1) breaking and scalar mesons in the Nambu-Jona-Lasinio model,''
Phys. Rev. C \textbf{53}, 1383-1396 (1996)
%doi:10.1103/PhysRevC.53.1383

%\cite{Das:2019crc}
\bibitem{Das:2019crc}
A.~Das, D.~Kumar and H.~Mishra,
%``Chiral susceptibility in the Nambu–Jona-Lasinio model: A Wigner function approach,''
Phys. Rev. D \textbf{100}, no.9, 094030 (2019)
%doi:10.1103/PhysRevD.100.094030
[arXiv:1907.12332 [hep-ph]].

%\cite{Duarte:2018kfd}
\bibitem{Duarte:2018kfd}
D.~C.~Duarte, R.~L.~S.~Farias and R.~O.~Ramos,
%``Regularization issues for a cold and dense quark matter model in $\beta-$equilibrium,''
Phys. Rev. D \textbf{99}, no.1, 016005 (2019)
%doi:10.1103/PhysRevD.99.016005
[arXiv:1811.10598 [hep-ph]].

%\cite{Boomsma:2009eh}
\bibitem{Boomsma:2009eh}
J.~K.~Boomsma and D.~Boer,
%``The High temperature CP-restoring phase transition at theta = pi,''
Phys. Rev. D \textbf{80}, 034019 (2009)
%doi:10.1103/PhysRevD.80.034019
[arXiv:0905.4660 [hep-ph]].

%%%%%%%Arpan ends%%%%%%%%%%%


%%%%%%%%Ranjita begins%%%%%%%%
%no references
%%%%%%%%Ranjita ends%%%%%%%%


%%%%%%%%Jayanta begins%%%%%%%%

\bibitem{Rafelski:1975rf} 
J.~Rafelski and B.~Muller,
%``Magnetic Splitting of Quasimolecular Electronic States in Strong Fields,''
Phys.\ Rev.\ Lett.\  {\bf 36}, 517 (1976).
%doi:10.1103/PhysRevLett.36.517
%%CITATION = doi:10.1103/PhysRevLett.36.517;%%
  
\bibitem{Tong:2016kpv}
D.~Tong,
%``Lectures on the Quantum Hall Effect,''
[arXiv:1606.06687 [hep-th]].

%\cite{Farias:2014eca}
\bibitem{Farias:2014eca}
R.~L.~S.~Farias, K.~P.~Gomes, G.~I.~Krein and M.~B.~Pinto,
%``Importance of asymptotic freedom for the pseudocritical temperature in magnetized quark matter,''
Phys. Rev. C \textbf{90}, 025203 (2014)
%doi:10.1103/PhysRevC.90.025203
[arXiv:1404.3931 [hep-ph]].

%%%%%%%%Jayanta ends%%%%%%%%


%%%%%%%%Deeptak start%%%%%%%%

%\cite{Lee:1974ma}
\bibitem{Lee:1974ma}
T.~D.~Lee and G.~C.~Wick,
%``Vacuum Stability and Vacuum Excitation in a Spin 0 Field Theory,''
Phys. Rev. D \textbf{9}, 2291-2316 (1974)
%doi:10.1103/PhysRevD.9.2291

%\cite{Collins:1974ky}
\bibitem{Collins:1974ky}
J.~C.~Collins and M.~J.~Perry,
%``Superdense Matter: Neutrons Or Asymptotically Free Quarks?,''
Phys. Rev. Lett. \textbf{34}, 1353 (1975)
%doi:10.1103/PhysRevLett.34.1353

%\cite{Itoh:1970uw}
\bibitem{Itoh:1970uw}
N.~Itoh,
%``Hydrostatic Equilibrium of Hypothetical Quark Stars,''
Prog. Theor. Phys. \textbf{44}, 291 (1970)
%doi:10.1143/PTP.44.291

%\cite{Luzum:2008cw}
\bibitem{Luzum:2008cw}
M.~Luzum and P.~Romatschke,
%``Conformal Relativistic Viscous Hydrodynamics: Applications to RHIC results at s(NN)**(1/2) = 200-GeV,''
Phys. Rev. C \textbf{78}, 034915 (2008)
%doi:10.1103/PhysRevC.78.034915
[arXiv:0804.4015 [nucl-th]].

%\cite{Luzum:2009sb}
\bibitem{Luzum:2009sb}
M.~Luzum and P.~Romatschke,
%``Viscous Hydrodynamic Predictions for Nuclear Collisions at the LHC,''
Phys. Rev. Lett. \textbf{103}, 262302 (2009)
%doi:10.1103/PhysRevLett.103.262302
[arXiv:0901.4588 [nucl-th]].

%\cite{Song:2010mg}
\bibitem{Song:2010mg}
H.~Song, S.~A.~Bass, U.~Heinz, T.~Hirano and C.~Shen,
%``200 A GeV Au+Au collisions serve a nearly perfect quark-gluon liquid,''
Phys. Rev. Lett. \textbf{106}, 192301 (2011)
%doi:10.1103/PhysRevLett.106.192301
[arXiv:1011.2783 [nucl-th]].

%\cite{Luzum:2010ag}
\bibitem{Luzum:2010ag}
M.~Luzum,
%``Elliptic flow at energies available at the CERN Large Hadron Collider: Comparing heavy-ion data to viscous hydrodynamic predictions,''
Phys. Rev. C \textbf{83}, 044911 (2011)
%doi:10.1103/PhysRevC.83.044911
[arXiv:1011.5173 [nucl-th]].

%\cite{Schenke:2011tv}
\bibitem{Schenke:2011tv}
B.~Schenke, S.~Jeon and C.~Gale,
%``Anisotropic flow in $\sqrt{s}=2.76$ TeV Pb+Pb collisions at the LHC,''
Phys. Lett. B \textbf{702}, 59-63 (2011)
%doi:10.1016/j.physletb.2011.06.065
[arXiv:1102.0575 [hep-ph]].

%\cite{Gale:2012rq}
\bibitem{Gale:2012rq}
C.~Gale, S.~Jeon, B.~Schenke, P.~Tribedy and R.~Venugopalan,
%``Event-by-event anisotropic flow in heavy-ion collisions from combined Yang-Mills and viscous fluid dynamics,''
Phys. Rev. Lett. \textbf{110}, no.1, 012302 (2013)
%doi:10.1103/PhysRevLett.110.012302
[arXiv:1209.6330 [nucl-th]].

%\cite{Bhalerao:2015iya}
\bibitem{Bhalerao:2015iya}
R.~S.~Bhalerao, A.~Jaiswal and S.~Pal,
%``Collective flow in event-by-event partonic transport plus hydrodynamics hybrid approach,''
Phys. Rev. C \textbf{92}, no.1, 014903 (2015)
%doi:10.1103/PhysRevC.92.014903
[arXiv:1503.03862 [nucl-th]].

%\cite{Landau:1953gs}
\bibitem{Landau:1953gs}
L.~D.~Landau,
%``On the multiparticle production in high-energy collisions,''
Izv. Akad. Nauk Ser. Fiz. \textbf{17}, 51-64 (1953)

%\cite{Murray:2004gh}
\bibitem{Murray:2004gh}
M.~J.~Murray [BRAHMS],
%``Scanning the phases of QCD with BRAHMS,''
%doi:10.1088/0954-3899/30/8/004
[arXiv:nucl-ex/0404007 [nucl-ex]].

%\cite{Bearden:2004yx}
\bibitem{Bearden:2004yx}
I.~G.~Bearden \textit{et al.} [BRAHMS],
%``Charged meson rapidity distributions in central Au+Au collisions at s(NN)**(1/2) = 200-GeV,''
Phys. Rev. Lett. \textbf{94}, 162301 (2005)
%doi:10.1103/PhysRevLett.94.162301
[arXiv:nucl-ex/0403050 [nucl-ex]].

%\cite{Murray:2007cy}
\bibitem{Murray:2007cy}
M.~Murray [BRAHMS],
%``Flavor dynamics,''
%doi:10.1088/0954-3899/35/4/044015
[arXiv:0710.4576 [nucl-ex]].

%\cite{Steinberg:2004wx}
\bibitem{Steinberg:2004wx}
P.~Steinberg,
%``Bulk dynamics in heavy ion collisions,''
Nucl. Phys. A \textbf{752}, 423-432 (2005)
%doi:10.1016/j.nuclphysa.2005.02.139
[arXiv:nucl-ex/0412009 [nucl-ex]].

%\cite{Steinberg:2007iv}
\bibitem{Steinberg:2007iv}
P.~Steinberg,
%``Entropy Production at High Energy and mu(B),''
PoS \textbf{CPOD2006}, 036 (2006)
%doi:10.22323/1.029.0036
[arXiv:nucl-ex/0702019 [nucl-ex]].

%\cite{Wong:2008ex}
\bibitem{Wong:2008ex}
C.~Y.~Wong,
%``Landau Hydrodynamics Revisited,''
Phys. Rev. C \textbf{78}, 054902 (2008)
%doi:10.1103/PhysRevC.78.054902
[arXiv:0808.1294 [hep-ph]].

%\cite{Srivastava:1992xb}
\bibitem{Srivastava:1992xb}
D.~K.~Srivastava, J.~Alam and B.~Sinha,
%``Rapidity distribution of secondaries in ultrarelativistic heavy ion collisions using Landau's hydrodynamic model,''
Phys. Lett. B \textbf{296}, 11-17 (1992)
%doi:10.1016/0370-2693(92)90796-7

%\cite{Srivastava:1992cg}
\bibitem{Srivastava:1992cg}
D.~K.~Srivastava, J.~e.~Alam, S.~Chakrabarty, B.~Sinha and S.~Raha,
%``Hydrodynamics of ultrarelativistic heavy ion collisions: Considerations of boost noninvariance and stopping,''
Annals Phys. \textbf{228}, 104-145 (1993)
%doi:10.1006/aphy.1993.1089

%\cite{Srivastava:1992gh}
\bibitem{Srivastava:1992gh}
D.~K.~Srivastava, J.~Alam, S.~Chakrabarty, S.~Raha and B.~Sinha,
%``Boost noninvariant hydrodynamics in ultrarelativistic heavy ion collisions,''
Phys. Lett. B \textbf{278}, 225-230 (1992)
%doi:10.1016/0370-2693(92)90185-7

%\cite{Mohanty:2003va}
\bibitem{Mohanty:2003va}
B.~Mohanty and J.~e.~Alam,
%``Velocity of sound in relativistic heavy ion collisions,''
Phys. Rev. C \textbf{68}, 064903 (2003)
%doi:10.1103/PhysRevC.68.064903
[arXiv:nucl-th/0301086 [nucl-th]].

%\cite{Hama:2004rr}
\bibitem{Hama:2004rr}
Y.~Hama, T.~Kodama and O.~Socolowski, Jr.,
%``Topics on hydrodynamic model of nucleus-nucleus collisions,''
Braz. J. Phys. \textbf{35}, 24-51 (2005)
%doi:10.1590/S0103-97332005000100003
[arXiv:hep-ph/0407264 [hep-ph]].

%\cite{Aguiar:2000hw}
\bibitem{Aguiar:2000hw}
C.~E.~Aguiar, T.~Kodama, T.~Osada and Y.~Hama,
%``Smoothed particle hydrodynamics for relativistic heavy ion collisions,''
J. Phys. G \textbf{27}, 75-94 (2001)
%doi:10.1088/0954-3899/27/1/306
[arXiv:hep-ph/0006239 [hep-ph]].

%\cite{Pratt:2008jj}
\bibitem{Pratt:2008jj}
S.~Pratt,
%``A Co-moving Coordinate System for Relativistic Hydrodynamics,''
Phys. Rev. C \textbf{75}, 024907 (2007)
%doi:10.1103/PhysRevC.75.024907
[arXiv:nucl-th/0612010 [nucl-th]].

%\cite{Bialas:2007iu}
\bibitem{Bialas:2007iu}
A.~Bialas, R.~A.~Janik and R.~B.~Peschanski,
%``Unified description of Bjorken and Landau 1+1 hydrodynamics,''
Phys. Rev. C \textbf{76}, 054901 (2007)
%doi:10.1103/PhysRevC.76.054901
[arXiv:0706.2108 [nucl-th]].

%\cite{Csorgo:2006ax}
\bibitem{Csorgo:2006ax}
T.~Csorgo, M.~I.~Nagy and M.~Csanad,
%``A New family of simple solutions of perfect fluid hydrodynamics,''
Phys. Lett. B \textbf{663}, 306-311 (2008)
%doi:10.1016/j.physletb.2008.04.038
[arXiv:nucl-th/0605070 [nucl-th]].

%\cite{Beuf:2008vd}
\bibitem{Beuf:2008vd}
G.~Beuf, R.~Peschanski and E.~N.~Saridakis,
%``Entropy flow of a perfect fluid in (1+1) hydrodynamics,''
Phys. Rev. C \textbf{78}, 064909 (2008)
%doi:10.1103/PhysRevC.78.064909
[arXiv:0808.1073 [nucl-th]].

%\cite{Osada:2008cn}
\bibitem{Osada:2008cn}
T.~Osada and G.~Wilk,
%``Nonextensive perfect hydrodynamics: A Model of dissipative relativistic hydrodynamics?,''
Central Eur. J. Phys. \textbf{7}, 432-443 (2009)
%doi:10.2478/s11534-008-0163-5
[arXiv:0810.3089 [hep-ph]].

%\cite{Landau_book} 
\bibitem{Landau_book} 
L.~D.~Landau and E.~M.~Lifshitz, 
{\it Fluid Mechanics}
(Butterworth-Heinemann, Oxford, 1987).

%\cite{Biswas:2019wtp}
\bibitem{Biswas:2019wtp}
D.~Biswas, K.~Deka, A.~Jaiswal and S.~Roy,
%``Viscosity, non-conformal equation of state and sound velocity in Landau hydrodynamics,''
[arXiv:1910.13368 [hep-ph]].

%\cite{Klay:2003zf}
\bibitem{Klay:2003zf}
J.~L.~Klay \textit{et al.} [E-0895],
%``Charged pion production in 2 to 8 agev central au+au collisions,''
Phys. Rev. C \textbf{68}, 054905 (2003)
%doi:10.1103/PhysRevC.68.054905
[arXiv:nucl-ex/0306033 [nucl-ex]].

%\cite{Alt:2007aa}
\bibitem{Alt:2007aa}
C.~Alt \textit{et al.} [NA49],
%``Pion and kaon production in central Pb + Pb collisions at 20-A and 30-A-GeV: Evidence for the onset of deconfinement,''
Phys. Rev. C \textbf{77}, 024903 (2008)
%doi:10.1103/PhysRevC.77.024903
[arXiv:0710.0118 [nucl-ex]].

%\cite{Afanasiev:2002mx}
\bibitem{Afanasiev:2002mx}
S.~V.~Afanasiev \textit{et al.} [NA49],
%``Energy dependence of pion and kaon production in central Pb + Pb collisions,''
Phys. Rev. C \textbf{66}, 054902 (2002)
%doi:10.1103/PhysRevC.66.054902
[arXiv:nucl-ex/0205002 [nucl-ex]].

%\cite{Gazdzicki:2010iv}
\bibitem{Gazdzicki:2010iv}
M.~Gazdzicki, M.~Gorenstein and P.~Seyboth,
%``Onset of deconfinement in nucleus-nucleus collisions: Review for pedestrians and experts,''
Acta Phys. Polon. B \textbf{42}, 307-351 (2011)
%doi:10.5506/APhysPolB.42.307
[arXiv:1006.1765 [hep-ph]].

%\cite{Gao:2015mha}
\bibitem{Gao:2015mha}
L.~N.~Gao and F.~H.~Liu,
%``On Distributions of Emission Sources and Speed of Sound in Proton-proton (Proton-antiproton) Collisions,''
Adv. High Energy Phys. \textbf{2015}, 641906 (2015)
%doi:10.1155/2015/641906
[arXiv:1509.09034 [nucl-th]].

%%%%%%%%Deeptak end%%%%%%%%


%%%%%%%%Sumana begins%%%%%%%%
\bibitem{Gupt:1983rq} 
C.~Gupt, R.~K.~Shivpuri, N.~S.~Verma and A.~P.~Sharma,
%``Quark Anti-quark Recombination In The Low Transverse Momentum Region,''
Nuovo Cim.\ A {\bf 75}, 408 (1983).

\bibitem{Fries:2003vb} 
R.~J.~Fries, B.~Muller, C.~Nonaka and S.~A.~Bass,
%``Hadronization in heavy ion collisions: Recombination and fragmentation of partons,''
Phys.\ Rev.\ Lett.\  {\bf 90}, 202303 (2003)
%doi:10.1103/PhysRevLett.90.202303
[nucl-th/0301087].

\bibitem{Kolb:2000sd} 
P.~F.~Kolb, J.~Sollfrank and U.~W.~Heinz,
%``Anisotropic transverse flow and the quark hadron phase transition,''
Phys.\ Rev.\ C {\bf 62}, 054909 (2000)
%doi:10.1103/PhysRevC.62.054909
[hep-ph/0006129].

\bibitem{Voloshin:2002wa} 
S.~A.~Voloshin,
%``Anisotropic flow,''
Nucl.\ Phys.\ A {\bf 715}, 379 (2003)
%doi:10.1016/S0375-9474(02)01450-1
[nucl-ex/0210014].

\bibitem{Lin:2002rw} 
Z.~w.~Lin and C.~M.~Ko,
%``Flavor ordering of elliptic flows at high transverse momentum,''
Phys.\ Rev.\ Lett.\  {\bf 89}, 202302 (2002)
%doi:10.1103/PhysRevLett.89.202302
[nucl-th/0207014].

\bibitem{H. Grad} H. Grad, Comm. Pure Appl. Math. \textbf{2}, 331 (1949).

%\cite{Romatschke:2009im}
\bibitem{Romatschke:2009im}
P.~Romatschke,
%``New Developments in Relativistic Viscous Hydrodynamics,''
Int. J. Mod. Phys. E \textbf{19}, 1-53 (2010)
%doi:10.1142/S0218301310014613
[arXiv:0902.3663 [hep-ph]].

%\cite{Jaiswal:2013npa}
\bibitem{Jaiswal:2013npa} 
A.~Jaiswal,
%``Relativistic dissipative hydrodynamics from kinetic theory with relaxation time approximation,''
Phys.\ Rev.\ C {\bf 87}, 051901 (2013)
%doi:10.1103/PhysRevC.87.051901
[arXiv:1302.6311 [nucl-th]].
%%CITATION = doi:10.1103/PhysRevC.87.051901;%%

%\cite{Jaiswal:2013vta}
\bibitem{Jaiswal:2013vta} 
A.~Jaiswal,
%``Relativistic third-order dissipative fluid dynamics from kinetic theory,''
Phys.\ Rev.\ C {\bf 88}, 021903 (2013)
%doi:10.1103/PhysRevC.88.021903
[arXiv:1305.3480 [nucl-th]].
%%CITATION = doi:10.1103/PhysRevC.88.021903;%%

%\cite{Song:2009gc}
\bibitem{Song:2009gc}
H.~Song,
%``Causal Viscous Hydrodynamics for Relativistic Heavy Ion Collisions,''
[arXiv:0908.3656 [nucl-th]].

%%%%%%%%Sumana ends%%%%%%%%


%%%%%%%%Sreekanth begins%%%%%%%%

%\cite{Romatschke:2017ejr}
\bibitem{Romatschke:2017ejr}
  P.~Romatschke and U.~Romatschke,
  %``Relativistic Fluid Dynamics In and Out of Equilibrium,''
  %doi:10.1017/9781108651998
  arXiv:1712.05815 [nucl-th].
  %%CITATION = doi:10.1017/9781108651998;%%

%\cite{Biro:1993qt}
\bibitem{Biro:1993qt}
  T.~S.~Biro, E.~van Doorn, B.~Muller, M.~H.~Thoma and X.~N.~Wang,
  %``Parton equilibration in relativistic heavy ion collisions,''
  Phys.\ Rev.\ C {\bf 48} (1993) 1275
  %doi:10.1103/PhysRevC.48.1275
  [nucl-th/9303004].
  %%CITATION = doi:10.1103/PhysRevC.48.1275;%%
  
%\cite{Muronga:2001zk}
\bibitem{Muronga:2001zk}
  A.~Muronga,
  %``Second order dissipative fluid dynamics for ultrarelativistic nuclear collisions,''
  Phys.\ Rev.\ Lett.\  {\bf 88} (2002) 062302
   Erratum: [Phys.\ Rev.\ Lett.\  {\bf 89} (2002) 159901]
  %doi:10.1103/PhysRevLett.89.159901, 10.1103/PhysRevLett.88.062302
  [nucl-th/0104064].
  %%CITATION = doi:10.1103/PhysRevLett.89.159901, 10.1103/PhysRevLett.88.062302
  
%\cite{Jaiswal:2013fc}
\bibitem{Jaiswal:2013fc} 
  A.~Jaiswal, R.~S.~Bhalerao and S.~Pal,
  %``Complete relativistic second-order dissipative hydrodynamics from the entropy principle,''
  Phys.\ Rev.\ C {\bf 87}, 021901 (2013)
  %doi:10.1103/PhysRevC.87.021901
  [arXiv:1302.0666 [nucl-th]].
  %%CITATION = doi:10.1103/PhysRevC.87.021901;%%
  
%\cite{Florkowski:2015lra}
\bibitem{Florkowski:2015lra} 
  W.~Florkowski, A.~Jaiswal, E.~Maksymiuk, R.~Ryblewski and M.~Strickland,
  %``Relativistic quantum transport coefficients for second-order viscous hydrodynamics,''
  Phys.\ Rev.\ C {\bf 91}, 054907 (2015)
  %doi:10.1103/PhysRevC.91.054907
  [arXiv:1503.03226 [nucl-th]].
  %%CITATION = doi:10.1103/PhysRevC.91.054907;%%
  
  %\cite{Bhatt:2009zg}
\bibitem{Bhatt:2009zg}
  J.~R.~Bhatt and V.~Sreekanth,
  %``Photon emission from out of equilibrium dissipative parton plasma,''
  Int.\ J.\ Mod.\ Phys.\ E {\bf 19} (2010) 299
  %doi:10.1142/S0218301310014765
  [arXiv:0901.1363 [hep-ph]].
  %%CITATION = doi:10.1142/S0218301310014765;%%

%\cite{Bhatt:2011kr}
\bibitem{Bhatt:2011kr}
  J.~R.~Bhatt, H.~Mishra and V.~Sreekanth,
  %``Shear viscosity, cavitation and hydrodynamics at LHC,''
  Phys.\ Lett.\ B {\bf 704} (2011) 486
  %doi:10.1016/j.physletb.2011.09.052
  [arXiv:1103.4333 [hep-ph]].
  %%CITATION = doi:10.1016/j.physletb.2011.09.052;%%
  
  %\cite{Rajagopal:2009yw}
\bibitem{Rajagopal:2009yw}
  K.~Rajagopal and N.~Tripuraneni,
  %``Bulk Viscosity and Cavitation in Boost-Invariant Hydrodynamic Expansion,''
  JHEP {\bf 1003} (2010) 018
  %doi:10.1007/JHEP03(2010)018
  [arXiv:0908.1785 [hep-ph]].
  %%CITATION = doi:10.1007/JHEP03(2010)018;%%
 
%\cite{Bhatt:2010cy}
\bibitem{Bhatt:2010cy}
  J.~R.~Bhatt, H.~Mishra and V.~Sreekanth,
  %``Thermal photons in QGP and non-ideal effects,''
  JHEP {\bf 1011} (2010) 106
  %doi:10.1007/JHEP11(2010)106
  [arXiv:1011.1969 [hep-ph]].
  %%CITATION = doi:10.1007/JHEP11(2010)106;%%
  
%\cite{Bhatt:2011kx}
\bibitem{Bhatt:2011kx}
  J.~R.~Bhatt, H.~Mishra and V.~Sreekanth,
  %``Cavitation and thermal dilepton production in QGP,''
  Nucl.\ Phys.\ A {\bf 875} (2012) 181
  %doi:10.1016/j.nuclphysa.2011.11.012
  [arXiv:1101.5597 [hep-ph]].
  %%CITATION = doi:10.1016/j.nuclphysa.2011.11.012;%%

  %\cite{Shen:2011eg}
\bibitem{Shen:2011eg}
  C.~Shen, U.~Heinz, P.~Huovinen and H.~Song,
  %``Radial and elliptic flow in Pb+Pb collisions at the Large Hadron Collider from viscous hydrodynamic,''
  Phys.\ Rev.\ C {\bf 84} (2011) 044903
  %doi:10.1103/PhysRevC.84.044903
  [arXiv:1105.3226 [nucl-th]].
  %%CITATION = doi:10.1103/PhysRevC.84.044903;%%
  
  %\cite{Sreekanth:2019}
\bibitem{Sreekanth:2019}
  V.~Sreekanth, \textit{manuscript under preparation}

%%%%%%%%Sreekanth ends%%%%%%%%



%%%%Samapan begins%%%%%%%%%

\bibitem{Mitra:2018akk} 
S.~Mitra and V.~Chandra,
%``Covariant kinetic theory for effective fugacity quasiparticle model and first order transport coefficients for hot QCD matter,''
Phys.\ Rev.\ D {\bf 97}, no. 3, 034032 (2018)
% doi:10.1103/PhysRevD.97.034032
[arXiv:1801.01700 [nucl-th]].
%%CITATION = doi:10.1103/PhysRevD.97.034032;%%

%\cite{Chandra:2011en}
\bibitem{Chandra:2011en} 
V.~Chandra and V.~Ravishankar,
%``A quasi-particle description of $(2+1)$- flavor lattice QCD equation of state,''
Phys.\ Rev.\ D {\bf 84}, 074013 (2011)
% doi:10.1103/PhysRevD.84.074013
[arXiv:1103.0091 [nucl-th]].
%%CITATION = doi:10.1103/PhysRevD.84.074013;%%

\bibitem{Anderson_Witting}
J.~L.~Anderson and H.~R.~Witting
%A relativistic relaxation-time for the Boltzmann equation. 
Physica \textbf{74}, 466 (1974).

%\cite{Bluhm:2011xu}
\bibitem{Bluhm:2011xu} 
M.~Bluhm, B.~Kampfer and K.~Redlich,
%``Ratio of bulk to shear viscosity in a quasigluon plasma: from weak to strong coupling,''
Phys.\ Lett.\ B {\bf 709}, 77 (2012)
[arXiv:1101.3072 [hep-ph]].
%%CITATION = doi:10.1016/j.physletb.2012.01.069;%%

%\cite{Bhadury:2019xdf}
\bibitem{Bhadury:2019xdf}
S.~Bhadury, M.~Kurian, V.~Chandra and A.~Jaiswal,
%``First order dissipative hydrodynamics and viscous corrections to the entropy four-current from an effective covariant kinetic theory,''
J. Phys. G \textbf{47}, no.8, 085108 (2020)
%doi:10.1088/1361-6471/ab907b
[arXiv:1902.05285 [hep-ph]].

%%%%Samapan end%%%%%%%%%



%%%%%%%Avdhesh start %%%%%%%

%\cite{STAR:2017ckg}
\bibitem{STAR:2017ckg}
L.~Adamczyk \textit{et al.} [STAR],
%``Global $\Lambda$ hyperon polarization in nuclear collisions: evidence for the most vortical fluid,''
Nature \textbf{548}, 62-65 (2017)
%doi:10.1038/nature23004
[arXiv:1701.06657 [nucl-ex]].

%\cite{Adam:2018ivw}
\bibitem{Adam:2018ivw}
J.~Adam \textit{et al.} [STAR],
%``Global polarization of $\Lambda$ hyperons in Au+Au collisions at $\sqrt{s_{_{NN}}}$ = 200 GeV,''
Phys. Rev. C \textbf{98}, 014910 (2018)
%doi:10.1103/PhysRevC.98.014910
[arXiv:1805.04400 [nucl-ex]].

%\cite{Karpenko:2016jyx}
\bibitem{Karpenko:2016jyx}
I.~Karpenko and F.~Becattini,
%``Study of $\Lambda $ polarization in relativistic nuclear collisions at $\sqrt{s_\mathrm {NN}}=7.7$ –200 GeV,''
Eur. Phys. J. C \textbf{77}, no.4, 213 (2017)
%doi:10.1140/epjc/s10052-017-4765-1
[arXiv:1610.04717 [nucl-th]].

%\cite{Becattini:2016gvu}
\bibitem{Becattini:2016gvu}
F.~Becattini, I.~Karpenko, M.~Lisa, I.~Upsal and S.~Voloshin,
%``Global hyperon polarization at local thermodynamic equilibrium with vorticity, magnetic field and feed-down,''
Phys. Rev. C \textbf{95}, 054902 (2017)
%doi:10.1103/PhysRevC.95.054902
[arXiv:1610.02506 [nucl-th]].

%\cite{Becattini:2017gcx}
\bibitem{Becattini:2017gcx}
F.~Becattini and I.~Karpenko,
%``Collective Longitudinal Polarization in Relativistic Heavy-Ion Collisions at Very High Energy,''
Phys. Rev. Lett. \textbf{120}, no.1, 012302 (2018)
%doi:10.1103/PhysRevLett.120.012302
[arXiv:1707.07984 [nucl-th]].

%\cite{Niida:2018hfw}
\bibitem{Niida:2018hfw}
T.~Niida [STAR],
%``Global and local polarization of $\Lambda$ hyperons in Au+Au collisions at 200 GeV from STAR,''
Nucl. Phys. A \textbf{982}, 511-514 (2019)
%doi:10.1016/j.nuclphysa.2018.08.034
[arXiv:1808.10482 [nucl-ex]].

%\cite{Becattini:2018duy}
\bibitem{Becattini:2018duy}
F.~Becattini, W.~Florkowski and E.~Speranza,
%``Spin tensor and its role in non-equilibrium thermodynamics,''
Phys. Lett. B \textbf{789}, 419-425 (2019)
%doi:10.1016/j.physletb.2018.12.016
[arXiv:1807.10994 [hep-th]].

%\cite{Florkowski:2017ruc}
\bibitem{Florkowski:2017ruc}
W.~Florkowski, B.~Friman, A.~Jaiswal and E.~Speranza,
%``Relativistic fluid dynamics with spin,''
Phys. Rev. C \textbf{97}, no.4, 041901 (2018)
%doi:10.1103/PhysRevC.97.041901
[arXiv:1705.00587 [nucl-th]].

%\cite{Florkowski:2017dyn}
\bibitem{Florkowski:2017dyn}
W.~Florkowski, B.~Friman, A.~Jaiswal, R.~Ryblewski and E.~Speranza,
%``Spin-dependent distribution functions for relativistic hydrodynamics of spin-1/2 particles,''
Phys. Rev. D \textbf{97}, no.11, 116017 (2018)
%doi:10.1103/PhysRevD.97.116017
[arXiv:1712.07676 [nucl-th]].

%\cite{Florkowski:2017njj}
\bibitem{Florkowski:2017njj}
W.~Florkowski, B.~Friman, A.~Jaiswal and E.~Speranza,
%``Relativistic hydrodynamics of particles with spin 1/2,''
Acta Phys. Polon. Supp. \textbf{10}, 1139 (2017)
%doi:10.5506/APhysPolBSupp.10.1139
[arXiv:1708.04035 [hep-ph]].

%\cite{Florkowski:2018ual}
\bibitem{Florkowski:2018ual}
W.~Florkowski, B.~Friman, A.~Jaiswal, R.~Ryblewski and E.~Speranza,
%``Fluid dynamics for relativistic spin-polarized media,''
Acta Phys. Polon. Supp. \textbf{11}, 507 (2018)
%doi:10.5506/APhysPolBSupp.11.507
[arXiv:1810.01709 [nucl-th]].

%\cite{Florkowski:2018ahw}
\bibitem{Florkowski:2018ahw}
W.~Florkowski, A.~Kumar and R.~Ryblewski,
%``Thermodynamic versus kinetic approach to polarization-vorticity coupling,''
Phys. Rev. C \textbf{98}, no.4, 044906 (2018)
%doi:10.1103/PhysRevC.98.044906
[arXiv:1806.02616 [hep-ph]].

%\cite{Florkowski:2018fap}
\bibitem{Florkowski:2018fap}
W.~Florkowski, A.~Kumar and R.~Ryblewski,
%``Relativistic hydrodynamics for spin-polarized fluids,''
Prog. Part. Nucl. Phys. \textbf{108}, 103709 (2019)
%doi:10.1016/j.ppnp.2019.07.001
[arXiv:1811.04409 [nucl-th]].

%\cite{Florkowski:2019qdp}
\bibitem{Florkowski:2019qdp}
W.~Florkowski, A.~Kumar, R.~Ryblewski and R.~Singh,
%``Spin polarization evolution in a boost invariant hydrodynamical background,''
Phys. Rev. C \textbf{99}, no.4, 044910 (2019)
%doi:10.1103/PhysRevC.99.044910
[arXiv:1901.09655 [hep-ph]].

%\cite{Becattini:2013fla}
\bibitem{Becattini:2013fla}
F.~Becattini, V.~Chandra, L.~Del Zanna and E.~Grossi,
%``Relativistic distribution function for particles with spin at local thermodynamical equilibrium,''
Annals Phys. \textbf{338}, 32-49 (2013)
%doi:10.1016/j.aop.2013.07.004
[arXiv:1303.3431 [nucl-th]].

%\cite{Vasak:1987um}
\bibitem{Vasak:1987um}
D.~Vasak, M.~Gyulassy and H.~T.~Elze,
%``Quantum Transport Theory for Abelian Plasmas,''
Annals Phys. \textbf{173}, 462-492 (1987).

%%%%%%%%avdesh end%%%%%%%%%%


%%%%%%%%Sunil begins%%%%%%%%

%\cite{Florkowski:2017olj}
\bibitem{Florkowski:2017olj} 
  W.~Florkowski, M.~P.~Heller and M.~Spalinski,
  %``New theories of relativistic hydrodynamics in the LHC era,''
  Rept.\ Prog.\ Phys.\  {\bf 81}, no. 4, 046001 (2018)
  %doi:10.1088/1361-6633/aaa091
  [arXiv:1707.02282 [hep-ph]].
  %%CITATION = doi:10.1088/1361-6633/aaa091;%%

%\cite{Heinz:2001xi}
\bibitem{Heinz:2001xi} 
  U.~W.~Heinz and P.~F.~Kolb,
  %``Early thermalization at RHIC,''
  Nucl.\ Phys.\ A {\bf 702}, 269 (2002)
  %doi:10.1016/S0375-9474(02)00714-5
  [hep-ph/0111075].
  %%CITATION = doi:10.1016/S0375-9474(02)00714-5;%%
  
  %\cite{Romatschke:2017vte}
\bibitem{Romatschke:2017vte} 
  P.~Romatschke,
  %``Relativistic Fluid Dynamics Far From Local Equilibrium,''
  Phys.\ Rev.\ Lett.\  {\bf 120}, no. 1, 012301 (2018)
  %doi:10.1103/PhysRevLett.120.012301
  [arXiv:1704.08699 [hep-th]].
  %%CITATION = doi:10.1103/PhysRevLett.120.012301;%%
  
%\cite{Hiscock:1985zz}
\bibitem{Hiscock:1985zz} 
  W.~A.~Hiscock and L.~Lindblom,
  %``Generic instabilities in first-order dissipative relativistic fluid theories,''
  Phys.\ Rev.\ D {\bf 31}, 725 (1985).
  %doi:10.1103/PhysRevD.31.725
  %%CITATION = doi:10.1103/PhysRevD.31.725;%%

%\cite{Muller:1967zza}
\bibitem{Muller:1967zza} 
  I.~Muller,
  %``Zum Paradoxon der Warmeleitungstheorie,''
  Z.\ Phys.\  {\bf 198}, 329 (1967).
  %doi:10.1007/BF01326412
  %%CITATION = doi:10.1007/BF01326412;%%
  
 %\cite{Israel:1979wp}
\bibitem{Israel:1979wp} 
  W.~Israel and J.~M.~Stewart,
  %``Transient relativistic thermodynamics and kinetic theory,''
  Annals Phys.\  {\bf 118}, 341 (1979).
  %doi:10.1016/0003-4916(79)90130-1
  %%CITATION = doi:10.1016/0003-4916(79)90130-1;%%
  
  %\cite{Maxwell:1867}
\bibitem{Maxwell:1867} 
  J.~C.~Maxwell, 
  Phil.\ Trans.\ R.\ Soc.\ {\bf 157:49} (1867).
  %10.1098/rstl.1867.0004.
  
 %\cite{Heller:2015dha}
\bibitem{Heller:2015dha} 
  M.~P.~Heller and M.~Spalinski,
  %``Hydrodynamics Beyond the Gradient Expansion: Resurgence and Resummation,''
  Phys.\ Rev.\ Lett.\  {\bf 115}, no. 7, 072501 (2015)
  %doi:10.1103/PhysRevLett.115.072501
  [arXiv:1503.07514 [hep-th]].
  %%CITATION = doi:10.1103/PhysRevLett.115.072501;%%
  
%\cite{Baier:2006um}
\bibitem{Baier:2006um} 
  R.~Baier, P.~Romatschke and U.~A.~Wiedemann,
  %``Dissipative hydrodynamics and heavy ion collisions,''
  Phys.\ Rev.\ C {\bf 73}, 064903 (2006)
  %doi:10.1103/PhysRevC.73.064903
  [hep-ph/0602249].
  %%CITATION = doi:10.1103/PhysRevC.73.064903;%%
  
 %\cite{Denicol:2012cn}
\bibitem{Denicol:2012cn} 
  G.~S.~Denicol, H.~Niemi, E.~Molnar and D.~H.~Rischke,
  %``Derivation of transient relativistic fluid dynamics from the Boltzmann equation,''
  Phys.\ Rev.\ D {\bf 85}, 114047 (2012)
  Erratum: [Phys.\ Rev.\ D {\bf 91}, no. 3, 039902 (2015)]
  %doi:10.1103/PhysRevD.85.114047, 10.1103/PhysRevD.91.039902
  [arXiv:1202.4551 [nucl-th]].
  %%CITATION = doi:10.1103/PhysRevD.85.114047, 10.1103/PhysRevD.91.039902;%%
  
 %\cite{Jaiswal:2019cju}
\bibitem{Jaiswal:2019cju} 
  S.~Jaiswal, C.~Chattopadhyay, A.~Jaiswal, S.~Pal and U.~Heinz,
  %``Exact solutions and attractors of higher-order viscous fluid dynamics for Bjorken flow,''
  Phys.\ Rev.\ C {\bf 100}, no. 3, 034901 (2019)
  %doi:10.1103/PhysRevC.100.034901
  [arXiv:1907.07965 [nucl-th]].
  %%CITATION = doi:10.1103/PhysRevC.100.034901;%%

 %\cite{Denicol:2017lxn}
\bibitem{Denicol:2017lxn} 
  G.~S.~Denicol and J.~Noronha,
  %``Analytical attractor and the divergence of the slow-roll expansion in relativistic hydrodynamics,''
  Phys.\ Rev.\ D {\bf 97}, no. 5, 056021 (2018)
  %doi:10.1103/PhysRevD.97.056021
  [arXiv:1711.01657 [nucl-th]].
  %%CITATION = doi:10.1103/PhysRevD.97.056021;%%
  
%\cite{Chattopadhyay:2018pwe}
\bibitem{Chattopadhyay:2018pwe}
C.~Chattopadhyay, A.~Jaiswal, S.~Jaiswal and S.~Pal,
%``Analytical solutions of causal relativistic hydrodynamic equations for Bjorken and Gubser flows,''
Nucl. Phys. A \textbf{982}, 911-914 (2019)
%doi:10.1016/j.nuclphysa.2018.12.012
[arXiv:1807.05544 [nucl-th]].
  
 %\cite{Strickland:2018ayk}
\bibitem{Strickland:2018ayk} 
  M.~Strickland,
  %``The non-equilibrium attractor for kinetic theory in relaxation time approximation,''
  JHEP {\bf 1812}, 128 (2018)
  %doi:10.1007/JHEP12(2018)128
  [arXiv:1809.01200 [nucl-th]].
  %%CITATION = doi:10.1007/JHEP12(2018)128;%%
  
  %\cite{Behtash:2018moe}
\bibitem{Behtash:2018moe} 
  A.~Behtash, C.~N.~Cruz-Camacho, S.~Kamata and M.~Martinez,
  %``Non-perturbative rheological behavior of a far-from-equilibrium expanding plasma,''
  Phys.\ Lett.\ B {\bf 797}, 134914 (2019)
  %doi:10.1016/j.physletb.2019.134914
  [arXiv:1805.07881 [hep-th]].
  %%CITATION = doi:10.1016/j.physletb.2019.134914;%%
  
 %\cite{Kurkela:2019set}
\bibitem{Kurkela:2019set} 
  A.~Kurkela, U.~A.~Wiedemann and B.~Wu,
  %``What attracts to attractors?,''
  arXiv:1907.08101 [hep-ph].
  %%CITATION = ARXIV:1907.08101;%%

%%%%%%%%Sunil ends%%%%%%%%


%%%%%%%%Deependra starts%%%%%%%%

%\cite{Gross:1973ju}
\bibitem{Gross:1973ju}
D.~J.~Gross and F.~Wilczek,
%``Asymptotically Free Gauge Theories - I,''
Phys. Rev. D \textbf{8}, 3633-3652 (1973).

\bibitem{Politzer:1973fx}
H.~D.~Politzer,
%``Reliable Perturbative Results for Strong Interactions?,''
Phys. Rev. Lett. \textbf{30}, 1346-1349 (1973).
%doi:10.1103/PhysRevLett.30.1346

\bibitem{Nambu:1974zg} 
Y.~Nambu,
%``Strings, Monopoles and Gauge Fields,''
Phys.\ Rev.\ D {\bf 10}, 4262 (1974).
%doi:10.1103/PhysRevD.10.4262

\bibitem{tHooft:1981bkw} 
G.~'t Hooft,
%``Topology of the Gauge Condition and New Confinement Phases in Nonabelian Gauge Theories,''
Nucl.\ Phys.\ B {\bf 190}, 455 (1981).
%doi:10.1016/0550-3213(81)90442-9

\bibitem{Pandey:2000bt} 
V.~P.~Pandey and H.~C.~Chandola,
%``Nonperturbative aspects of dual QCD vacuum and confinement,''
Phys.\ Lett.\ B {\bf 476}, 193 (2000).
%doi:10.1016/S0370-2693(00)00033-2

\bibitem{Chandola:2009zz} 
H.~C.~Chandola and D.~Yadav,
%``Thermo-field dynamics and quark-hadron phase transition in QCD,''
Nucl.\ Phys.\ A {\bf 829}, 151 (2009).
%doi:10.1016/j.nuclphysa.2009.08.006, 10.1016/j.nuclphysa2009.08.006

\bibitem{Rawat:2018mxg} 
D.~S.~Rawat, H.~C.~Pandey, H.~C.~Chandola and D.~Yadav,
%``Flux Tubes, Field Configuration and Non-Perturbative Dynamics of QCD,''
Springer Proc.\ Phys.\  {\bf 203}, 625 (2018).

\bibitem{Chandola:2019xwo} 
H.~C.~Chandola, D.~Singh Rawat, H.~C.~Pandey, D.~Yadav and H.~Dehnen,
%``Non perturbative and thermal dynamics of confined fields in dual QCD,''
Adv. High Energy Phys. \textbf{2020}, 4240512 (2020),
arXiv:1904.11714 [hep-th].

\bibitem{Cho:1979nv} 
Y.~M.~Cho,
%``A Restricted Gauge Theory,''
Phys.\ Rev.\ D {\bf 21}, 1080 (1980).
%doi:10.1103/PhysRevD.21.1080

\bibitem{Cho:1980nx} 
Y.~M.~Cho,
%``Extended Gauge Theory and Its Mass Spectrum,''
Phys.\ Rev.\ D {\bf 23}, 2415 (1981).
%doi:10.1103/PhysRevD.23.2415

\bibitem{Cho:2012pq} 
Y.~M.~Cho, F.~H.~Cho and J.~H.~Yoon,
%``Dimensional Transmutation by Monopole Condensation in QCD,''
Phys.\ Rev.\ D {\bf 87}, no. 8, 085025 (2013)
%doi:10.1103/PhysRevD.87.085025
[arXiv:1206.6936 [hep-th]].

\bibitem{Nielsen:1973cs} 
H.~B.~Nielsen and P.~Olesen,
%``Vortex Line Models for Dual Strings,''
Nucl.\ Phys.\ B {\bf 61}, 45 (1973).
%doi:10.1016/0550-3213(73)90350-7

\bibitem{Cea:2015wjd} 
P.~Cea, L.~Cosmai, F.~Cuteri and A.~Papa,
%``Flux tubes at finite temperature,''
JHEP {\bf 1606}, 033 (2016)
%doi:10.1007/JHEP06(2016)033
[arXiv:1511.01783 [hep-lat]].

%%%%%%%%Deependra ends%%%%%%%%

\end{thebibliography}
